\documentclass[12pt]{report}
\usepackage[titletoc]{appendix}
\usepackage{verbatim}
\usepackage{cite}
\usepackage{graphicx}
\usepackage{color}
\usepackage[fleqn]{amsmath}
\usepackage{amssymb}
\usepackage{slashed}
\usepackage{epsfig}
\usepackage{textcomp}
\usepackage{hyperref}
 \usepackage{geometry}
\newcommand{\nn}{\nonumber}
\newcommand{\p}{p_}
\newcommand{\eps}{\epsilon}
\newcommand{\veps}{\varepsilon}
\newcommand{\ieps}{i\eps}
\newcommand{\gmu}{\gamma^{\mu}}
\newcommand{\gnu}{\gamma^{\nu}}
\newcommand{\gro}{\gamma^{\rho}}
\newcommand{\gsig}{\gamma^{\sigma}}
\newcommand{\gal}{\gamma^{\alpha}}

\newcommand{\gfive}{\gamma^5}
\newcommand{\gVV}{g g \to  V V^\prime g}
\newcommand{\gVVpp}{p p  \to V V^\prime j+X}
\newcommand{\gpp}{g g  \to \gamma \gamma g}
\newcommand{\gpZ}{g g \to  \gamma Z g}
\newcommand{\gpZpp}{p p  \to \gamma Z j+X}
\newcommand{\gZZ}{g g \to  Z Z g}

\newcommand{\gWW}{g g \to  W^{+} W^{-} g}

\newcommand{\VV}{g g \to  V V^\prime }

\newcommand{\BG}{g g \to  B G_{\rm KK} }
\newcommand{\HG}{g g \to  H G_{\rm KK} }
\newcommand{\pG}{g g \to  \gamma G_{\rm KK} }
\newcommand{\ZG}{g g \to  Z G_{\rm KK} }
\newcommand{\BGpp}{p p \to  B G_{\rm KK} +X}

\newcommand{\T}{\rule{0pt}{3.0ex}}
\newcommand{\B}{\rule[-2.2ex]{0pt}{0pt}}

\begin{document}
\pagenumbering{roman}
\thispagestyle{empty}

\begin{center}
%     {\bf \large One Loop Study of Gluon Fusion Processes at Hadron Colliders within the SM and Beyond }
      {\bf \Large GLUON FUSION PROCESSES AT ONE-LOOP WITHIN THE STANDARD MODEL AND BEYOND }
\end{center}

\vspace{0.3 cm}
\begin{center}
    {\it \large By}
\end{center}

\begin{center}
%   {\bf {\Large Ambresh Kumar Shivaji} \\ (Enrollment No. : PHYS07200604037)}
    {\bf {\Large Ambresh Kumar Shivaji} \\ PHYS07200604037}
\end{center}

\begin{center}
    {\large Institute of Physics, Bbhubaneswar }
%\vskip 1.0 cm
%\epsfig{file=IOP_logo.pdf, width=2.5cm, height= 2.5cm}\\
\end{center}

\vskip 2.0 cm
\begin{center}
\large
\emph{
 A thesis submitted to the \\
Board of Studies in Physical Sciences \\
\vspace{0.2cm}
In partial fulfillment of requirements \\
For the Degree of } \\
{\bf DOCTOR OF PHILOSOPHY} \\
\emph{of} \\
{\bf HOMI BHABHA NATIONAL INSTITUTE}
\vskip 2.0 cm
\epsfig{file=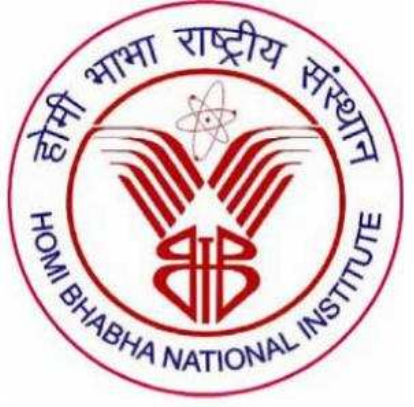, width=3.0cm, height= 3.0cm}\\
\vskip 1.0 cm
{\bf April, 2013}
\end{center}

% \begin{comment}
\newpage
\thispagestyle{empty}

\begin{center}
\large \bf
STATEMENT BY AUTHOR
\end{center}

This dissertation has been submitted in partial fulfillment of 
requirements for an advanced degree at Homi Bhabha National Institute 
(HBNI) and is deposited in the Library to be made available to 
borrowers under rules of the HBNI. Brief quotations from this 
dissertation are allowable without special permission,
provided that accurate acknowledgement of source is made. Requests 
for permission for extended quotation from or reproduction of this 
manuscript in whole or in part may be granted by the Competent Authority 
of HBNI when in his or her judgment the
proposed use of the material is in the interests of scholarship. In all 
other instances, however, permission must be obtained from the author.
\vskip 3.0cm
\noindent
\hspace*{10.5cm} (Ambresh Kumar Shivaji)\\
\noindent

\newpage
\thispagestyle{empty}

\begin{center}
\large \bf
DECLARATION
\end{center}

I, Ambresh Kumar Shivaji, hereby declare that the investigations presented in
the thesis have been carried out by me. The matter embodied in the thesis is
original and has not been submitted earlier as a whole or in part for a 
degree/diploma at this or any other Institution/University.

\vskip 3.0cm
\noindent
\hspace*{10.5cm} (Ambresh Kumar Shivaji)\\
\noindent

% Dedications
\newpage
\thispagestyle{empty}
\vspace*{3.0in}
\hspace*{2.0in}
\begin{center}
{\Large \emph{ To My Brother}\\}
\hrule
\end{center}

\newpage
\thispagestyle{empty}
\begin{center}
\large \bf
ACKNOWLEDGEMENTS
\end{center}
\vspace*{-0.1in}
{ \noindent
I would like to acknowledge the direct financial support, in the form 
of my fellowship, from the Department of Atomic Energy, Government 
of India. The financial support from other funding agencies for attending
various schools and conferences in India and abroad, including short academic
visits, is also greatly 
acknowledged. I would also like to thank all the active members of the Indian 
high energy physics community for continuing the effort of making 
it one of the most organized research communities in India. 
Special thanks to the organizers and participants of the \emph{SERC Schools in Theoretical High 
Energy Physics} which I attended as a beginning graduate student. I should 
also thank the countable members of Indian RADCOR group for organizing an 
\emph{Advanced School on Radiative Corrections} and the \emph{RADCOR 2011 Symposium} 
first time in India and fortunately during my PhD career.
}

Although Pankaj is my official supervisor, I have always seen him as a good 
friend and collaborator. I would like to thank him for whatever academic and 
non-academic support he could provide me in last five years. I wish to thank 
other supporting members of the Institute also. 
I have extensively used the small cluster-computing facility of the Institute 
computer centre. I give special thanks to all those who kept 
the facility alive during my crucial PhD projects. I have gone through many wonderful 
experiences during my stay in the campus, and I wish to thank all the people who 
were responsible for them.    

During my PhD career, I have been greatly benefited by interacting with many 
individuals. I must thank my all such teachers which include friends, professional 
teachers and professional researchers for showing great patience with me and
helping me out in difficult situations. The unconditional love and support of 
my family members; my parents, my brother, my sisters and their lovely kids 
have continuously helped in reaching this milestone.

\vskip 2.0cm
%\vspace{1.5in}
{\bf Date:}
\hspace{3.2in} {\bf Ambresh Kumar Shivaji}
 \hspace*{0.32cm}
%\hrule

% \begin{abstract}
% We have studied certain processes, within the Standard Model (SM) and beyond,
% which are initiated by gluon-gluon fusion at Hadron colliders. We consider
% di-vector boson production in association with a jet via gluon fusion in the
% SM. We have taken the model of large extra space-dimensions as an example for the 
% model of new physics. Within this model we consider the associated production of 
% a boson and the KK-gravitons via gluon fusion. All these processes proceed via 
% quark loop diagrams at the leading order itself. Total cross-section calculation
% and various kinematic distributions constitute the main results of these studies. 
% \end{abstract}

% \end{comment}

\numberwithin{equation}{chapter}
\numberwithin{figure}{chapter}
\numberwithin{table}{chapter}
%%%%%%%%%%%%%%%%%%%%%%%%%%%%%%%%%%%%%%%%%%%%%%%%%%%%%%%%%%%%%%%%%%%%%%
% \newpage
% \setcounter{page}{1}
% \pagenumbering{roman}
\tableofcontents

% Symopsis

\newpage
\chapter*{Synopsis}
\addcontentsline{toc}{chapter}{Synopsis}
\vskip 0.5cm

The {\it Standard Model} (SM) of particle physics is our present understanding of 
Nature. It is a framework in which properties and behavior of the fundamental 
particles and forces can be studied. It offers a unified
description of the electromagnetic and weak forces along with the strong
 force in the microscopic domain. At the fundamental level, these forces act 
among {\it quarks} and {\it leptons} (the matter particles) through {\it gauge bosons} 
(the mediator particles). The model has been remarkably  successful in explaining a 
huge amount of data collected so far at various high energy particle collider experiments 
throughout the world. Despite its success, 
there are direct as well as indirect evidences which indicate that the SM cannot be a 
complete story of the Universe. It does not include the fourth force of Nature, 
the Gravity. It does not account correctly for the dominance of matter over antimatter as 
observed in the Universe. It does not have any suitable candidate for the proposed 
{\it Dark matter} particles. 

Apart from these direct evidences, there are certain theoretical inconsistencies associated with 
the model. It predicts the existence of a fundamental scalar, the {\it Higgs boson} as a result 
of the {\it electroweak symmetry breaking} (EWSB) through the SU(2) doublet of scalar fields. 
The recent discovery of a fundamental boson ($\simeq$ 125 GeV) at the {\it Large Hadron Collider} (LHC) 
experiments might confirm the existence of this last missing piece of the SM particle jungle soon. 
Within the model, the quantum correction to the Higgs mass is not stable; it is 
quadratically divergent and requires a very fine tuning of parameters. This is known as a 
{\it fine tuning} or {\it naturalness problem } in the literature.
Another problem, more of a philosophical nature and closely related to the naturalness problem, 
is the unexplained hierarchy between the electroweak symmetry breaking scale ($\sim 100 \; \rm{GeV}$) 
and the fundamental scale of Gravity in 4 dimensions, the {\it Planck  scale} ($\sim 10^{19}\; \rm{GeV}$). 
The desirable unification of the fundamental forces is also not feasible within the model.  
All these lead us to believe that the SM is only a low energy description of a deeper reality. 
We expect new physics and new particles to appear at higher energy scales ($\sim$ TeV), where 
the SM predictions are not yet tested. \\

There are many candidates for new physics, with their own merits and demerits, which address 
many of the above mentioned problems. Models based on {\it supersymmetry}, models of 
{\it new forces} and models of {\it extra dimensions} are few among many. Their predictions 
can be tested at the present day high energy colliders such as the LHC and this requires 
calculations to be done in both the SM and new physics models. Any deviation from the SM 
predictions is a signature of new physics. The problem of large hierarchy between the 
electroweak and the Planck  scales can be addressed within the models of extra space dimensions. 
Among various models of extra dimensions, the ADD model of large extra dimensions 
is one of the very first successful attempts in this direction.
In the ADD model, the number of space-time dimensions is taken $4+\delta$. 
The SM degrees of freedom live on a $(3 + 1)$-dimensional {\it brane}, 
while the Gravity can access the full $4+\delta$ dimensions. 
The extra space dimensions 
are supposed to be compact. In this model, the 4-dimensional Planck  scale is only an 
effective scale and the fundamental scale of Gravity, $M_S$ can be near TeV scale. 
In 4 dimensions, the $4+\delta$ dimensional graviton appears as an infinite tower of 
{\it Kaluza-Klein} (KK) modes and it couples to the energy-momentum tensor ($\cal{T}_{\mu\nu}$) 
of the SM fields. The direct production of the KK-gravitons gives rise to missing energy 
signals in the detector. \\

At {\it hadron colliders} the fundamental interactions take place among its constituents, 
the quarks and the gluons. Collectively these are called {\it partons} and they carry a 
certain momentum fraction of the parent hadron. Although the coupling of the strong interaction 
($\alpha_s$) acting among the partons is quite large at low energies, due to the property of 
{\it asymptotic freedom} it is possible to apply perturbative methods at the parton level  
at higher energies. The {\it color confinement} forces the partons not to be seen as free 
particles and therefore, perturbative calculations at the parton level may appear meaningless. 
The {\it factorization theorem} provides a means to calculate hadronic 
cross sections in terms of partonic ones. Predictions obtained this way 
have been verified at the hadron colliders such as the {\it Tevatron} and the LHC. The $p\bar p$ 
collider facility at the Tevatron has been recently shutdown. The LHC is a {\it proton-proton} 
collider with the centre-of-mass (c.m.) energy in the multi-TeV range. It is presently running 
at 8 TeV c.m. energy and more than $ 10 fb^{-1} $ of data has been collected so far at the 
two general purpose detectors, the ATLAS ({\it A Toroidal LHC Apparatus}) and the CMS 
({\it Compact Muon Solenoid} ). At higher energies the gluon distribution functions are 
quite important, and therefore, gluon fusion processes can contribute significantly towards the SM 
as well as new physics predictions. \\

In this thesis, I have studied certain gluon fusion processes at the hadron colliders. We 
have considered the {\it di-vector boson production in association with a jet via gluon fusion}  
within the SM. We have also considered {\it the direct production of the KK-gravitons in 
association with a boson via gluon fusion} in the ADD model. All these processes share a 
common feature that they proceed via the {\it quark loop diagrams} at the leading order (LO) 
itself and being LO contributions these are expected to be finite. Total cross sections and 
phenomenologically relevant kinematic distributions constitute the main results of these 
studies. The amplitude calculation is done at the parton level. One of the most difficult 
parts of the calculation is the reduction of one-loop {\it tensor integrals} into a suitable 
set of one-loop {\it scalar integrals}. In our projects, we have one-loop five-point tensor 
integrals of rank five as the most complicated tensor structures. We have worked with 
two different codes for the one-loop tensor reduction in $n$ dimensions. An analytical tensor 
reduction code in FORM was developed during the ADD model projects. 
It is based on the reduction method suggested by Denner-Dittmaier, and the reduction of any 
one-loop tensor integral up to four-rank, four-point function can be performed using it. 
We have used it successfully in our ADD model projects to make various 
checks on amplitudes. For actual numerical calculations we have used a numerical code written 
in FORTRAN following the one-loop tensor reduction method of Oldenborgh-Vermaseren. 
Any one-loop amplitude in 4 dimensions, after the tensor reduction, 
can be expressed in terms of the scalar integrals of {\it box}, {\it triangle}, {\it bubble} 
and {\it tadpole} types and an additional piece called the {\it Rational part}. Thus the 
singularity structure of any one-loop amplitude is dictated by those of the scalar integrals. 
The rational part is an artifact of the regularization of {\it ultra-violet} (UV) divergences 
of one-loop tensor integrals. We have derived the one-loop scalar integrals, required in this 
thesis, following the method of 't Hooft-Veltman. UV singularities 
are regularized in $4-2\epsilon$ dimensions while {\it infra-red} (IR) singularities are 
regularized by giving a small mass to the quarks in the loop. Due to the numerical instability 
these scalar integrals are used only for making finiteness checks on amplitudes. Actual numerical 
results are obtained using the scalar integrals from the LoopTools and the OneLOop 
packages.\\

In the ADD model, the process $\BG$ contributes to  $\BGpp$ at the next-to-leading order 
 (NLO) in $\alpha_s$. These processes proceed via quark loop diagrams of the triangle and 
the box types. We work with all the six quark flavors and except the top quark all others are 
treated as massless. We find that the amplitude for $\pG$ vanishes at the LO. This can be 
shown using {\it Furry's Theorem} and the charge conjugation property of the graviton. Due to 
the same reason the vector part of the $\ZG$ amplitude does not contribute. Since the axial 
part of the amplitude is proportional to the $T^3_q$ value, it does not receive any 
contribution from the first two generations. We find that due to the nature of the graviton 
coupling with the quarks, both the triangle and box diagrams are linearly divergent 
and give rise to the {\it anomaly}. We have studied the intricate relationship between the 
anomaly and the rational terms in linearly divergent fermion loop amplitudes. We find that 
in fermion loop amplitudes (including those plagued with the chiral anomaly), correct rational 
terms can be obtained utilizing the {\it Decoupling theorem}. The 
process $\HG$ receives dominant contribution from the top quark in the loop. It is a leading 
non-zero contribution to $ p p \rightarrow H G_{KK} + X$, if we neglect the bottom quark mass. 
In all these cases, we have studied the variation of the total cross section with respect 
to the collider c.m. energy. We observe a significant cancellation between the triangle and 
the box contributions at the amplitude level. We find that at the typical LHC energy the 
cross section is only few fb. We give the transverse momentum ($p_T$) and the rapidity 
($\eta$) distributions of the bosons ($Z/H$) and examine the effect of changing the 
{\it renormalization} and the {\it factorization} scales. In the direct production processes 
of the KK-gravitons, all the kinematically allowed modes are produced and therefore we also 
discuss their distribution. We have studied changes in our results as the ADD model parameters 
$\delta$ and $M_S$ are varied. We have also checked the effect of choosing different sets of 
the {\it parton distribution functions} (PDFs). \\

In the SM, the process $g g \rightarrow V V' g$ contributes to $ p p \rightarrow V V'j +X$ 
at the next-to-next-to-leading order in $\alpha_s$. Like the corresponding di-vector boson 
production cases, these are important backgrounds to the Higgs boson production as well as 
new physics scenarios. In particular, we have considered the production of $\gamma\gamma g$, 
$\gamma Zg$, $ZZg$ and $W^{+}W^{-}g$ at the LHC. 
The process $g g \rightarrow \gamma \gamma g$ has been calculated 
in the past. We have updated its cross section and have reconfirmed the importance of 
this processes at the LHC.
In the $ZZg$ and $W^{+}W^{-}g$ cases, we have ignored the Higgs boson interference effects. 
These proceed via the quark loop diagrams of the box 
and the pentagon types. The box diagrams give only the vector contribution while the pentagon 
diagrams give both the vector and the axial-vector contributions. We work in the limit of 
the decoupling of the top quark. We have studied the variation of the hadronic cross section 
with respect to the collider c.m. energy for all the four processes. We find that these 
one-loop contributions are in the range of $4-15 \%$ of the corresponding tree-level 
contributions. We also give some important kinematic distributions common to all the four processes.

In a more complete study of the $\gamma Zg$ production process, 
we explicitly check the decoupling of the 
top quark at the amplitude level as well as at the level of the total cross section.
We have quantified the contributions of the vector and the axial-vector 
parts of the amplitude towards the total cross section. We also study the effects of 
changing the renormalization and the factorization scales and the effect of choosing 
the PDF sets. We have considered the decay of the $Z$ boson, and 
a comparison with the corresponding LO and NLO calculation is also made. 
We briefly discuss the method adopted to deal with the numerical instabilities in 
our calculations.

\newpage
\addcontentsline{toc}{chapter}{List of Publications/Preprints}
\begin{center}
 \Large {\bf List of Publications/Preprints}
\end{center}

% \newpage
% \pagestyle{empty}
\begin{center}
\underline{\bf Conference Proceedings}
\end{center}

%\vspace*{1.6cm}
%\hspace*{0.32cm}
\hrule {A (\textbf{$^{\ddagger}$}) indicates papers on which this
thesis is based.}

\addcontentsline{toc}{chapter}{List of Figures}
 \listoffigures
 \listoftables
\addcontentsline{toc}{chapter}{List of Tables}

% \chapter{Synopsis}
\newpage
\setcounter{page}{1}
\pagenumbering{arabic}

\chapter{Introduction}\label{chapter:intro}
% \numberwithin{equation}{chapter}
It is an exciting period for all of us who have been looking beyond the well established 
fundamental laws of Nature. Our current understanding of fundamental particles and their 
interactions is dubbed as the Standard Model (SM) of particle physics. In the past four decades
or so, its predictions have been tested time and again in a wide-variety of experiments, and 
the model has become a paradigm. However, the story does not end here. We now know that the SM,
despite its success, does not address many of the important issues about which we have become 
aware over the years. We do not have a satisfactory quantum theory of Gravity, the fourth 
fundamental force of Nature. Although the SM has ingredients to generate matter-antimatter asymmetry, 
the amount of asymmetry predicted by it does not explain the asymmetry observed in the Universe. 
We know that the SM particle spectrum describes only 4 $\%$ of the Universe. Investigating the 
nature of dark matter and dark energy, which make up 96 $\%$ of the Universe, remains a big challenge. 
Few experimental measurements like the forward-backward asymmetry of the top quark pair 
production at the Tevatron, and the muon $g-2$ measurement at the BNL, seem to be 
inconsistent with the SM calculations~\cite{Aaltonen:2011kc,Bennett:2006fi}. The well 
known hierarchy between the Planck scale and the scale of electroweak symmetry breaking also seeks a 
natural explanation. All these and many other issues suggest the existence of new physics beyond the 
SM. There have been many proposals such as supersymmetric models, extra-dimensional models and
models of grand unification to address some of these questions, but their confirmation can come only 
from the experiments. For a review on the topics related to the SM and possible new physics, one 
may refer to~\cite{Beringer:1900zz}.  

The discovery of new physics (a signal) is always complemented by the knowledge of known physics 
(the background). Every new discovery in an experiment becomes a background for future experiments. 
To find a signal, a precise knowledge of the background is essential. The Tevatron, a proton-antiproton 
collider at the Fermilab, was in service for almost 28 years before it was shut down on September 
30, 2011. It has tested and refined the SM through many fundamental discoveries and measurements, 
such as the discovery of the top quark, the observation of direct CP violation in the decay of neutral 
kaons, the direct observation of the tau neutrino, precision measurements of weak interaction parameters,
the observation of single top quark production and the discovery of many mesons and baryons~\cite{TEVATRON:2011xx}. 
It also played a crucial role in narrowing down the mass range for the SM Higgs boson, discovery of which is expected 
to be confirmed soon at the Large Hadron Collider (LHC)~\cite{TEVNPH:2012ab}. The collider delivered more 
than 10 fb$^{-1}$ of data before going for a full stop. The data analysis will keep experimentalists 
engaged for several years. After the Tevatron, in a collider type experiment, the LHC is the next big 
step in achieving the goal of testing the SM predictions at higher energies and uncovering new physics.

The LHC, a proton-proton collider at CERN, seems to have achieved a lot within a short span of its ongoing 
career. It is a fact of amusement and amazement that the whole SM of particle physics was 
rediscovered at the LHC withing a year of its full operation~\cite{Green:2011zza}.
Very recently, the discovery of a Higgs-like particle at both the general purpose LHC experiments, the 
CMS and the ATLAS, was announced~\cite{cern:2012xx,ATLAS:2012gk,CMS:2012gu}. It is believed that this 
Higgs-like particle could be the very Higgs boson of the SM. Further analysis is going on in this 
direction. Before the discovery of this Higgs-like particle, the observation of a new quarkonium state
and the observation of a new bottom baryon was also reported~\cite{Aad:2011ih,Chatrchyan:2012ni}. 
Although their discovery is not as crucial as that of the Higgs boson, their existence and measured 
properties are definitely consistent with our beloved SM. So far, we do not have a conclusive evidence of 
any new physics~\cite{Jakobs:2012qn,Nahn:2012ey,Adams:2012xx,Rolandi:2012xd}. However, with ever growing data 
and its ongoing analysis, a large parameter space in many potential candidates of new physics, 
has already been ruled out.

At the LHC, two beams of protons are collided with an unprecedented amount of energy. At present 
the collider is operating at 8 TeV centre-of-mass energy. Future plans to upgrade it to 
14 TeV and beyond are already under considerations. The proton is a composite
particle and at very high energy it can be seen as a collection of quarks, antiquarks and gluons. 
For a given set of final state particles, any process at the proton-proton collider may proceed 
through various channels like quark-quark, quark-gluon and gluon-gluon channels. Due to the 
possibility of a large gluon flux at high energy hadron colliders, we have considered a specific 
class of processes which are initiated by the gluon-gluon channel. We have divided our study of 
gluon fusion processes into two parts -- the background processes and the signal processes. 
We have considered di-vector boson and a jet production  
via gluon fusion within the SM. These processes have never been seen earlier, so it will be a test for the 
SM itself. They may have significant cross sections at the LHC and 
therefore are important backgrounds to the Higgs boson and many new physics signals. We have also
considered the gluon-gluon contribution to the associated production of an electroweak boson with 
the KK-gravitons in the ADD (Arkani-Hamed, Dimopoulos and Dvali) model. 
The model offers an explanation to the SM hierarchy problem. 
% The direct production of KK-gravitons 
% will give rise to a missing energy signal in the detector. 
All these gluon fusion processes proceed
via quark loop diagrams at the leading order (LO). Unlike in the case of radiative corrections to 
a tree-level process, there is no issue of renormalization/ factorization because these LO one-loop 
processes are finite. However, the calculation of multiparticle one-loop amplitudes itself is quite difficult. 
Although the method of calculating one-loop amplitudes has evolved quite a lot lately, we have 
taken the traditional route of one-loop tensor reduction to calculate them. In addition, the fermion loop amplitudes, 
due to their special properties, provide a unique opportunity to understand the structure of one-loop 
amplitudes in general. 

This thesis is organized as follows. Within this chapter, in the next few sections we will give
a brief overview of the Standard Model and the ADD model. We will also
argue that the gluon fusion processes at high energy hadron colliders are important. 
In chapter~\ref{chapter:oneloop}, we will describe the general structure of one-loop amplitudes using 
the methods of tensor reduction. Some special properties of fermion loop amplitudes will also be discussed. 
The production of a pair of electroweak vector bosons in association with a jet via gluon fusion will
be studied in chapter~\ref{chapter:sm}. We will present the cross section calculations and important 
kinematic distributions at the LHC. Next, in chapter~\ref{chapter:add}, we will consider the 
associated production of an electroweak boson and KK-gravitons via gluon fusion at the LHC. The dependence 
of cross sections and various kinematic distributions on the ADD model parameters will also be studied. We will 
summarize the thesis in chapter~\ref{chapter:summary}. Many complementary topics are included in 
the appendix.
% 
% 
%%%%%%%%%%%%%%%%%%%%%%%%%%%%%%%%%%%%%%%%%%%%%%%%%%%%%%%%%%%%%%%%%%%%%%%%%%%%%%%%%%%%%%%%%%%%%%%%%%%%%
\section{The Standard Model}\label{section:sm}
The SM is a renormalizable local quantum field theory, based on the gauge group 
$SU(2)_L \times U(1)_Y \times SU(3)_c$. It is a gauge theory meaning that 
the allowed interactions among fundamental particles are fixed by demanding the 
invariance of the free field Lagrangian under the local gauge transformation of 
fields~\cite{cheng:2007xx,Abers:1973qs}. The (spin-1/2) matter particles, leptons
and quarks, live in the fundamental representation while the (spin-1) mediator 
particles, gauge bosons, live in the adjoint representation of the gauge groups. 
The $SU(2)_L \times U(1)_Y$ component of the SM can be considered a unified 
description of the electromagnetic and weak interactions. The electroweak classical 
Lagrangian consistent with the local gauge invariance is given by
\begin{eqnarray}\label{eq:LEW}
 {\cal L}_{EW} &=& -\frac{1}{4} (W^a_{\mu\nu})^2 -\frac{1}{4} (B_{\mu\nu})^2 + i \bar{\psi}\slashed{D} \psi,            
\end{eqnarray}
where the field strength tensor of the $SU(2)$ gauge fields $W^a_\mu(x)\; (a=1,2,3)$ 
and that of the $U(1)_Y$ gauge field $B_\mu(x)$ are
\begin{eqnarray}
 W^a_{\mu\nu} &=& \partial_\mu W^a_\nu - \partial_\nu W^a_\mu + g_w\; \veps^{abc} W^b_\mu W^c_\nu, \\
 B_{\mu\nu} &=& \partial_\mu B_\nu - \partial_\nu B_\mu.
\end{eqnarray}
Here $g_w$ is the $SU(2)$ gauge coupling parameter. Physical fields of the electroweak 
interaction are linear combinations of the above fields,
\begin{eqnarray}
 W^{\pm}_\mu &=& \frac{1}{\sqrt{2}} (W^1_\mu \mp i W^2_\mu), \\ 
 \left( \begin{array}{c} Z_\mu \\ A_\mu \end{array} \right) &=& 
\begin{pmatrix} {\rm cos}\theta_w & -{\rm sin}\theta_w \\ 
                {\rm sin}\theta_w & {\rm cos}\theta_w \end{pmatrix} \; 
 \left( \begin{array}{c} W^3_\mu \\ B_\mu \end{array} \right).
\end{eqnarray}
The angle $\theta_w$ is known as the {\it weak mixing angle} and it is defined by 
\begin{eqnarray}
 {\rm cos}\theta_w = \frac{g_w}{\sqrt{g_w^2+g'^2}}, \;\;\; {\rm sin}\theta_w = \frac{g'}{\sqrt{g_w^2+g'^2}},
\end{eqnarray}
where $g'$ is the $U(1)_Y$ gauge coupling parameter. Though all the gauge fields
are massless at this stage, their triple and quartic couplings are already present 
in the Lagrangian; a hallmark of non-Abelian gauge theories. The coupling of the 
fermions with the gauge bosons is defined through the covariant derivative,
\begin{eqnarray}\label{eq:c-derivative}
D_\mu &\equiv& \partial_\mu -i g_w W^a_\mu T^a  - i g' B_\mu Y \\
      &=&  \partial_\mu -i \frac{g_w}{\sqrt{2}} (W^+_\mu T^+ + W^-_\mu T^-) 
        - i \frac{g_w}{{\rm cos}\theta_w} Z_\mu (T^3 - {\rm sin}^2\theta_w Q) \nn \\ && - \;i e A_\mu Q, \\
{\rm with} \; \; T^{\pm} &=& T^1 \pm i T^2, \;\;\ Q = T^3 + Y \;\; {\rm and} \;\; 
e = g_w \;{\rm sin}\theta_w = g' \;{\rm cos}\theta_w.
\end{eqnarray}
$Q$ and $e$ are the charge and the coupling parameter of the electromagnetic interaction, 
respectively. $Y$ is the $U(1)_Y$ charge called the {\it hypercharge}.
The $\{T^a\}$ are the generators of the $SU(2)$ gauge group and their algebra is given by,
\begin{eqnarray}
 [T^a,T^b] = i \veps^{abc} T^c \;\;{\rm and} \;\; [T^a,Y] = 0.
\end{eqnarray}
In Eq.~\ref{eq:c-derivative}, these generators appear in the fundamental representation. 
We see here that all the couplings of the electroweak vector bosons are described by two parameters, 
$e$ and $\theta_w$. However, a shift from the set $\{g_w,\;g'\} \to \{e,\; \theta_w\}$, does 
not imply true unification of the weak and the electromagnetic interactions. Existence of 
the $Z$ boson, and therefore the existence of weak neutral currents was a very successful 
prediction of the model. The fact that weak interactions differentiate left and right-handed 
fermions, is implemented in the theory by grouping left-handed fermions into $SU(2)$ doublets 
and taking right-handed fermions as $SU(2)$ singlets. That explains the subscript `$L$' 
in the $SU(2)_L$. For the first generation of fermions, the $\psi$ field in Eq.~\ref{eq:LEW}, 
represents the following set of fermion fields:
$$
 \left( \begin{array}{c} \nu_e \\ e \end{array} \right)_L, \; e_R, \; 
 \left( \begin{array}{c} u \\ d \end{array} \right)_L, \; u_R, \; d_R.
$$
Quantum numbers of these fermion fields are shown in Table~\ref{tbl:EWQN}. There are 
two more copies of the above set in the SM and all of them have been observed 
experimentally. Fermions of the first generation are sufficient for this brief 
introduction. Due to historical reasons, the $\nu_{eR}$ field is not included 
in the above set. 
% Note that we do not have field $\nu_{eR}$ in this set which is consistent with 
% the observation that we have seen only left-handed neutrinos and right-handed 
% anti-neutrinos in experiments. 

It is clear that mass terms for both the gauge bosons and fermions violate gauge 
invariance. But in Nature, we do see massive fermions and massive gauge bosons of 
weak interaction. This implies that the electroweak symmetry must be broken to meet 
the reality. In the standard electroweak model, this symmetry breaking is achieved 
by introducing a $SU(2)$ doublet of complex scalar fields with $Y=1/2$,
\begin{eqnarray}\label{eq:PHI-SU2}
 \Phi(x) =  \left( \begin{array}{c} \phi^+(x) \\ \phi^0(x) \end{array} \right).        
\end{eqnarray}
\begin{table}[h!]\label{tbl:EWQN}
\begin{center}
\begin{tabular}{|c|c|c|c|}
\hline
Fermions &$T^3$ &$Y$ &$Q$ \T \\
\hline
{$ \nu_{eL} $ } & $+\frac{1}{2}$ & $-\frac{1}{2}$ &  $\;\;\;0$ \T \B \\ 

{$ e_L $ }       & $-\frac{1}{2}$ & $-\frac{1}{2}$ & $-1$ \T \B \\  
 \hline
{$e_R$} & $\;\;\;0$ & $-1$ &  $-1$ \T \B \\ 

 \hline
{$ u_L $} & $+\frac{1}{2}$ & $+\frac{1}{6}$ &  $+\frac{2}{3}$ \T \B \\

{$ d_L $} & $-\frac{1}{2}$ & $+\frac{1}{6}$ & $-\frac{1}{3}$ \T \B \\
 \hline
{$u_R$} & $\;\;\;0$ & $+\frac{2}{3}$ &  $+\frac{2}{3}$ \T \B \\ 
 \hline
{$d_R$} & $\;\;\;0$ & $-\frac{1}{3}$ &  $-\frac{1}{3}$ \T \B \\ 
\hline
\end{tabular}
\end{center}
\caption{Quantum numbers of leptons and quarks of the first generation. The $Y$ value of fields
         is decided using the Gell-Mann--Nishijima relation, $Q = T^3 + Y$. This relation naturally 
         arises in the theory. Each quark field is threefold degenerate in electroweak interactions 
         due to its color content.}
\end{table}
The idea here is to couple the complex scalar doublet with the gauge fields and 
with the fermions in a gauge invariant manner. The scalar doublet couples with the gauge 
fields through the covariant derivative given in Eq.~\ref{eq:c-derivative}, and 
it couples with the fermion fields through Yukawa interactions,
\begin{eqnarray}
 {\cal L}_{scalar} &=& |D_\mu\Phi|^2 - V(\Phi) - y_d \;\bar{\psi_L}.\Phi \;\psi_R 
                                               - y_u \;\bar{\psi_L}.\tilde{\Phi}\; \psi_R + h.c., \; 
\label{eq:L-scalar} \\
           V(\Phi) &=& -\mu^2(\Phi^\dagger \Phi) + \lambda(\Phi^\dagger \Phi)^2, \label{eq:V-Higgs}\\ 
{\rm and} \; \; \tilde{\Phi} &=& i\sigma_2\Phi^\dagger = 
\left( \begin{array}{c} \phi^0 \\ -\phi^- \end{array} \right).\label{eq:phitilde}
\end{eqnarray}
$\psi_L$ and $\psi_R$ represent, respectively, the set of doublets and singlets of fermion fields 
introduced above. Only gauge invariant combinations of these fields can appear in Eq.~\ref{eq:L-scalar}. 
$y_u$ and $y_d$ are Yukawa couplings and they are related to the upper and lower components of 
the left-handed fermion doublets. In Eq.~\ref{eq:phitilde}, $\sigma_2$ is a Pauli matrix.
Note that, no new coupling parameter is introduced for the interaction of scalars with the 
gauge bosons. The first term in the Lagrangian includes gauge invariant terms, quadratic in 
gauge fields and coupled with the scalar doublet. Now the electroweak symmetry is broken 
spontaneously in the scalar sector, which then propagates into the gauge and fermion sector 
through its couplings with them, by giving a non-zero vacuum expectation value to the scalar
doublet, {\it i.e.},
\begin{eqnarray}\label{eq:VEV}
 <\Phi>_0 = \frac{1}{\sqrt{2}} \left( \begin{array}{c} 0 \\ v \end{array} \right).
\end{eqnarray}
Here $v = \sqrt{\mu^2/\lambda}$ is obtained by minimizing the scalar potential in 
Eq.~\ref{eq:V-Higgs}, provided $\lambda, \mu^2 > 0$. A comparison with the effective 
theory of weak interactions at low energy implies 
$v = (G_F\sqrt{2})^{-1/2} \simeq 246$ GeV, 
$G_F$ being the Fermi constant. Remember that the $\mu^2$-term in the potential is 
not the mass term as it appears with the incorrect sign in the Lagrangian. As a 
result of the spontaneous symmetry breaking, the quadratic gauge field terms become 
the mass term of the corresponding gauge bosons and the fermions become massive through 
the Yukawa terms. The masses of the electroweak bosons and fermions, in this model, 
are given by
\begin{eqnarray}
 M_W &=& g_w\; \frac{v}{2}, \; \; M_Z = \sqrt{g_w^2+g'^2}\;\frac{v}{2} = \frac{M_W}{{\rm cos}\theta_w}, \\
 M_\gamma &=& 0 \; \; {\rm and } \; \; m_f = \lambda_f\; \frac{v}{\sqrt{2}}.
\end{eqnarray}
Taking the experimental inputs for $v$ and $\theta_w$ (${\rm sin}\theta_w \simeq 0.23 $), the 
correct masses for the $W^\pm$ and $Z$ bosons were predicted by the model. On the other
hand, fermion masses are parameterized in terms of unknown Yukawa couplings. The observed
hierarchy of fermion masses does not find a natural explanation in this model.
The above mechanism of mass generation for fermions and gauge bosons is known as the 
{\it Higgs mechanism}~\cite{Higgs:1964ia,Higgs:1964pj,Englert:1964et,Guralnik:1964eu,Higgs:1966ev,Kibble:1967sv}.
The fact that the photon remains massless implies that the subgroup $U(1)_{em}$ is unbroken. 
Note that neutrinos are massless in the SM as defined above. The neutrino oscillation experiments 
suggest that neutrinos are massive, however, whether they are Dirac particles or 
Majorana particles is not yet clear~\cite{Beringer:1900zz,Fukuda:1998mi,GonzalezGarcia:2007ib}. 
In the SM, a Dirac mass term for neutrinos can be generated via Higgs mechanism by introducing a 
$\nu_{eR}$ field.

Since the true vacuum is defined by a non-zero expectation value, we must reparametrize 
the complex scalar doublet taken in Eq.~\ref{eq:PHI-SU2}. In the {\it unitary gauge}, 
there are no unphysical degrees of freedom and the scalar doublet can be taken as
\begin{eqnarray}
 \Phi(x) = \frac{1}{\sqrt{2}} \left( \begin{array}{c} 0 \\ v+H(x) \end{array} \right).
\end{eqnarray}
The real scalar field $H(x)$, also called the Higgs field, is the only physical scalar 
field left after the symmetry breaking has taken place. The other three real scalar 
fields of the scalar doublet are the goldstone bosons resulting from the spontaneous 
symmetry breaking. They provide longitudinal degrees of freedom to weak bosons making 
them massive. Replacing $\Phi(x)$ in Eq.~\ref{eq:L-scalar} with the above choice, we 
not only generate masses of various fields as described above, we also obtain interactions 
of the Higgs field with the gauge and fermion fields. The scalar potential contains the 
Higgs boson mass term, with the mass given by
\begin{eqnarray}
 M_H = \sqrt{2 \mu^2} = \sqrt{2\lambda}\;v,
\end{eqnarray}
along with its self interactions.
The SM does not provide any direct information on the value of $M_H$ because  
the parameter $\mu^2$ of the scalar potential is {\it a priori} unknown. Direct 
search experiments and various theoretical studies advocate a light Higgs boson~\cite{Beringer:1900zz}.
The recent discovery of a fundamental boson at the LHC, in the mass range 125-126 GeV, might 
confirm this last missing piece of the SM~\cite{ATLAS:2012gk,CMS:2012gu}. 
The electroweak model of $SU(2)_L\times U(1)_Y$
gauge group supplemented with the spontaneous electroweak symmetry breaking, through
 a complex scalar doublet, is also known as the Glashow-Weinberg-Salam (GWS) 
model~\cite{Glashow:1961tr,Weinberg:1967tq,salam:1968xx}. 
The renormalizability of the GWS model was proved by 't Hooft~\cite{'tHooft:1971fh}. 

The SM does not put any constraint on the number of fermion generations $n_g$, but to avoid 
{\it anomalies} (violation of gauge symmetries of classical Lagrangian due to quantum effects), 
it does require that the number of quark generations must be equal to the number of lepton generations.
Introduction of more than one generation of fermions in the model leads to the mixing of fields in 
the quark sector. This is related to the fact that weak interaction eigenstates are different 
from the mass eigenstates. If the number of fermion generations is greater than two, which is 
the case in the SM ($n_g=3$), CP violation can occur in weak interactions~\cite{Kobayashi:1973fv}. 
A very small CP violation (1 part in 1000) has been observed in kaon and B-meson 
decays~\cite{Nakada:2009zz,Beringer:1900zz}. CP violation is 
necessary to explain the matter-antimatter asymmetry in the Universe, but the amount of CP
violation present in the SM is not enough to quantify the observed asymmetry. A primer 
on the electroweak sector of the SM can be found in Ref.~\cite{Novaes:1999yn}. \\

% QCD
The $SU(3)_c$ sector of the SM represents the strong interaction among colored fields, 
the quarks and the gluons. Leptons are immune to the color force. The interactions of quarks 
and gluons are governed by the classical Lagrangian
\begin{eqnarray}\label{eq:LQCD}
 {\cal L}_{QCD} &=& -\frac{1}{4} (G^a_{\mu\nu})^2 + \bar{q}_i (i \slashed{D}-m_q)_{ij} q_j,
\end{eqnarray}
where the field strength of massless gluon fields $G^a_\mu \;(a=1\to 8)$, and the covariant 
derivative acting on the quark fields $q_i\;(i=1,2,3)$, are
\begin{eqnarray}
 G^a_{\mu\nu} &=& \partial_\mu G^a_\nu - \partial_\nu G^a_\mu + g_s \; f^{abc} G^b_\mu G^c_\nu, \\
 (D_\mu)_{ij} &=& \delta_{ij} \partial_\mu - i g_s G^a_\mu t^a_{ij}.\label{eq:Dmu}
\end{eqnarray}
Here $g_s$ is the coupling parameter of the strong interaction and $f^{abc}$ are the structure
constants of the $SU(3)$ group. The generators $\{t^a\}$ of the group 
satisfy, $[t^a, t^b] = i f^{abc} \; t^c$. In Eq.~\ref{eq:Dmu}, they are in the fundamental 
representation.
Like the photon in QED, gluons -- the force carriers of the strong interaction,
remain massless due to the local gauge invariance. On the other hand, 
the quark mass term is gauge invariant. Unlike the photon, the gluons are charged under the 
gauge group and they interact among themselves. The quantization of massless gauge field theories 
also require a gauge-fixing term in the Lagrangian~\cite{Faddeev:1967fc}. 
It is not possible to define the propagator for massless gauge fields without making a gauge choice. 
A class of covariant gauges can be introduced in Eq.~\ref{eq:LQCD}, to fix the gauge
\begin{eqnarray}
 {\cal L}_{GF} = - \frac{1}{2\xi} (\partial^\mu G^a_\mu)^2 ,
\end{eqnarray}
where $\xi$ is the gauge-fixing parameter. In non-Abelian massless gauge theories, to cancel the effects of 
the unphysical timelike and longitudinal polarization states of the gauge bosons, the covariant 
gauge-fixing term must be supplemented with the Faddeev-Popov ghost Lagrangian
\begin{eqnarray}
 {\cal L}_{FP} &=& \bar{\eta}^a(-\partial^\mu D^{ab}_\mu) \eta^b, \\
{\rm with} \;\;\; (D_\mu)^{ab} &=& \delta^{ab} \partial_\mu - g f^{abc} G^c_\mu. 
\end{eqnarray}
The ghost fields $\eta^a$ are complex scalar fields which obey Fermi-Dirac statistics. 
Independence of physical amplitudes on the gauge-fixing parameter $\xi$, provides a powerful check on calculations. 
For practical reasons, the gauge fixing principle is also useful in the quantization of electroweak
Lagrangian.

An essential feature of a renormalizable QFT is the scale dependence of parameters of the theory. The 
scale variation of coupling parameter $\alpha_s = g_s^2/4\pi$, is determined by the 
renormalization group equation~\cite{Callan:1970yg,Symanzik:1970rt},
\begin{eqnarray}
 Q^2 \frac{\partial}{\partial Q^2} \alpha_s(Q^2) = \beta(\alpha_s).
\end{eqnarray}
The $\beta$ function has a perturbative expansion in $\alpha_s$
\begin{eqnarray}
 \beta(\alpha_s) = - \alpha_s^2 (b_0 + b_1 \alpha_s + b_2 \alpha_s^2 + ...),
\end{eqnarray}
where $b_i$ is the $(i+1)$-loop contribution to the $\beta$ function. At one-loop, the 
solution to the renormalization group equation is 
\begin{eqnarray}
 \alpha_s(Q^2) = \frac{\alpha_s(\mu_r^2)}{1+b_0 \alpha_s(\mu_r^2){\rm ln}(Q^2/\mu_r^2)},
\end{eqnarray}
where $\mu_r$ is the renormalization scale and $b_0 = (33-2n_f)/12\pi$ with $n_f$, the number
of light quark flavours. For $n_f < 16$, $b_0 > 0$ and therefore, unlike QED, in QCD the 
coupling parameter decreases with increasing $Q$. This is the property of {\it asymptotic freedom}
which makes perturbative methods useful in strong interactions at high energy~\cite{Gross:1973id,Politzer:1973fx}. 
The fact that 
isolated quarks and gluons are not observed and the only finite energy asymptotic states of the 
theory are color singlets, is known as {\it color confinement}. This can be seen as a consequence 
of the growth of the coupling at low $Q$ ($\lesssim$ 1 GeV). A formal discussion on perturbative 
QCD can be found in Refs.~\cite{muta:2010xx,Brock:1993sz}. 
% The SM Feynman rules used in this thesis are taken from Ref.~\cite{peskin:2005xx,ellis:2003xx}. 
% 
%%%%%%%%%%%%%%%%%%%%%%%%%%%%%%%%%%%%%%%%%%%%%%%%%%%%%%%%%%%%%%%%%%%%%%%%%%%%%%%%%%%%%%%%%%%%%%%%%%%%%
\section{The ADD Model: An Example of New Physics}\label{section:add}
The SM of particle physics, although very successful, is conceptually incomplete. In 
absence of any new physics, the most fundamental scale in Nature is the 4-dimensional {\it Planck} scale,
$M_P\;(=\frac{1}{\sqrt{8\pi G_N}}) \sim 10^{19}$ GeV. It is the scale around which the quantum effects of 
4-dimensional Gravity can no longer be ignored. On the other hand, the electroweak symmetry breaking
required to generate masses for fermions and gauge bosons takes place at the scale of about 246 GeV. 
In the SM, the unexplained hierarchy between the electroweak scale and the 4-dimensional Planck scale 
is considered unnatural. The evolution of gauge couplings of the SM at higher energy scales 
($\sim 10^{16}$ GeV) provides a hint of a true unification (in the sense of a single coupling parameter) 
of strong, weak and electromagnetic interactions. This scale, also known as the {\it Grand Unification} 
scale, may be considered a natural scale of new physics but it does not resolve the hierarchy problem. 
The Higgs boson mass is not stable against the quantum corrections. Unlike fermions and gauge bosons, 
the Higgs boson mass is not protected by any symmetry present in the SM. The one-loop quantum correction 
to the Higgs mass is quadratically divergent and therefore, it strongly depends on the new physics scale. 
Provided there is no new physics scale between the electroweak and the Grand Unification scales, 
the Higgs boson mass cannot be kept low (around the electroweak symmetry breaking scale) naturally. 
In other words, for a stable and natural mass of the SM Higgs boson, a very fine tuning of various 
parameters in the theory would be required. This is known as the {\it naturalness} or {\it fine tuning} 
problem of the SM. These, so to speak problems of the SM, are central to many beyond the SM physics 
proposals~\cite{Bustamante:2009us}. Various studies suggest that any new physics beyond the SM might 
be operative around TeV scale. The LHC is fully capable of testing the SM predictions very accurately at 
the TeV scale; therefore, it can find out signatures of possible new physics. 

There are many models of new physics which have been proposed in the last 30 years, and they come 
with their own set of pros and cons. Supersymmetric models and extra-dimensional models are among 
many which have been studied extensively in the literature~\cite{Beringer:1900zz}. Both the 
supersymmetry and the extra space dimensions are important ingredients in the construction of 
the String theory, a promising theory of the quantum Gravity. In the supersymmetric models, the 
quadratic divergence of the Higgs mass is smoothened by introducing a symmetry between fermions 
and bosons. In models of extra dimensions, Gravity plays a direct role in fixing the TeV scale 
physics. In this thesis, we will consider a particular model of large extra dimensions, as an 
example of new physics scenario beyond the SM. \\

The problem of large hierarchy between the electroweak and Planck scales can be addressed 
within the models of extra space dimensions. Among various models of extra dimensions the
 model of $large$ extra dimensions, proposed by Arkani-Hamed, Dimopoulos and Dvali (ADD),
 is one of the very first successful attempts in this direction 
\cite{Antoniadis:1990ew,ArkaniHamed:1998rs,Antoniadis:1998ig,ArkaniHamed:1998nn}\footnote{Another 
very popular model of extra dimensions, in the context of the
SM hierarchy problem, is the model of $warped$ extra dimensions, first proposed by 
Randall and Sundrum (RS)~\cite{Randall:1999ee}. The model considers 5-dimensional 
space-time where the fifth dimension is curved. In this model, the size of the extra 
dimension can be very small and the fundamental scale of Gravity (the 5-dimensional 
Planck scale) can be close to the 4-dimensional Planck scale. Warping of the fifth 
dimension plays the trick of resolving the hierarchy problem. For a review on various 
other models of extra dimensions, one may refer to~\cite{Ponton:2012bi}.}. 
The predictions of this model can be probed at present day high energy colliders.
In the ADD model, the number of space-time dimensions is assumed to be $4+\delta$. 
 The SM degrees 
of freedom live on a $(3+1)$-dimensional brane, while Gravity is allowed
to propagate in full $4+\delta$ dimensions. To avoid any direct conflict with the
present day observations, these extra dimensions are assumed to be compact. 
For simplicity, we consider compactification of these extra dimensions on a 
$\delta$-dimensional torus ($T^\delta$) with radius $R/2\pi$. The space is thus factorized
meaning that the 4-dimensional part of the metric does not depend on the 
extra-dimensional coordinates. We assume that the energy density on the brane 
does not affect the space-time curvature and the extra space dimensions
are flat. The fundamental scale of Gravity 
$M_S$, {\it i.e.}, the (4+$\delta$)-dimensional Planck scale is related to the 
4-dimensional Planck scale by,
\begin{eqnarray}\label{eq:plank-scales}
 M_P^2 \approx M_S^{\delta+2}\: R^{\delta}.
\end{eqnarray}
From this relation one can argue that if $R$ (in general the volume
of the compact extra dimensions, $V_\delta$) is large enough, the fundamental scale of
 Gravity in $4+\delta$ dimensions can be as small as few TeV. The ADD model thus resolves 
the hierarchy problem of the SM by proposing the fundamental scale of Gravity in the TeV range.
However, it introduces another hierarchy (unexplained within the model) 
between the scale for the size of the extra dimensions ($1/R$) and the fundamental scale of 
Gravity ($M_S$). Even then it stands as a good phenomenological model to study. 
In 4 dimensions, the $(4+\delta)$-dimensional graviton appears as an infinite 
tower of KK-modes which includes one spin-2 state, $(n-1)$ spin-1 states and $n(n-1)/2$
spin-0 states. All these states are mass-degenerate and mass of the $n^{th}$ mode
is given by, 
\begin{eqnarray}\label{eq:KK-mass}
 m_{\vec{n}}^2 = \frac{4 \pi^2 \vec{n}^2}{R^2}.
\end{eqnarray}
This mass spectrum is cutoff at the scale $M_S$. The $\vec{n}=0$ mode 
corresponds to the 4-dimensional massless graviton, $U(1)$ gauge 
bosons and scalars. The vector KK-modes ($A_{\mu i}^{\vec{n}}$) decouple 
from the theory and the scalar KK-modes ($\phi_{ij}^{\vec{n}}$)
couple through their trace. In this thesis, we are interested in the 
interaction of KK-gravitons. The interaction of spin-$2$ component
of these KK-modes ($h_{\mu\nu}^{(\vec{n})}$) with the SM is given by an 
effective Lagrangian
\begin{eqnarray}\label{eq:LADD}
\mathcal{L}_{int} \sim \frac{1}{M_P} \sum_{\vec{n}}  h_{\mu\nu}^{(\vec{n})}(x) \: T^{\mu\nu}(x), 
\end{eqnarray} 
where $T^{\mu\nu}$ is the energy-momentum tensor for the SM fields. $M_S$ plays the role
of an ultraviolet (UV) cutoff of the effective theory. The Feynman rules
of the model are derived in \cite{Giudice:1998ck,Han:1998sg}. We have followed Han et al. 
\cite{Han:1998sg} and the ADD model Feynman rules, required in this thesis, are listed in 
the appendix~\ref{appendix:feynrules}.

The collider signatures of the model can be studied by looking at processes in which
KK-gravitons may appear as virtual states, or they may be produced directly in the 
final state~\cite{Giudice:1998ck}. In this thesis, we will be interested in the direct production of KK-gravitons.
The effective interaction in Eq.~\ref{eq:LADD}, suggests that the coupling of a single 
KK-graviton with the SM particles is suppressed by the 4-dimensional 
Planck scale. In the direct production processes, all the kinematically allowed KK-modes 
are produced and their contributions should be added up to obtain the inclusive 
cross sections. Since each KK-graviton mode is degenerate and the mass separation between 
modes is very small $(\sim 1/R)$, we can work in the continuum limit to 
sum the contributions of kinematically allowed KK-modes. In the continuum limit,
the inclusive cross section is
\begin{eqnarray}
 \sigma = \int dm \; \rho(m)\; \frac{d\sigma_m}{dm}. 
\end{eqnarray}
Here $\rho(m)$ is the density of KK states and it is given by,
\begin{eqnarray}\label{eq:KK-density}
 \rho(m) = 2 \frac{R^\delta m^{\delta-1}}{(4\pi)^{\delta /2} \Gamma(\delta /2)}.
\end{eqnarray}
From Eqs.~\ref{eq:plank-scales} and \ref{eq:KK-density}, we can see that
although the cross section for the production of a single KK-graviton ($\sigma_m$)
is proportional to $1/M_P^2$, the inclusive cross section, obtained by 
summing over all KK-mode contributions, is only $1/M_S^{\delta+2}$ suppressed. Thus, if $M_S$ is in the 
TeV range, it can have observable effects at the LHC. Since the coupling of each KK-mode with
the SM particles is very small, the direct production of KK-gravitons gives rise to missing energy 
signal in the detector. More details on the model and its phenomenology can be found in 
\cite{Giudice:1998ck,Mirabelli:1998rt,Han:1998sg}.
% 
%%%%%%%%%%%%%%%%%%%%%%%%%%%%%%%%%%%%%%%%%%%%%%%%%%%%%%%%%%%%%%%%%%%%%%%%%%%%%%%%%%%%%%%%%%%%%%%%%%%%%
 \section{Physics at Hadron Colliders}\label{section:hc}
High energy particle colliders are powerful microscopes which help us in revealing the 
secrets of Nature 
at the very small length scales ($\sim 10^{-19}$ m). We have come a long way since the first 
collider ADA ({\it Anello Di Accumulazione}) collided $e^+$ and $e^-$ beams at 500 MeV centre-of-mass 
energy and recorded single photon production\footnote{http://www.lnf.infn.it/acceleratori/ada/}.
We have already entered into an era of multi-TeV hadron colliders 
such as the Tevatron and the LHC, and have seen some of the very important discoveries in 
particle physics in recent times. 

\begin{figure}[ht]
 \begin{minipage}[b]{0.5\linewidth}
\centering
\includegraphics[width=\textwidth]{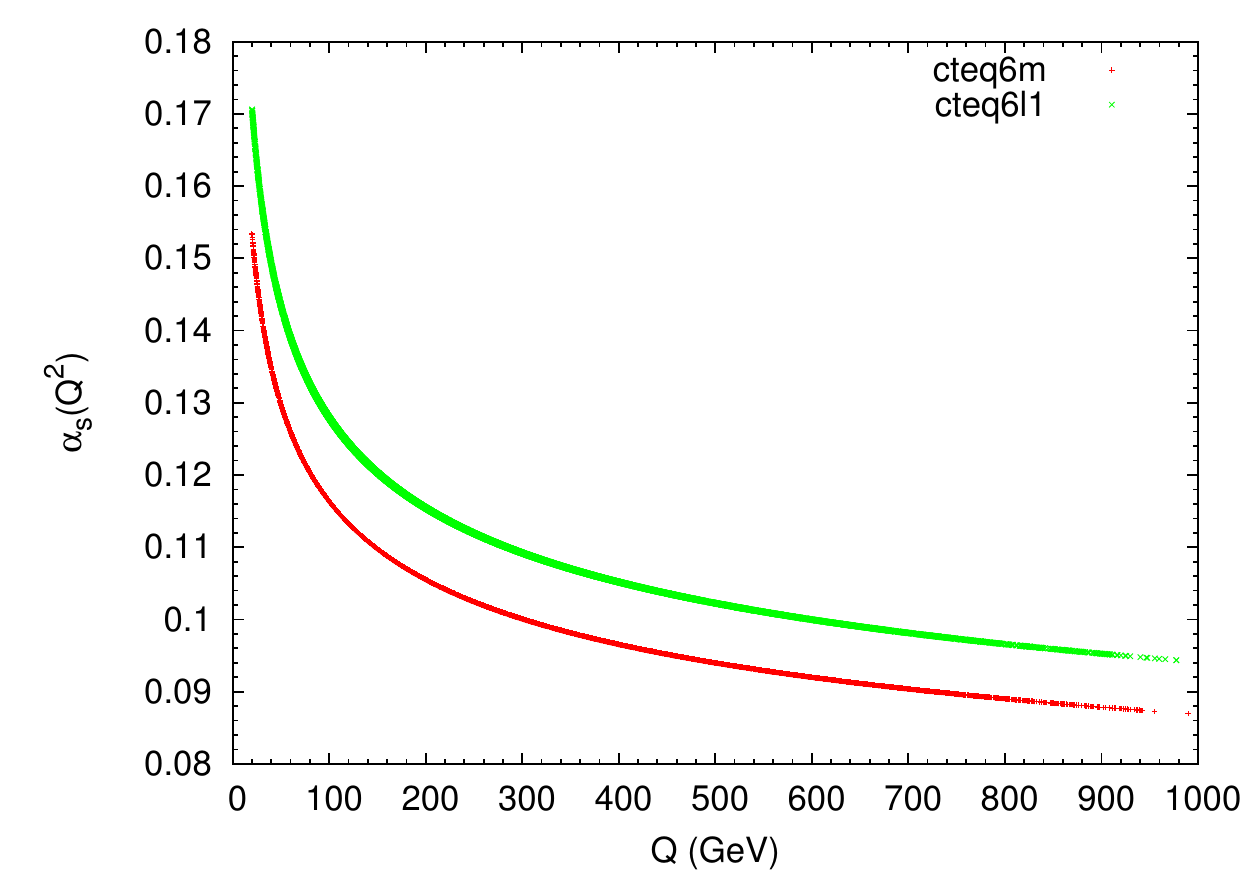}
\caption{Running of QCD coupling with the scale.}
\label{fig:alpha_s}
 \end{minipage}
 \hspace{0.2cm}
 \begin{minipage}[b]{0.5\linewidth}
\centering
 \includegraphics[width=\textwidth]{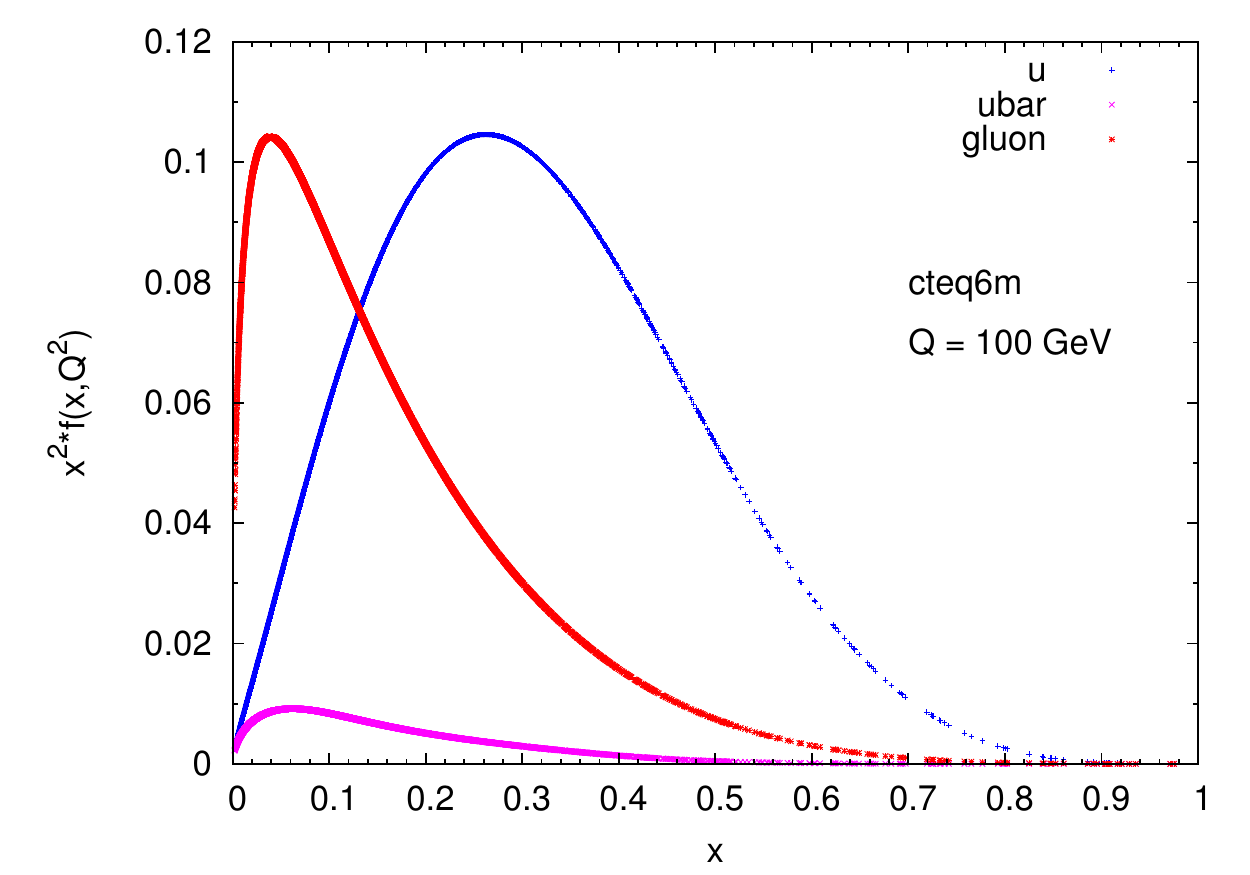}
\caption{Parton Distribution Functions of quark and gluon.}
\label{fig:pdf6m}
 \end{minipage}
\end{figure}

Deep inelastic lepton-hadron scatterings, performed at SLAC\footnote{Stanford Linear Accelerator Center} 
and HERA\footnote{Hadron Electron Ring Accelerator}, did confirm the parton model 
picture of hadrons~\cite{Close:1979xx}. According to the parton model, hadrons are composite objects 
and they are made up of 
partons, the quarks and the gluons. At hadron colliders, the fundamental interactions take place 
among these partons and they carry a certain momentum fraction of the parent hadron. The coupling
of the strong interaction, $\alpha_s(Q^2)$ is quite large at hadronic mass scales. However, due to
the property of {\it asymptotic freedom}, it is possible to apply perturbative methods at the 
parton level at higher energies. In Fig.~\ref{fig:alpha_s}, we can see the running of the strong
coupling parameter with the scale of interaction at the LO and the next-to-leading order (NLO).
In hadronic collisons, the partons undergo many soft subprocesses (processes involving low momentum
transfer) before reaching the interaction point 
of a hard scattering. These subprocesses do affect the cross section predictions but they cannot be 
calculated perturbatively due to a large $\alpha_s(Q^2)$ value at low $Q$. These contributions can be factorized 
into parton distribution functions (PDFs) at the level of cross section. Hadronic cross sections can be obtained 
in terms of partonic ones using the {\it factorization theorem} to all orders in perturbation 
theory~\cite{Collins:1989gx}. 
At the LHC for a $2 \to n$ process, the hadronic cross section is given by
\begin{flalign}                         
 \sigma(p+p\rightarrow k_1 + k_2 + ... + k_n + X) = &   
 \int_0^1 dx_1 dx_2 \, \sum_{a,b} f_a(x_1,Q^2)\times f_b(x_2,Q^2)\times \nn \\
& \hat{\sigma}(a+b\rightarrow k_1 + k_2 + ... + k_n; Q^2).
\end{flalign}  
The quantity $f_{a/b}(x,Q^2)$, known as a parton distribution function (PDF), defines the probability that the 
parton `$a/b$' inside a proton carries a momentum fraction $x$. The scale dependence in the partonic
cross section ($\hat{\sigma}$) may enter through higher order corrections due to the renormalization 
of the ultraviolet 
singularities and/or due to the factorization of the collinear singularities.
The scale dependence of the parton distribution function is a prediction of perturbative QCD 
and it has been verified, experimentally.   
The parton distribution functions contain the information on the 
soft/nonperturbative part of the hadronic interaction and they are universal. Although they cannot be calculated from 
the first principle, their evolution with the scale is governed by the perturbative QCD. Their evolution 
equation, known as the {\it Dokshitzer-Gribov-Lipatov-Altarelli-Parisi} (DGLAP) equation, is analogous to the QCD
$\beta$ function \cite{peskin:2005xx,ellis:2003xx}.
% %  
\begin{figure}[ht]
 \begin{minipage}[b]{0.5\linewidth}
\centering
\includegraphics[width=\textwidth]{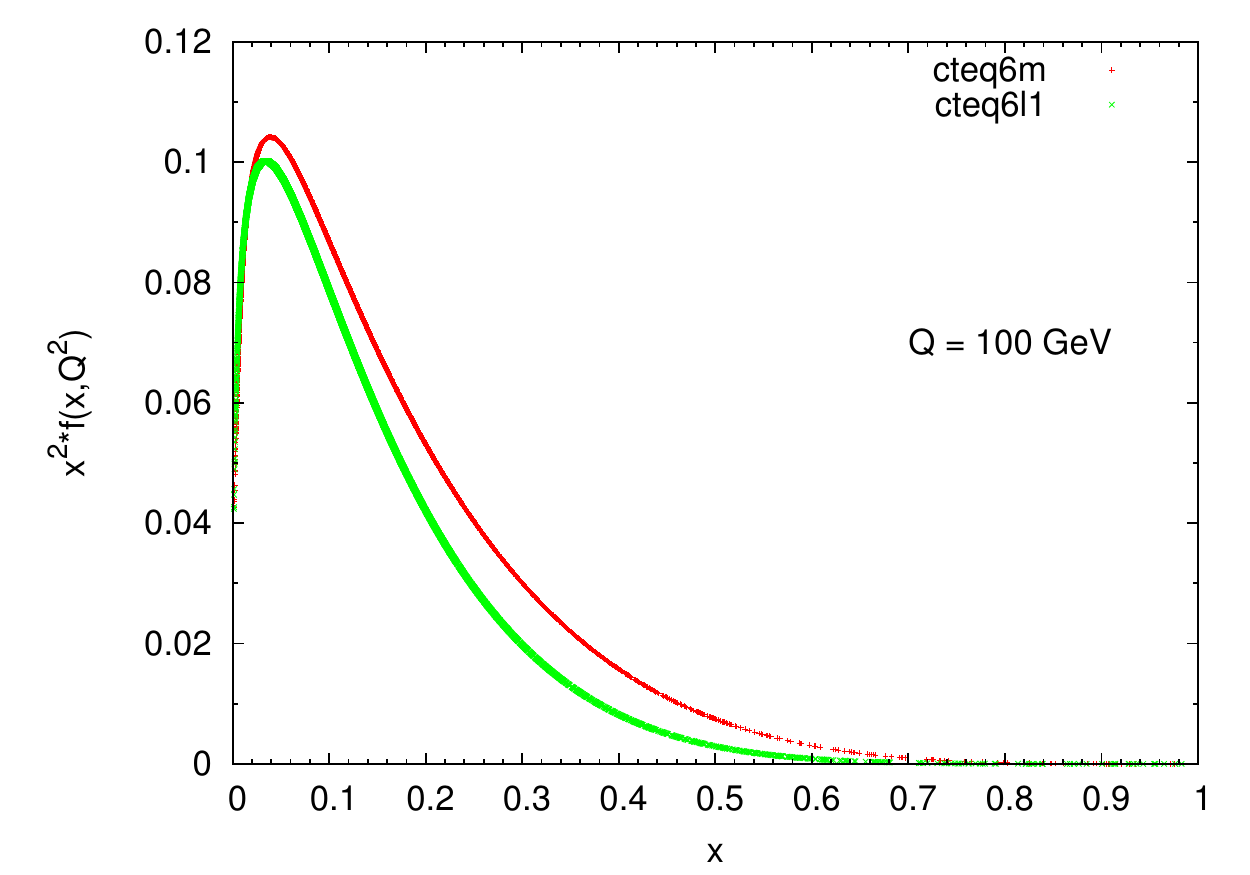}
\caption{Gluon Distribution Functions at LO and NLO.}
\label{fig:pdf-glu}
 \end{minipage}
 \hspace{0.2cm}
 \begin{minipage}[b]{0.5\linewidth}
\centering
 \includegraphics[width=\textwidth]{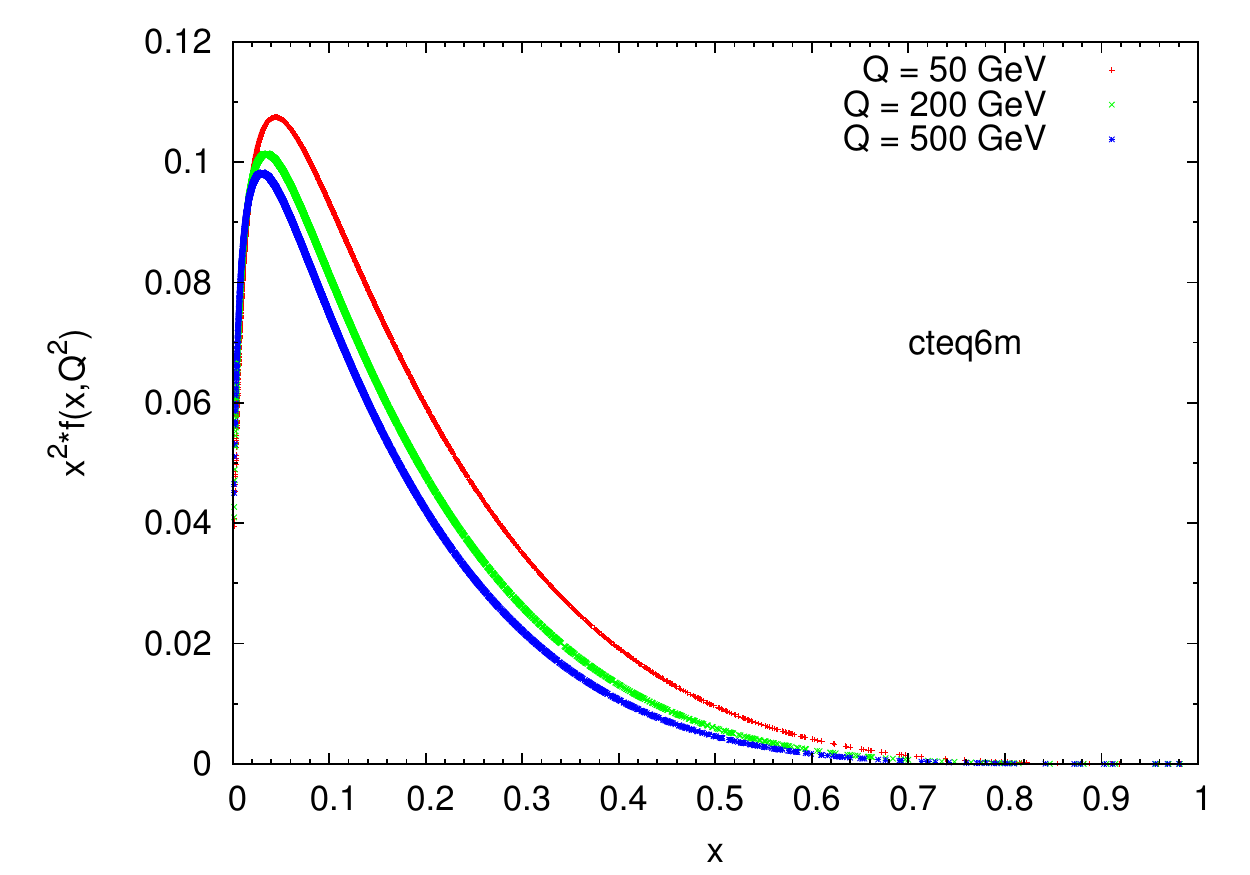}
\caption{Gluon Distribution Functions at various scales.}
\label{fig:pdfQ-glu}
 \end{minipage}
\end{figure}

In this thesis, we will be interested in gluon-gluon initiated processes at hadron colliders. 
In Fig.~\ref{fig:pdf6m}, we have plotted quark and gluon distribution functions at $Q=100$ GeV 
using the CTEQ6M PDF set. Note that we have multiplied the distributions by the momentum
fraction squared to reduce the large variation. It can be seen that the gluon distribution 
function dominates over the quark distribution function in the low-$x$ region. As the collider 
centre-of-mass energy increases, partons with smaller $x$ values may also contribute towards 
a physical process. Therefore, at the hadron collider such as the LHC, the gluon-gluon fusion
processes are quite important. The amplitudes of gluon fusion processes considered in this thesis
are finite, therefore, no renormalization and/or factorization proceedure is required. 
Thus the scale dependence in the cross section predictions enters only through the QCD coupling 
parameter, $\alpha_s(Q^2)$ and the PDFs, $f_a(x,Q^2)$. In Fig.~\ref{fig:pdf-glu},
we have plotted the gluon distributions at the LO and NLO using CTEQ6L1 and CTEQ6M PDF sets. The 
scale dependence of gluon distributions can be seen in Fig.~\ref{fig:pdfQ-glu}. A comprehensive 
introduction of particle phenomenology at hadron colliders can be referred 
to~\cite{barger:1996xx,ellis:2003xx,green:2009xx}.
% 

%%%%%%%%%%%%%%%%%%%%%%%%%%%%%%%%%%%%%%%%%%%%%%%%%%%%%%%%%%%%%%%%%%%%%%%%%%%%%%%%%%%%%%%%%%%%%%%%%%%%%
%%%%%%%%%%%%%%%%%%%%%%%%%%%%%%%%%%%%%%%%%%%%%%%%%%%%%%%%%%%%%%%%%%%%%%%%%%%%%%%%%%%%%%%%%%%%%%%%%%%%%
% \newpage
% 
% \bibliographystyle{utcaps}
% \bibliography{thesis}
% 
% \end{document}

\setcounter{equation}{0}
\setcounter{figure}{0}
\newpage
\chapter{One-loop Tensor Reduction}\label{chapter:oneloop}
\begin{figure}[h!]
\begin{center}
\includegraphics [angle=0,width=0.7\linewidth] {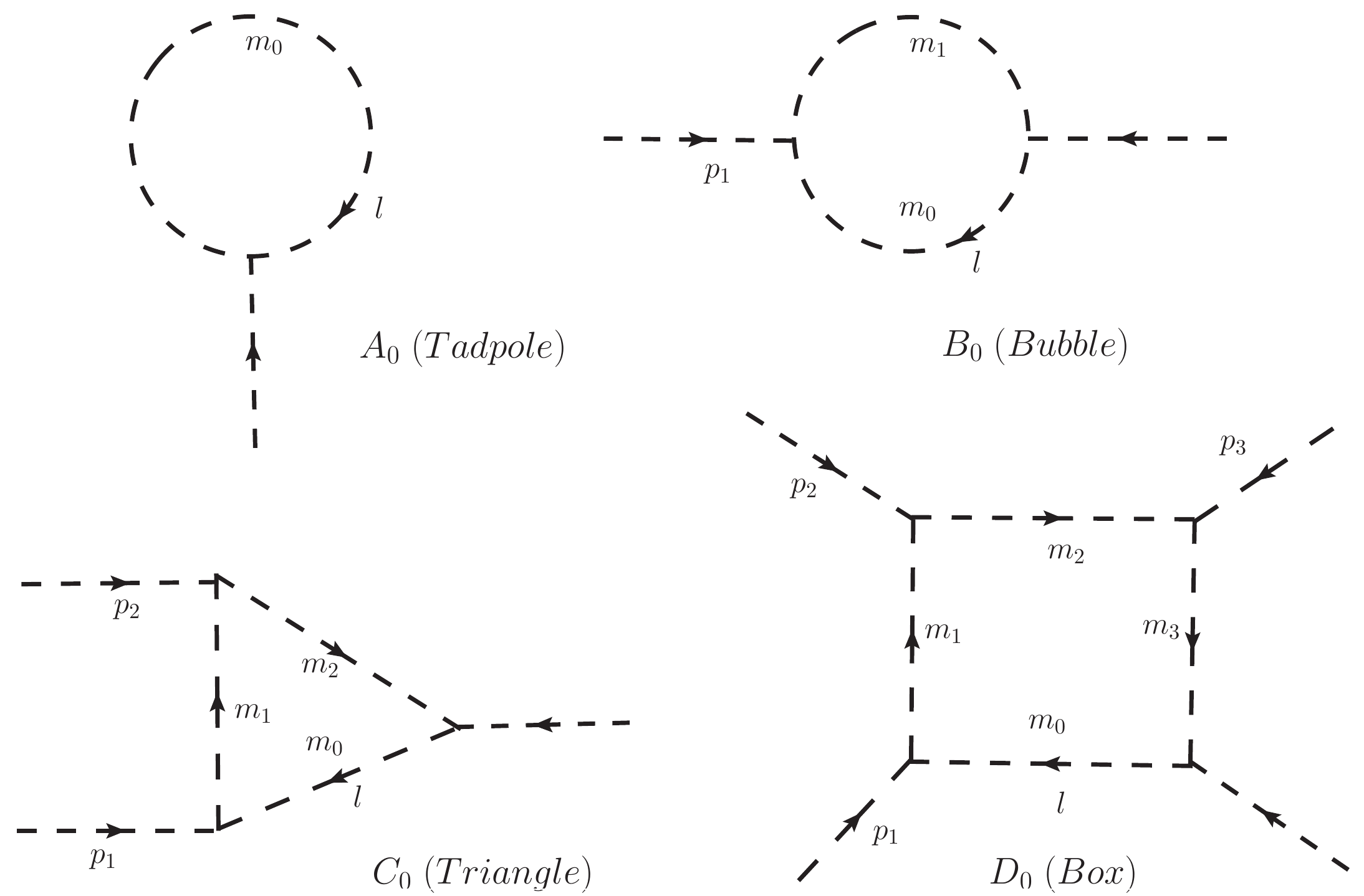}
\caption{Standard one-loop scalar integrals.}
\label{fig:scalars}
\end{center}
\end{figure}

% \numberwithin{equation}{chapter}
% 
We will now describe the reduction of one-loop tensor integrals, a very important ingredient in the 
calculation of any one-loop amplitude.
Calculation of scattering amplitudes at one-loop level involves {\it ill-defined} integration over loop 
momentum.The simplest example of loop integrals appear in $\phi^3$-theory. These are called 
{\it scalar integrals}. Depending upon the number of propagators in the loop, these integrals can be 
classified into various classes. Basic one-loop scalar integrals are shown in Fig.~\ref{fig:scalars}\footnote{
In this thesis, all the Feynman diagrams are drawn using Jaxodraw package~\cite{Binosi:2008ig}.}. 
Following the nomenclature of 't Hooft and Veltman~\cite{'tHooft:1978xw}, these integrals are
\begin{eqnarray}
 A_0 &\equiv& I_0^1(m_0)  = (\mu^2)^{2-n/2} \int \frac{d^nl}{(2\pi)^n} \frac{1}{d_0} , \\
\nn \\
 B_0 &\equiv& I_0^2(p_1;m_0,m_1)  = (\mu^2)^{2-n/2} \int \frac{d^nl}{(2\pi)^n} \frac{1}{d_0 \; d_1}, \\
\nn \\
 C_0 &\equiv& I_0^3(p_1,p_2;m_0,m_1,m_2)  = (\mu^2)^{2-n/2} \int \frac{d^nl}{(2\pi)^n} \frac{1}{d_0 \; d_1 \; d_2}, \\
\nn \\
 D_0 &\equiv& I_0^4(p_1,p_2,p_3;m_0,m_1,m_2,m_3)  = (\mu^2)^{2-n/2} \int \frac{d^nl}{(2\pi)^n} \frac{1}{d_0 \; d_1 \; d_2 \; d_3}.
\end{eqnarray}
Here, $d_i = (l+q_i)^2 - m_i^2 +  \ieps$ with $q_i = \sum_{j=1}^i p_j$ and $q_0 = 0$. We take all
the external momenta as incoming.
 We have introduced a mass scale $\mu$ to keep the natural mass-dimension of scalar integrals intact 
in $n$ dimensions. Hereafter, we will not show this scale dependence explicitly in $n$-dimensional 
loop integrals. 
In this thesis, we will be interested in one-loop diagrams with all internal lines of 
equal masses, {\it i.e.}, $m_i^2  = m^2$.

Thus, a general $N$-point scalar integral will have $N$ such denominators. In 4 dimensions, 
any $N\geq5$-point scalar integral can be always written in terms of box scalar integrals. In 
particular, a 5-point (pentagon) scalar integral can be reduced to a sum of five box scalar 
integrals \cite{vanNeerven:1983vr,Bern:1992em,Bern:1993kr},
\begin{eqnarray}\label{eq:E0}
 E_0 = \sum_{i=0}^4\;c_i \; D_0(i),
\end{eqnarray}
where $D_0(i)$ is the box scalar integral obtained after removing the $i^{th}$ propagator in $E_0$.
This derivation is included in the appendix~\ref{appendix:E0}. 
These scalar integrals are well known in the literature \cite{Ellis:2007qk}. 
We have derived all the scalar integrals for the case of massless internal lines,
required in this thesis, using the method of 't Hooft and Veltman \cite{'tHooft:1978xw}, and 
they are listed in the appendix~\ref{appendix:scalars}. \\
% Scalar integrals for massive internal lines are calculated 
% using OneLOop package developed by van Hameren \cite{vanHameren:2010cp}.\\
% 

In gauge theories, one generally comes across loop integrals with loop momentum appearing 
in the numerator. These are called {\it tensor integrals}. To set our notation, we write down
an $N$-point tensor integral of rank-m,
\begin{eqnarray}\label{eq:N-point}
 I^N_{\mu_1,\mu_2,....,\mu_m}(p_1,....,p_{N-1};m_0,....,m_{N-1}) =  \int \frac{d^nl}{(2\pi)^n} 
                              \frac{l_{\mu_1}l_{\mu_2}....l_{\mu_m}}{d_0 \; d_1 \; d_2 ..... d_{N-1}}.
\end{eqnarray}
In renormalizable gauge theories, $m \le N$.
Tensor integrals can also be evaluated in the manner of scalar integrals. However, with increasing power 
of loop momentum in the numerator, the straightforward evaluation of tensor integrals is not very 
appealing. A systematic approach towards the reduction of tensor integrals into the scalar integrals 
was presented by Passarion and Veltman (PV) \cite{Passarino:1978jh}. It is based on writing down the most general 
tensor structure possible using the set of external momenta $\{p_i\}$ and the metric tensor $g_{\mu\nu}$. 
For example,
the two-point (bubble) tensor integrals of the rank one and two, which appear in the self-energy correction to 
photon, are given by
\begin{eqnarray}
 B_\mu &=& p_\mu \; B_1 , \\
B_{\mu\nu} &=& g_{\mu\nu} \; B_{00} + p_\mu p_\nu \; B_{11} .
\end{eqnarray}
The tensor coefficients $B_1, B_{00}$ and $B_{11}$ can be written in terms of the tadpole and bubble 
scalar integrals. For the case of $p^2 \ne 0$,
\begin{eqnarray}
 B_1 &=& - \frac{1}{2} \; B_0,\\
 B_{00} &=& \frac{1}{(n-1)} \left[\frac{1}{2} A_0(0) + m^2 B_0 + \frac{1}{2} p^2 B_1\right], \\
 B_{11} &=& \frac{1}{p^2}\; \left[\frac{(n-2)}{2(n-1)}A_0(0) -\frac{1}{(n-1)}m^2 B_0 - 
                          \frac{n}{2(n-1)} p^2 B_1\right]. 
\end{eqnarray}
Here $A_0(0)$ is obtained from $B_0$, dropping the $d_0$ propagator. We will use this {\it missing propagator
notation} extensively, in the reduction of tensor integrals.  
Except for the special case of $m^2=0$, the above expression for $B_{11}$ cannot be used if $p^2$=0. 
In that case, we can evaluate the integral separately and the result is,
\begin{eqnarray}
 B_{\mu\nu}(p^2=0;m^2) = \left[ \frac{1}{(n-2)}m^2g_{\mu\nu} + \frac{1}{3} p_\mu p_\nu \right] B_0(p^2=0;m^2).
\end{eqnarray}

The reduction of higher point and higher rank tensor integrals can be carried out in a similar manner. 
Note that we have performed this reduction in $n$ dimensions. The reduction can of course be done in 
4 dimensions, but then one has to be very careful in shifting the loop momentum in UV divergent
integrals. In linearly (or worse) divergent loop integrals, any shift in the loop momentum
leads to a finite contribution. It is convenient, therefore, to perform the reduction in $n=4-2\eps$ 
dimensions.   
A detailed discussion on the reduction of higher point tensor integrals is given in the next 
section. The upshot is that any one-loop amplitude, regularized properly in 4 dimensions, 
can be written in the basis of tadpole, bubble, triangle and box scalar integrals, {\it i.e.},
 \begin{eqnarray}\label{eq:1loop}
 {\cal M}^{1-loop} = \sum_i \left(d_i D_0^i\right) +  \sum_i \left(c_i C_0^i\right) 
+  \sum_i \left(b_i B_0^i\right) +  \sum_i \left(a_i A_0^i\right)  + {\cal R}.  
\end{eqnarray}
In this equation, ${\cal R}$ is known as the {\it rational part} and it is an artifact of the UV 
regularization of tensor integrals. 
It is clear from the above relation that singularities of one-loop amplitudes are encoded in a set of basic
scalar integrals. In this thesis, numerical results are presented using a FORTRAN implementation of the tensor 
reduction method developed by Oldenborgh and Vermaseren (OV) \cite{vanOldenborgh:1989wn}. We have also implemented 
an analytic PV reduction of tensor integrals in FORM \cite{Vermaseren:2000nd}, following Denner and Dittmaier 
\cite{Denner:2002ii}. The analytic reduction code has been extremely useful in understanding the structure of 
one-loop amplitudes. We have used it to make various checks on our amplitudes in the ADD model projects. 
In recent times, there have been many new developments in the direction of one-loop calculations.  
A good review on the subject of one-loop calculations can be found in Ref.~\cite{Ellis:2011cr}.   

%%%%%%%%%%%%%%%%%%%%%%%%%%%%%%%%%%%%%%%%%%%%%%%%%%%%%%%%%%%%%%%%%%%%%%%%%%%%%%%%%%%%%%%%%%%%%%%%%%%%%% 
\section{Reduction of Standard Three and Four-point Tensor Integrals}
In this section, we will describe important features of the two reduction methods that we have
worked with in our projects. Before considering the actual reduction of tensor integrals we would 
like to construct some useful objects and study their properties. 

\subsection{Projection operators for tensor reduction}\label{section:projection}
For given two linearly independent vectors $p_1$ and $p_2$, we can always construct their 
{\it dual vectors} $v_1$ and $v_2$ such that,
\begin{eqnarray}
 v_i.p_j = \delta_{ij}.
\end{eqnarray}
Clearly, $v_i$s are linear combinations of $p_1$ and $p_2$. Let $ v_1^\mu = a\;p_1^\mu + b\;p_2^\mu$, then 
the coefficients $a$ and $b$ can be obtained by solving the following matrix equation,
\begin{eqnarray}
\left( \begin{array}{c} 1 \\ 0 \end{array} \right) &=& \begin{pmatrix} p_1.p_1 & p_1.p_2 \\ 
                                                                                   p_1.p_2 & p_2.p_2 \end{pmatrix} \; 
                                                     \left( \begin{array}{c} a \\ b \end{array} \right).
\end{eqnarray}
The $2\times2$ symmetric matrix on the right hand side is known as the {\it Gram matrix} of $p_1$ and $p_2$ 
and we will 
denote it by $X_2[p_1,p_2]$. The solution for $a$ and $b$, and therefore the existence of $v_1$ is related to 
the existence of the inverse of the Gram matrix. For that, the determinant of the Gram matrix, {\it i.e.}, 
the {\it Gram determinant} $\Delta_2(p_1,p_2)$, should not vanish and this will always be the case as long as 
$p_1$ and $p_2$ are linearly independent. The inverse of the Gram matrix of $p_1$ and $p_2$ is given by
\begin{eqnarray} \label{eq:GD2IN}
 X_2^{-1}[p_1,p_2] = \frac{1}{\Delta_2(p_1,p_2)} \begin{pmatrix} p_2.p_2 & -p_1.p_2 \\
                                                                -p_1.p_2 & p_1.p_1 \end{pmatrix},
\end{eqnarray}
with, $\Delta_2 = p_1^2 p_2^2 - (p_1.p_2)^2$.
Solving for $a$ and $b$ we get,
\begin{eqnarray}
 v_1^\mu = \frac{p_2.p_2}{\Delta_2}\; p_1^\mu - \frac{p_1.p_2}{\Delta_2}\; p_2^\mu = 
\delta^{\mu p_2}_{p_1p_2}/\Delta_2.
\end{eqnarray}
Here we have introduced the {\it generalized Kronecker delta notation} given in \cite{vanNeerven:1983vr}. 
In this notation, 
\begin{eqnarray}
\delta^{\mu_1 \mu_2}_{\nu_1 \nu_2} &=&  
 \left|\begin{array}{cc} \delta^{\mu_1}_{\nu_1} & \delta^{\mu_1}_{\nu_2} \\ 
\delta^{\mu_2}_{\nu_1} & \delta^{\mu_2}_{\nu_2} \end{array}\right|  \nn \\
  {\rm and} \; \; \delta^{p_1 p_2}_{q_1q_2} &=& \delta^{\mu_1 \mu_2}_{\nu_1 \nu_2} 
\;p_{1\mu_1}\;p_{2\mu_2}\;q_1^{\nu_1}\;q_2^{\nu_2}.
\end{eqnarray}
Therefore, the Gram determinant $\Delta_2 = \delta^{p_1 p_2}_{p_1p_2} $.
Similarly we can write,
\begin{eqnarray}
  v_2^\mu = -\frac{p_1.p_2}{\Delta_2}\; p_1^\mu + \frac{p_1.p_1}{\Delta_2}\; 
  p_2^\mu = \delta^{p_1 \mu}_{p_1p_2}/\Delta_2.
\end{eqnarray}
Note that the inverse of the Gram matrix of $p_1$ and $p_2$, shown in Eq.~\ref{eq:GD2IN}, 
is the Gram matrix of $v_1$ and $v_2$, {\it i.e.},
\begin{eqnarray}
X_2^{-1}[p_1,p_2] = X_2[v_1,v_2]. 
\end{eqnarray}

This analysis can now easily be extended for constructing the duals of three and four linearly independent vectors. 
For three linearly independent vectors $p_1,p_2$ and $p_3$, the duals are
\begin{eqnarray}
v_1^\mu = \delta^{\mu p_2p_3}_{p_1p_2p_3}/\Delta_3, \; v_2^\mu = \delta^{p_1 \mu p_3}_{p_1p_2p_3}/\Delta_3\; 
{\rm and } \; v_3^\mu = \delta^{p_1p_2 \mu}_{p_1p_2p_3}/\Delta_3, 
\end{eqnarray}
with, $\Delta_3 = \delta^{p_1p_2 p_3}_{p_1p_2p_3}$. In the case of four linearly independent vectors 
$p_1,p_2,p_3$ and $p_4$, the duals are
\begin{eqnarray}\label{eq:v4}
v_1^\mu &=& \delta^{\mu p_2p_3p_4}_{p_1p_2p_3p_4}/\Delta_4, \; 
v_2^\mu = \delta^{p_1 \mu p_3p_4}_{p_1p_2p_3p_4}/\Delta_4, \nn \\
v_3^\mu &=& \delta^{p_1p_2 \mu p_4}_{p_1p_2p_3p_4}/\Delta_4\;{\rm and } \; 
v_4^\mu = \delta^{p_1p_2 p_3 \mu}_{p_1p_2p_3p_4}/\Delta_4,
\end{eqnarray}
with, $\Delta_4 = \delta^{p_1p_2 p_3p_4}_{p_1p_2p_3p_4}$. In general, we have
\begin{eqnarray}
 v_i^\mu &=& \sum_j (v_i.v_j)\;p_j^\mu \;\; {\rm and} \;\; 
p_i^\mu = \sum_j (p_i.p_j)\;v_j^\mu \\
\Rightarrow  \delta_{ik}  &=& \sum_j (v_i.v_j)\;(p_j.p_k).
\end{eqnarray}

In integer space-time dimensions where we can define
the antisymmetric $\veps$-tensors, these dual vectors can be rewritten using,
\begin{eqnarray}
 \delta^{p_1p_2....p_m}_{q_1q_2....q_m} = \veps^{p_1p_2....p_m}\veps_{q_1q_2....q_m} .
\end{eqnarray}
For example, in the case of two linearly independent vectors $p_1$ and $p_2$, the dual vectors become
\begin{eqnarray}
 v_1^\mu = \frac{\veps^{\mu p_2}}{\veps^{p_1p_2}}\;\; {\rm and} \;\;  v_2^\mu = \frac{\veps^{p_1 \mu}}{\veps^{p_1p_2}}.
\end{eqnarray}
In the context of one-loop tensor reduction, the set of dual vectors is known as {\it van 
Neerven-Vermaseren basis vectors}.
To regulate the UV singularity of tensor integrals, we will be carrying out the reduction in $n = 4-2\eps$
dimensions and therefore the Kronecker delta representation of dual vectors will be more useful. It 
is easy to show that in $n$ dimensions,
\begin{eqnarray}
  \delta^{p_1p_2....p_m \mu}_{q_1q_2....q_m \mu} = (n-m) \; \delta^{p_1p_2....p_m}_{q_1q_2....q_m}.
\end{eqnarray}

Now we consider an another very useful object required in the reduction of tensor integrals, 
the {\it projective tensor}
\begin{eqnarray} \label{eq:omega-mn}
 \omega^\mu_\nu &=&   \delta^{p_1p_2....p_m \mu}_{p_1p_2....p_m \nu}/\Delta_m \nonumber \\
                &=&   \left(\delta^\mu_\nu - \sum_{i=1}^{m}v_i^\mu p_{i\nu}\right) 
                =   \left(\delta^\mu_\nu - \sum_{i=1}^{m}v_{i\nu} p_i^\mu\right),
\end{eqnarray}
for $m$ linearly independent vectors defined in $n$ dimensions . The  projective tensor satisfies 
following identities
\begin{eqnarray}
 \omega^\mu_\mu  = n-m; \; \omega^\mu_\nu \; p_{i \mu} = \omega^\mu_\nu \; p_i^\nu = 0; \; 
 \omega^\mu_\nu \; v_{i \mu} = \omega^\mu_\nu \; v_i^\nu = 0 \;\; {\rm and} \;\; 
\omega^\mu_\nu\omega^\nu_\rho = \omega^\mu_\rho.
\end{eqnarray}
Therefore, it is a projection operator into the space orthogonal to the vector space of $p_i$s 
(the transverse space). The defining equation of the projective tensor can be seen as a decomposition 
of the full metric tensor into two orthogonal spaces, {\it i.e.},
\begin{eqnarray}
 g^{\mu\nu} = \hat{g}^{\mu\nu} + \omega^{\mu\nu},
\end{eqnarray}
where $\hat{g}^{\mu\nu} = \sum_{i=1}^{m}v_i^\mu p_i^\nu$, is the metric tensor of the vector space of $p_i$s
and it satisfies $\hat{g}^\mu_\mu = m; \; \hat{g}^{\mu\nu}p_{i\mu} = p_{i\nu}$ \& 
$\hat{g}^{\mu\nu}v_{i\mu} = v_{i\nu}$. 
Note that $(n-m)$ is the dimensionality of the transverse space and therefore the projective tensor exists only for
$m < n$. Like $\hat{g}^{\mu\nu}$, $\omega^{\mu\nu}$ can also be written in terms 
of orthonormal basis vectors of the transverse space, say $\tilde{p}_i$s. Thus any vector in $n$ dimensions  can be 
completely expressed in terms of $p_i$s and $\tilde{p}_i$s. This is particularly useful in decomposing
the loop momentum appearing in the numerator of tensor integrals. From Eq.~\ref{eq:omega-mn},
\begin{eqnarray}\label{eq:l-mu}
 l^\mu &=&  \sum_{i=1}^{m}(l.v_i)\;p_i^\mu + \omega^\mu_l \; \; {\rm (Conventional\; decomposition)}\nonumber \\
       &=&  \sum_{i=1}^{m}(l.p_i)\;v_i^\mu + \omega^\mu_l \; \; {\rm (van \; Neerven-Vermaseren \;decomposition)}.
\end{eqnarray}
For the particular case of $n = m+1$, the transverse space is one dimensional and 
\begin{eqnarray}
  \omega^\mu_\nu =   \delta^{p_1p_2....p_m \mu}_{p_1p_2....p_m \nu}/\Delta_m = \tilde{p}^\mu \tilde{p}_\nu,
\end{eqnarray}
where $\tilde{p}^\mu = \veps^{p_1p_2....p_m \mu}/\veps^{p_1p_2....p_m}$, is the one and only basis vector in the 
transverse space. The explicit form of $\tilde{p}^\mu$ can be used in the reduction of box tensor integrals ($m =3 $) 
in 4 dimensions. Clearly for $m \ge n$, $\omega^\mu_\nu = 0$ and the loop momentum is completely expressible in $p_i$s.
We have used this fact to reduce the pentagon scalar and tensor integrals ($m=4$) in 4 dimensions. We will 
heavily use the ideas developed here in the reduction of tensor integrals~\cite{Beenakker:1989xx}.

\subsection{Passarino-Veltman (PV) reduction}
Let's first consider the PV method of reducing tensor integrals. The 3-point 
functions have two linearly independent momenta, $\p1$ and $\p2$. We can either use $p_i$s 
(the $p$-basis) or their linear combinations $q_i$s (the $q$-basis), to 
 decompose the tensor integrals into Lorentz-covariant structures. In the $q$-basis,
\begin{eqnarray}
 C^\mu &=& \sum_{i=1}^2 q_i^\mu C_i ,\;\;  C^{\mu\nu} = g^{\mu\nu}C_{00} + \sum_{i,j=1}^2 q_i^\mu q_j^\nu C_{ij}, \\
 C^{\mu\nu\rho} &=& \sum_{i=1}^2 g^{[\mu\nu} q_i^{\rho]} C_{00i} + \sum_{i,j,k=1}^2 q_i^\mu q_j^\nu q_k^\rho C_{ijk} .
\end{eqnarray}
The Lorentz indices within the square bracket represent a sum of all tensor structures with a cyclic permutation 
of these indices. Note that the tensor coefficients are symmetric under the permutation of all indices.
We can extract tensor coefficients by multiplying these expressions with appropriate combinations
of the projection operators $u_i$ (see Eq.~\ref{eq:appendixTR}) and $\omega^{\mu\nu}$ defined above. The tensor coefficients of rank-1, 3-point
function are given by,
\begin{eqnarray} 
  C_i &=& u_i^\mu C_\mu = \overline{l.u_i} \\
      &=& \sum_{n=1}^2(u_i.u_n)\overline{l.q_n} \nn \\
      &=& \sum_{n=1}^2 (X_2^{-1})_{in} R^1_n, 
\end{eqnarray}
where
\begin{eqnarray}
R^1_n &=& \overline{l.q_n} = \frac{1}{2}\overline{(d_n-d_0-q_n^2)} \nn \\
      &=& \frac{1}{2}\left[B_0(n) - B_0(0) - r_n C_0\right];\;r_n = q_n^2.
\end{eqnarray}
The overline notation implies a division by $d_0d_1d_2$ and the $n$-dimensional loop integration consistent with
Eq.~\ref{eq:N-point}. The tensor coefficients of rank-2, 3-point function are given by, 
\begin{eqnarray}
  C_{00} &=& \frac{1}{(n-2)}\omega^{\mu\nu} C_{\mu\nu} = \frac{1}{(n-2)}\;\overline{\omega^{ll}} \\
         &=& \frac{1}{(n-2)}\left[\overline{l^2} - \sum_{i=1}^2\overline{l.q_il.u_i}\right] \nn \\
         &=& \frac{1}{(n-2)}\left[B_0(0) + m^2C_0 -\frac{1}{2}\sum_{i=1}^2\overline{(d_i-d_0 -q_i^2)l.u_i}\right] \nn \\
         &=& \frac{1}{(n-2)}\left[B_0(0) + m^2C_0 -\frac{1}{2}\left(B_0(0) - \sum_{i=1}^2 q_i^2 C_i\right)\right] \nn \\
\Rightarrow C_{00} &=& \frac{1}{(n-2)}\left[ m^2C_0 +\frac{1}{2}B_0(0) + \frac{1}{2}\sum_{i=1}^2 r_i C_i\right].
\end{eqnarray}
Note that, $\overline{d_il.u_i} = 0$ and $\overline{d_0\sum_{i=1}^2l.u_i} = -\overline{d_0} = -B_0(0)$.
\begin{eqnarray}
  C_{ij} &=& (u_i^\mu u_j^\nu - \frac{u_i.u_j}{(n-2)}\omega^{\mu\nu})C_{\mu\nu} = \overline{l.u_i l.u_j} - u_i.u_j C_{00} \\
       &=& \sum_{n=1}^2u_i.u_n\overline{l.q_nl.u_j} - u_i.u_j C_{00} \nn \\
       &=& \sum_{n=1}^2u_i.u_n\left[\overline{l.q_nl.u_j}{} - \delta_{nj} C_{00}\right] \nn \\
       &=& \sum_{n=1}^2(X_2^{-1})_{in}\left[R^2_{nj} - \delta_{nj} C_{00}\right],\label{eq:cij1}
\end{eqnarray}
where
\begin{eqnarray}
 R^2_{nj} &=& \overline{l.q_n l.u_j} = \frac{1}{2}\overline{(d_n-d_0-q_n^2)l.u_j} \nn \\
          &=& \frac{1}{2}\left[ B_\mu(n) - B_\mu(0) -r_nC_\mu \right]u_j^\mu \nn \\
          &=& \frac{1}{2}\left[ B_1(n)\bar{\delta_{nj}} - B_j(0) -r_nC_j \right],\label{eq:cij2}
\end{eqnarray}
with $\bar{\delta_{nj}} = (1- \delta_{nj})$. $B_j(0)$ are the coefficients of,
\begin{eqnarray}
 B^\mu(0) = \overline{l^\mu} = B_1(0)q_1^\mu + B_2(0)q_2^\mu. 
\end{eqnarray}
The overline notation used here includes only the propagators $d_1$ and $d_2$.
The tensor coefficients of rank-3, 3-point function are given by,
\begin{eqnarray}
  C_{00i} &=&  \frac{1}{(n-2)} \omega^{\mu\nu}u_i^\rho C_{\mu\nu\rho}= \frac{1}{(n-2)}\overline{\omega^{ll} \;l.u_i} \\
          &=& \frac{1}{(n-2)}\sum_{n=1}^2 (u_i.u_n) \overline{\omega^{ll}\;l.q_n} \nn \\
          &=& \sum_{n=1}^2 (X_2^{-1})_{in} R^3_{n00},
\end{eqnarray}
where
\begin{eqnarray}
R^3_{n00} &=& \frac{1}{(n-2)}\overline{l.q_n\;\omega^{ll}} = \frac{1}{2(n-2)}\overline{(d_n-d_0-q_n^2)\omega^{ll}} \nn \\
          &=& \frac{1}{2}\left[B_{\mu\nu}(n) - B_{\mu\nu}(0) -r_n C_{\mu\nu}\right]\;\frac{\omega^{\mu\nu}}{(n-2)} \nn \\
          &=& \frac{1}{2}\left[B_{00}(n) - B_{00}(0) -r_n C_{00}\right].
\end{eqnarray}
Instead of expanding $l.q_i$ if we expand $\omega^{ll}$ and use the above result, we get an alternative
expression for $C_{00i}$,
\begin{eqnarray}
 C_{00i} = \frac{1}{(n-1)} \left[ m^2C_i +\frac{1}{2}B_i(0) + \frac{1}{2}\sum_{j=1}^2 r_j C_{ij}\right].
\end{eqnarray}
\begin{eqnarray}
  C_{ijk} &=&  u_i^\mu (u_j^\nu u_k^\rho - \frac{u_j.u_k}{(n-2)}\omega^{\nu\rho})C_{\mu\nu\rho} - u_i.u_j C_{00k} -u_i.u_k C_{00j} \\
          &=& \overline{l.u_i (l.u_j l.u_k-\frac{u_j.u_k}{(n-2)}\omega^{ll})} - u_i.u_j C_{00k} -u_i.u_k C_{00j} \nn \\
          &=& \sum_{n=1}^2 (u_i.u_n) \overline{l.q_n (l.u_j l.u_k-\frac{u_j.u_k}{(n-2)}\omega^{ll})} - u_i.u_j C_{00k} -u_i.u_k C_{00j} \nn \\
          &=& \sum_{n=1}^2 (X_2^{-1})_{in} \left[R^3_{njk} - \delta_{nj} C_{00k} - \delta_{nk} C_{00j}\right], 
\end{eqnarray}
where
\begin{eqnarray}
R^3_{njk} &=& \frac{1}{2}\overline{(d_n-d_0-q_n^2) (l.u_j l.u_k-\frac{u_j.u_k}{(n-2)}\omega^{ll})} \nn\\
          &=& \frac{1}{2} \left[ B_{\mu\nu}(n)- B_{\mu\nu}(0) -r_n C_{\mu\nu}\right]
              (u_j^\mu u_k^\nu - \frac{u_j.u_k}{(n-2)}\omega^{\mu\nu}) \nn \\
          &=& \frac{1}{2} \left[ B_{11}(n)\bar{\delta_{nj}}\bar{\delta_{nk}}- B_{jk}(0) -r_n C_{jk}\right].
\end{eqnarray}
$B_{00}(0)$ and $B_{ij}(0)$ are the coefficients of,
\begin{eqnarray}
B^{\mu\nu}(0) = \overline{l^\mu l^\nu} = B_{00}(0)g^{\mu\nu}+ \sum_{i,j=1}^2 q_i^\mu q_j^\nu B_{ij}(0). 
\end{eqnarray}

Notice that we required at maximum one $l.u_i$ to be written in terms of $l.q_n$. Thus a higher point 
tensor coefficient is expressed in terms of lower point tensor and scalar coefficients. The same analysis
can be extended to the reduction of 4-point functions up to rank-3 ($D_{\mu\nu\rho}$) with appropriate 
definitions of $u_i$ and $\omega_{\mu\nu}$. For completeness, we give here the reduction of rank-4,
4-point function,
\begin{eqnarray}
  D^{\mu\nu\rho\sigma} &=&  g^{[\mu\nu} g^{\rho]\sigma} D_{0000} +
                         \sum_{i,j=1}^3 \left(g^{[\mu\nu} q_i^{\rho]} q_j^\sigma + 
                                              g^{\sigma[\mu} q_i^\nu q_j^{\rho]} \right) D_{00ij} \nn \\ 
                         && \;+ \sum_{i,j,k,m=1}^3 q_i^\mu q_j^\nu q_k^\rho q_m^\sigma D_{ijkm} .
\end{eqnarray}
The tensor coefficients are again extracted using suitable combinations of $u_i$ and $\omega_{\mu\nu}$,
\begin{eqnarray}
 D_{0000} &=& \frac{1}{(n-1)(n-3)} \omega^{\mu\nu}\omega^{\rho\sigma} D_{\mu\nu\rho\sigma} \\
          &=& \frac{1}{(n-1)} \left[ m^2D_{00} +\frac{1}{2}C_{00}(0) + \frac{1}{2}\sum_{i=1}^3 r_i D_{00i}\right].
\end{eqnarray}
\begin{eqnarray}
 D_{00ij} &=& \frac{\omega^{\mu\nu}}{(n-3)}\left[u_i^\rho u_j^\sigma - u_i.u_j\frac{\omega^{\rho\sigma}}{(n-1)}\right]D_{\mu\nu\rho\sigma} \\
          &=& \sum_{n=1}^3(X_2^{-1})_{in}\left[S^4_{n00j} - \delta_{nj} D_{0000}\right],
\end{eqnarray}
where
\begin{eqnarray}
S^4_{n00j} &=& \frac{1}{2}\left[C_{\mu\nu\rho}(n) - C_{\mu\nu\rho}(0) -r_n D_{\mu\nu\rho}\right]\;
                             \frac{\omega^{\mu\nu}}{(n-3)} u_j^\rho \nn \\
          &=& \frac{1}{2}\left[C_{00j_n}(n)\bar{\delta_{nj}} - C_{00j}(0) -r_n D_{00j}\right].
\end{eqnarray}
with $j_n = j\; \Theta(n-j) + (j-1)\; \Theta(j-n)$. Noting that,
\begin{eqnarray}\label{eq:d00ij}
D_{00ij} = \frac{\omega^{\mu\nu}}{(n-3)}\left[u_i^\rho u_j^\sigma - 
             u_i.u_j\frac{\omega^{\rho\sigma}}{(n-3)}\right]D_{\mu\nu\rho\sigma} + 2 u_i.u_j D_{0000},
\end{eqnarray}
an another equivalent form can be obtained,
\begin{eqnarray}
D_{00ij} = \frac{1}{(n-1)}\left[m^2 D_{ij} + \frac{1}{2} C_{ij}(0) + \frac{1}{2}\sum_{n=1}^3 r_n D_{ijn} \right].
\end{eqnarray}
\begin{flalign}
  D_{ijkm} =&\; u_i^\mu\Big[u_j^\nu u_k^\rho u_m^\sigma - u_j.u_k\frac{\omega^{\nu\rho}}{(n-3)}u_m^\sigma
                                                        - u_j.u_m\frac{\omega^{\nu\sigma}}{(n-3)}u_k^\rho 
                                                        - u_k.u_m\frac{\omega^{\rho\sigma}}{(n-3)}u_j^\nu 
                      \Big]\times \nn \\ & D_{\mu\nu\rho\sigma} 
            - u_i.u_j D_{00km} - u_i.u_k D_{00jm} - u_i.u_m D_{00jk} \\
            =& \; \sum_{n=1}^3(X_2^{-1})_{in}\left[S^4_{njkm} - \delta_{nj} D_{00km}
                                                          - \delta_{nk} D_{00jm}
                                                          - \delta_{nm} D_{00jk} \right],
\end{flalign}
where
\begin{eqnarray}
S^4_{njkm} &=& \frac{1}{2}\left[C_{\mu\nu\rho}(n) - C_{\mu\nu\rho}(0) -r_n D_{\mu\nu\rho}\right]\times \nn \\ &&
\left[u_j^\mu u_k^\nu u_m^\rho - u_j.u_k\frac{\omega^{\mu\nu}}{(n-3)}u_m^\rho
                               - u_j.u_m\frac{\omega^{\mu\rho}}{(n-3)}u_k^\nu
                               - u_k.u_m\frac{\omega^{\nu\rho}}{(n-3)}u_j^\mu\right] \nn \\
          &=& \frac{1}{2}\left[C_{j_nk_nm_n}(n)\bar{\delta_{nj}}\bar{\delta_{nk}}\bar{\delta_{nm}} - C_{jkm}(0) -r_n D_{jkm}\right]. 
\end{eqnarray}
$C_{00i}(0)$ and $C_{ijk}(0)$ are the coefficients of, 
\begin{eqnarray}
 C^{\mu\nu\rho}(0) = \overline{l^\mu l^\nu l^\rho} = 
\sum_{i=1}^3 g^{[\mu\nu} q_i^{\rho]} C_{00i}(0) + \sum_{i,j,k=1}^3 q_i^\mu q_j^\nu q_k^\rho C_{ijk}(0) .
\end{eqnarray}

A complete list of tensor coefficients related to 1-,2-,3- and 4-point functions is given in the appendix B of 
Ref.~\cite{Denner:2002ii}. We have maintained the $n$ dependence in writing these tensor coefficients. 
In the above, one could have also used $p-$basis to expand tensor integrals. It is just a matter of personal taste. 

\subsection{Oldenborgh-Vermaseren (OV) reduction}\label{section:OV}
Now we turn to a tensor reduction method which is equivalent to the above but avoids explicit computation 
of tensor coefficients. Introduction of artificial virtualities for internal lines and the use of 
generalized Kronecker deltas allow compact formulae for the tensor reduction. We will describe all the important
aspects of this method by applying it to the reduction of 3-point functions. We define $s-$vectors related to the 
internal lines by $ p_1 = s_1-s_0, \; p_2 = s_2-s_1, \; p_3 = s_0-s_2 $, such that 
$ s_i^2 = m_i^2$, and $p_1+p_2+p_3 =0$. Recall the {\it van Neerven-Vermaseren} decomposition of the 
loop momentum for a three-point function,
\begin{eqnarray} \label{eq:Cl-mu}
l^\mu &=&  {\cal P}^\mu + \omega^\mu_l ; \; {\cal P}^\mu = \sum_{i=1}^{2}(l.p_i)\;v_i^\mu, \\
{\rm where} \;\;
v_1^\mu &=& \delta^{\mu p_2}_{p_1p_2}/\Delta_2, \; v_2^\mu = \delta^{p_1 \mu}_{p_1p_2}/\Delta_2 \;\; {\rm and} \;\;
\omega^\mu_\nu = \delta^{p_1p_2\mu}_{p_1p_2\nu}/\Delta_2.
\end{eqnarray}
Writing $l.p_i$ in terms of propagators,
\begin{eqnarray}
 {\cal P}^\mu = \frac{1}{\Delta_2}\left[ \delta^{s_0\alpha}_{p_1p_2}\delta^{\mu\alpha}_{p_1p_2} 
                                       -\frac{1}{2}\left(d_0 \delta^{\mu p_2}_{p_1p_2}
                                                        +d_1 \delta^{\mu p_3}_{p_1p_2}
                                                        +d_2 \delta^{\mu p_1}_{p_1p_2}\right)  \right].
\end{eqnarray}
Also note that $\omega^\mu_\nu p_i^\nu = \omega^\mu_\nu {\cal P}^\nu = 0$. 
We are now set to do the tensor reduction. We will again use the overline notation
introduced previously. The rank-1, 3-point function is give by,
\begin{eqnarray}
 C^\mu &=& \overline{l^\mu} \\
       &=& \overline{{\cal P}^\mu} + \overline{\omega^\mu_l}  \\
       &=& \frac{1}{\Delta_2}\left[ \delta^{s_0\alpha}_{p_1p_2}\delta^{\mu\alpha}_{p_1p_2} C_0
                                       -\frac{1}{2}\left( \delta^{\mu p_2}_{p_1p_2} B_0(0)
                                                        + \delta^{\mu p_3}_{p_1p_2} B_0(1)
                                                        + \delta^{\mu p_1}_{p_1p_2} B_0(2) \right)  \right].
\end{eqnarray}
Since $\overline{l^\nu}$ can only be a linear combination of $p_1$ and $p_2$, it implies
$\overline{\omega^\mu_l} = 0$. This holds true for any odd combination of $\omega^\mu_l$. 
In the reduction of higher rank tensors we replace one $l$ at a time using Eq.~\ref{eq:Cl-mu}.
 The rank-2, 3-point function is given by,
\begin{flalign}
 C^{\mu\nu} =& \;\overline{l^\mu l^\nu} \\
       =&\; \overline{{\cal P}^\mu l^\nu} + \overline{\omega^\mu_l l^\nu}  \\
       =&\; \frac{1}{\Delta_2}\Big[ \delta^{s_0\alpha}_{p_1p_2}\delta^{\mu\alpha}_{p_1p_2} C_\nu
                                       -\frac{1}{2}\Big( \delta^{\mu p_2}_{p_1p_2} B_\nu(0)
                                                        + \delta^{\mu p_3}_{p_1p_2} B_\nu(1) 
                                                         +\; \delta^{\mu p_1}_{p_1p_2} B_\nu(2) \Big) \Big] \nn \\
         & + \omega^\mu_\nu C_{00}, \\
{\rm where}\;\; C_{00} =&\; \frac{1}{(n-2)}\overline{\omega^l_l} = \frac{1}{(n-2)}\overline{(l^2-{\cal P}^2)} \label{eq:C00} \\
{\rm and}\;\; \overline{(l^2-{\cal P}^2)} =& \;\frac{1}{\Delta_2}\Big[ \delta^{s_0p_1p_2}_{s_0p_1p_2} C_0 +
                                        \frac{1}{2}\delta^{s_1p_2}_{p_1p_2}B_0(0) +
                                        \frac{1}{2}\delta^{s_0p_3}_{p_1p_2}B_0(1) + 
                                        \frac{1}{2}\delta^{s_0p_1}_{p_1p_2}B_0(2)\Big].
\end{flalign}

Similarly for the rank-3, 3-point function,
\begin{flalign}
 C^{\mu\nu\rho} =&\; \overline{l^\mu l^\nu l^\rho} \\
       =&\; \overline{{\cal P}^\mu l^\nu l^\rho} + \overline{\omega^\mu_l l^\nu l^\rho}\\
       =&\; \frac{1}{\Delta_2}\left[ \delta^{s_0\alpha}_{p_1p_2}\delta^{\mu\alpha}_{p_1p_2} C_{\nu\rho}
                                       -\frac{1}{2}\left( \delta^{\mu p_2}_{p_1p_2} B_{\nu\rho}(0)
                                                        + \delta^{\mu p_3}_{p_1p_2} B_{\nu\rho}(1)
                                                        + \delta^{\mu p_1}_{p_1p_2} B_{\nu\rho}(2) \right)  \right] \nn \\
       &\;+ \overline{\omega^\mu_l {\cal P}^\nu l^\rho} + \overline{\omega^\mu_l \omega^\nu_l l^\rho}
\end{flalign}
\begin{flalign} 
\Rightarrow C^{\mu\nu\rho}       =&\; \frac{1}{\Delta_2}\left[ \delta^{s_0\alpha}_{p_1p_2}\delta^{\mu\alpha}_{p_1p_2} C_{\nu\rho}
                                       -\frac{1}{2}\left( \delta^{\mu p_2}_{p_1p_2} B_{\nu\rho}(0)
                                                        + \delta^{\mu p_3}_{p_1p_2} B_{\nu\rho}(1)
                                                        + \delta^{\mu p_1}_{p_1p_2} B_{\nu\rho}(2) \right)  \right] \nn \\
       &+ \frac{\omega^\mu_\rho}{\Delta_2}\left[ \delta^{s_0\alpha}_{p_1p_2}\delta^{\nu\alpha}_{p_1p_2} C_{00}
                                       -\frac{1}{2}\left( \delta^{\nu p_2}_{p_1p_2} B_{00}(0)
                                                        + \delta^{\nu p_3}_{p_1p_2} B_{00}(1)
                                                        + \delta^{\nu p_1}_{p_1p_2} B_{00}(2) \right)  \right] \nn \\
       &+ \overline{\omega^\mu_l \omega^\nu_l {\cal P}^\rho} + \overline{\omega^\mu_l \omega^\nu_l \omega^\rho_l}
\end{flalign}
\begin{flalign}
\Rightarrow C^{\mu\nu\rho}       =&\; \frac{1}{\Delta_2}\left[ \delta^{s_0\alpha}_{p_1p_2}\delta^{\mu\alpha}_{p_1p_2} C_{\nu\rho}
                                       -\frac{1}{2}\left( \delta^{\mu p_2}_{p_1p_2} B_{\nu\rho}(0)
                                                        + \delta^{\mu p_3}_{p_1p_2} B_{\nu\rho}(1)
                                                        + \delta^{\mu p_1}_{p_1p_2} B_{\nu\rho}(2) \right)  \right] \nn \\
       &+ \frac{\omega^\mu_\rho}{\Delta_2}\left[ \delta^{s_0\alpha}_{p_1p_2}\delta^{\nu\alpha}_{p_1p_2} C_{00}
                                       -\frac{1}{2}\left( \delta^{\nu p_2}_{p_1p_2} B_{00}(0)
                                                        + \delta^{\nu p_3}_{p_1p_2} B_{00}(1)
                                                        + \delta^{\nu p_1}_{p_1p_2} B_{00}(2) \right)  \right] \nn \\
       &+ \frac{\omega^\mu_\nu}{\Delta_2}\left[ \delta^{s_0\alpha}_{p_1p_2}\delta^{\rho\alpha}_{p_1p_2} C_{00}
                                       -\frac{1}{2}\left( \delta^{\rho p_2}_{p_1p_2} B_{00}(0)
                                                        + \delta^{\rho p_3}_{p_1p_2} B_{00}(1)
                                                        + \delta^{\rho p_1}_{p_1p_2} B_{00}(2) \right)  \right].
\end{flalign}
$B_{00}(N)$ is the coefficient of $B_{\mu\nu}(N)$. Although we can reduce 2-point functions also
using this method, we stick to the PV type reduction for $B_\mu$ and $B_{\mu\nu}$ described above.
One can perform the reduction of 4-point functions along the same line with appropriately 
defined ${\cal P}^\mu$ and $\omega^\mu_\nu$. For the sake of completeness we list all
the important relations required in the recursive reduction of 4-point functions up to 
rank-4. For the reduction of box tensor integrals, 
\begin{eqnarray}
l^\mu &=&  {\cal P}^\mu + \omega^\mu_l ; \; {\cal P}^\mu = \sum_{i=1}^{3}(l.p_i)\;v_i^\mu, \\ \nn \\
{\rm where} \;\;
v_1^\mu &=& \delta^{\mu p_2p_3}_{p_1p_2p_3}/\Delta_3, \; v_2^\mu = \delta^{p_1 \mu p_3}_{p_1p_2p_3}/\Delta_3, \nn \\
v_3^\mu &=& \delta^{p_1 p_2\mu }_{p_1p_2p_3}/\Delta_3 \;\;
{\rm and} \;\; \omega^\mu_\nu = \delta^{p_1p_2p_3\mu}_{p_1p_2p_3\nu}/\Delta_3.
\end{eqnarray}
\begin{eqnarray}
 {\cal P}^\mu = \frac{1}{\Delta_3}\left[ \frac{1}{2}\delta^{s_0\alpha\beta}_{p_1p_2p_3}\delta^{\mu\alpha\beta}_{p_1p_2p_3} 
                                       -\frac{1}{2}\left(d_0 \delta^{\mu p_2p_3}_{p_1p_2p_3}
                                                        -d_1 \delta^{\mu p_3p_4}_{p_1p_2p_3}
                                                        +d_2 \delta^{\mu p_4p_1}_{p_1p_2p_3}
                                                        -d_3 \delta^{\mu p_1p_2}_{p_1p_2p_3}\right)  \right].
\end{eqnarray}
\begin{flalign}
\overline{(l^2-{\cal P}^2)} 
        =&\; \frac{1}{\Delta_3}\Big[ \delta^{s_0p_1p_2p_3}_{s_0p_1p_2p_3} D_0 
                                       + \frac{1}{2}\delta^{s_1p_2p_3}_{p_1p_2p_3}C_0(0) 
                                       - \frac{1}{2}\delta^{s_0p_3p_4}_{p_1p_2p_3}C_0(1) \nn \\
                                 &~~~~~+ \frac{1}{2}\delta^{s_0p_4p_1}_{p_1p_2p_3}C_0(2)
                                       - \frac{1}{2}\delta^{s_0p_1p_2}_{p_1p_2p_3}C_0(3) \Big].
\end{flalign}
\begin{flalign}
\overline{(l^2-{\cal P}^2)^2} 
        =&\; \frac{1}{\Delta_3}\Big[ \delta^{s_0p_1p_2p_3}_{s_0p_1p_2p_3} \overline{(l^2-{\cal P}^2)}
                                       + \frac{n-3}{2}\delta^{s_1p_2p_3}_{p_1p_2p_3}C_{00}(0)  
                                       - \frac{n-3}{2}\delta^{s_0p_3p_4}_{p_1p_2p_3}C_{00}(1) \nn \\ 
                                 &~~~~~+ \frac{n-3}{2}\delta^{s_0p_4p_1}_{p_1p_2p_3}C_{00}(2)
                                       - \frac{n-3}{2}\delta^{s_0p_1p_2}_{p_1p_2p_3}C_{00}(3) \Big].
\end{flalign}
$C_{00}(N)$ can be obtained following Eq.~\ref{eq:C00}. The tensor reduction method 
of Oldenborgh and Vermaseren can be seen as a reduction in dual vector basis. 
This method of reduction, due to its compactness, is very well suited for numerical 
implementation. More details on the method can be found in Ref.~\cite{vanOldenborgh:1989wn}. 

\section{Reduction of Five-point Tensor Integrals} 
The SM processes considered in this thesis have five external particles, therefore the reduction of 
5-point functions is also required. The most complicated tensor structure that occur in these processes is 
the rank-5, 5-point function $E_{\mu\nu\rho\sigma\alpha}$. Since these integrals are UV finite, we can 
perform the tensor reduction in 4 dimensions. The PV type reduction can be applied to them, but at this 
level it becomes very cumbersome. Also the tensor decomposition of $E_{\mu\nu}$ and of higher rank 
5-point functions is overcomplete because in 4 dimensions, 
\begin{eqnarray}
 g^{\mu\nu} = \sum_{i=1}^4 p_i^\mu v_i^\nu = \sum_{i,j=1}^4 (v_i.v_j)\; p_i^\mu p_j^\nu,
\end{eqnarray}
 {\it i.e.}, the four vectors $p_i^\mu$ span the $n=4$ dimensional Minkowski space. Thus for the 
5-point functions,
\begin{eqnarray}
 l^\mu &=& {\cal P}^\mu = \sum_{i=1}^{4}(l.p_i)\;v_i^\mu \nn \\
       &=& \sum_{i=1}^{4}v_i^\mu\;\left[\frac{1}{2}d_i -\frac{1}{2}d_{i-1} + s_0.p_i\right],
\end{eqnarray}
and we can use $\veps$-tensor representation of dual vectors, given by
\begin{eqnarray}
 v_1^\mu &=& \veps^{\mu p_2p_3p_4}/ \veps^{p_1p_2p_3p_4},\;  v_2^\mu = \veps^{p_1\mu p_3p_4}/ \veps^{p_1p_2p_3p_4},\; \nn \\
 v_3^\mu &=& \veps^{p_1p_2\mu p_4}/ \veps^{p_1p_2p_3p_4},\;  v_4^\mu = \veps^{p_1p_2p_3\mu}/ \veps^{p_1p_2p_3p_4}.
\end{eqnarray}
Therefore,
\begin{flalign}
 {\cal P}^\mu =&\;  \sum_{i=1}^{4} s_0.p_i\;v_i^\mu -\frac{1}{2}\big( d_0\; \veps^{\mu p_2p_3p_4} 
                   + d_1\; \veps^{\mu p_3p_4p_5} + d_2\; \veps^{\mu p_4p_5p_1} \nn \\
&~~~~~~~~~~~~~~~~~~~~+ d_3\; \veps^{\mu p_5p_1p_2} + d_4\; \veps^{\mu p_1p_2p_3} \big)/\veps^{p_1p_2p_3p_4}.
\end{flalign}
This relation now can be used to write down 5-point tensor integrals in terms of 5-point
and 4-point functions of lower rank. For example, the rank-5, 5-point function is given by,
\begin{flalign}
 E^{\mu\nu\rho\sigma\alpha} =&\; \overline{l^\mu l^\nu l^\rho l^\sigma l^\alpha} 
                             = \overline{l^\mu l^\nu l^\rho l^\sigma {\cal P}^\alpha} \\
  =&\; E^{\mu\nu\rho\sigma}\;\sum_{i=1}^{4} s_0.p_i\;v_i^\alpha - 
      \Big[ D^{\mu\nu\rho\sigma}(0)\; \veps^{\alpha p_2p_3p_4} + D^{\mu\nu\rho\sigma}(1)\; \veps^{\alpha p_3p_4p_5} \nn \\
  &+ D^{\mu\nu\rho\sigma}(2)\; \veps^{\alpha p_4p_5p_1} + D^{\mu\nu\rho\sigma}(3)\; \veps^{\alpha p_5p_1p_2}
   + D^{\mu\nu\rho\sigma}(4)\; \veps^{\alpha p_1p_2p_3}\Big]{\Big /}\veps^{p_1p_2p_3p_4}
\end{flalign}
and so on for lower rank 5-point functions. The reduction of 4-point functions appearing
in the above pentagon tensor reduction must be done in $n$ dimensions. This ends our discussion on the 
reduction of one-loop tensor integrals considered in this thesis and we confirm 
the most general structure of one-loop amplitudes given in Eq.~\ref{eq:1loop}. 

{\bf Remark}: 
In a dimensionally regulated tensor coefficient/integral, replacing $n=4-2\eps$ at the end of the
complete reduction leads to ${\cal O}(\eps)$ terms in its expression. Whenever these ${\cal O}(\eps)$ 
terms hit $1/\eps$ pole of UV divergent integrals ($A_0$ and $B_0$ in the complete reduction), the 
tensor coefficient gets finite contributions. These finite contributions are called rational terms. 
The rational part ${\cal R}$ in Eq.~\ref{eq:1loop}, is the collection of all such terms in a one-loop 
amplitude. We see that the rational terms appear due to the UV regularization of tensor integrals.
However, this does not mean that those tensor coefficients/integrals which are UV finite will not have 
rational terms, or tensor coefficients which are UV divergent will necessarily have rational terms. 
For example, all rank-one tensor coefficients (including $B_1$ which is UV divergent) do not have 
rational terms because the $n$ dependence enters through $g_{\mu\nu}$, only when rank $\ge 2$. On 
the other hand, the tensor coefficients $C_{ij}$, $D_{ijk}$, $D_{00ij}$ and $D_{ijkl}$, all are UV 
finite, but they do have rational terms. Unlike the rational terms of a UV divergent tensor coefficient, 
those of a UV finite tensor coefficient are implicit, that is, the rational terms in a UV finite
tensor coefficient arise only from the UV divergent tensor coefficients.  
For example, see the expression of $C_{ij}$ in Eqs.~\ref{eq:cij1} and \ref{eq:cij2}. This also implies 
that the $n$ dependence in $D_{00ij}$ (Eq.~\ref{eq:d00ij}) is of no consequence and we can take $n=4$.
The same is true for the UV finite tensor coefficients of 5-point functions with rank $\le 5$, and
the reduction of 5-point functions into 4-point functions using the 4-dimensional Schouten identity is 
completely justified. Note that the rational terms in tensor coefficients of $N$-point functions with 
$N>2$ are independent of internal line masses if $m_i^2=m^2$. The explicit expressions for the rational parts of basic tensor coefficients are given in Ref.~\cite{Binoth:2006hk}.
% 
%%%%%%%%%%%%%%%%%%%%%%%%%%%%%%%%%%%%%%%%%%%%%%%%%%%%%%%%%%%%%%%%%%%%%%%%%%%%%%%%%%%%%%%%%%%%%%%%%%%%%
\section{Fermion Loop Amplitudes}
In this thesis, as mentioned in the introduction, we will be dealing with the fermion loop amplitudes. 
The fermion loop diagrams/amplitudes have many special properties which can serve as an important 
check on calculations and can even simplify certain calculations sometimes. In this section, we will 
discuss some of the properties which we have come across while calculating fermion loop diagrams.
\subsection{IR finiteness of fermion loop diagrams }\label{section:IRfinite}
In the mass regularization, the infrared singular part of a massless fermion loop 
diagram appears as, 
\begin{eqnarray}
&\sim& A \; {\rm ln}^2(m^2) + B \;  {\rm ln}(m^2) + \mathcal{O}(m^2),
\end{eqnarray}
where $A\;{\rm and}\; B$ are complex functions of kinematic invariants. 
The ln$^2(m^2)$ piece refers to an overlapping of the soft and collinear singularities. 
The classification of IR singularities at one-loop is given in appendix~\ref{appendix:IR-structure}.
In the following, we will show that for a given (individual) fermion loop diagram, $ A=B=0$, that is, 
it is IR finite. In general, IR finiteness is expected to hold for a gauge invariant 
combination of fermion loop diagrams like in the case of their UV finiteness.
The proof is obvious for the case of massive fermions in the loop. 
Also, if all the external legs are massive/offshell, the diagram would be IR finite even for 
massless fermions in the loop. So we need to consider only those fermion loop diagrams 
in which fermions are massless or more correctly their masses can be neglected and
at least one external leg is massless.

\begin{figure}[h]
\begin{center}
\includegraphics [angle=0,width=0.4\linewidth] {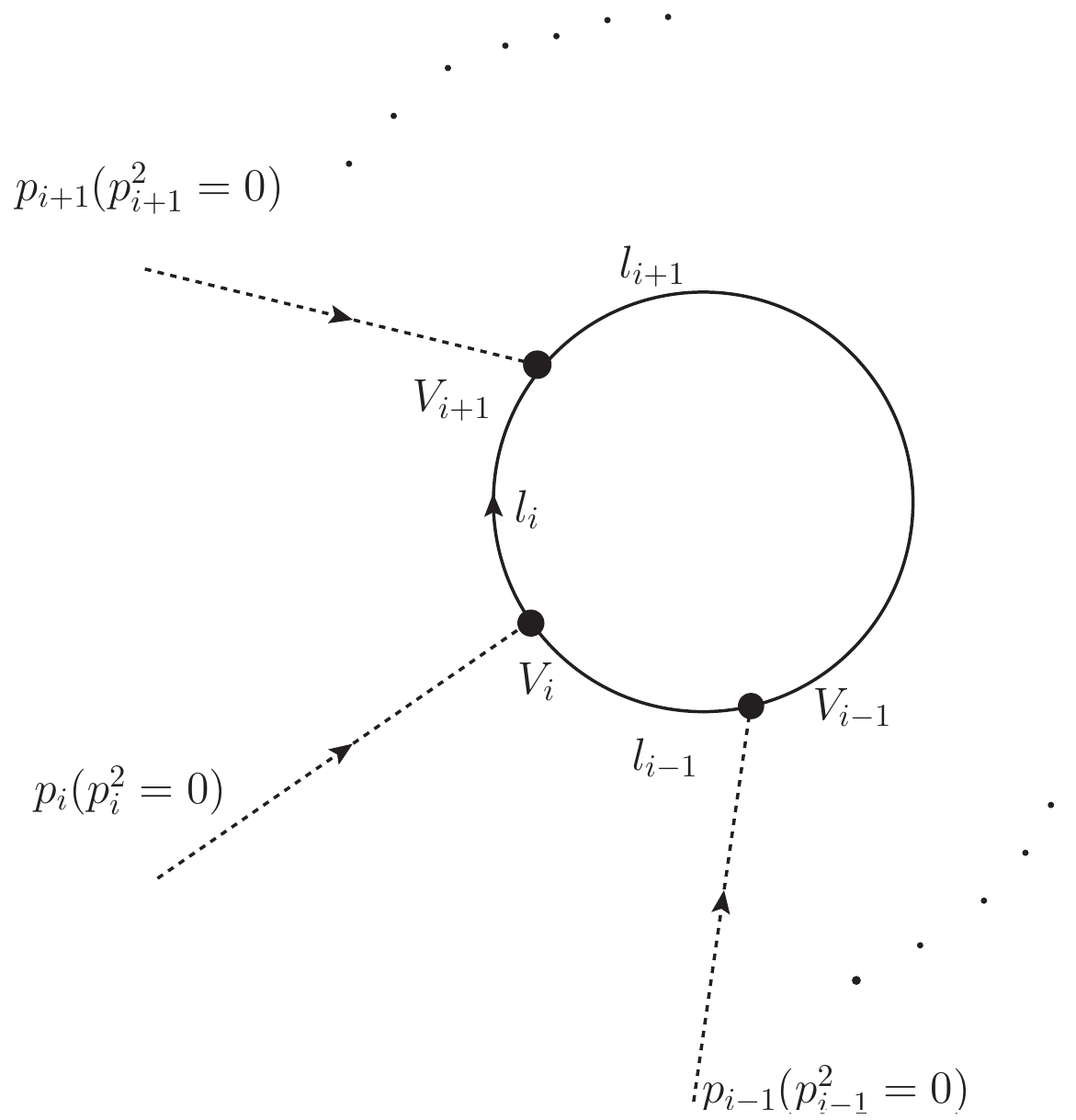}
\caption{ Massless fermion loop diagram. Dotted external lines represent any suitable massless particle.}
\label{fig:fermionloop}
\end{center}
\end{figure}

\begin{figure}[h]
\begin{center}
\includegraphics [angle=0,width=0.8\linewidth] {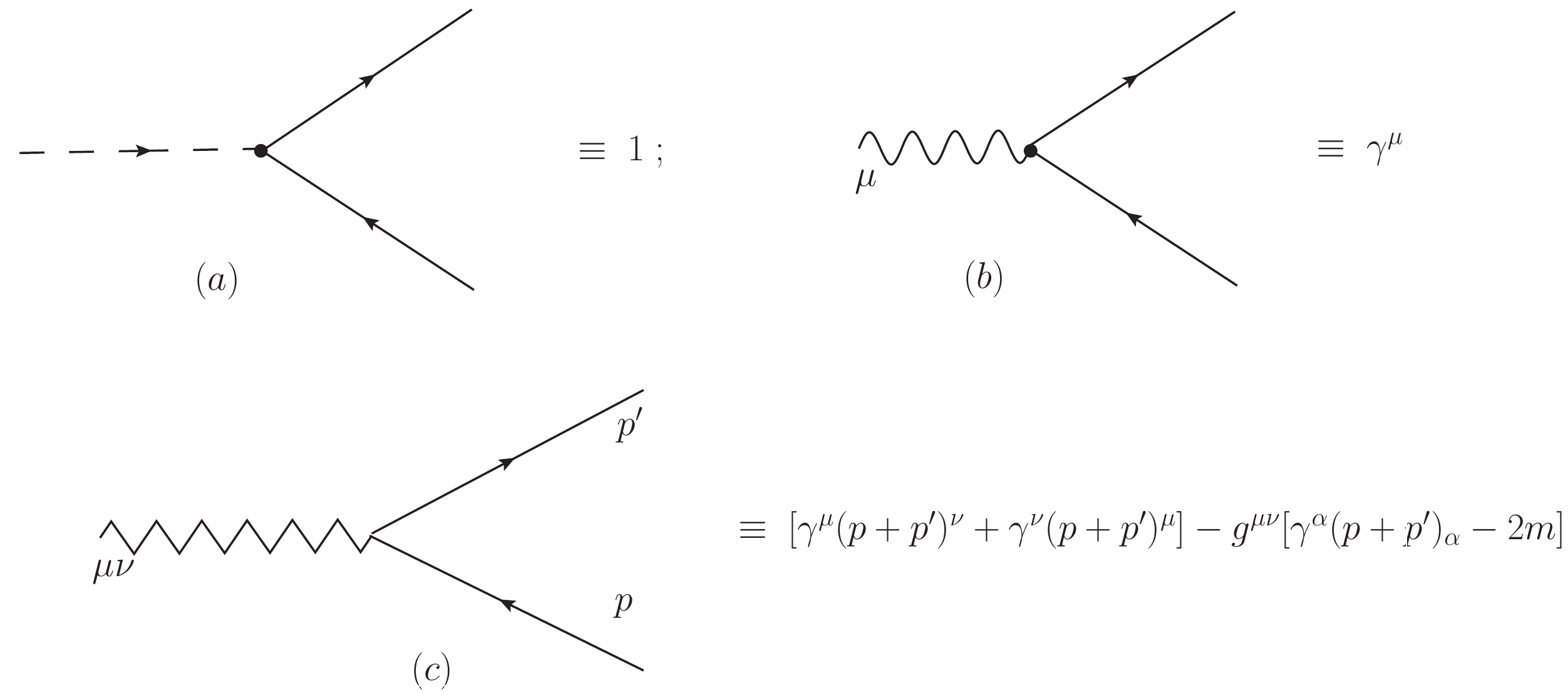}
\caption{Feynman rules: (a) scalar-fermion-fermion vertex (b) vector boson-fermion-fermion vertex 
        (c) graviton-fermion-fermion vertex }
\label{fig:vertices}
\end{center}
\end{figure}

In general, we can have scalars, gauge bosons and gravitons attached to a fermion loop. 
With one massless external particle we expect only collinear singularity while for two adjacent 
external massless particles, the soft and collinear singularities and their overlap may develop. 
IR finiteness of 
a fermion loop diagram can be shown by showing its soft finiteness and collinear finiteness. This 
automatically takes care of its finiteness in overlapping regions. At one-loop, the general fermion 
loop integral has the following form (see Fig.~\ref{fig:fermionloop}),
\begin{eqnarray}\label{eq:finite-I}
 I &\simeq& \int d^n l \; {{\rm tr}(....\slashed l_{i-1}V_i\slashed l_iV_{i+1}\slashed l_{i+1}....)\over
....l_{i-1}^2\;l_i^2\;l_{i+1}^2....},
 \end{eqnarray}
where $V_i$s are vertex factors for a given massless external particle attached to the fermion loop. 
From the momentum conservation at each vertex, it is clear that $l_i = l_{i-1} + p_i, \; l_{i+1} = l_i + p_{i+1}$ 
etc. The Feynman rules for needed vertices are given in Fig.~\ref{fig:vertices}. For simplification,
factors of `$i$' and coupling constants etc. are dropped in writing those rules. 
 We did not consider graviton-graviton-fermion-fermion 
vertex, since it does not correspond to a soft or collinear configuration described above. We will closely 
follow the analysis of IR singular structure of one-loop diagrams described in the appendix 
\ref{appendix:IR-structure}. \\

%%%%%%%%%%%%%%%%%%%%%%%%%%%%%%%%%%%%% Soft finiteness %%%%%%%%%%%%%%%%%%%%%%%%%%%%%%%%%%%%%%
%%%%%%%%%%%%%%%%%%%%%%%%%%%%%%%%%%%%%%%%%%%%%%%%%%%%%%%%%%%%%%%%%%%%%%%%%%%%%%%%%%%%%%%%%%%%%

First we would like to see the behavior of this integral as any of the internal lines becomes soft,
{\it i.e.}, its momentum vanishes. Without loss of generality, we consider softness of $l_i$ and take 
$l_i = \eps$. We see that in the limit $\eps \rightarrow 0$, the denominators which vanish in general 
are
\begin{eqnarray}\label{eq:soft-l}
 l_{i-1}^2 &=& \eps^2 - 2\eps \cdot p_i, \nonumber \\
l_i^2 &=& \eps^2 \nonumber \\
{\rm and} \;\; l_{i+1}^2 &=& \eps^2 + 2\eps \cdot p_{i+1},
\end{eqnarray}
where we have used on mass-shell conditions for $p_i$ and $p_{i+1}$. Neglecting $\eps^2$ with respect 
to $\eps\cdot p_i$ and $\eps\cdot p_{i+1}$ in Eq.~\ref{eq:soft-l}, we see that the integral in 
Eq.~\ref{eq:finite-I}, in the soft limit, behaves as
\begin{eqnarray}
 I &\sim& \int d^n\eps \; \frac{\slashed \eps}{\eps\cdot p_i\; \eps^2\; \eps\cdot p_{i+1}} \sim \eps^{n-3},
\end{eqnarray}
and it vanishes in $n=4$ dimensions. Thus each fermion loop diagram is soft finite, independent 
of the kind of massless external particles attached to it. We should mention here that the soft finiteness 
of fermion loop diagrams is indirectly shown by Kinoshita in \cite{Kinoshita:1962ur}.

%%%%%%%%%%%%%%%%%%%%%%%%%%% Collinear Finiteness %%%%%%%%%%%%%%%%%%%%%%%%%%%%%%%%%%%
%%%%%%%%%%%%%%%%%%%%%%%%%%%%%%%%%%%%%%%%%%%%%%%%%%%%%%%%%%%%%%%%%%%%%%%%%%%%%%%%%%%%

Next we consider the behavior of the fermion loop integral in collinear region. We take 
{$ l_i = x p_{i+1}+\epsilon_\perp$}, ($x\ne 0,-1$ as it corresponds to the softness of 
$l_i$ and $l_{i+1}$, respectively) with $\eps_\perp\cdot p_{i+1} = 0 $. Note that in 
$\eps_\perp \rightarrow 0$ limit, this condition implies collinearity of $l_i$ with 
$p_{i+1}$. In this collinear limit, only vanishing denominators are $l_i^2 = \eps_\perp^2$ 
and $l_{i+1}^2 = \eps_\perp^2$. Thus the integral in Eq.~\ref{eq:finite-I} reads,
\begin{eqnarray}\label{eq:collinear-I}
 I &\simeq& \int {d^n\epsilon_\perp \over\epsilon_\perp^4} \; {\rm tr} (...\slashed l_{i-1} 
\; V_i \; \slashed p_{i+1} \; V_{i+1} \; \slashed p_{i+1}... ).
\end{eqnarray}
We need to make the above substitution for $l_i$ in $V_i$s also, in case they may depend 
on the loop momentum, {\it e.g.}, in the graviton-fermion-fermion vertex. Since the vertex factors 
are different for different kind of external particles (see Fig.~\ref{fig:vertices}),
we will consider three separate cases to see the behavior of fermion 
loop integral in the collinear limit.
\begin{enumerate}
 \item 
{\bf Scalars:} \\
In this case, the vertex factor is simply  $V_{i+1} = 1$,
 and since $\slashed p_{i+1}\slashed p_{i+1} = p_{i+1}^2 = 0$, the integral 
in Eq.~\ref{eq:collinear-I} is
\begin{eqnarray}
 I &\simeq& \int {d^n\epsilon_\perp \over\epsilon_\perp^4} \; 
{\rm tr} (...\slashed l_{i-1} \slashed p_{i+1}\slashed p_{i+1}... ) = 0. 
\end{eqnarray}
 \item
{\bf Vector bosons:} \\
The vertex factor in this case, is
\begin{eqnarray}
 V_{i+1} &=& \gamma_\mu e_{i+1}^\mu = \slashed e_{i+1}.
\end{eqnarray}
 Here $e_{i+1}^\mu$ is the polarization vector of the vector boson with momentum $p_{i+1}$. Using 
the transversality and on-shell conditions for the vector boson, we see that
\begin{eqnarray}
 \slashed p_{i+1}\slashed e_{i+1}\slashed p_{i+1} &=& 2 \slashed p_{i+1} e_{i+1}\cdot p_{i+1}-p_{i+1}^2 
 \slashed e_{i+1} =   0.
\end{eqnarray}
and therefore the fermion loop integral in collinear limit, Eq.~\ref{eq:collinear-I}, vanishes.
 \item
{\bf Gravitons:} \\
The graviton-fermion vertex factor is given by
\begin{eqnarray}\label{eq:Gff-vertex}
 V_{i+1} &=& \big[\gamma_\mu(2l_i+p_{i+1})_\nu + \gamma_\nu(2l_i+p_{i+1})_\mu  \nn \\
          && - g_{\mu\nu}(2\slashed l_i+\slashed p_{i+1}-2m)\big] e_{i+1}^{\mu\nu},
\end{eqnarray}
where $e_{i+1}^{\mu\nu}$ is the polarization tensor for graviton of momentum $p_{i+1}$. It has 
following well known properties,
\begin{eqnarray}
 e_{i+1}^{\mu\nu}g_{\mu\nu} = ({e_{i+1}})^\mu_\mu &=& 0 \; {\rm (traceless \;condition)}, \nonumber\\
p_{i+1}^\mu ({e_{i+1}})_{\mu\nu} &=& 0 \; {\rm (transverse \; condition)}.
\end{eqnarray}

Using these properties the vertex factor in Eq.~\ref{eq:Gff-vertex} becomes
\begin{eqnarray}
 V_{i+1} &=& 4 \gamma_\mu e_{i+1}^{\mu\nu}(l_i)_\nu.
\end{eqnarray}
In the collinear limit, taken above, it is
\begin{eqnarray}
 V_{i+1} &=& 4 x \gamma_\mu e_{i+1}^{\mu\nu}(p_{i+1})_\nu = 0,
\end{eqnarray}
due to the transverse condition. Therefore the fermion loop diagram with external gravitons, 
like the cases of scalars and vector bosons, is also collinear finite. 
\end{enumerate}

Combining all the above results 
of this section, we see that a general fermion loop diagram is soft as well as collinear finite.
To summarize, the soft finiteness of fermion loop diagrams follows from simple power counting 
in vanishing loop momentum, while their collinear finiteness results utilizing various properties 
of massless external particles attached to the loop. The result holds even for axial coupling
of external particles with the fermion. Though we have not considered any flavor change in the loop, 
it should be clear from the above analysis that our result remains true for any possible flavor 
changing interaction vertex in the loop. \\

We can utilize the above fact regarding an individual fermion loop diagram to show,
that the IR structure of any one-loop amplitude with $N\ge 3$ can be fixed completely in terms
of the IR structure of 3-point functions only. Since we have learned quite a bit about 
one-loop amplitudes, it would be sufficient for us to show that the IR singularities
of box scalar integrals can be expressed in terms of those of reduced triangle scalars.   
\begin{figure}[h]
\begin{center}
\includegraphics [angle=0,width=0.4\linewidth] {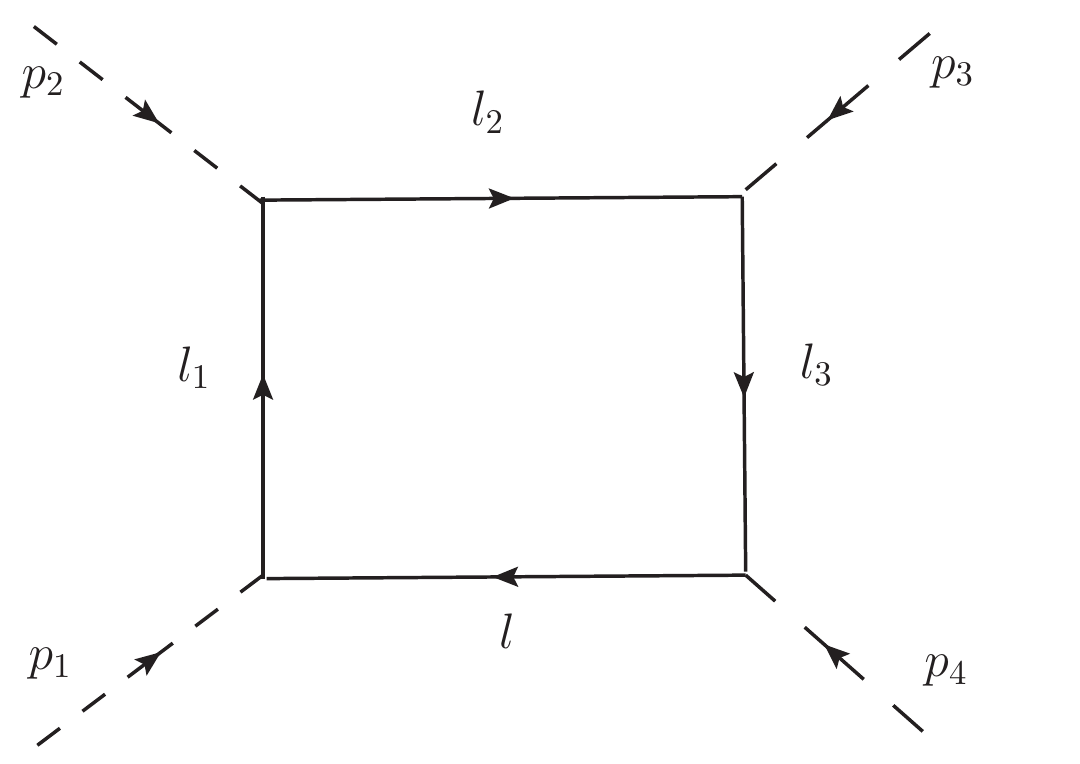}
\caption{ Interaction of 4 scalars in the Yukawa theory of scalars and fermions. }
\label{fig:phi-4}
\end{center}
\end{figure}
The amplitude of the fermion loop diagram, shown in Fig.~\ref{fig:phi-4}, is\footnote{Note 
that there are two more independent diagrams which contribute to the $\phi^4$-amplitude 
in the Yukawa theory. We do not require them here.} 
\begin{eqnarray}
 {\cal M}_{1234}(\phi^4) &=& 
\int \frac{d^n l}{(2\pi)^n}\; \frac{{\rm tr}({\slashed l} {\slashed l_3} 
                                             {\slashed l_2} {\slashed l_1} )}{l^2\; l_1^2\; l_2^2\; l_3^2} 
\nn \\
&=&\int \frac{d^n l}{(2\pi)^n}\; \frac{4(l.l_3\;l_2.l_1 - l.l_2\;l_3.l_1 + l.l_1\;l_2.l_3)}
{l^2\; l_1^2\; l_2^2\; l_3^2}.
\end{eqnarray}
Rearranging the terms in the numerator, we can write
\begin{eqnarray}
 {\cal M}_{1234}(\phi^4) &=& (p_1^2 p_3^2 + p_2^2 p_4^2 - st)\; D_0 + (s - p_1^2 - p_2^2)\; C_0(3) \nn \\
                         && + (t - p_1^2 - p_4^2)\; C_0(2) + (s - p_3^2 - p_4^2)\; C_0(1) \nn \\
                         && + (t - p_2^2 - p_3^2)\; C_0(0) + 2\; B_0(0,2) + 2\; B_0(1,3),
\end{eqnarray}
where $s=(p_1+p_2)^2$ and $t=(p_2+p_3)^2$. $C_0(i)$ and $B_0(i,j)$ are reduced triangles and bubbles
written in the missing propagator notation introduced earlier in this chapter. Since the bubble scalars
are IR finite, the IR finiteness of the above fermion loop amplitude implies                                                                                                 
\begin{eqnarray}\label{eq:IR-D0toC0}
 D_0|_{IR} &=& \frac{1}{(st - p_1^2 p_3^2 - p_2^2 p_4^2)}\Big[ (t - p_2^2 - p_3^2)\; C_0(0)|_{IR} +
                                                             (s - p_3^2 - p_4^2)\; C_0(1)|_{IR} \nn \\
&     & + (t - p_1^2 - p_4^2)\; C_0(2)|_{IR} + (s - p_1^2 - p_2^2)\; C_0(3)|_{IR} \Big],
\end{eqnarray}
which is the desired result. It can be compared with the identity, derived in \cite{Bern:1993kr} 
\begin{eqnarray}
 D_0^n = \frac{1}{2}\left( \sum_{i=0}^3 \alpha_i C_0^n(i) + (3-n)\beta D_0^{n+2}\right),
\end{eqnarray}
where $\alpha_i = \sum_{i=0}^3 (Y^{-1})_{i+1\;j+1}$ and $\beta = \sum_{i=0}^3 \alpha_i$. $Y_{ij}$
is the modified Cayley matrix of the box integral. Since the six-dimensional box integral is finite,
the IR structure of box integrals is completely decided by that of triangle integrals in 4 dimensions.
We have used the identity in Eq.~\ref{eq:IR-D0toC0}, to obtain the IR singular terms in some of the 
complicated box scalar integrals listed in the appendix~\ref{appendix:scalars}. 
The IR finiteness of an individual Feynman diagram can also serve a good check on tensor reduction 
routines. 

\subsection{Rational terms, Gauge invariance, Anomaly $\&$ Decoupling in fermion loop amplitudes}
\label{section:rational}
We know that the rational part $\mathcal{R}$ in Eq.~\ref{eq:1loop}, is a result of 
the UV regularization of tensor integrals. In a UV finite fermion loop amplitude, ${\cal R}$
is independent of masses of internal lines \footnote{To our knowledge, there is no proof
of this statement and it can be taken as a conjecture. From the point-of-view of the validity
of this statement, any overall factor of fermion mass in the amplitude is irrelevant.}. 
We have verified this in many UV finite fermion loop 
amplitudes including those considered in this thesis. It should be 
clear from the general structure of one-loop amplitudes 
that in a gauge invariant amplitude, $\mathcal{R}$ should be 
separately gauge invariant. In the case of one-loop calculation of 
fermion loop diagrams, in dimensional regularization,  the
issue of chiral anomaly is related to the presence of odd number of $\gamma^5$ 
in a linearly (or worse) divergent integrals. There is always an 
ambiguity in carrying out the full calculation in dimensional regularization
as $\gamma^5$ is strictly a 4-dimensional object. However, his ambiguity 
arises only in fixing the rational part of the amplitude. The non-rational finite part
of the amplitude is strictly 4-dimensional. Therefore, any $\mathcal{O}(\eps)$ 
structure of $\gamma^5$, if it at all exists in $n= 4-2\eps$ dimensions, should
contribute to the rational part only. Thus anomalies affect 
only the rational part of the amplitude. They are also independent of the fermion mass. 
It is well known that anomalies may affect the gauge invariance of the amplitude. 
In the presence of anomaly, any violation of the gauge invariance of an amplitude 
should occur only in the rational part of the amplitude -- the non-rational part should 
remain gauge invariant. 

In a fermion loop amplitude containing both the vector ($\gamma^\mu$) and axial-vector 
($\gamma^\mu \gamma^5$) type couplings, 
it is reasonable to expect that the chiral anomaly, being associated with $\gamma^5$,
 should spoil only the axial-vector
current conservation and the conservation of vector current must hold.
However, it is not very clear if this feature can always be ensured by treating $\gamma^5$ 
in 4 dimensions. The 4-dimensional $\gamma^5$ may 
lead to spurious anomalies in the amplitude, resulting in the non-conservation of vector currents.
To ensure the vector current conservation in the amplitude,
it is advisable to use an appropriate prescription for $\gamma^5$ in $n$ dimensions. 
It is important to note that if we do not regulate anomaly by 
using a suitable $n$-dimensional $\gamma^5$ prescription, various other symmetries of the amplitude
may not hold. Even the relation between charge-conjugated fermion loop diagrams does not hold and 
the violation, as we now expect, is only in the rational part. Thus by using an appropriate prescription for
$\gamma^5$ in $n$ dimensions, we generate a rational part consistent with various symmetries of the amplitude.
The treatment of $\gamma^5$ in $n$ dimensions is discussed in the appendix~\ref{appendix:g5-nd}. 

Now we will discuss how one can obtain the correct rational part of a fermion loop amplitude
by doing the full calculation just in 4 dimensions.
The decoupling of heavy fermions is in general expected in any UV finite fermion loop 
amplitude involving couplings which are not proportional to the fermion mass in the loop \cite{Appelquist:1974tg}.
This is a very nice feature of fermion loop amplitudes. It can be utilized to fix the rational 
part of the amplitude without any ambiguity. It is particularly important in calculating 
fermion loop amplitudes suffering from the anomaly and where an $n$-dimensional $\gamma^5$ prescription
is needed to generate the rational part consistent with various symmetries of the amplitude.
Since the amplitudes are UV finite and the decoupling 
of heavy fermions does hold, it implies that there must be a cancellation between the mass 
independent $(n-4)$-dimensional rational part and the 4-dimensional non-rational finite 
part of the amplitude in Eq.~\ref{eq:1loop} as large fermion 
mass limit is taken. Therefore, we do not need to calculate the rational terms explicitly. 
The rational part $\mathcal{R}$ is simply the negative of the non-rational part of the amplitude 
in large fermion mass limit. Though the decoupling does not hold if a Higgs boson is attached to the fermion
loop, the structure of the fermion loop amplitude (with Higgs bosons) suggests that one can
still fix the rational part by taking the large fermion mass limit. For that, one should take away the 
overall (explicit) factor of the fermion mass before taking the limit. Once again the amplitude
goes to zero indicating a cancellation between the rational part and the non-rational part of the amplitude.
In the appendix~\ref{appendix:ggB}, we have given the expressions of $ggH$ and $ggZ$ quark
loop amplitudes treating both the gluons off-shell. One can see there that, indeed, after 
removing any overall factor of quark mass from the amplitude, the rational terms are independent of quark mass
and in the $m^2\to \infty$ limit, the amplitude vanishes due to a cancellation between rational and non-rational 
parts of the amplitude. This knowledge, about the rational terms in UV finite fermion loop amplitudes, also gives an 
opportunity to examine the validity of the reduction of 5-point functions into 4-point functions using
the 4-dimensional identity. One may worry that carrying out these reductions also in $n$ dimensions may
give additional contribution to the rational part of the amplitude. In our SM projects described in 
chapter~\ref{chapter:sm}, we have confirmed by making separate gauge invariance checks on the rational part
and also by checking the decoupling theorem in UV finite amplitudes, that the $(n=4-2\eps)$-dimensional reduction
of 5-point functions does not differ from the 4-dimensional one in $\eps \to 0$ limit. We would like to emphasize that in a UV
finite and gauge invariant fermion loop amplitude, decoupling theorem check is more powerful 
than the gauge invariance check, in ensuring the correctness of the rational part.  

{\bf Remark}: Fermion loop amplitudes involve the trace of a string of $\gamma$-matrices over Dirac indices. 
After the trace evaluation a fermion loop amplitude is, in general, a collection of the scalar and tensor integrals. 
It is interesting to note that in a fermion loop diagram containing at maximum one spin-0 particle leg,
the amplitude can always be cast in such a form that it does not explicitly depend on the fermion mass. 
The fermion mass terms appear explicitly after tensor reduction. They are also present in the scalar integrals.
In all our fermion loop processes including the one involving a Higgs boson, we have observed this feature. 

%%%%%%%%%%%%%%%%%%%%%%%%%%%%%%%%%%%%%%%%%%%%%%%%%%%%%%%%%%%%%%%%%%%%%%%%%%%%%%%%%%%%%%%%%%%%%%%%%%%
%%%%%%%%%%%%%%%%%%%%%%%%%%%%%%%%%%%%%%%%%%%%%%%%%%%%%%%%%%%%%%%%%%%%%%%%%%%%%%%%%%%%%%%%%%%%%%%%%%%

% \newpage
% 
% \bibliographystyle{utcaps}
% \bibliography{thesis}
% 
% \end{document}
 
\setcounter{equation}{0}
\setcounter{figure}{0}
\chapter{Gluon-Gluon Contribution to Di-vector Boson + Jet Production}
\label{chapter:sm}
% \numberwithin{equation}{chapter}
% 
In this chapter, we will study within the SM, the production of a pair of electroweak vector bosons
in association with a hard jet via gluon fusion. In particular, we will consider leading order 
$\gpp$, $\gpZ$, $\gZZ$ and $\gWW$ processes 
at hadron colliders such as the LHC. These processes proceed via quark loop diagrams at the 
leading order itself. In fact, the process $\gVV$ contributes to the hadronic process $\gVVpp$ at the 
next-to-next-leading order (NNLO) in 
$\alpha_s$. Being leading order contributions, these are finite and therefore their contributions towards the 
hadronic cross sections can be calculated separately. We will compute the hadronic cross sections of these gluon 
initiated process  at various collider centre-of-mass energies and compare them with the corresponding tree-level 
hadronic processes. We will also make a comparison with the corresponding $\VV$ processes. Some 
important kinematic distributions common to them, will also be given. We will present a detailed study of $\gpZ$ process 
which is central to all other processes. We will make comments on the issue of numerical instabilities which 
commonly arises in such calculations. This chapter is based on the work reported in 
\cite{Agrawal:2012sq,Agrawal:2012df,Agrawal:2012as}. 

%%%%%%%%%%%%%%%%%%%%%%%%%%%%%%%%%%%%%%%%%%%%%%%%%%%%%%%%%%%%%%%%%%%%%%
% \section{Introduction}

   The search for new physics at the LHC is in progress and the collider is delivering
   data presently at 8 TeV centre-of-mass energy. The discovery of a fundamental
   scalar particle (most probably a Higgs boson) of mass around 125 GeV has received a 
   lot of world-wide attention \cite{cern:2012xx,ATLAS:2012gk,CMS:2012gu}. 
   We expect more good news from the experiments at 
   the LHC before the collider goes for a two year long pause. So far, the 
   SM of particle physics seems to be in excellent agreement with the collected data 
   (more than $10\;{\rm fb}^{-1}$). 
  There have been searches for the hints of physics beyond the SM such as supersymmetry, large extra 
  dimensions, etc. But, as of now, there is no clear evidence~\cite{Jakobs:2012qn,Nahn:2012ey,Adams:2012xx,Rolandi:2012xd}. The process of 
  identifying the discovered fundamental scalar particle as \emph{the Higgs boson} is also continuing. 
  Due to the lack of signals for beyond the SM scenarios, there is a need to look for the SM 
  processes that were not accessible earlier at the Tevatron. Most of such processes have several 
  particles in the final state, and/or occur at the higher order. Calculations of multiparticle processes not
  only provide tests of the SM, they can also contribute to the background to new physics signals. One such 
  class of processes is multi-vector boson production in association with one or more jets.

\begin{figure}[h!]
\begin{center}
 \includegraphics[width=0.5\textwidth]{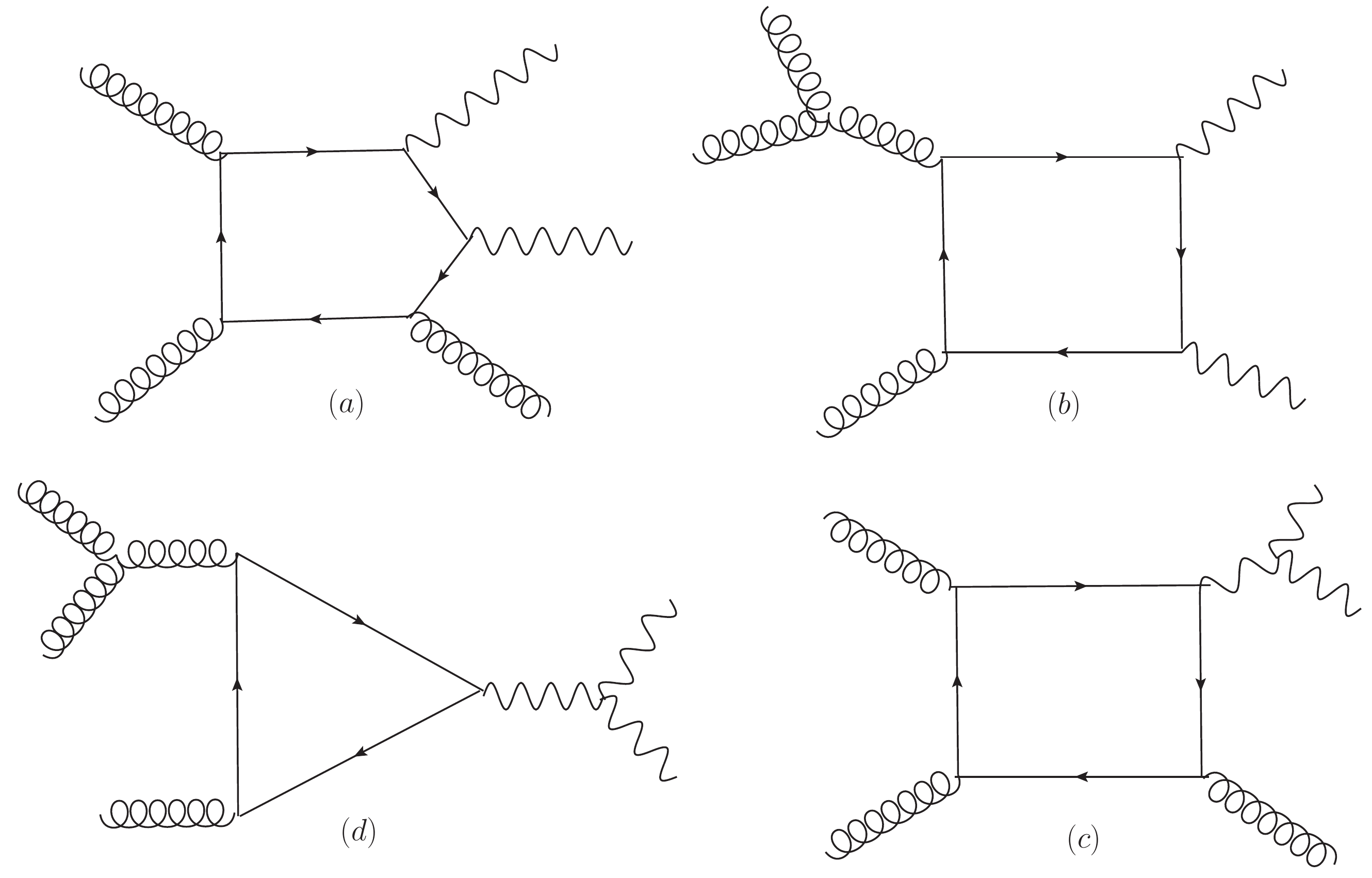}
\caption{The prototype diagrams for the processes $gg \to V V^\prime g$. 
The wavy lines represent the appropriate combination of
the $\gamma, Z {\rm or}\; W$ boson. The last two classes $(c)$ and $(d)$ are relevant to $WWg$
production only. We do not consider diagrams involving Higgs boson for the $ZZg$ and $WWg$ cases.}
\label{fig:VVg}
\end{center}
\end{figure}

    At the LHC centre-of-mass energy, the collider has another useful feature.
   In the proton-proton collisions, the gluon luminosity can be quite significant.
   It can even dominate over the quark luminosity in certain kinematic domains.
   Therefore, at the LHC, loop mediated gluon fusion processes can be important.
   Di-vector boson production via gluon fusion have been studied by many authors 
   \cite{Dicus:1987dj,Dicus:1987fk,vanderBij:1988fb,Glover:1988rg,Glover:1988fe,Campbell:2011bn,Campbell:2011cu}.
   We consider another class of processes $gg \to V V^\prime g$, where $V$ and $V^\prime$
   can be any allowed combination of electroweak vector bosons. These processes can be
   a background to the Higgs boson production as well as new physics scenarios such as the
   \emph{Technicolor}. At the leading order, these processes
   receive contribution from quark loop diagrams. The prototype diagrams are displayed
   in Fig.~\ref{fig:VVg}. The calculations for the process $\gpp$ have already been performed 
   \cite{deFlorian:1999tp,Agrawal:1998ch}. Preliminary results for $\gpZ$ were presented in \cite{Agrawal:2012sq}.
    Recently, {\it Melia et al.} have presented calculations for $\gWW$ \cite{Melia:2012zg}. In the next section,
    we give details on the structure of the amplitudes. In section 3.2, we describe the method of calculation and 
    important numerical checks. Numerical results are presented in section 3.3 and a discussion on the issue 
    of numerical instability in our calculations follows in section 3.4.

%%%%%%%%%%%%%%%%%%%%%%%%%%%%%%%%%%%%%%%%%%%%%%%%%%%%%%%%%%%%%%%%%%%%%%%%%%%%%%%%%%%%%%%%%%%%%%%%%%%%%
\section{The Amplitudes}
   The processes $\gamma \gamma g$, $\gamma Z g$ and $ZZg$ receive contribution from two main classes of quark loop 
   diagrams -- pentagon and box types, as shown in $(a)\;{\rm and}\;(b)$ of Fig.~\ref{fig:VVg}. 
   The box class of diagrams are due to the triple gluon vertices and they can be further divided 
   into three subclasses. This subclassification has its own physical importance. These subclasses are 
   separately gauge invariant with respect to the electroweak vector bosons. Other diagrams can 
   be obtained by a suitable permutation of external legs. For each quark flavor, there are 24 
   pentagon-type and 18 box-type diagrams. However, due to Furry's theorem, only half of the 42
   diagrams are independent. The $\gamma\gamma g$ amplitude is purely vector type. In $\gamma Zg$ and 
   $ZZg$ cases, the pentagon diagrams give both the vector as well as axial-vector 
   contributions, while the box diagrams give only vector contribution. We work with five massless 
   quark flavors and expect decoupling of the top quark. 
%We will check this in the $\gamma\gamma g$ and $\gamma Zg$ amplitudes.

   In the case of $WWg$ process, instead of a single quark flavor, two quark flavors of a single generation 
   contribute to the above discussed pentagon and box diagrams. To keep the matter simple, for this process, 
   we work with the first two generations of massless quarks. It is expected that the contribution from the 
   third generation will not be significant in low $p_T$ region \cite{Campbell:2011cu}. There are additional 
   box and triangle classes of diagrams due to $\gamma/Z\;W W$ vertex for each quark flavor. These are shown 
   in $(c)$ and $(d)$ of Fig.~\ref{fig:VVg}. Due to Furry's theorem, the triangle diagrams with $\gamma W W$ 
   coupling do not contribute and only axial-vector part of the triangle diagrams with $ZWW$ coupling contribute 
   to the amplitude. Since the axial-vector coupling of the $Z$ boson to a quark is proportional to the $T^3_q$ 
   value, the axial-vector contributions from additional triangle and box diagrams, when summed over a massless 
   quark generation, vanish. The vector contribution from the additional box-type diagrams is separately gauge 
   invariant. Because of its color structure, it interferes with the axial-vector part of the pentagon amplitude.
   We have explicitly checked that its contribution towards the cross section is very small; therefore we have 
   dropped this contribution. Thus, effectively we are left with the $ZZg$-like contributions for $WWg$. The 
   Higgs boson interference effects for the cases of $ZZg$ and $WWg$ are ignored in the present calculation.  
   Our one-loop processes, being the leading order processes, are expected to be finite, {\it i.e.}, free from
   ultraviolet (UV) and infrared (IR) divergences.
  The amplitudes of our processes has the following general structure:
\begin{eqnarray}
{\cal M}^{abc}(\gVV) & = & i\; \frac{f^{abc}}{2} {\cal M}_V(V V^\prime g) 
                           + \frac{d^{abc}}{2} {\cal M}_A(V V^\prime g), \label{eq:amp-gVV} \nn \\
{\cal M}_V(V V^\prime g) & = & -\; e^2 g_s^3 \; C_V(V V^\prime g)\;
                                \left( {\cal P}_V - {\cal B}_V \right),  \nn \\
{\cal M}_A(V V^\prime g) & = & -\; e^2 g_s^3 \; C_A(V V^\prime g)\;\left( {\cal P}_A \right).
\end{eqnarray}
This structure is explained in detail in~\cite{Agrawal:2012sq}. Here 
${\cal M}_{V,A}$ are amplitudes for the vector and axial-vector parts of the full amplitude 
under consideration. Because of the Bose symmetry ${\cal M}_V \rightarrow -{\cal M}_V$ under 
the exchange of any two external gluons while ${\cal M}_A$ remains same. ${\cal B}_V\;{\rm and}\; 
{\cal P}_{V,A}$ are the box and pentagon contributions from a single flavor (single generation for 
the $WWg$ case) of quarks. The structure of the amplitude suggests that the vector and axial-vector 
contributions, in the $\gamma Zg, ZZg$ and $WWg$ cases, should be separately gauge invariant. 
Moreover, due to the color structure when 
we square the amplitude, the interference between the vector part and the axial-vector part vanishes, 
{\it i.e.},
\begin{eqnarray}
|{\cal M}(\gVV)|^2 = \left( 6 |{\cal M}_V|^2 + \frac{10}{3} |{\cal M}_A|^2 \right).
\end{eqnarray}
     Therefore the cross section of any of these processes, is an incoherent sum of the vector and axial-vector contributions. 
The couplings $C_{V,A}$ for various cases are listed below and contributions from all the relevant quark flavors
(described above) are included appropriately. 
\begin{eqnarray}\label{eq:couplings-gvv}
C_V(\gamma\gamma g) &=& \frac{11}{9}, \;\; C_A(\gamma\gamma g) = 0, \nn \\
C_V(\gamma Z g) & = & \frac{1}{{\rm sin}\theta_w {\rm cos}\theta_w} \left( \frac{7}{12} - \frac{11}{9} {\rm sin}^2\theta_w \right),
\nonumber \\ 
C_A(\gamma Z g) & = & \frac{1}{{\rm sin}\theta_w {\rm cos}\theta_w} \left( -\frac{7}{12} \right),\nonumber  \\
C_V(Z Z g) & = & \frac{1}{{\rm sin}^2\theta_w {\rm cos}^2\theta_w} \left( \frac{5}{8} - \frac{7}{6} {\rm sin}^2\theta_w +
                 \frac{11}{9} {\rm sin}^4\theta_w \right), \nonumber \\ 
C_A(Z Z g) & = & \frac{1}{{\rm sin}^2\theta_w {\rm cos}^2\theta_w} \left( -\frac{5}{8} + 
                 \frac{7}{6} {\rm sin}^2\theta_w \right), \nonumber \\
C_V(W W g) & = & \frac{1}{{\rm sin}^2\theta_w} \left( \frac{1}{2} \right),\nonumber \\ 
C_A(W W g) & = & \frac{1}{{\rm sin}^2\theta_w} \left( -\frac{1}{2} \right).
\end{eqnarray}

% In addition to the above three processes, we will briefly discuss $\gpp$ and $\ppZ$ processes at the LHC. Like the 
% vector part of the $\gamma Z g$ amplitude, the $\gamma\gamma g$ amplitude receives contributions from
% the pentagon and the box classes of diagrams. On the other hand, the $\gamma\gamma Z$, like the axial-vector
% part of the $\gamma Z g$ amplitude, gets its contribution only from the pentagon class of diagrams.
% The amplitudes of these processes are: 
% \begin{eqnarray}
% {\cal M}^{abc}(\gpp) & = & -i e^2g_s^3 \; \frac{f^{abc}}{2} \; 
% \left[\frac{11}{9} \left( {\cal P}_V - {\cal B}_V \right) \right], \label{eq:gpp}\\
% {\cal M}^{ab}(\ppZ) & =  & \frac{e^3g_s^2}{{\rm sin}{\theta_w} {\rm cos}{\theta_w}}\;\frac{\delta^{ab}}{2} \;
% \left[\frac{5}{36} \left( {\cal P}_A \right) \right]. \label{eq:ppZ}
% \end{eqnarray}
We would also like to see if the top quark loop contributes significantly in $\gamma \gamma g$ and $\gamma Zg$
processes. The top quark coupling in these amplitudes can be found in~\cite{Agrawal:2012as}.   

\section{Calculation and Numerical Checks}
 For each class of diagrams, we write down the prototype amplitudes using the SM Feynman rules \cite{peskin:2005xx}. 
 The amplitude of all other diagrams are generated by appropriately permuting the external momenta and 
 polarizations in our code. The quark loop traces without $\gamma_5$ are calculated in $n$ dimensions, while
 those with $\gamma_5$ are calculated in 4 dimensions using FORM \cite{Vermaseren:2000nd}. We do not need
 any $n$-dimensional prescription for $\gamma_5$ as the pentagon diagrams, which are the only contributions to the 
axial-vector part of the amplitude, are UV finite and free of anomaly. 
 The amplitude contains one-loop tensor integrals. In the case of pentagon-type diagrams, the most complicated 
integral
 is rank-5 tensor integral ($E^{\mu \nu \rho \sigma \delta}$), while for the box-type diagrams, rank-4 tensor 
 integral ($D^{\mu \nu \rho \sigma}$) is the most complicated one. 
Five point tensor and scalar integrals are written in terms of box tensor and scalar integrals using 
4-dimensional Schouten Identity. 
The box tensor integrals are reduced, in $n=4-2\eps$ dimensions, into the standard scalar integrals
 -- $A_0$, $B_0$, $C_0$ and $D_0$
 using FORTRAN routines that follows from the reduction scheme
developed by Oldenborgh and Vermaseren \cite{vanOldenborgh:1989wn}.
 We require box scalar integrals with two massive external legs at the most.
The scalar integrals with massive internal lines are calculated using OneLOop library \cite{vanHameren:2010cp}.
Because of a very large and complicated expression of the amplitude, we calculate the amplitude numerically before 
squaring it. This requires numerical evaluation of the polarization vectors of the gauge bosons.
We choose the real basis, instead of the helicity basis, for the polarization vectors to calculate the amplitude.
This is to reduce the size of the compiled program and the time taken in running the code.
We use RAMBO in our Monte Carlo integration subroutine to generate three particle phase 
space for our processes \cite{Kleiss:1985gy}.
% Ward Identity
\begin{figure}[h!]
\begin{center}
\includegraphics [angle=0,width=0.9\linewidth] {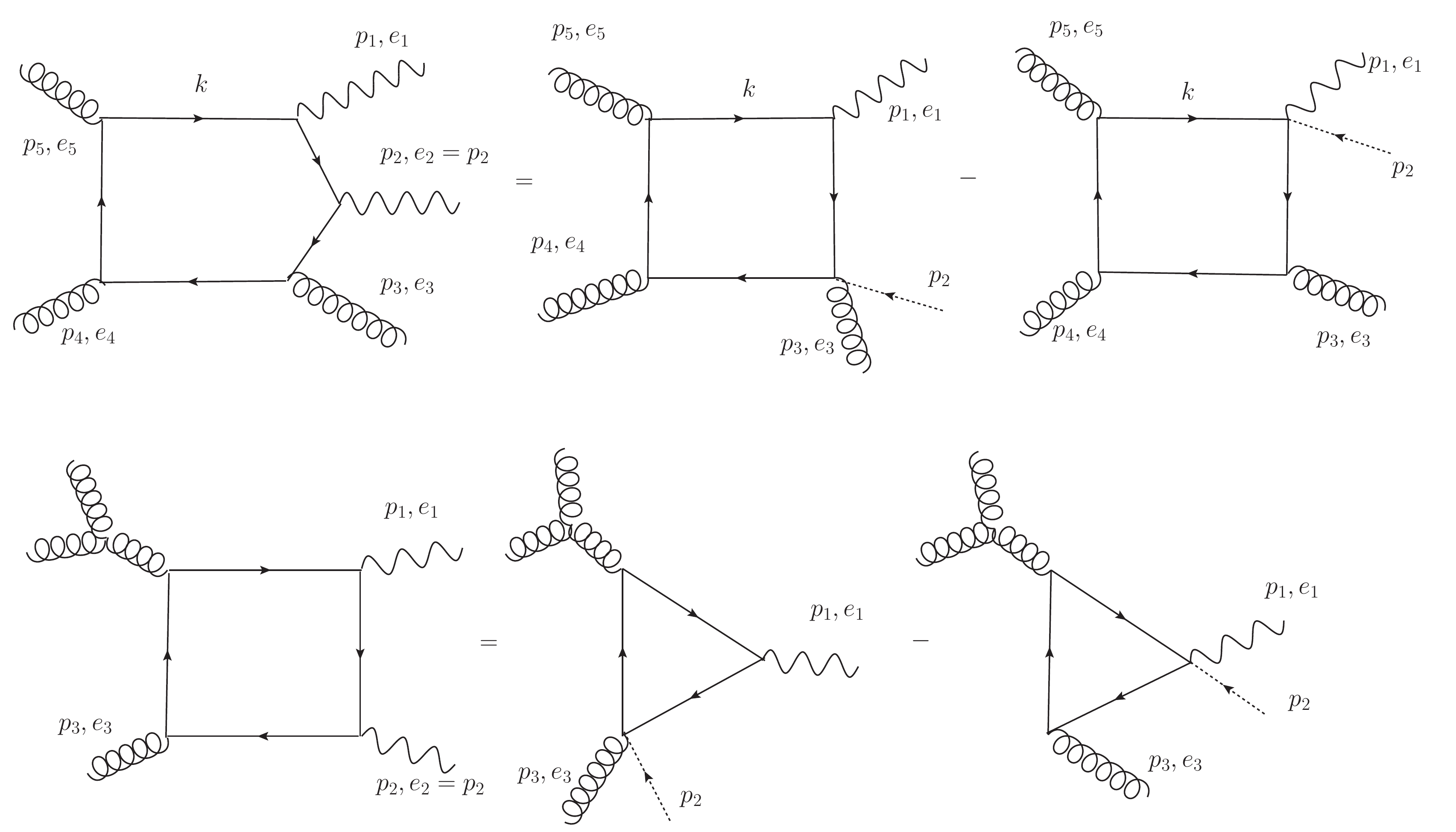}
\caption{Ward identities for $\gpp$ pentagon and box diagrams. All the momenta are taken incoming. The 
 dotted lines take care of momentum insertion at relevant vertices.}
\label{fig:WI}
\end{center}
\end{figure}

  Since the processes $\gVV$ are leading order one-loop processes, the general amplitude in Eq.~\ref{eq:amp-gVV} 
  should be both UV as well as IR finite. However, individual diagrams may be UV and/or IR divergent. The IR 
  divergence is relevant to only light quark cases. All these singularities are encoded in various scalar 
  integrals. To make UV and IR finiteness checks on our amplitude we have derived all the required scalar 
  integrals and they are listed in the appendix~\ref{appendix:scalars}. 
% We regulate the UV divergences of the scalar 
% integrals using the dimensional regularization and the IR singularities by using a small quark mass, 
% the mass regularization. 
Following are the details of various checks made on our amplitude given in 
  Eq.~\ref{eq:amp-gVV}.

\begin{figure}[h!]
\begin{center}
\includegraphics [angle=0,width=0.6\linewidth] {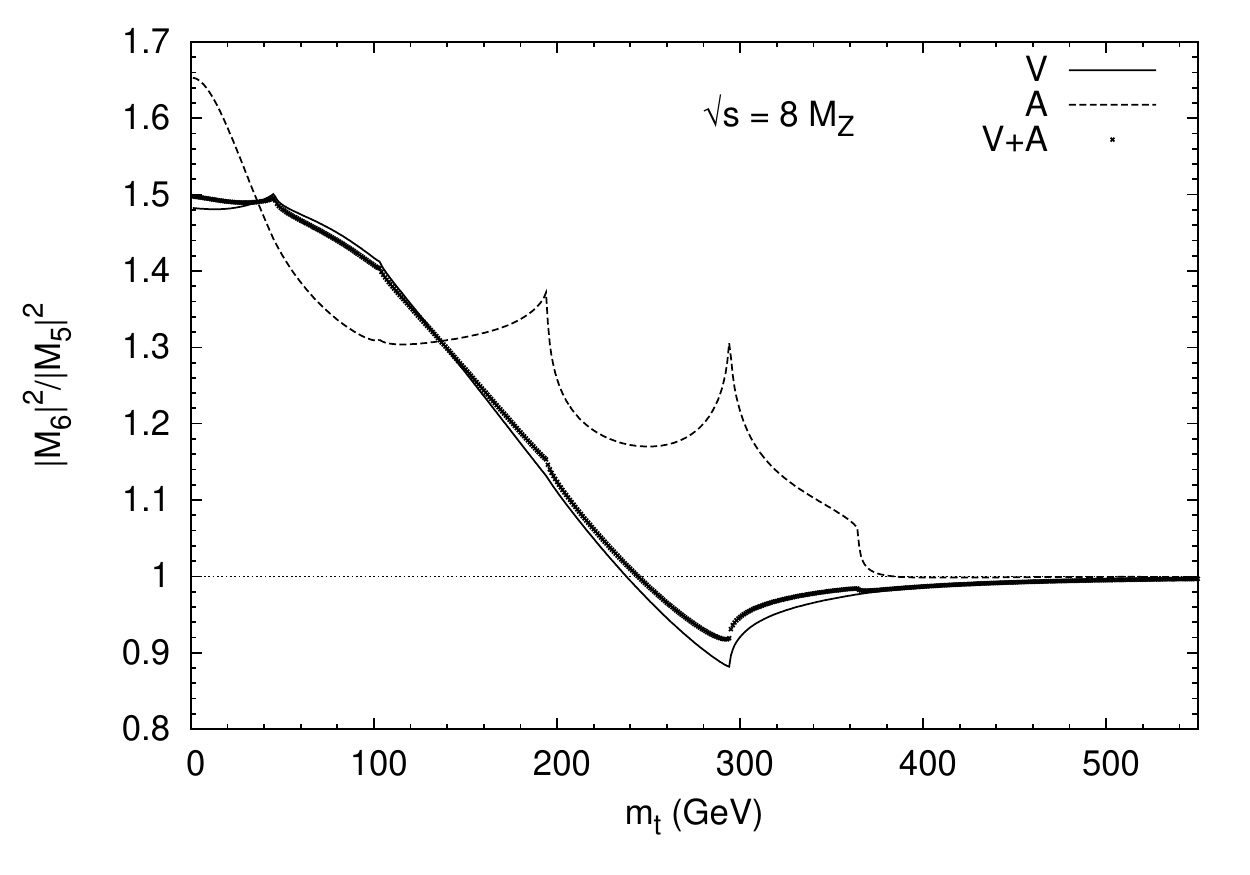}
 \caption{Decoupling of the top quark in $\gpZ$ amplitude. The vector and axial-vector contributions 
are shown separately. }
\label{fig:ampsq_mt}
\end{center}
\end{figure}

\begin{enumerate}
\item {\bf UV Finiteness:}
The tadpole and bubble scalar integrals ($A_0$ and $B_0$) are the only sources of UV singularity 
in any one-loop amplitude. For the case of massless internal lines, $A_0$s do not appear 
in the tensor reduction. For both the massive and massless quark contributions, we have 
verified that the amplitude is UV finite. The amplitude of each pentagon diagram has only 
UV finite tensor integrals. Therefore, each pentagon diagram is UV finite by itself as 
expected from a naive power counting. The box diagrams individually are not UV finite. 
Therefore, the cancellation of the divergence in the sum of the box diagrams is an 
important check.  We find that the three classes of box diagrams are separately UV finite. 
\item {\bf IR Finiteness:}
 The diagrams with massless internal quarks have mass singularities. Even in the case of a quark of small
 mass, like the bottom quark, these diagrams may have large logarithms which should cancel 
 for the finiteness of the amplitude. There can be ${\rm ln}^2(m_q^2)$ and ${\rm ln}(m_q^2)$ 
 types of mass singular terms. We have checked explicitly that such terms are absent from
 the amplitude. Moreover, we have verified that the IR finiteness holds for each fermion loop 
 diagram, confirming the general result proved in Sec.~\ref{section:IRfinite}.
\item {\bf Ward Identities:}
  Certain mathematical identities can be obtained by replacing a polarization vector by its 
  4-momentum in any of the pentagon/box amplitudes. This way a pentagon amplitude can be written 
  as a difference of two (reduced) box amplitudes and also a box amplitude can be written 
  as a difference of two (reduced) triangle amplitudes. For example, if we replace 
  the polarization vector of one of the photons by its 4-momentum in the prototype pentagon 
  and box amplitudes for $\gpp$, we get
\begin{flalign}
{\cal M}_P(p_1,p_2,p_3,p_4;e_1,e_2=p_2,e_3,e_4,e_5) =& \;  
{\cal M}_B(p_1,p_2+p_3,p_4;e_1,e_3,e_4,e_5) \nn \\
&\; - {\cal M}_B(p_1+p_2,p_3,p_4;e_1,e_3,e_4,e_5), \label{eq:WI-penta} \\ 
{\cal M}_B(p_1,p_2,p_3;e_1,e_2=p_2,e_3,e_{45}) =& \;{\cal M}_T(p_1,p_2+p_3;e_1,e_3,e_{45}) \nn \\ 
&\; - {\cal M}_T(p_1+p_2,p_3;e_1,e_3,e_{45}). \label{eq:WI-box}
\end{flalign}
  Here $e_i$s are polarization vectors of the gauge bosons and $e_{45}$ denotes the triple gluon vertex 
  attached to the box/triangle diagram. These identities are shown diagrammatically in Fig.~\ref{fig:WI}.
  We have verified them numerically. These Ward identities are important checks on individual 
  diagrams and a set of these identities can also be utilized for a systematic study of numerical instabilities
  in the tensor reduction, near exceptional phase space points. An exceptional phase space point corresponds 
  to the vanishing of a partial sum of the external momenta.
\item {\bf Gauge Invariance:}
As we have seen, because of the color structure, the vector and axial-vector parts of the amplitude
do not interfere and they are separately gauge invariant. The vector part of the amplitude has gauge 
invariance with respect to the three gluons and the electroweak bosons. This has been checked by replacing 
the polarization vector of any of these gauge particles by its momentum ($\veps^\mu(p_i)\to p^\mu_i$) 
which makes the amplitude vanish. As one would expect the pentagon and the three classes of box contributions 
are separately gauge invariant with respect to the electroweak bosons. For each gluon, one of the three 
classes of box amplitudes is separately gauge invariant and further cancellation takes place among the
pentagon and the other two box contributions. The axial-vector part of the amplitude is separately gauge 
invariant with respect to all the three gluons and the photon. We verify that due to explicit breaking 
of the chiral symmetry in the presence of a quark mass, the axial-vector part of the amplitude, in 
$\gamma Zg, ZZg$ and $WWg$ cases, vanishes on 
replacing the $Z/W$ boson polarization by its 4-momentum only in the $m_q \rightarrow 0$ limit.
\item{\bf Decoupling of heavy quarks: }
As a consistency check, we have also verified that the vector and axial-vector parts of the amplitude vanish
in the large quark mass limit \cite{Appelquist:1974tg}. This feature of the amplitude is very closely related
to its UV structure. The decoupling theorem holds for each pentagon amplitude and also for each class of box 
amplitudes. For the process $\gpZ$, in Fig.~\ref{fig:ampsq_mt}, we have plotted the ratio of the squared-amplitudes
 for five and six quark flavor contributions as a function of the top quark mass. The phase space point 
corresponds to a fixed partonic centre-of-mass energy, $\sqrt{s} = 8 M_Z$. The vector and axial-vector 
contributions are plotted separately. The  $m_t = 4 M_Z$ corresponds to the scale above which the top 
quarks in the loop cannot go on-shell. Various slope changes shown in this plot correspond to the possibilities 
of producing one or more final state particles via on-shell $t\bar t$ annihilation. 
\end{enumerate}
\section{Numerical Results}
  Based on the procedure outlined above, we can now compute the hadronic cross sections and examine various 
  features of our processes. As we have already mentioned, we compute amplitudes numerically using the 
  real polarization vectors of the gauge bosons. There are 32 polarized amplitudes for the case of 
  $\gamma \gamma g$, 48 for the case of $\gamma Zg$ and 72 for the cases of 
  $ZZg$ and $WWg$. Given the number of diagrams, the number of polarization combination and the 
  length of the amplitude, the computation becomes very time consuming. Each phase space point 
  evaluation takes about 1.3 seconds on a single machine that we use. We use ifort compiler on Intel 
  Xeon CPU 3.20GHz machines. We, therefore, run the code in a parallel environment using AMCI package, 
  a PVM implementation of the VEGAS algorithm~\cite{Veseli:1997hr,PVM:1994xx}. We have used more than 
  30 cores to run the code in the parallel environment. Still it takes more than 12 hours to get 
  suitable cross section which includes both the massive and massless quark contributions. 
  We will now divide our numerical results into two parts. In the first part, we will do a comparative 
  study of the $\gamma Z g$, $ZZg$ and $WWg$ processes. Also we will update the results for $\gamma\gamma g$ 
  at 8 TeV LHC. The second part will deal with the numerical results for the $\gamma Zg$ process in greater 
  detail.
\begin{figure}[ht]
 \begin{minipage}[b]{0.5\linewidth}
\centering
 \includegraphics[width=\textwidth]{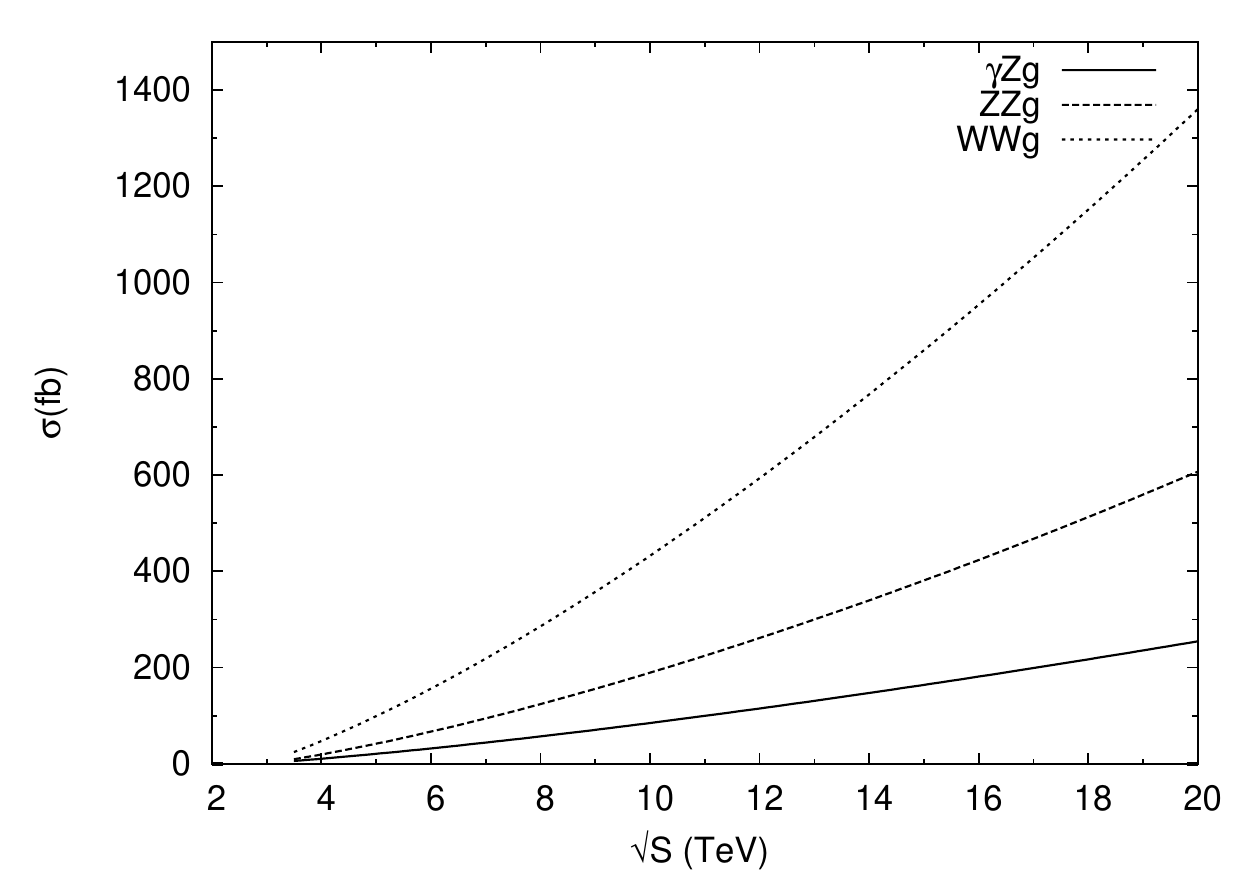}
\caption{Variation of the cross sections with the collider centre-of-mass energy for $\gVV$.}
\label{fig:sigma_cmeVVg}
 \end{minipage}
 \hspace{0.5cm}
 \begin{minipage}[b]{0.5\linewidth}
\centering
\includegraphics[width=\textwidth]{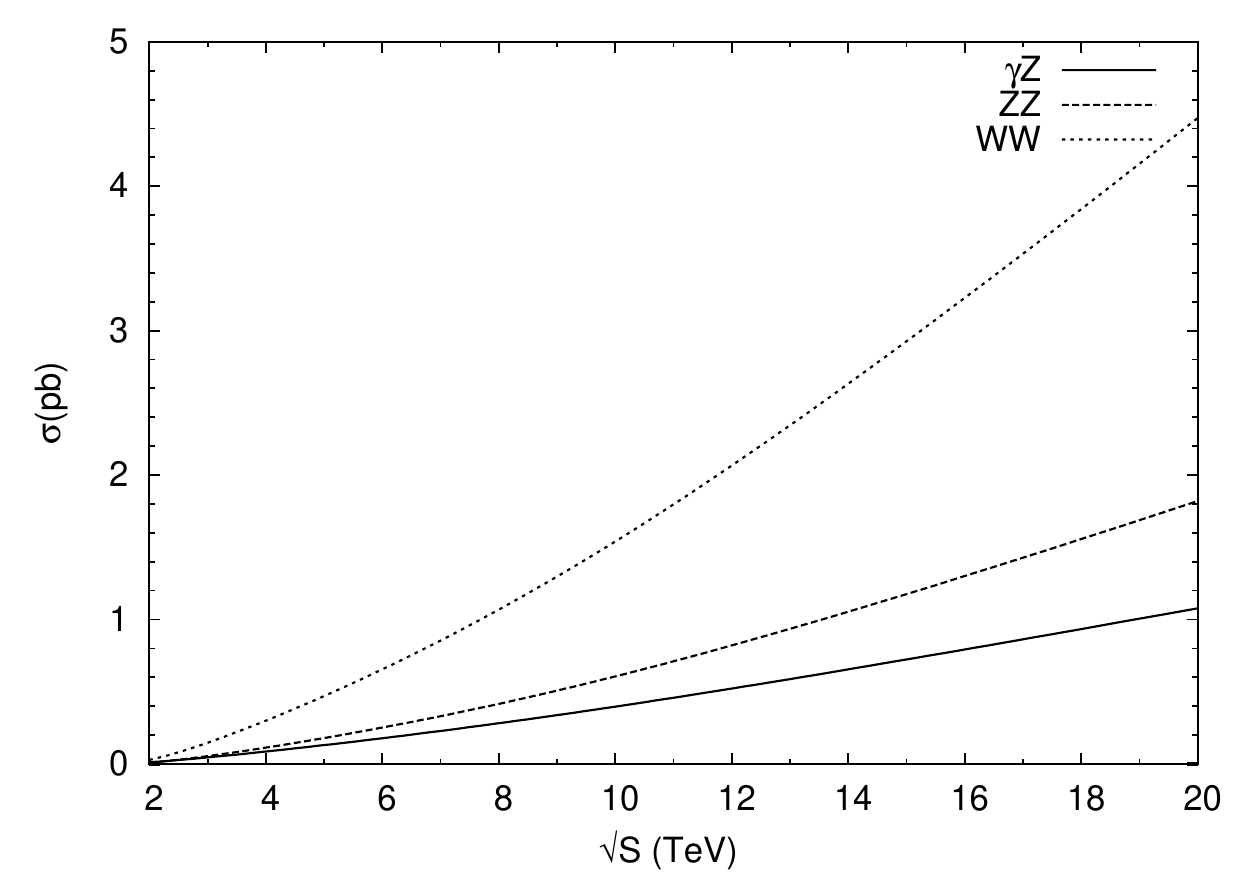}
\caption{Variation of the cross sections with the collider centre-of-mass energy for $\VV$.}
\label{fig:sigma_cmeVV}
\end{minipage}
\end{figure}
%%%%%%%%%%%%%%%%%%%%%%%%%%%%%%%%%%%%%%%%%%%%%%%%%%%%%%%%%%%%%%%%%%%%%%%%%%%%%%%%%%%%%%%%%%%%%%%%%%%%%
\subsection{Numerical results for $VV'g$}
Since the processes $\gamma Z g$, $ZZg$ and $WWg$ have both the vector and the axial-vector 
contributions, we can put them in one class and study their relative behavior. Their results 
do not include the top quark-loop contribution.
The comparative study, presented in this section, uses following kinematic cuts: 
\begin{eqnarray}
p_T^{\gamma,Z,W,j} > 30 \; {\rm GeV },\: |\eta^{\gamma,Z,W,j}| < 2.5,\; ,R(\gamma,\gamma/j) > 0.6,
\end{eqnarray}
where $R(\gamma,j)=\sqrt{\Delta\eta_{\gamma j}+\Delta\phi_{\gamma j}}$, is the isolation cut between 
the photon and the jet. 
In addition to this, we have chosen the factorization and the renormalization scales as
$\mu_f = \mu_r = p_T^{\gamma/Z/W}$, as appropriate. In this section, we have used CTEQ6M parton distribution 
functions to obtain the results~\cite{Nadolsky:2008zw}.        
In Fig.~\ref{fig:sigma_cmeVVg}, we present the results of the cross section calculation
for the three processes. We note that at typical LHC energy, the cross sections are of the order of 
100 fb. For example, at the centre-of-mass energy of the 8 TeV, the cross sections are
46.7 fb, 95.5 fb, and 225.2 fb, respectively for the  $\gamma Z g, Z Z g \; {\rm and}\; W W g$
production processes. Therefore, one may expect a few thousand of such events at the end of the present run.
But a $W/Z$ boson is observed through its decay channels. If we consider the case when all the
$W/Z$ bosons are seen through their decays to the electron/muon only, then, with $20 \; {\rm fb}^{-1}$
integrated luminosity, the number of events for these processes will be approximately 60, 10, and
220 respectively. However, if we allow one of the $W/Z$ boson to decay hadronically, then the
number of events would increase significantly. At the 14 TeV centre-of-mass energy, the numbers
will be about a factor of three larger.
 The relative behavior of cross sections of the three processes,
as the centre-of-mass energy varies, is quite similar to the case of di-vector boson production 
via gluon fusion as shown in Fig.~\ref{fig:sigma_cmeVV}. This common behavior is mainly due to 
the relative couplings of the processes listed above in Eq.~\ref{eq:couplings-gvv}. We find that at 
14 TeV the cross sections
for our processes are $20-30 \%$ of those for the corresponding di-vector boson production 
(without jet) processes. We can also compare the contributions of these loop processes
with those of the corresponding tree-level processes. We find that the processes $g g \to \gamma Z g,
WWg$ make about $4-5 \%$ contribution to the processes $ p p \to \gamma Z j,
WWj$, while $g g \to ZZ g$ makes a contribution of about $10-15 \%$ to the
$p p \to ZZ j$ process. Here `$j$' stands for a jet. This is quite similar to the case 
of the di-vector boson production. Tree-level estimates are obtained using MadGraph~\cite{Alwall:2011uj}.
% M12 and ptg distributions 
\begin{figure}[ht]
 \begin{minipage}[b]{0.5\linewidth}
\centering
 \includegraphics[width=\textwidth]{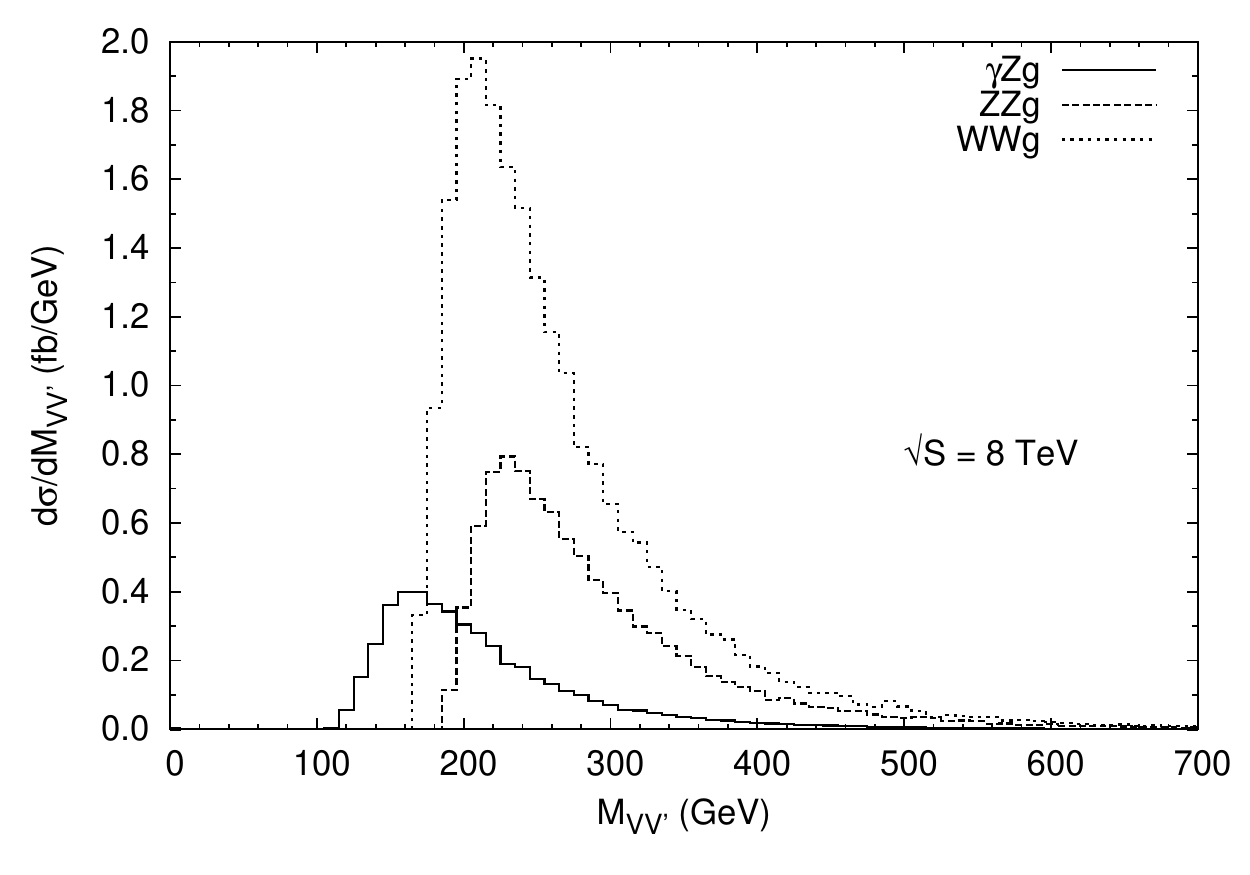}
\caption{Invariant mass distributions of the pair of vector boson at 8 TeV, for $\gVV$.}
\label{fig:m12-8}
 \end{minipage}
 \hspace{0.5cm}
 \begin{minipage}[b]{0.5\linewidth}
\centering
 \includegraphics[width=\textwidth]{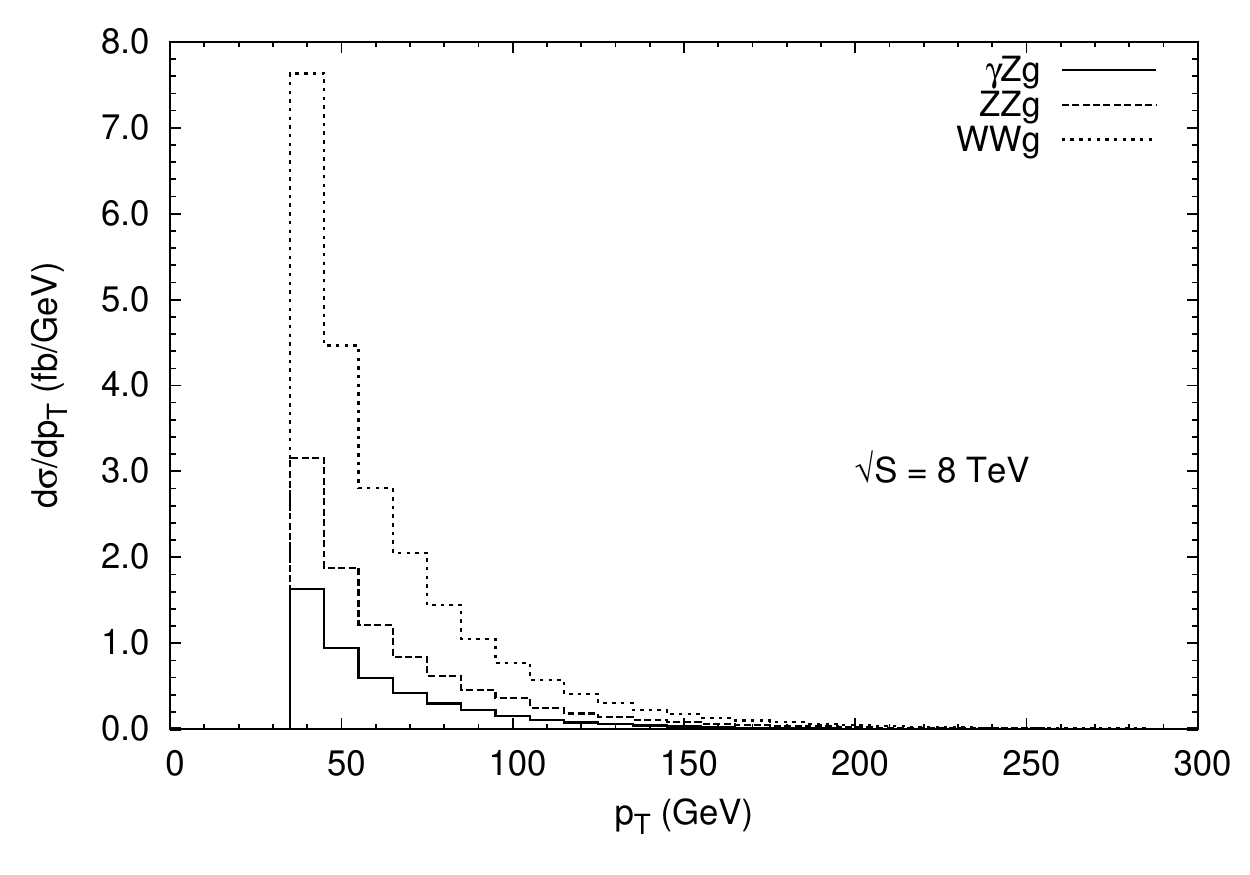}
\caption{Transverse momentum distributions of the gluon jet at 8 TeV, for $\gVV$.}
\label{fig:ptg-8}
 \end{minipage}
\end{figure}

We have also compared our results for the $WW j$ production case
with those of {\it Melia et al.}~\cite{Melia:2012zg}. Though they have considered the leptonic 
decays of the $W$ bosons and the kinematic cuts, choice of scales and parton distributions etc. are 
quite different, the percentage contribution of gluon-gluon channel as compared to the LO cross section 
is same within the allowed range of uncertainty, {\it i.e.}, $4-5 \%$. The values of the cross section 
are more strongly dependent on the values of parameters and kinematic cuts; still two results are similar
if we take into account quoted uncertainties and branching ratios. The contribution of these gluon fusion 
processes can be even larger in appropriate kinematic regime.

We now discuss few kinematic distributions at 8 TeV collider centre-of-mass energy. 
These distributions remain same, characteristically, at 14 TeV centre-of-mass energy.
The invariant mass distributions for the pair of vector boson are shown in Fig.~\ref{fig:m12-8}.
The positions of the peaks are related to the masses of the vector bosons in each case. Of course, 
the $WW$ invariant mass cannot be measured experimentally. 
In Fig.~\ref{fig:ptg-8}, the transverse momentum distributions for gluon jet is given for the three processes.
The major contribution to the cross section comes from low $p_T$ region, as we expect. The cross section 
is very sensitive to the $p_T$ cut on gluon as it is radiated from one of the incoming gluons in the 
box diagrams.
\begin{figure}[ht]
\begin{minipage}[b]{0.5\linewidth}
\centering
\includegraphics[width=\textwidth]{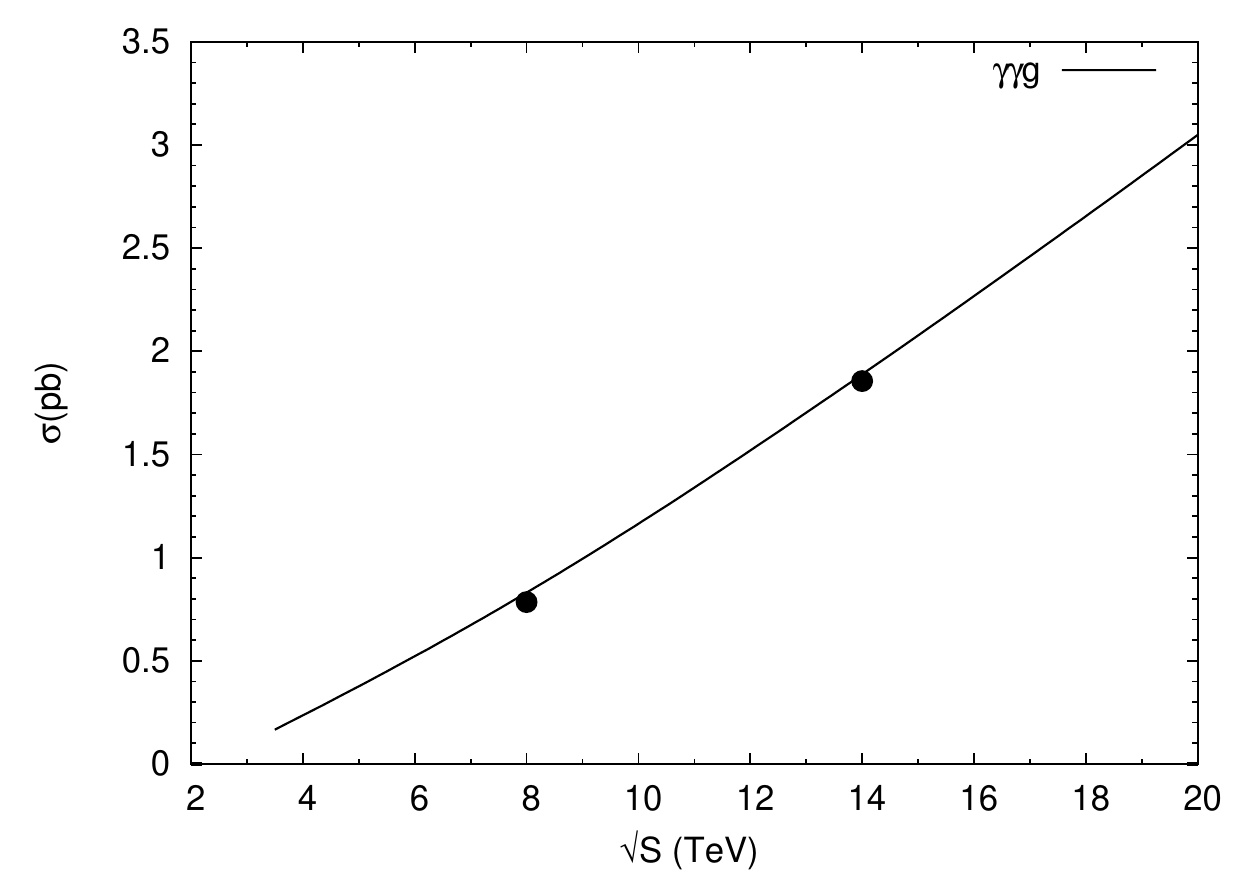}
 \caption{The collider energy dependence of the cross section for $\gpp$. }
\label{fig:sigma_cme-ppg}
\end{minipage}
\hspace{0.5cm}
\begin{minipage}[b]{0.5\linewidth}
\centering
\includegraphics[width=\textwidth]{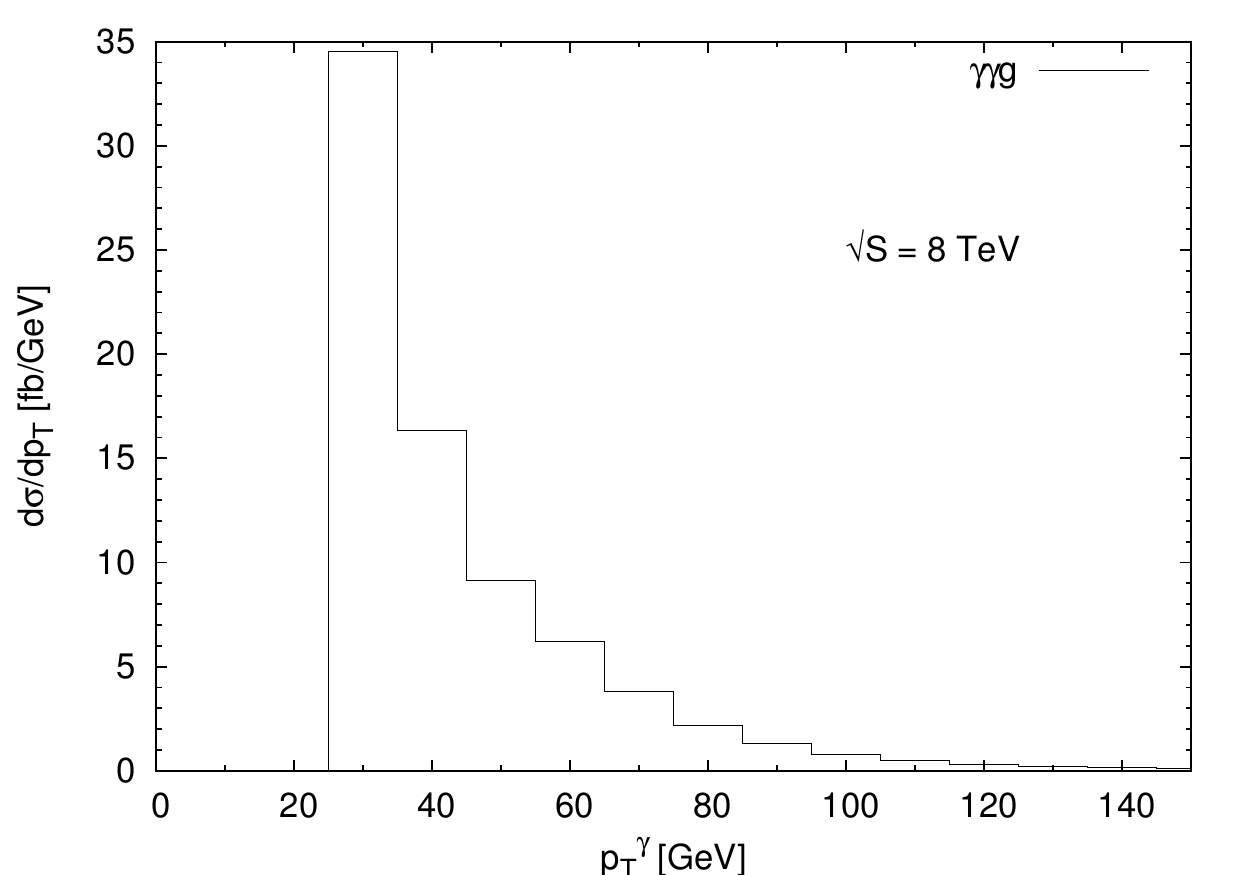}
 \caption{Transverse momentum distributions of the photon in $\gpp$. }
\label{fig:sigma8_ptp-ppg}
\end{minipage}
\end{figure}
\begin{figure}[ht]
\begin{minipage}[b]{0.5\linewidth}
\centering
\includegraphics[width=\textwidth]{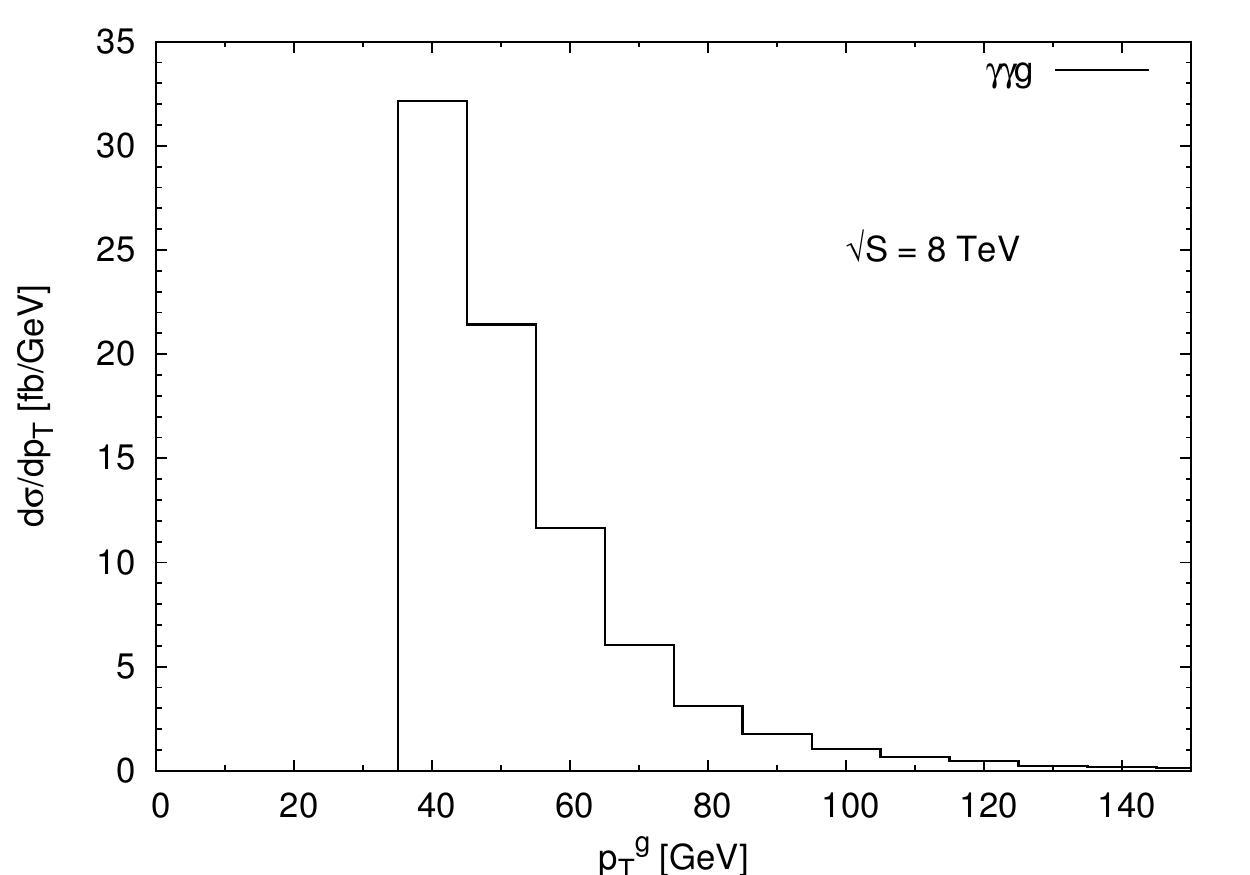}
 \caption{Transverse momentum distribution of the gluon jet in $\gpp$. }
\label{fig:sigma8_ptg-ppg}
\end{minipage}
\hspace{0.5cm}
\begin{minipage}[b]{0.5\linewidth}
\centering
\includegraphics[width=\textwidth]{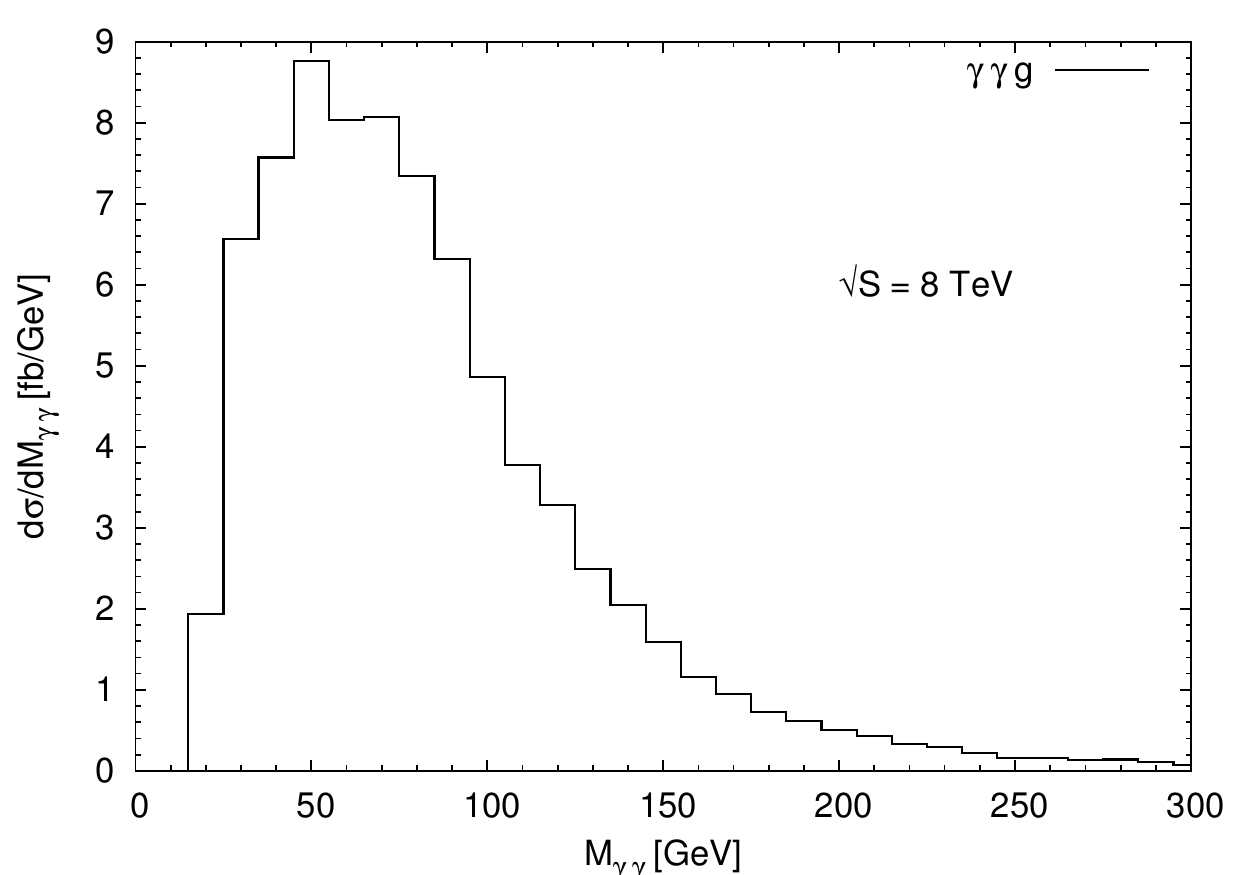}
 \caption{Invariant mass distribution of the two photons in $\gpp$. }
\label{fig:sigma8_m12-ppg}
\end{minipage}
\end{figure}

The variation of the hadronic cross section with the collider centre-of-mass energy, for $\gamma \gamma g$ 
production via gluon fusion, is given in Fig.~\ref{fig:sigma_cme-ppg}. The result is with the above kinematic 
cuts and choices except that, $p_T^\gamma > 20$ GeV. We reconfirm the importance of $\gpp$ process at 
the LHC. We have checked that the top quark contribution to the cross section is very small. Therefore
the results include light quark loop contributions only. The cross 
section at the 8 TeV (14 TeV) centre-of-mass energy is about 0.78 pb (1.86 pb), 
leading to several hundred (thousand) events with even 
1 fb$^{-1}$ of integrated luminosity. Few important kinematic distributions related to this process are 
given at 8 TeV LHC in Figs.~\ref{fig:sigma8_ptp-ppg},\ref{fig:sigma8_ptg-ppg} and \ref{fig:sigma8_m12-ppg}. 

\subsection{Numerical results for $\gamma Z g$}
\begin{figure}[h!]
\begin{center}
\includegraphics [angle=0,width=0.6\linewidth] {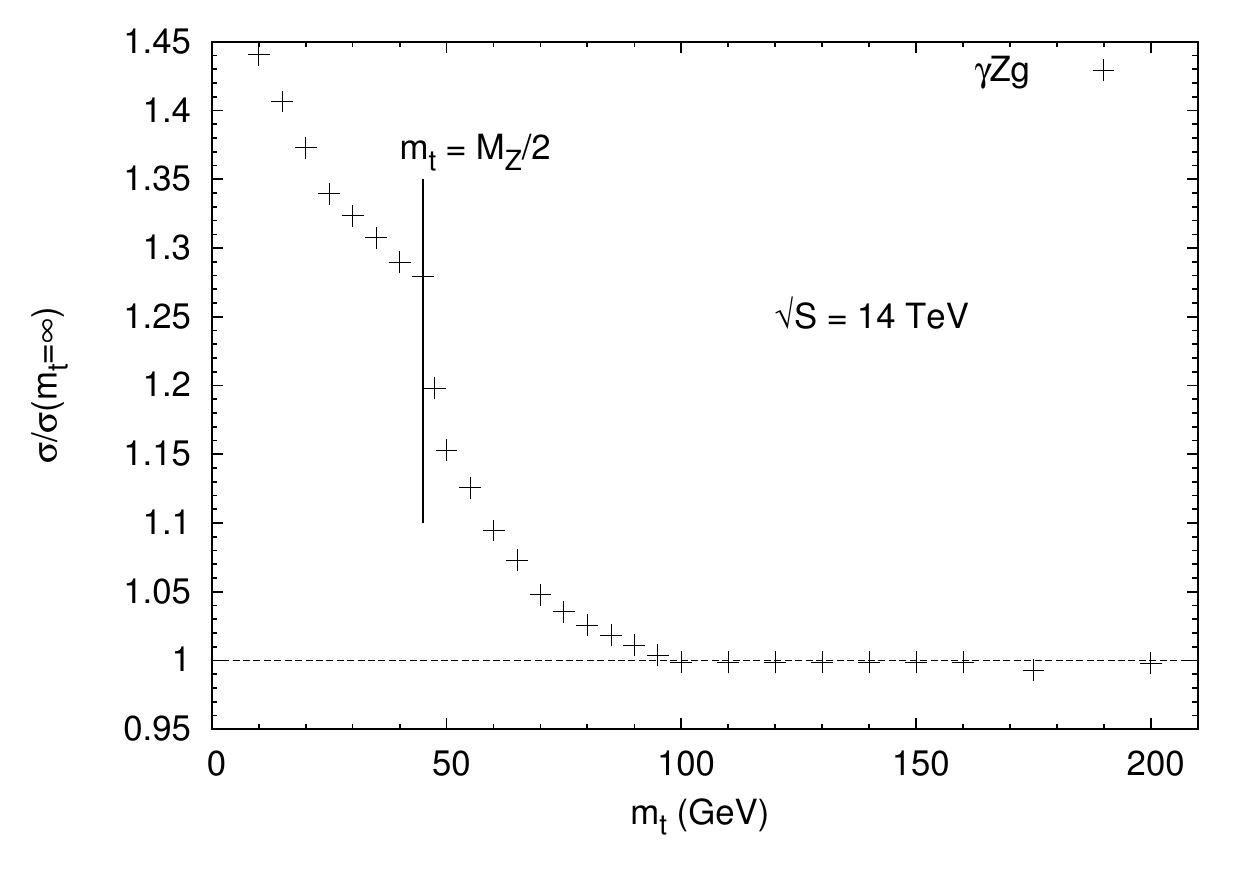}
\caption{Decoupling of the top quark in the cross section calculation of $\gpZ$.}
\label{fig:sigma_mt}
\end{center}
\end{figure}

Now we will have a detailed discussion on the results for the $\gpZ$ process.
We first study the importance of the diagrams with the top quark in the loop. In Fig.~\ref{fig:sigma_mt}, 
we see that the contribution of the top quark ($m_t = 175 $ GeV) to the hadronic cross section is negligible. 
We also see a knee in the plot at $m_t = {M_Z \over 2}$. This corresponds to the $Z$ boson production via 
$t{\bar t}$ annihilation. The top quark decouples at around 100 GeV. We have, therefore, not included its 
contribution in our results presented below. The run time of our code is also reduced by $50 \%$. It is not 
surprising that the top quark decouples at such a low value for our processes. This is simply because there 
are four/five quark propagators in each box/pentagon diagram, leading to a large power of the top quark mass 
in the denominator.

\begin{figure}[ht]
\begin{minipage}[b]{0.5\linewidth}
\centering
\includegraphics[width=\textwidth]{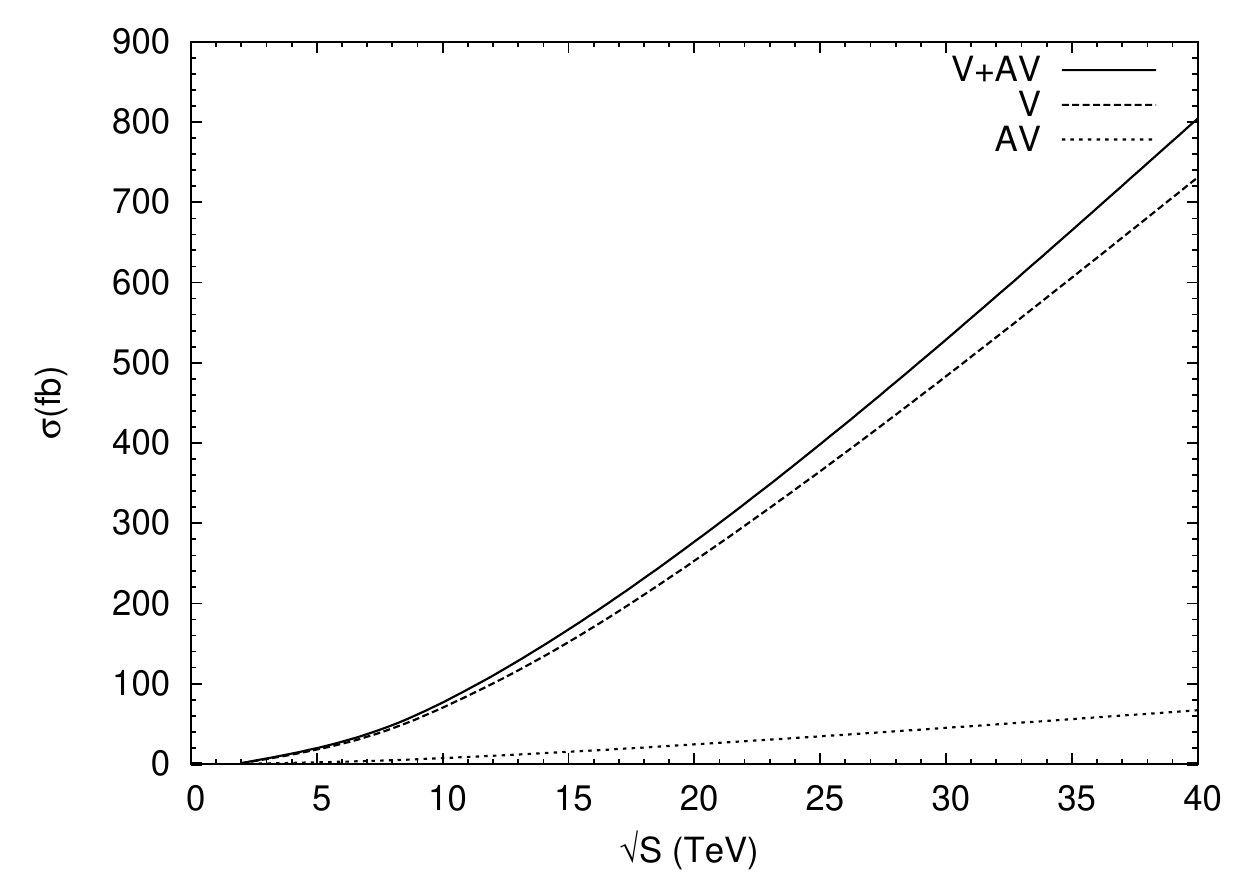}
 \caption{The vector and axial-vector contributions of the hadronic cross section for $\gpZ$.}
\label{fig:sigma_cme-V+A}
\end{minipage}
\hspace{0.5cm}
\begin{minipage}[b]{0.5\linewidth}
\centering
\includegraphics[width=\textwidth]{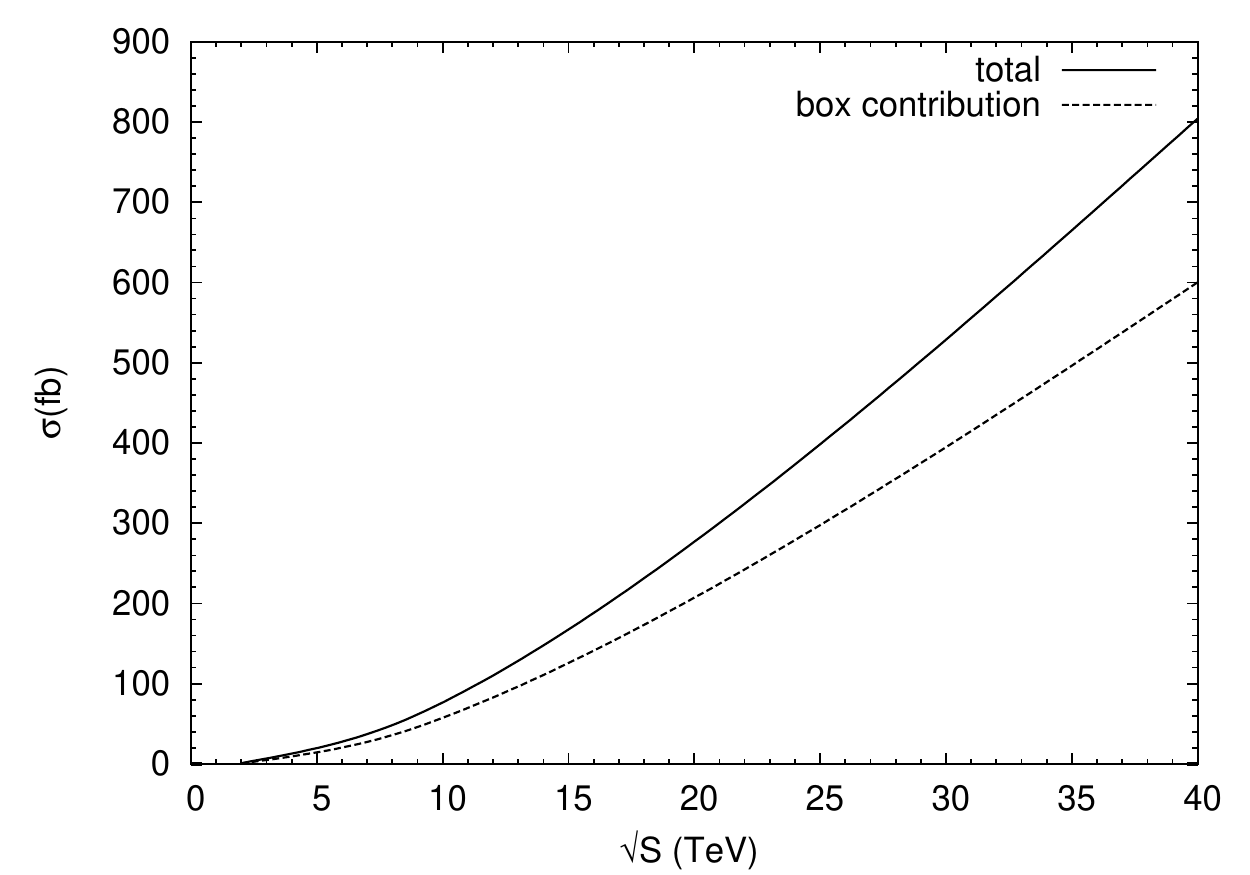}
 \caption{Contribution of the box-amplitude towards the hadronic cross section for $\gpZ$.}
\label{fig:sigma_cme-box}
\end{minipage}
\end{figure}

\begin{figure}[ht]
\begin{minipage}[b]{0.5\linewidth}
\centering
\includegraphics[width=\textwidth]{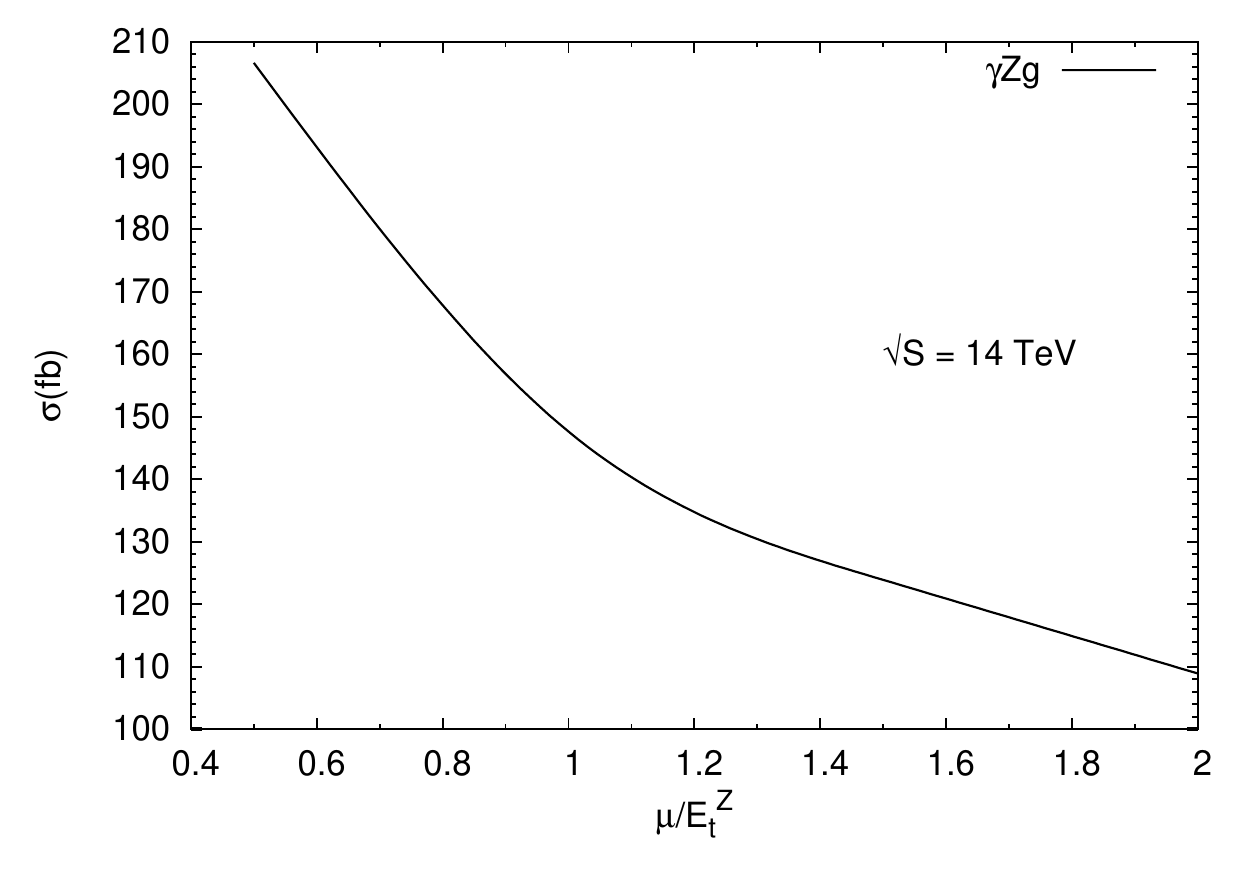}
 \caption{Variation of the cross section for $\gpZ$ with the scale, $\mu = \mu_r = \mu_f$ at 14 TeV.}
\label{fig:sigma14Q}
\end{minipage}
\hspace{0.5cm}
\begin{minipage}[b]{0.5\linewidth}
\centering
\includegraphics[width=\textwidth]{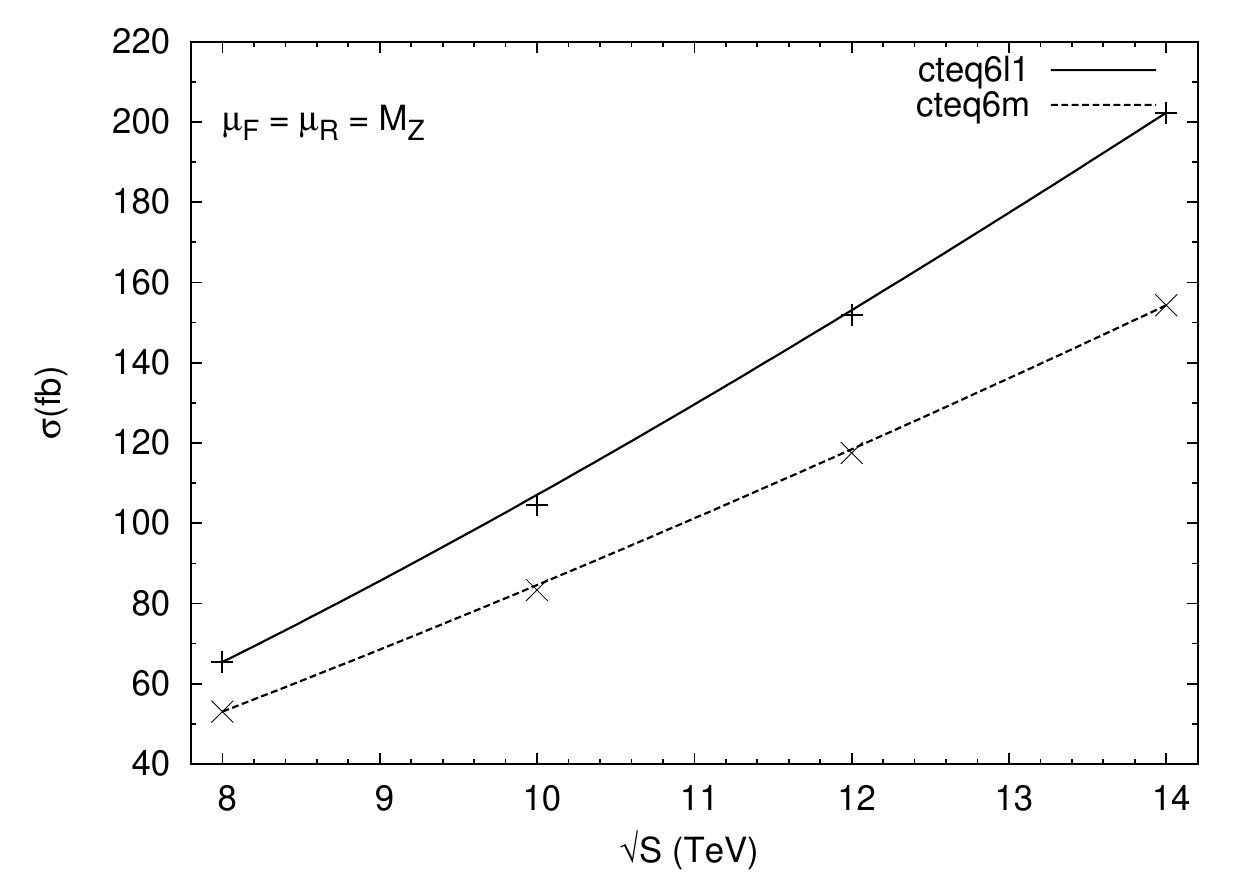}
 \caption{Dependence of the cross section on the collider energy, for $gg\to \gamma Z(\to l^+l^-)g$.}
\label{fig:sigma_cme-gpz}
\end{minipage}
\end{figure}

We can divide our numerical results that are presented in this section into two categories.
 We first discuss theoretical results, related to the structure of the amplitude,
keeping the $Z$ boson on-shell. 
Theoretical results presented below include following kinematic cuts:
\begin{eqnarray}
p_T^{j} > 30 \; {\rm GeV},\; p_T^{\gamma,Z} > 20 \; {\rm GeV},\:|\eta^{\gamma,Z,j}| < 2.5,\; R(\gamma,j) > 0.6 .
\end{eqnarray}
We have chosen the factorization and the renormalization scales as
$\mu = \mu_f = \mu_r = E_T^Z (=\sqrt{M_Z^2+(p_T^Z)^2})$, the transverse energy of the $Z$ boson. Results are obtained using CTEQ6M 
PDFs~\cite{Nadolsky:2008zw}. In Fig.~\ref{fig:sigma_cme-V+A},
we give the dependence of the cross section on the collider centre-of-mass energy  to see
the effect of a large gluon luminosity at higher energies. We have already seen 
that the vector and axial-vector parts of the amplitude are separately gauge invariant.
Their contributions towards the cross section are also included in the figure.
The axial-vector contribution is only about $10 \%$ of the total cross section;
this contribution comes from the pentagon class of diagrams. Although the
box-contribution to the cross section is not separately gauge invariant with respect to the gluons, it 
is gauge invariant with respect to the $\gamma$ and the $Z$ boson. We find that more than
$70 \%$ of the total cross section is due to the box-amplitude only, see Fig.~\ref{fig:sigma_cme-box}.
The scale variation of the cross section about the central value $\mu_0 = E_t^Z$, is shown
in Fig.~\ref{fig:sigma14Q}. On increasing the scale by a factor of 2, the cross section 
decreases by about $25 \%$; it increases by about $40 \%$ on decreasing the scale by a factor of 2. 
These large variations are expected because our calculation is effectively 
LO as far as the $\mu$ dependence is concerned. 
We see that the cross section falls as we increase the scale $\mu$.
This is because an increase in the factorization scale increases the cross section due to the increase
in the gluon luminosity; but an increase in the renormalization scale decreases the cross section
because of the decrease in  the value of $\alpha_s(\mu)$. When we increase both the scales 
at the same time, the effect of the change in the renormalization scale is stronger. It leads to an
overall decrease in the cross section with the increase in the scale $\mu$.

\begin{figure}[ht]
\begin{minipage}[b]{0.5\linewidth}
\centering
\includegraphics[width=\textwidth]{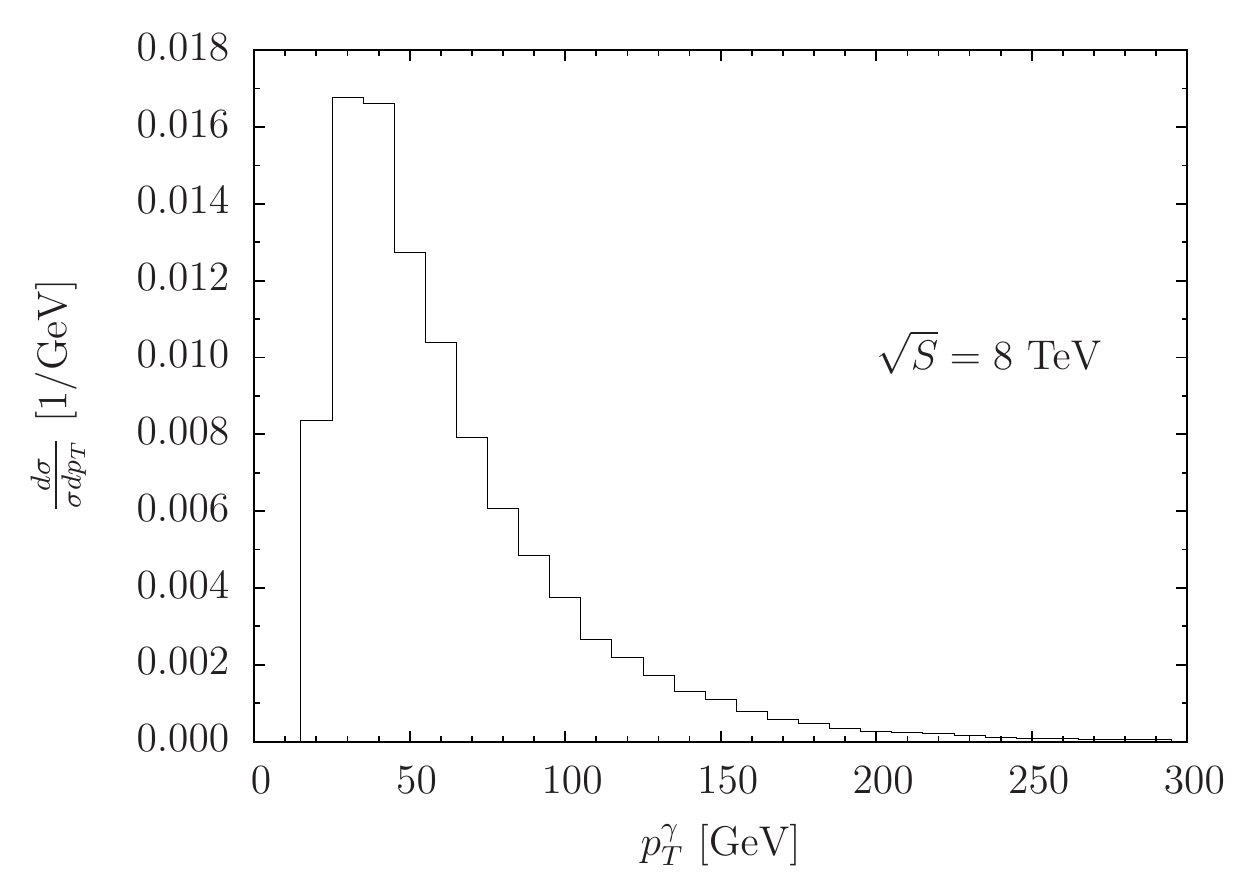}
 \caption{Transverse momentum distribution of the photon at 8 TeV centre-of-mass energy in $gg\to \gamma Z(\to l^+l^-)g$. }
\label{fig:ptp8}
\end{minipage}
\hspace{0.5cm}
\begin{minipage}[b]{0.5\linewidth}
\centering
\includegraphics[width=\textwidth]{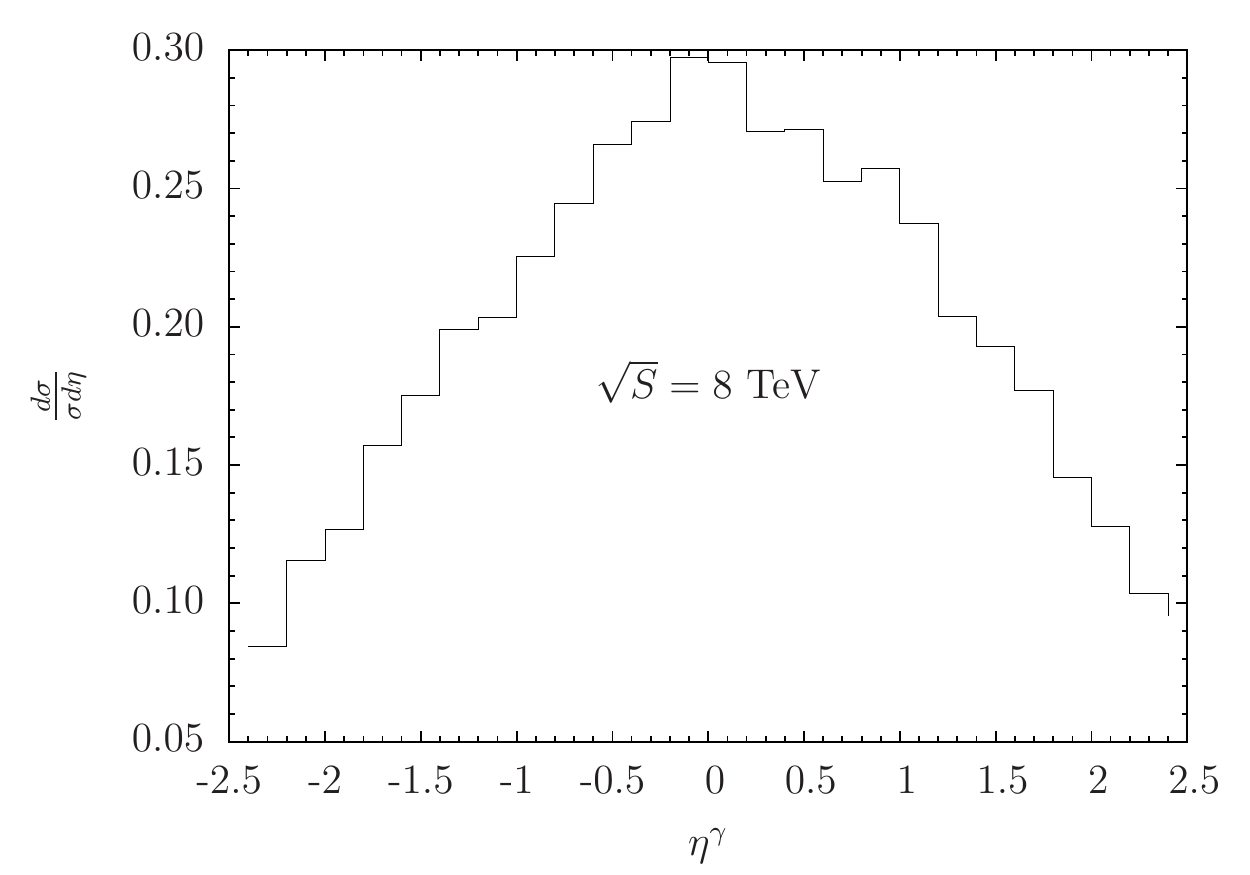}
 \caption{Rapidity distribution of the photon at 8 TeV centre-of-mass energy in $gg\to \gamma Z(\to l^+l^-)g$. }
\label{fig:etap8}
\end{minipage}
\end{figure}

\begin{figure}[ht]
\begin{minipage}[b]{0.5\linewidth}
\centering
\includegraphics[width=\textwidth]{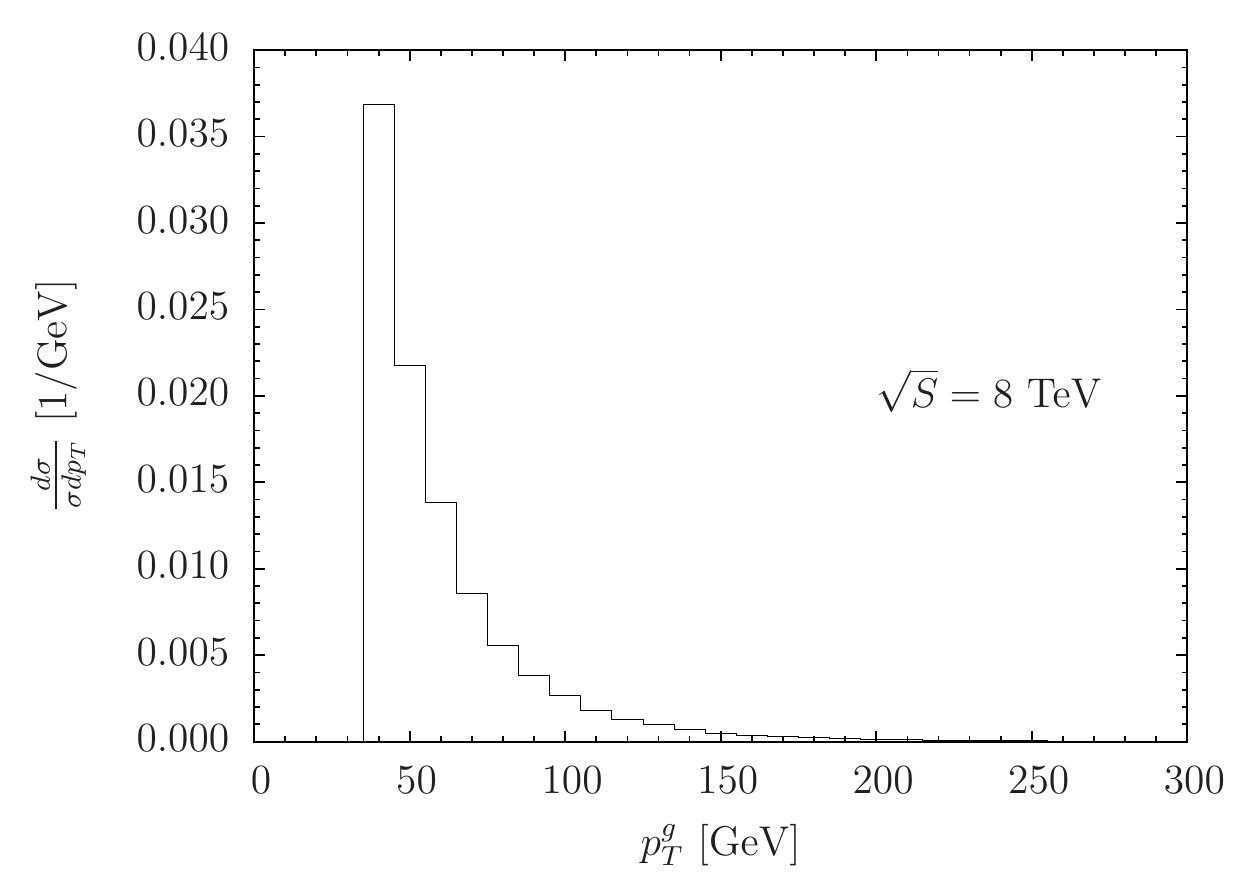}
 \caption{Transverse momentum distribution of the gluon jet at 8 TeV centre-of-mass energy in $gg\to \gamma Z(\to l^+l^-)g$. }
\label{fig:ptg8}
\end{minipage}
\hspace{0.5cm}
\begin{minipage}[b]{0.5\linewidth}
\centering
\includegraphics[width=\textwidth]{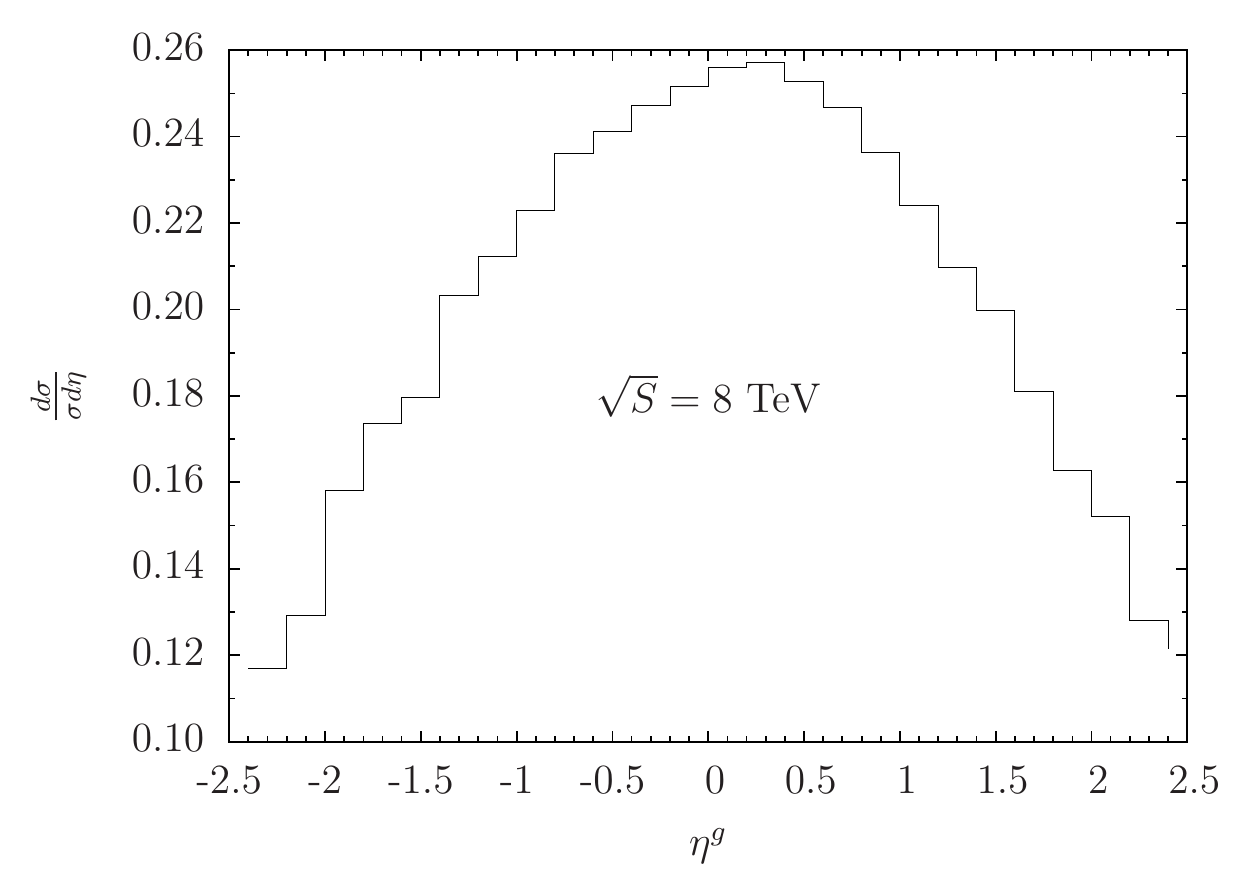}
 \caption{Rapidity distribution of the gluon jet at 8 TeV centre-of-mass energy in $gg\to \gamma Z(\to l^+l^-)g$. }
\label{fig:etag8}
\end{minipage}
\end{figure}

\begin{figure}[ht]
\begin{minipage}[b]{0.5\linewidth}
\centering
\includegraphics[width=\textwidth]{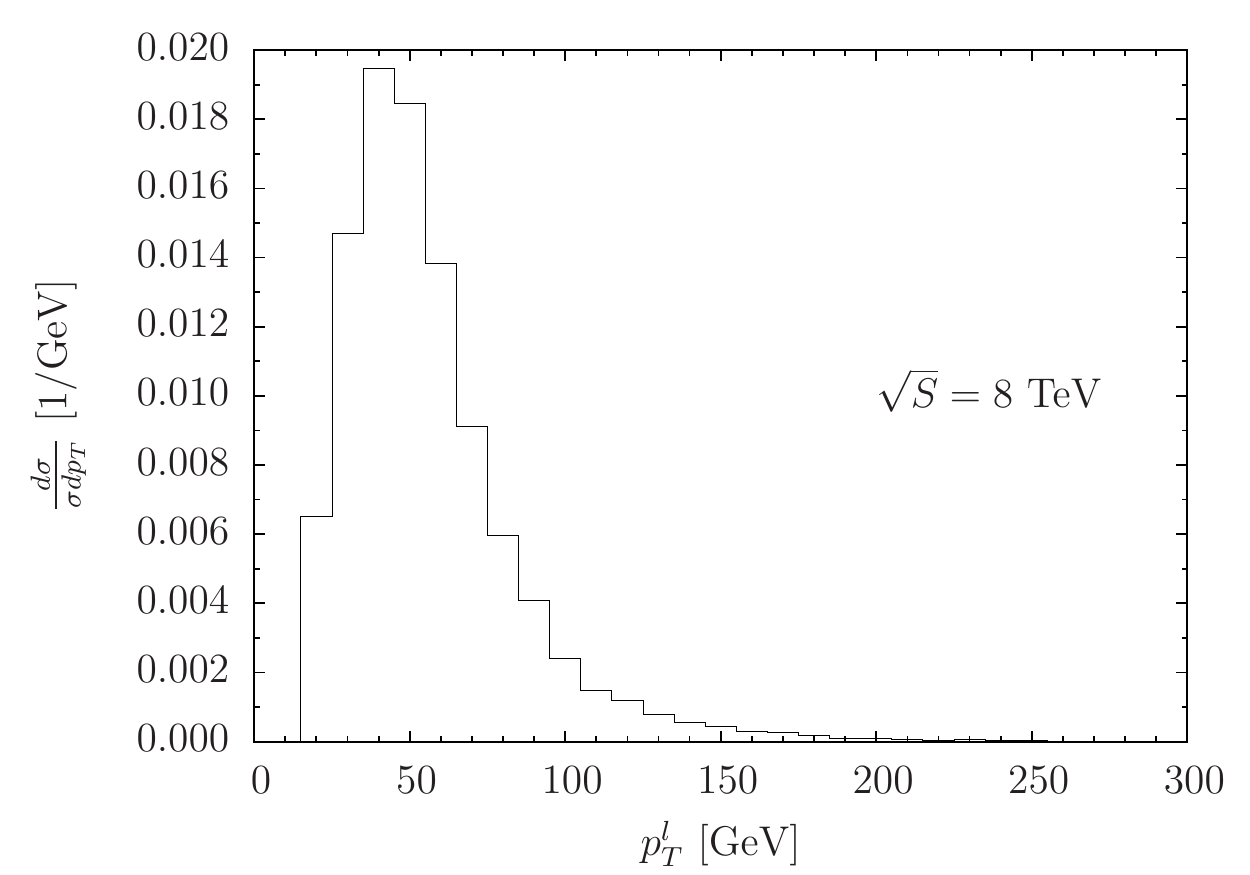}
 \caption{Transverse momentum distribution of lepton at 8 TeV centre-of-mass energy in $gg\to \gamma Z(\to l^+l^-)g$. }
\label{fig:ptl28}
\end{minipage}
\hspace{0.5cm}
\begin{minipage}[b]{0.5\linewidth}
\centering
\includegraphics[width=\textwidth]{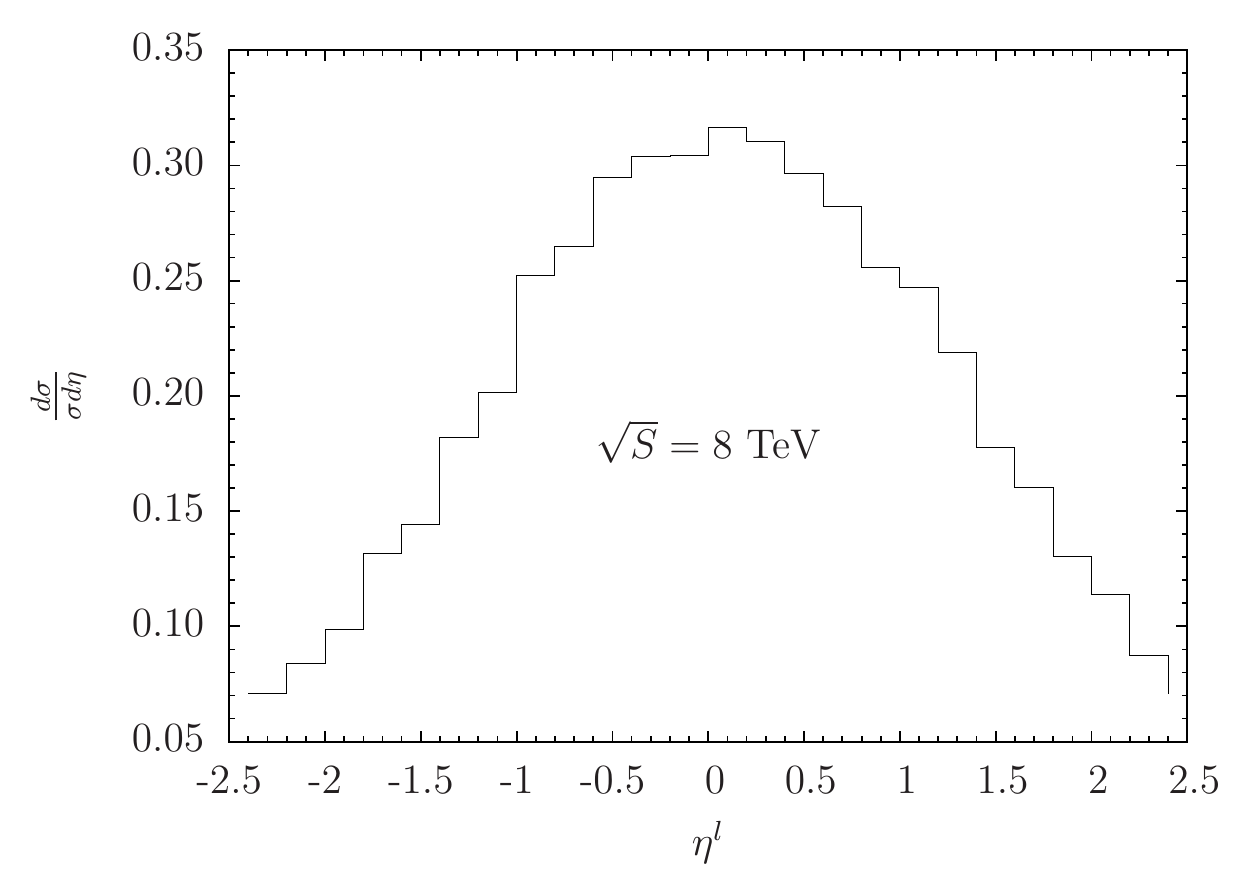}
 \caption{Rapidity distribution of lepton at 8 TeV centre-of-mass energy in $gg\to \gamma Z(\to l^+l^-)g$. }
\label{fig:etal28}
\end{minipage}
\end{figure}

 Next, we come to the discussion on our phenomenological results. These results include various kinematic 
distributions related to the final state particles. For phenomenological results, we work in the narrow width 
approximation. We allow the $Z$ boson to decay into two leptons in the phase space.
In this case, the kinematic cuts are
\begin{eqnarray}
p_T^{j} > 30 \; {\rm GeV},\; p_T^{\gamma} > 15 \; {\rm GeV},\; p_T^l > 10 \; {\rm GeV},
\:|\eta^{\gamma,l,j}| < 2.5, \; R(i_1,i_2) > 0.4.
\end{eqnarray}
Here $i_1$ and $i_2$ may represent any of the $\gamma/l/j$.
For convenience, we have chosen the scale $\mu = \mu_f = \mu_r = M_Z$.
In Fig. ~\ref{fig:sigma_cme-gpz}, we give the cross section variation in the range of 8 TeV to 14 
TeV centre-of-mass energies using both the CTEQ6l1 and CTEQ6M PDFs. 
These numbers do not include the branching ratio of $Z \to l^{+} l^{-}$.  In particular, 
the cross sections with CTEQ6l1 (CTEQ6M) parton distributions, are 65.4 (53.0) fb and 
202.4 (154.3) fb at 8 TeV and 14 TeV centre-of-mass energies respectively.
With these cross sections, number of $gg \to \gamma Z g$ events can be as large as 
20000 at the 14 TeV LHC, with 100 fb$^{-1}$ integrated luminosity. However, to observe these
events, one may have to look at $Z \to l^{+} l^{-}$ decay channel; here $l$ can be
an electron/muon. So including the branching ratios, one may expect more than
1000 events for $gg \to \gamma Z (\to l^{+} l^{-}) g$ process.
The transverse momentum and rapidity distributions for the
final state particles are shown in Figs.~\ref{fig:ptp8}-\ref{fig:etal28} at the 8 TeV centre-of-mass energy. 
We have given normalized distributions as they remain same for different choices of parton distributions
and/or scales. These distributions are characteristically similar at different collider centre-of-mass
energies, but at higher energies contribution coming from high $p_T$/rapidity region grows, while low $p_T$/rapidity
region contribution goes down. 
We note that $p_T^j$ is softer as compared to $p_T^\gamma$. It is because the cross section is dominated by the
box class of diagrams and in these diagrams, the gluon is emitted as a bremsstrahlung radiation, see 
Fig.~\ref{fig:sigma_cme-box}. Due to the same reason, {\it i.e.}, the gluon is emitted more collinearly, 
the rapidity distribution of the gluon jet is broader as compared to that of the photon. The lepton-$p_T$ 
distribution peaks around $M_Z/2$. On the other hand the rapidity distribution of the lepton is more central 
compared to the $\eta^\gamma$ distribution.
\begin{table}[h!]\label{tbl:gpz}
\begin{center}
\begin{tabular}{|c|c|c|c|c|c|c|}
\hline
$\sqrt{\rm S}$ &$p^{\gamma,min}_T$ &$\sigma^{\rm LO}$ &$\sigma^{\rm NLO}$ 
&$\sigma^{\rm NNLO}_{gg}$  &$\sigma^{\rm NNLO}_{gg}/(\sigma^{\rm NLO}-\sigma^{\rm LO})$ 
&$\sigma^{\rm NNLO}_{gg}/\sigma^{\rm NLO}$ \T \\
(TeV) &(GeV) &(pb) &(pb) &(fb) &$(\%)$ &$(\%)$ \B \\
\hline
{8} & 30 & 2.202 & 3.391 & 46.05 (38.25) & 3.87 (3.22)  & 1.36 (1.13)  \T\B \\
    & 50 & 1.144 & 1.744 & 30.49 (25.61) & 5.08 (4.27) & 1.75 (1.47)  \T\B \\ 
\hline
{14} & 30 & 4.868 & 7.722 & 158.72 (124.48) & 5.56 (4.36) & 2.06 (1.61)  \T\B \\
     & 50 & 2.608 & 4.158 & 109.92 (86.61)  & 7.09 (5.59) & 2.64 (2.08)  \T\B \\ 
\hline
{35} & 30 & 14.973 & 23.548 & 854.09 (606.07) & 9.96 (7.07) & 3.63 (2.57)  \T\B \\
     & 50 &  8.220 & 13.514 & 607.35 (438.88) & 11.47 (8.29) & 4.49 (3.25)  \T\B \\ 
\hline
\end{tabular}
\end{center}
\caption{ Cross sections for the production of $\gpZpp$ at various collider centre-of-mass energies. We use
CTEQ6l1 PDF set at the LO and CTEQ6M PDF set at the NLO. The NNLO predictions are with CTEQ6l1(CTEQ6M) 
parton distribution. The factorization and renormalization scales are set to, $\mu_f = \mu_r = \mu_0 = M_Z$.}
\end{table}

 We have also compared results of this NNLO calculation with the LO and NLO predictions for 
$ p p \to \gamma (Z \to \nu \bar\nu) j + X$~\cite{Campbell:2012ft}. The LO and NLO results are 
obtained using parton-level next-to-leading order program MCFM\footnote{http://mcfm.fnal.gov/}.
The comparison is presented after removing the branching ratios in Table~3.1. 
The table includes results at three different centre-of-mass energies and for two values of the 
$p_{T}^{\gamma,min}$. We have included the centre-of-mass energy of 35 TeV, as it is proposed for 
the HE-LHC collider. The other kinematic cuts are: 
$p_T^{j} > 30 \; {\rm GeV},\; p_T^{miss} > 30 \; {\rm GeV },\:|\eta^{\gamma,j}| < 2.5,R(\gamma,j) > 0.4$.
This table illustrates two facts -- 1) the fraction of NNLO events increases with the increase in  
$p_{T}^{\gamma,min}$, 2) the NNLO process becomes more important as we increase the centre-of-mass 
energy. There is an increase in the NNLO fraction with an increase in $p_{T}^{\gamma,min}$ because, 
in the NLO events, photon is emitted from a quark line; a larger  $p_{T}^{\gamma,min}$ suppresses 
the NLO contribution more than the NNLO contribution. In Fig.~\ref{fig:sigma14ptp30}, we have compared 
the normalized $p_T^\gamma$-distributions at the NLO and NNLO, leading to the same conclusion. The importance 
of the NNLO process is more at higher centre-of-mass energy simply because of the increase in the 
gluon-gluon luminosity. At 8 TeV, the scale uncertainties in the NLO calculations, on changing 
the scale by a factor of two in both the directions of the central value, are in the range $7-8\%$, while 
the same in the NNLO calculations is $30-50\%$.

%%%%%%%%%%%%%%%%%%%%%%%%%%%%%%%%%%%%%%%%%%%%%%%%%%%%%%%%%%%%%%%%%%%%%%%%%%%%%%%%%%%%%%%%%%%%%%%%%%%%%
\section{The Issue of Numerical Instability}

\begin{figure}[ht]
\begin{minipage}[b]{0.5\linewidth}
\centering
\includegraphics[width=\textwidth]{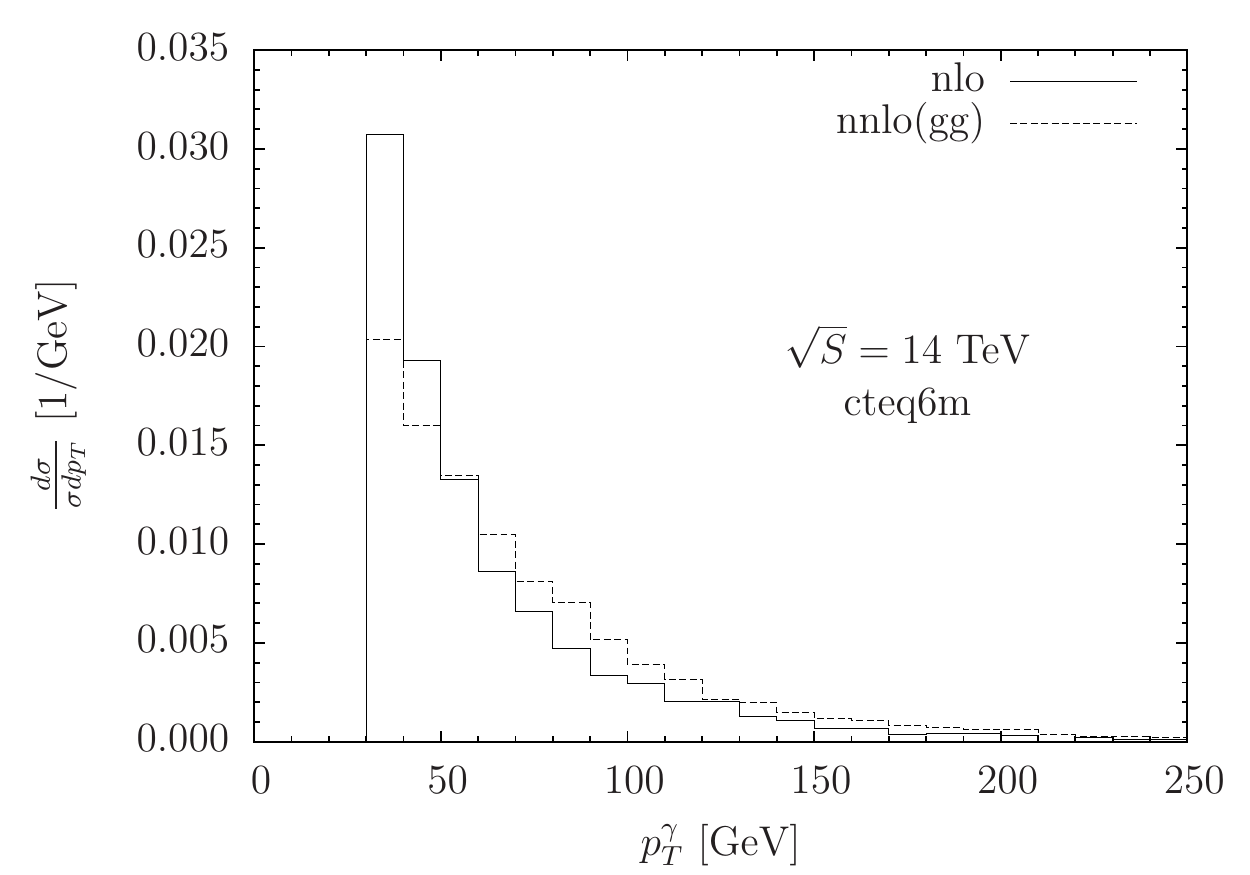}
 \caption{A comparison of the normalized $p_T-$distributions of $\gamma$ at NLO and NNLO. The NLO
          distribution is obtained using MCFM.}
\label{fig:sigma14ptp30}
\end{minipage}
\hspace{0.5cm}
\begin{minipage}[b]{0.5\linewidth}
\centering
\includegraphics[width=\textwidth]{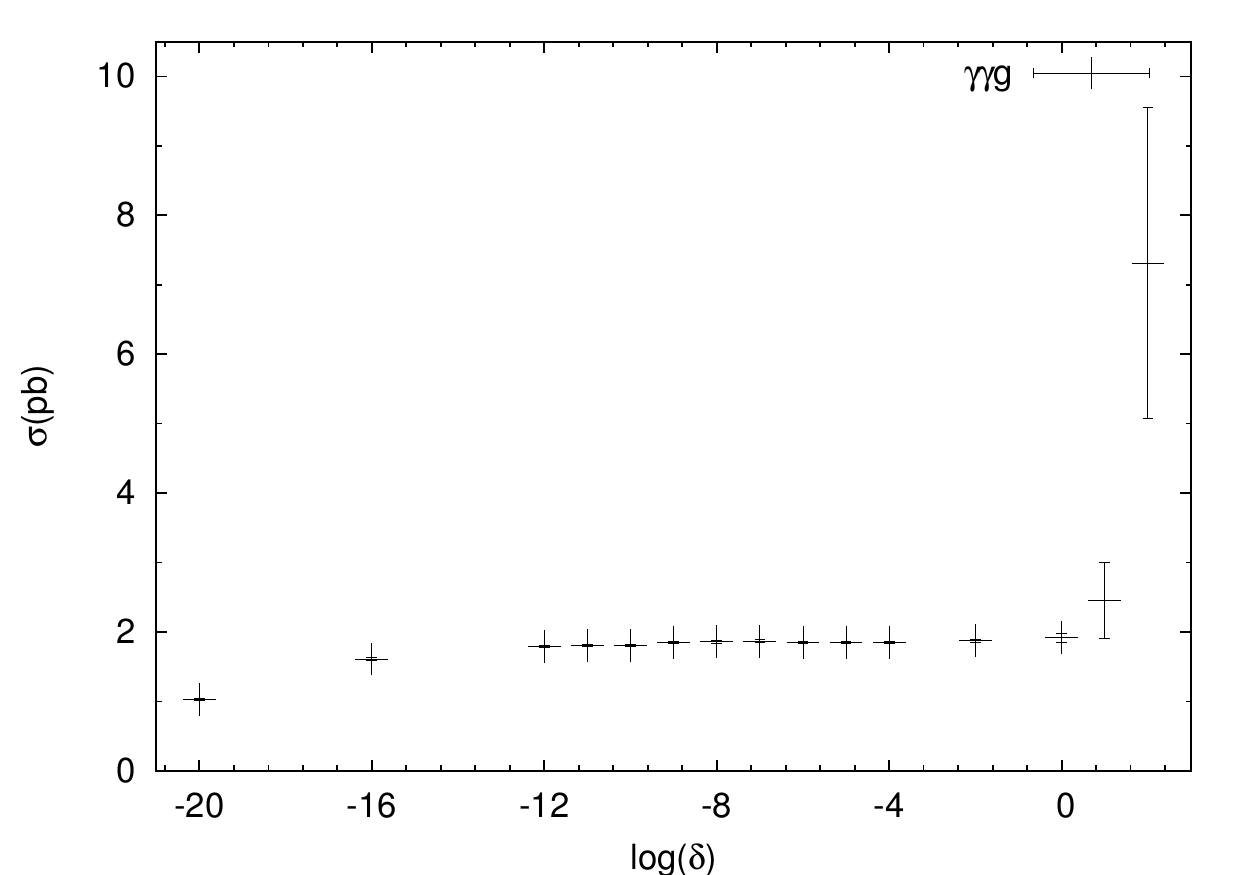}
\caption{Dependence of the total cross section on GI test cut-parameter ($\delta$) for $\gpp$ at 14 TeV. The
         error bars are shown explicitly.}
\label{fig:sigma-cut}
\end{minipage}
\end{figure}

 Like other calculations of our types, we have also faced the issue of numerical instability in our 
 calculations for certain phase space points. This is a well known issue in the reduction of one-loop 
 tensor integrals of higher rank and higher points. The issue of numerical instability may also occur 
 in the evaluation of the scalar integrals. We have taken care of this by using the OneLOop implementation 
 of the scalar integrals. We face numerical instabilities primarily in the evaluation of pentagon 
 tensor integrals. This is related to the inaccurate evaluation of the Gram determinants in those phase 
 space regions  where the linear independence of external momenta (modulo 4-momentum conservation) is 
 compromised, {\it i.e.}, near the exceptional phase space points. 
The inverse Gram determinants appear in the reduction of tensor integrals. Near exceptional 
 phase space points, the Gram determinants become very small and give rise to numerical problems. 
These numerical problems are result 
 of a loss of precision due to the large cancellations. This problem can be handled in several ways. One way is 
 to use higher precision for the tensor reduction and for the evaluation of scalar integrals. This certainly reduces 
 the number of exceptional phase space points but the code becomes enormously slow. Another approach 
 could be to use special expressions for the tensor reduction, near such phase space points~\cite{Denner:2005nn}. 
 It is important to mention here that none of these two approaches cure 
 the problem of numerical instability completely \cite{Campanario:2011cs}. A more economic and convenient way 
 to proceed in this situation is to judiciously ignore the  contributions from such phase space points. This one
    can do because we are not doing precision calculations and exceptional phase space points 
    are unlikely to give a significant contribution to the total cross section.
    We perform a gauge invariance (GI) test on the full amplitude for each phase space point.
In practice, we introduce a small cut-parameter $\delta$, and under GI we check, if $|{\cal M}|^2 < \delta$
holds true for each phase space point. 
% In practice, we put a GI test cut of, $\delta = 10^{-6}$ on the amplitude-squared. 
    We ignore all those points which fail to pass this test. In our code we have set $\delta=10^{-6}$. 
    With this value of $\delta$ and other kinematic
    cuts, the fraction of points ignored is below $2 \% $. However, with higher $p_T$ cut and/or less stringent 
    cut on the $\delta$, the number of such points can drop to a level of $0.01 \%$. Our Monte Carlo phase space 
integration subroutine is based on the VEGAS algorithm.
We believe that the adaptive nature of the VEGAS algorithm also affects the percentage of exceptional 
phase space points that one may come across in such calculations.
We sample about 0.4-0.5 
    million phase space points to obtain the numerical results. Given the volume of phase space, the number of 
    exceptional phase space points is small and it is reasonable to assume that the cross section is not dominated 
    by this region of phase space. We find that our result depends on this cut very weakly and remains quite 
    stable over the range of $10^{-4}-10^{-12}$ for the choice of the cut-parameter. This can be seen in 
    Fig.~\ref{fig:sigma-cut}. This stability reflects that the exceptional phase space points are few and make 
    small contribution at the level of total cross section.

    We have also checked our results for the cross section calculation by implementing 
    a set of Ward identity tests and its sensitivity on $\delta$-like cut-parameter.
    Examples of these identities are given in Eqs.~\ref{eq:WI-penta} and \ref{eq:WI-box}.  
    However, these identities should not be implemented in a straightforward manner. 
    For example, the pentagon ward identity in Eq.~\ref{eq:WI-penta} can be implemented
    in the following form,
\begin{eqnarray}\label{eq:WI-cut}
(1-\frac{{\cal M}_P}{{\cal M}_B^1-{\cal M}_B^2}) < \delta.
\end{eqnarray}
Numerically, due to the Ward identity, the ratio ${\cal M}_P/({\cal M}_B^1-{\cal M}_B^2) \sim {\cal O}(1)$. 
In Eq.~\ref{eq:WI-cut}, we compare two numbers of ${\cal O}(1)$, which gives precise information 
on the order of cancellation within the working precision.
    Although the total cross section is quite stable, various distributions, specially 
    the rapidity distributions near the edges, are quite sensitive to the variation in 
    $\delta$. Edges of distributions define the region where exceptional phase space 
    points may lie. Therefore, the inaccuracy of the distributions at the edges of
    the phase space is not surprising. We have seen that the exceptional phase space 
    points may defy the GI and/or Ward identity tests sometimes. One can make these
    $\delta$-like cuts more stringent to get more reliable distributions. One can also
    identify exceptional phase space points at  the level of Gram determinants which 
    may be more economical. Phase space points corresponding to 
    a large cancellation in the Gram determinants can be ignored without 
    putting stringent cuts on $\delta$ and again leading to more reliable distributions.
    A method to implement this criterion is discussed in \cite{Campanario:2011cs}.

\newpage
 
\vspace{3cm}

% for photon

% for gluon

% for e+/e-

% ppg and ppZ processes %%%%%%%%%%%%%%%%%%%%%%%%%%%%%%%%%%%%%%%%%%%%%%%%%%%%%%
% 
%%%%%%%%%%%%%%%%%%%%%%%%%%%%%%%%%%%%%%%%%%%%%%%%%%%%%%%%%%%%%%%%%%%%%%%%%%%%%%%%%%%%%%%%%%%%%%%%%%%%%
%%%%%%%%%%%%%%%%%%%%%%%%%%%%%%%%%%%%%%%%%%%%%%%%%%%%%%%%%%%%%%%%%%%%%%%%%%%%%%%%%%%%%%%%%%%%%%%%%%%%%
% \newpage
% 
% \bibliographystyle{utcaps}
% \bibliography{thesis} 
% \end{document}

\setcounter{equation}{0}
\setcounter{figure}{0}
\newpage
\chapter{Production of KK-gravitons with a Boson via Gluon Fusion}
\label{chapter:add}
% \numberwithin{equation}{chapter}
% 
In this chapter, we will consider the associated production of an electroweak boson and the KK-graviton modes
via gluon fusion in the ADD model. In particular, we will study leading order $\HG$, $\pG$ and $\ZG$
processes. Like the SM processes discussed earlier, these processes also proceed via quark loop diagrams 
at the leading order and are finite. These processes contribute to the hadronic processes $\BGpp$,
at the NLO in $\alpha_s$, where $B \in \{H, \gamma, Z\}$. We will calculate the inclusive cross sections 
of these gluon channel processes  at the LHC and discuss some important kinematic distributions. 
In this chapter, by inclusive we mean the contribution from all the kinematically accessible KK-modes. 
We will also study the dependence of these results on the model parameters $\delta$ and $M_S$.
This chapter is based on the work reported in~\cite{Shivaji:2011re,Mitra:2011sj,Shivaji:2011ww}.
A brief overview of the model is given in Sec.~\ref{section:add}.

Like the other new physics scenarios, 
the searches for the model of large extra dimensions or the ADD model is also continuing at the LHC. 
With the 7 TeV LHC data, the bounds on the model parameters have also improved quite 
significantly~\cite{Chatrchyan:2011fq,Chatrchyan:2012kc,
Chatrchyan:2012tea,Chatrchyan:2012me,Aad:2012fw,ATLAS:2012ky,Aad:2012cy,ATLAS:2012cb}.  
In the context of the ADD model, studies of processes involving the exchange of KK-gravitons and those 
in which KK-gravitons are produced directly, have been reported for both the Tevatron and the 
LHC~\cite{Eboli:1999aq,Cheung:1999wt,Mathews:2004xp,Kumar:2008pk,Karg:2009xk,Gao:2009pn,Kumar:2010kv,
Kumar:2010ca,Kumar:2011jq}. 
Due to a large gluon flux available at the LHC, the gluon initiated processes 
can be quite important. In this regard, we have investigated the KK-graviton production in 
association with a boson ($H/\gamma / Z$) via gluon fusion. Unlike the cases of 
$\gamma/ZG_{\rm KK}$ production, the $q{\bar q}$ initiated
tree-level $HG_{\rm KK}$ production process has very small cross section due to a vanishingly 
small coupling of the Higgs boson with light quarks. Even with a non-zero bottom quark mass, the tree-level
cross section can be at best of the order $10^{-3}$ fb at the LHC. The contribution of the gluon-gluon channel
is, therefore, expected to be relatively more important for the $HG_{\rm KK}$ production at the LHC.
The situation is analogous to the single Higgs boson production in the SM. There too the $gg\to H$ channel 
dominates the hadronic cross section. 
The rest of the chapter is organized as follows: In the next section, we give some details on the structure of 
the amplitudes. Various checks on our amplitudes and the method of computation are described in section 4.2 and 4.3. 
Numerical results are presented in section 4.4. An interesting issue related to the $ZG_{\rm KK}$ amplitude 
calculation is added in the end.

% 
%%%%%%%%%%%%%%%%%%%%%%%%%%%%%%%%%%%%%%%%%%%%%%%%%%%%%%%%%%%%%%%%%%%%%%%%%%%%%%%%%%%%%%%%%%%%%%%%%%%
\section{The Structure of Amplitudes}
\begin{figure}[h]
\begin{center}
\includegraphics[width=10cm]{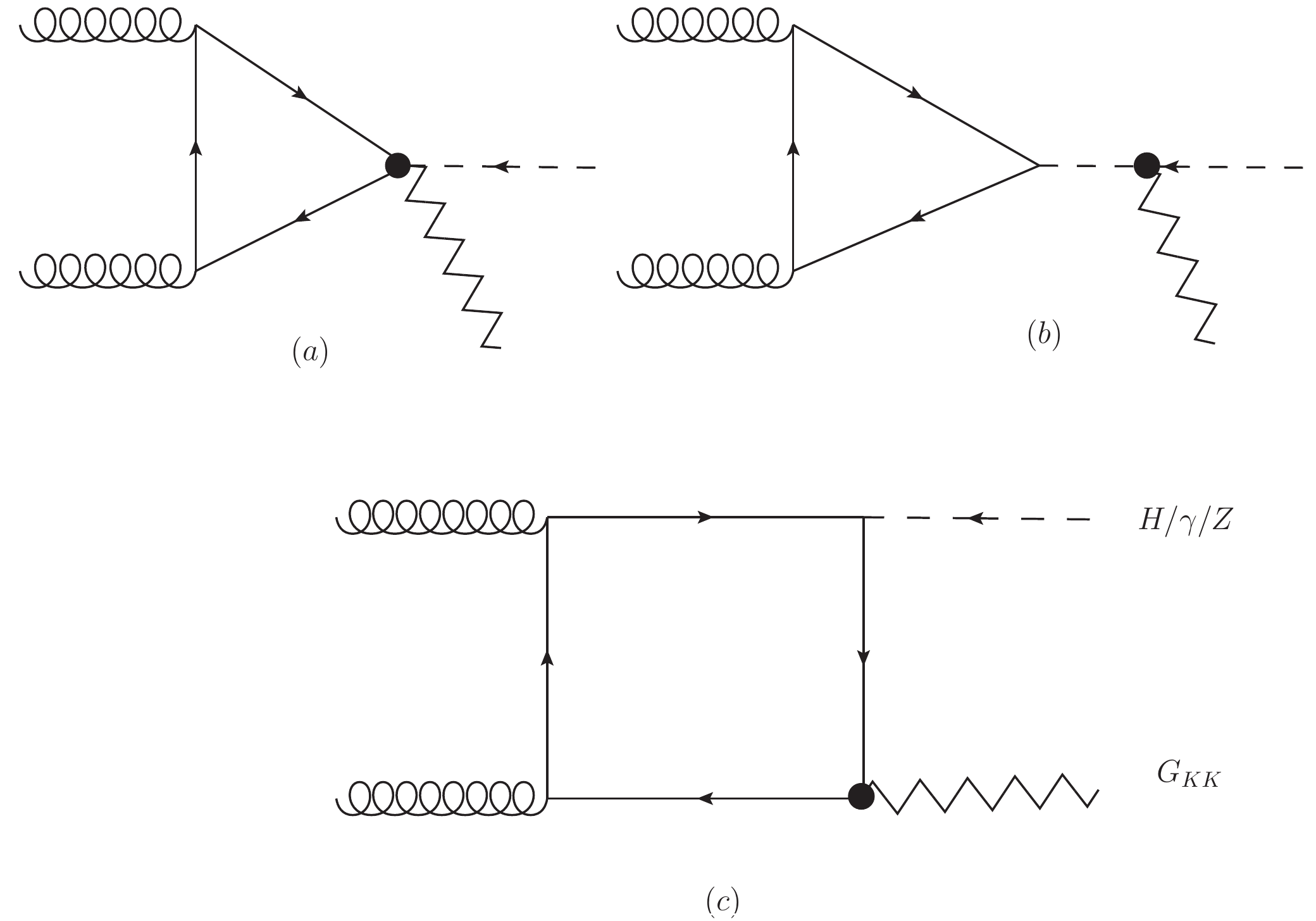}
\caption{Prototype Feynman diagrams for $ gg \to H/\gamma/Z+G_{\rm KK} $ in the ADD model.}
\label{fig:ggBG}
\end{center}
\end{figure}

At the leading order, the process $ \BG $ proceeds via quark loop diagrams. The allowed vertices and 
their Feynman rules, in the ADD model, are listed in Ref.~\cite{Han:1998sg}. 
Depending upon the coupling of the KK-graviton with the standard model particles, there are three 
classes of diagrams: a triangle class of diagrams due to $quark-boson-graviton$ coupling, another
triangle class of diagrams due to $boson-graviton$ coupling and  the box class of diagrams due 
to $quark-graviton$ coupling.
The prototype diagram in each class is shown in Fig.~\ref{fig:ggBG}. Other 
diagrams are obtained just by appropriate permutations of the external legs. There are total   
six box and twelve triangle diagrams for each quark flavor. However, due to the charge-conjugation 
property of the fermion loop diagrams, only half of the diagrams are independent. Since the coupling 
of the Higgs boson with quarks is proportional to the quark masses, in the $\HG$ case, we consider 
only bottom and top quark contributions. For a given massive quark in the loop, the $HG_{\rm KK}$ 
amplitude has the following structure:
\begin{eqnarray}
{\cal M}_q^{ab}(\HG) &=& 
\frac{1}{8}\; y_q \;g_s^2 \;\kappa \left(\frac{\delta^{ab}}{2}\right)\; {\cal A}(m_q), \nn \\
{\cal A}(m_q) &=& \left[ 2 \; \mathcal{A}_{tri}(m_q) - \mathcal{A}_{box}(m_q) \right].
\end{eqnarray}
Here the Yukawa coupling, $ y_q  = \frac{1}{2} g_w \left(m_q/M_W\right)$ and 
$\kappa = \sqrt{2}/M_P$. Furthermore, $a$ and $b$ are color indices of
the two gluons. $\mathcal{A}_{tri}$ is the net contribution from the two triangle
classes of diagrams, shown in Figs.~\ref{fig:ggBG} ($a$) and ($b$).

For $\pG$ case, we find that the amplitudes of the diagrams, related by charge-conjugation, 
are equal and opposite to each other. This implies that at the LO,
\begin{eqnarray}
 {\cal M}(\pG) = 0.
\end{eqnarray}
By introducing charge-conjugation transformation of the KK-graviton field and 
using charge-conjugation properties of the gluon and photon fields, it can be 
shown that the $\gamma G_{\rm KK}$ amplitude does vanish at the LO. This is just an 
implication of the extension of Furry's theorem in the presence of gravitons,
photons and gluons. The graviton field is considered even under charge-conjugation 
as it couples with the energy-momentum tensor only. This, we have explicitly verified 
using the ADD model Lagrangian \cite{Mitra:2011sj}. This result would remain valid to 
all orders if only QED and/or QCD radiative corrections are included. In the presence 
of the weak interaction, this result may not hold to all orders.
  
The $\ZG$ amplitude has both the vector 
and axial-vector contributions coming from the $Z$ boson coupling to the quarks.
The vector part of the amplitude is similar to the $\gamma G_{\rm KK}$ amplitude 
and therefore at the LO, the process receives contribution only from the axial-vector 
part of the amplitude. For a given quark flavor in the loop, the amplitude has the 
following structure:
\begin{eqnarray}
  \mathcal{M}^{ab}_q(\ZG) &=& g_z\; g_s^2\; \kappa\; \left(\frac{\delta^{ab}}{2}\right)\; c^q_A \; \mathcal{A}(m_q), \\
\mathcal{A}(m_q) &=& \mathcal{A}_{tri}(m_q)-\mathcal{A}_{box}(m_q).
\end{eqnarray}
Here $g_z = g_w/{\rm cos}\theta_w \; {\rm and} \; c_A^q = - T^3_q/2$.
$\mathcal{A}_{tri}$ includes the contributions from both types of triangle 
diagrams. We find 
that due to the nature of KK-graviton coupling with quarks, both the triangle and 
box diagrams are linearly divergent and therefore they will give rise to anomalous
contributions to the amplitude.
Of the six quark flavours, we treat the $u, d, s,$ and $c$ quarks as massless.
 Since the amplitude of the process is proportional to $T^3_q$ value, the first 
two generations do not contribute. Therefore, the full amplitude, including 
the contributions from all the six quarks, is
\begin{eqnarray}\label{eq:amp-ZG}
\sum_q  \mathcal{M}_q^{ab}(\ZG) = -\frac{1}{4} g_w g_s^2 \kappa 
\left(\frac{\delta^{ab}}{2}\right) \left[\mathcal{A}(m_t)-\mathcal{A}(m_b)\right].
\end{eqnarray}
We note that the cross section is of $\mathcal{O}(\alpha_s^2)$, and therefore this 
LO contribution can be included in $\sigma_{\rm NLO}(pp \rightarrow ZG_{\rm KK}+X)$ 
\cite{Kumar:2010kv, Kumar:2010ca}.
 
The one-loop amplitudes for $gg\to H/ZG_{\rm KK}$ are expected to be free of ultraviolet (UV) 
and infrared (IR) singularities for each quark flavour. The IR singularities (large logs in the 
mass regularization) are applicable to light quark cases only. We also expect gauge invariance
with respect to the gluon and the KK-graviton currents. However, in the $ZG_{\rm KK}$ case, 
due to the presence of anomalies, the amplitude for an individual quark flavour may not be gauge 
invariant with respect to the axial-vector current in the $m_q\to 0$ limit. Since the model 
is free from anomalies, we expect gauge invariance after summing over all the six
quark flavours. The confirmation of the cancellation of UV and IR
singularities and gauge invariance with respect to the vector and axial-vector currents 
 are powerful checks on our calculation. We make all these checks as described
in the next section.
%%%%%%%%%%%%%%%%%%%%%%%%%%%%%%%%%%%%%%%%%%%%%%%%%%%%%%%%%%%%%%%%%%%%%%%%%%%%%%%%%%%%%%%%%%%%%%%%%%%%%%%%
\section{Details of Calculation and Checks}
Our one-loop calculation is based on the traditional Feynman diagram method.
The amplitude of each diagram is written using the SM and ADD model Feynman rules. 
The ADD model Feynman rules we require are also listed in the appendix~\ref{appendix:feynrules}. 
However, we need not compute all the diagrams explicitly. We only need to compute
the prototype diagrams. All other diagrams can be obtained by suitable permutations of the external 
momenta and polarizations. This works quite well in the $HG_{\rm KK}$ case. However, due to the presence 
of $\gamma^5$ in the $ZG_{\rm KK}$ amplitude, one needs to take extreme care in making these 
permutations in $n$ dimensions. The permutation should not be across the $\gamma^5$ vertex. Such 
permuted diagrams need to be computed explicitly. Due to the presence of a quark loop, 
the amplitude of each diagram is proportional to the trace of a string of gamma matrices. 
We compute these traces using FORM \cite{Vermaseren:2000nd}. This is the most important
part of the calculation. The presence of 4-dimensional $\gamma^5$ in the trace, leads to spurious anomalies 
in the amplitude. We, therefore, need an appropriate $n$-dimensional treatment of $\gamma^5$. We 
have used Larin's prescription for $\gamma^5$ to calculate the trace in $n$ dimensions ~\cite{Larin:1992du}. 
According to this prescription,
\begin{eqnarray}\label{eq:g5-nd}
 \gamma_\mu \gamma^5 = -\frac{i}{6}\; \eps_{\mu\nu\rho\sigma} \gamma^\nu \gamma^\rho \gamma^\sigma.
\end{eqnarray}

After calculating the trace, we express the amplitude in terms of appropriate tensor 
integrals. The box amplitude has rank-4 tensor integrals, while the triangle amplitude 
has rank-2 tensor integrals at the most. The tensor reduction into scalars is done 
in $n=(4-2\eps)$ dimensions using the methods of Odenborgh and Vermaseren 
\cite{vanOldenborgh:1989wn}, also described in the Sec.~\ref{section:OV}. All the required 
scalar integrals for the massless quark case are listed in the appendix~\ref{appendix:scalars}. 
We need only the UV and IR singular pieces of these scalars to verify the cancellation of 
UV and IR singularities. For both the cases of bottom and top quarks, we use FF library 
to calculate the required scalars \cite{vanOldenborgh:1990yc}. 
Due to a very large and
complicated expression of the amplitude, we compute the amplitude numerically before squaring it. 
This requires computation of the polarization vectors for the gauge bosons and for the KK-graviton. 
We have chosen the helicity basis for them. It also helps in making additional checks on our 
calculation by verifying relations among helicity amplitudes. The KK-graviton polarization tensor
is constructed from the polarization vectors of two massive vector bosons as suggested in \cite{Han:1998sg}.
To obtain hadronic cross sections, we perform integrations over two body phase space and 
the gluon PDFs, using a Monte Carlo integration subroutine based on the VEGAS algorithm. 
Since the KK-gravitons are produced directly, we also require an additional integration over 
the KK-graviton mass parameter $M_{\rm KK}$, to obtain an inclusive cross section.  \\

As discussed in the previous section, our one-loop processes are expected to be finite. We verify that 
both the massive and massless 
contributions are UV finite. We observe that each triangle diagram is UV finite by itself, 
while the box amplitude is UV finite only after adding all the box contributions. 
As we discussed in the Sec.~\ref{section:IRfinite}, fermion loop 
diagrams are known to be IR finite, in both the massive and massless fermion cases, 
for any kind and any number of external particles attached to the loop. 
In the massless quark case, we check that each diagram is IR finite and therefore 
IR finiteness holds for the full amplitude. 
Finally, we check the gauge invariance of the amplitude with respect to the two gluons by replacing 
their polarizations with their respective momenta. In the $HG_{\rm KK}$ case, we observe that some 
of the triangle diagrams are separately gauge invariant with respect to both the gluons. 
To ensure the correctness of their contribution towards the full amplitude, we have also
performed a gauge invariance check with respect to the KK-graviton current.
In the $ZG_{\rm KK}$ case, we find that (only after using $\gamma^5$ prescription
in the trace) both the bottom and top quark contributions are separately gauge invariant 
with respect to the two gluons. In $ZG_{\rm KK}$-triangle class of diagrams, involving 
graviton-gauge boson coupling, we have chosen the gauge-fixing parameter $\xi =1 $ (the Feynman gauge)
for the gluon case and $\xi = \infty$ (the Unitary gauge) for the $Z$ boson case. 
As expected, the calculation does not depend on any specific choice of the gauge-fixing parameter.
We also check gauge invariance of the amplitude with respect to the $Z$ boson. Because of the anomaly, 
the two contributions are not separately gauge invariant. However, the 
total amplitude is gauge invariant up to  the top quark mass, as expected due 
to the explicit breaking of the chiral symmetry. All these checks on our amplitude have been 
made both numerically  as well as analytically. The issue of anomaly in $ZG_{\rm KK}$ amplitude 
is discussed in the end.

\begin{figure}[ht]
 \begin{minipage}[b]{0.5\linewidth}
\centering
 \includegraphics[width=\textwidth]{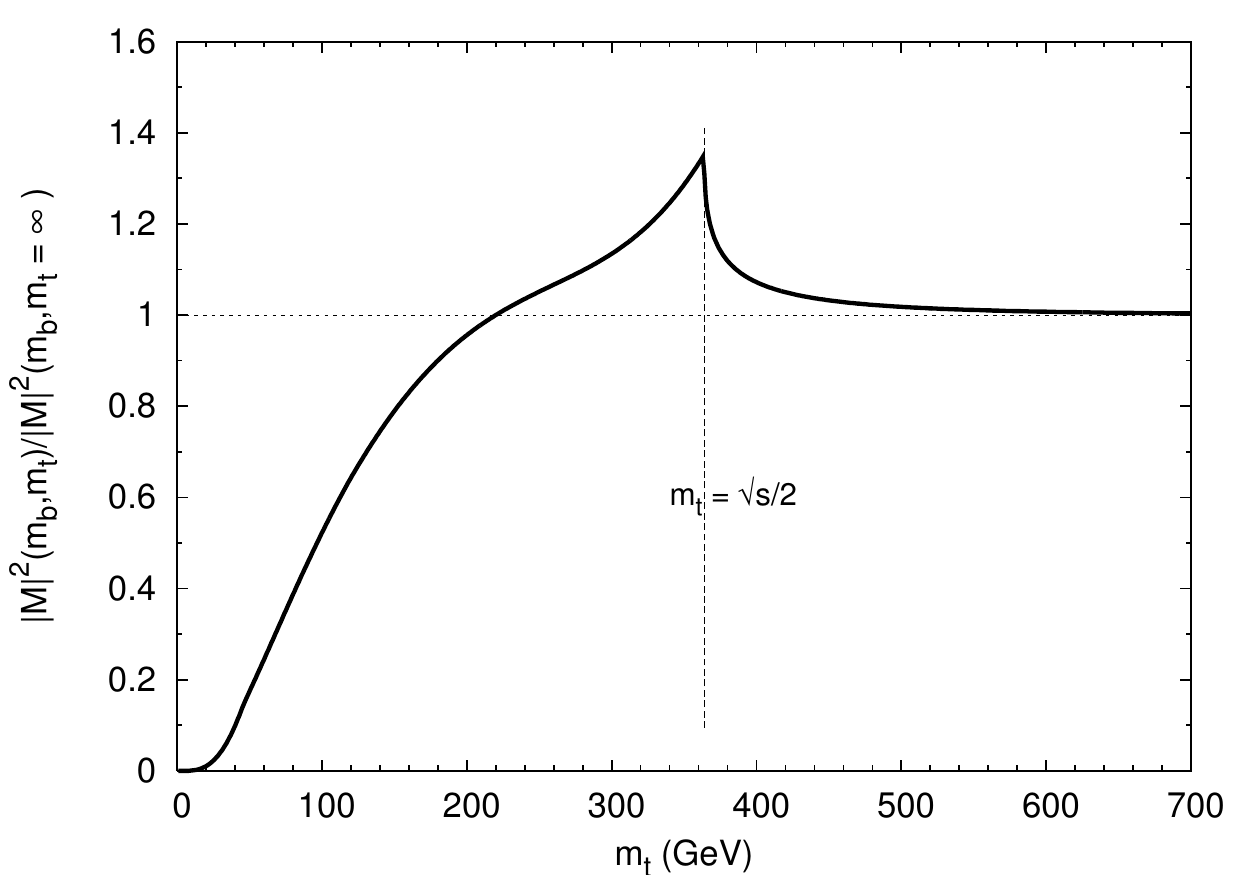}
\caption{Decoupling of the top quark as $m_t \rightarrow \infty$, in $\ZG$.}
\label{fig:ampsq_mtZG}
 \end{minipage}
 \hspace{0.5cm}
 \begin{minipage}[b]{0.5\linewidth}
\centering
\includegraphics[width=\textwidth]{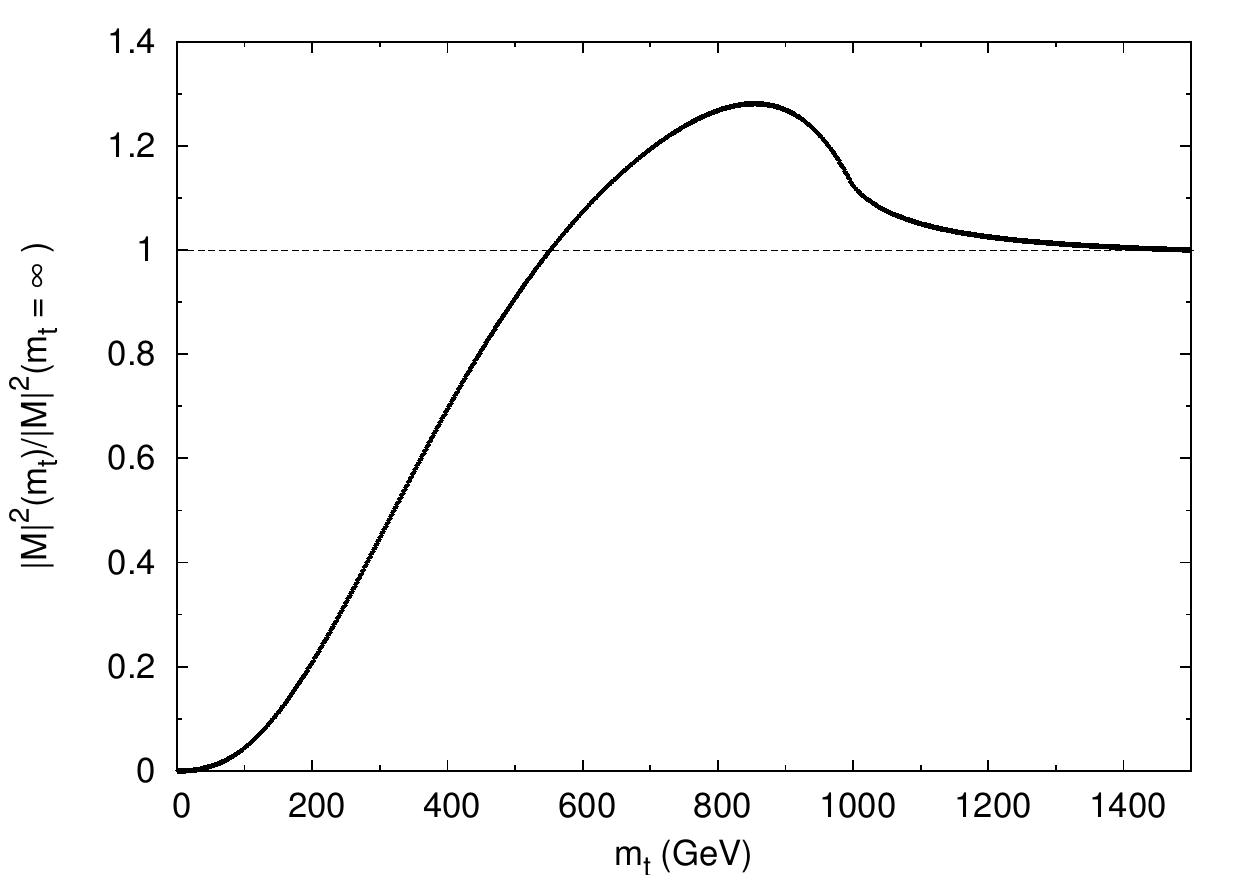}
\caption{Non-decoupling of the top quark as $m_t \rightarrow \infty$, in $\HG$.}
\label{fig:ampsq_mtHG}
\end{minipage}
\end{figure}

We have also cross-checked our calculations by taking the $m_t \to \infty$ limit. In the $ZG_{\rm KK}$
case, for a given phase space point, we vary the top quark mass and observe that 
the amplitude-squared (which includes both the bottom and top quark contributions) approaches 
a constant value (the bottom quark contribution) as $m_t \rightarrow \infty$. 
This implies the complete decoupling of the
top quark, {\it i.e.},  the top quark contribution of the amplitude goes to zero in large $m_t$ limit.
It is expected from the decoupling theorem \cite{Appelquist:1974tg}. This feature has been plotted in 
Fig.~\ref{fig:ampsq_mtZG}.
The change in slope around $m_t = \sqrt{s}/2 $ corresponds to a physical threshold after which 
the top quark propagators cannot go on-shell and the amplitude is real. 
It is well known that the decoupling theorem does not hold for fermion loop amplitudes
involving a Higgs boson. This is because, the Higgs boson coupling to fermions is proportional to
the fermion masses. Like in the case of $gg\to H$ amplitude, we do observe non-decoupling 
of the heavy top quark in the $\HG$ amplitude. In Fig.~\ref{fig:ampsq_mtHG}, the rise in the
curve, as $m_t$ increases in the beginning, is due to the explicit top quark mass dependence 
in the numerator ($m_t^2$ in the amplitude). As we approach larger values of $m_t$, the 
effective suppression ($\sim 1/m_t^2$ in the amplitude) due to the propagators dominates and the 
amplitude becomes independent of $m_t$. 
\section{$\HG$ Calculation in the Effective theory of Gluon-Higgs coupling}
\label{section:HG-effective}
\begin{figure}[h]
\begin{center}
\includegraphics[width=10cm]{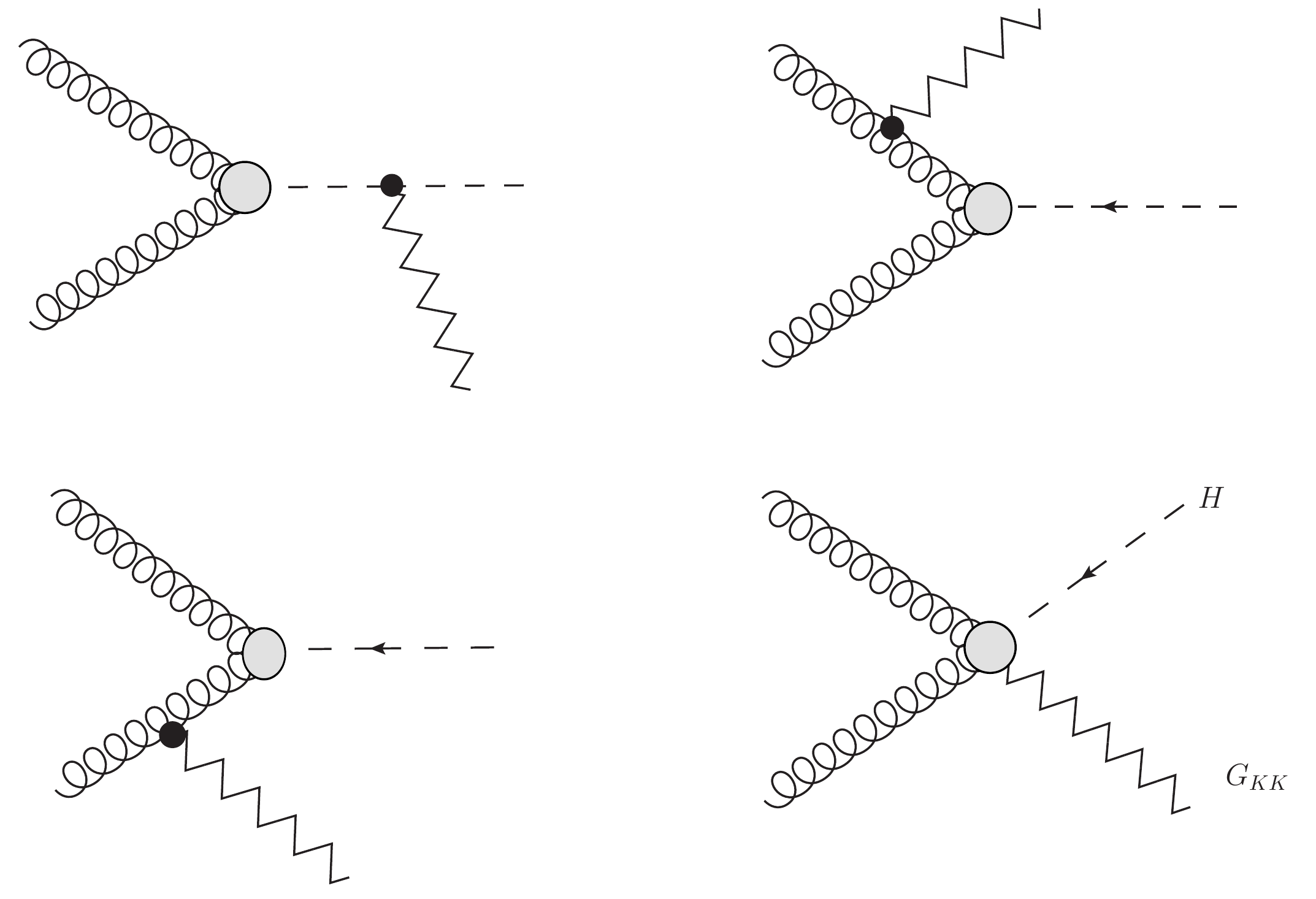}
\caption{Feynman diagrams for $\HG$ in the heavy top quark limit.}
\label{fig:HG-heft}
\end{center}
\end{figure}

In the heavy top quark limit ($m_t >> M_H/2$), the interaction of gluons with the Higgs boson
can be described by an effective Lagrangian~\cite{Kauffman:1996ix},
\begin{eqnarray}
 {\cal L}_{eff} = -\frac{1}{4} g_{eff} G^a_{\mu\nu} G^{a,\mu\nu}H,
\end{eqnarray}
where the effective coupling is
\begin{eqnarray}
 g_{eff} = \frac{\alpha_s}{3\pi v}\left[1+{\cal O}(M_H^2/4m_t^2)\right].
\end{eqnarray}
 It is known that the full calculation of the Higgs production via gluon fusion 
(including its radiative corrections), 
matches quite well with the calculation performed using this effective Lagrangian,
even for physical top quark mass $m_t=175$ GeV and $M_H =120$ GeV \cite{Dawson:1990zj,Djouadi:1991tka,Pak:2009dg}.
We can also use this Lagrangian, in the ADD model to calculate $\HG$ process. Furthermore, we
can compare this effective theory calculation with the full calculation. In the heavy top quark
limit, both calculations should be in complete agreement. In this case, the 
diagrams contributing to the process are displayed in the Fig.~\ref{fig:HG-heft}.
The Feynman rule for the last diagram is derived following Ref.~\cite{Han:1998sg} and it 
is added in the appendix~\ref{appendix:feynrules}.
In this effective theory, the amplitude-squared of $\HG$ process is 
% \begin{eqnarray}
\begin{flalign}
& \sum_{pol.}|\mathcal{M}(\HG)|^2 = \frac{1}{6 (M_H^2 - s)^2 t^2 u^2} 
 \Big[6 M_H^{12} (t + u)^2 + 7 s^2 t^2 u^2 (t + u)^2 - 12M_H^{10} \nn \\ 
  & ((t + u)^3 + 2 s (t^2 + t u + u^2)) +  6 M_H^8 ((t + u)^2 (t^2 + 4 t u + u^2) + 2 s^2 (3 t^2 + t u + 3 u^2) + \nn \\
  &   2 s (t + u) (3 t^2 + 4 t u + 3 u^2)) - 4 M_H^2 s t u (3 t u (t + u)^2 + 3 s^2 (t^2 + t u + u^2) + s (t + u) \nn \\
  &   (3 t^2 + 8 t u + 3 u^2)) - 12 M_H^6 (t u (t + u)^3 + 2 s^3 (t^2 + u^2) + 3 s^2 (t + u)(t^2 + t u + u^2) +  \nn \\
  &   s (t^4 + 7 t^3 u + 10 t^2 u^2 + 7 t u^3 + u^4)) + 2 M_H^4 (3 t^2 u^2 (t + u)^2 + 12 s t u (t + u)^3 +   3 s^4 \nn \\
  &   (t^2 + u^2) + 6 s^3 (t + u) (t^2 + t u + u^2) + s^2 (3 t^4 + 30 t^3 u + 44 t^2 u^2 + 30 t u^3 + 3 u^4))\Big], 
\end{flalign}
% \end{eqnarray}
where a summation over external polarizations is included. Here $s,t$ and $u$ are Mandelstam variables and 
they satisfy, $s+t+u = M_H^2 + M_{\rm KK}^2$. 

%%%%%%%%%%%%%%%%%%%%%%%%%%%%%%%%%%%%%%%%%%%%%%%%%%%%%%%%%%%%%%%%%%%%%%%%%%%%%%%%%%%%%%%%%%%%%%%%%%%%%%%%%%%%%%
% \new page
\section{Numerical Results}
\begin{figure}[ht]
\begin{center}
\includegraphics[width=8cm]{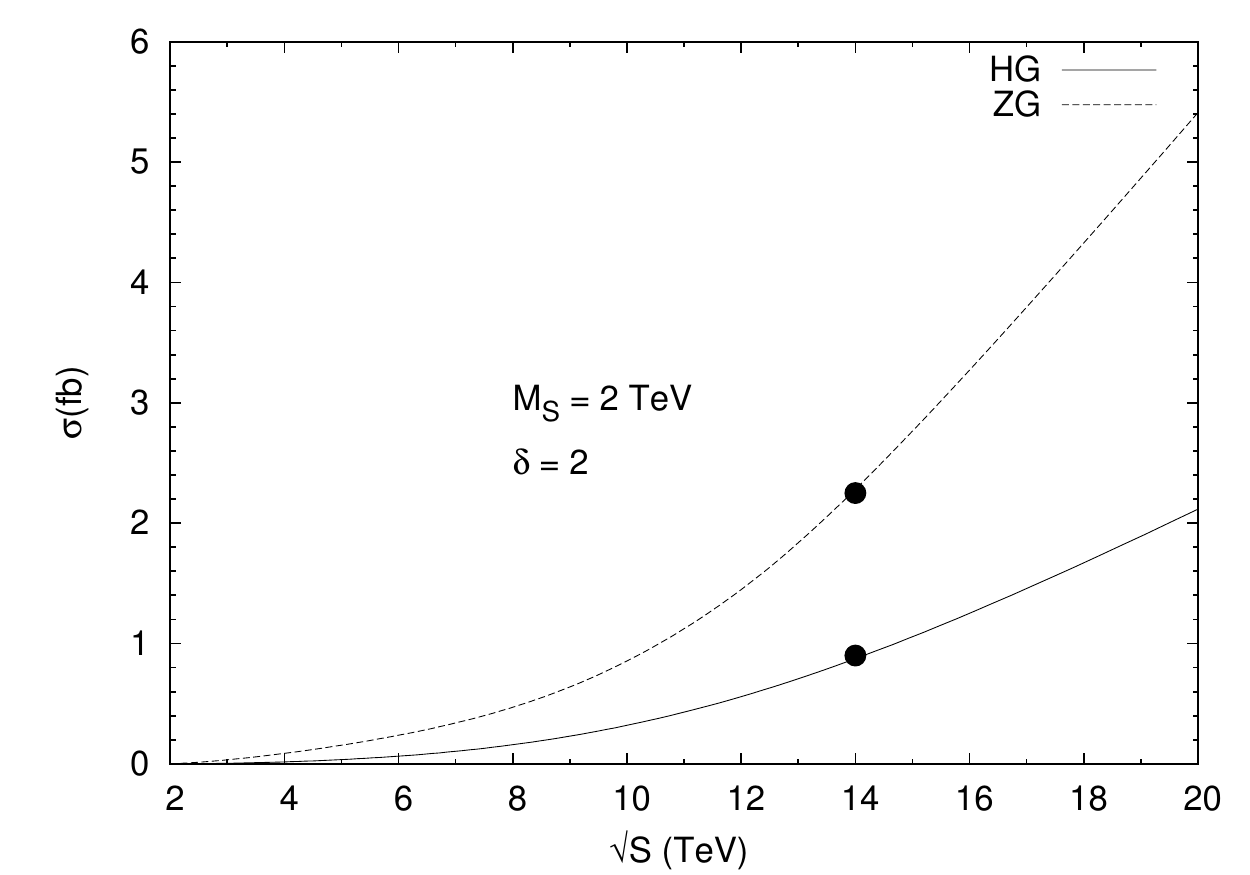}
\caption{Collider energy dependence of the hadronic cross sections for $gg\to H/Z G_{\rm KK}$ at the LHC.}
\label{fig:sigma_cme}
\end{center}
\end{figure}

In this section, we present results for $gg \to H/ZG_{\rm KK}$ processes at the LHC.
The results depend on the two parameters of the ADD model - (i) the number of extra-space 
dimensions $\delta$ and (ii) the fundamental scale of Gravity $M_S$. We will study this dependence and 
other features of our processes in the following. In Fig.~\ref{fig:sigma_cme}, we have plotted the hadronic
 cross sections as a function of the collider centre-of-mass energy for both the processes. Due to a large gluon flux
at higher energies, the cross sections also increase. Here we have 
chosen $\delta=2$ and $M_S = 2$ TeV. This combination of the model parameters has already been 
ruled out. Nevertheless, for a study of the qualitative features of many of the results presented here, 
this combination is as good as any other combination. In addition, we have applied following kinematic cuts:
\begin{eqnarray}
p_T^H > 20\;{\rm GeV},\;p_T^Z > 30\; {\rm GeV},\: |\eta^{H/Z}| < 2.5 ,\:\sqrt{s} < M_S.
\end{eqnarray}
\begin{figure}[h!]
 \begin{minipage}[b]{0.5\linewidth}
\centering
\includegraphics[width=\textwidth]{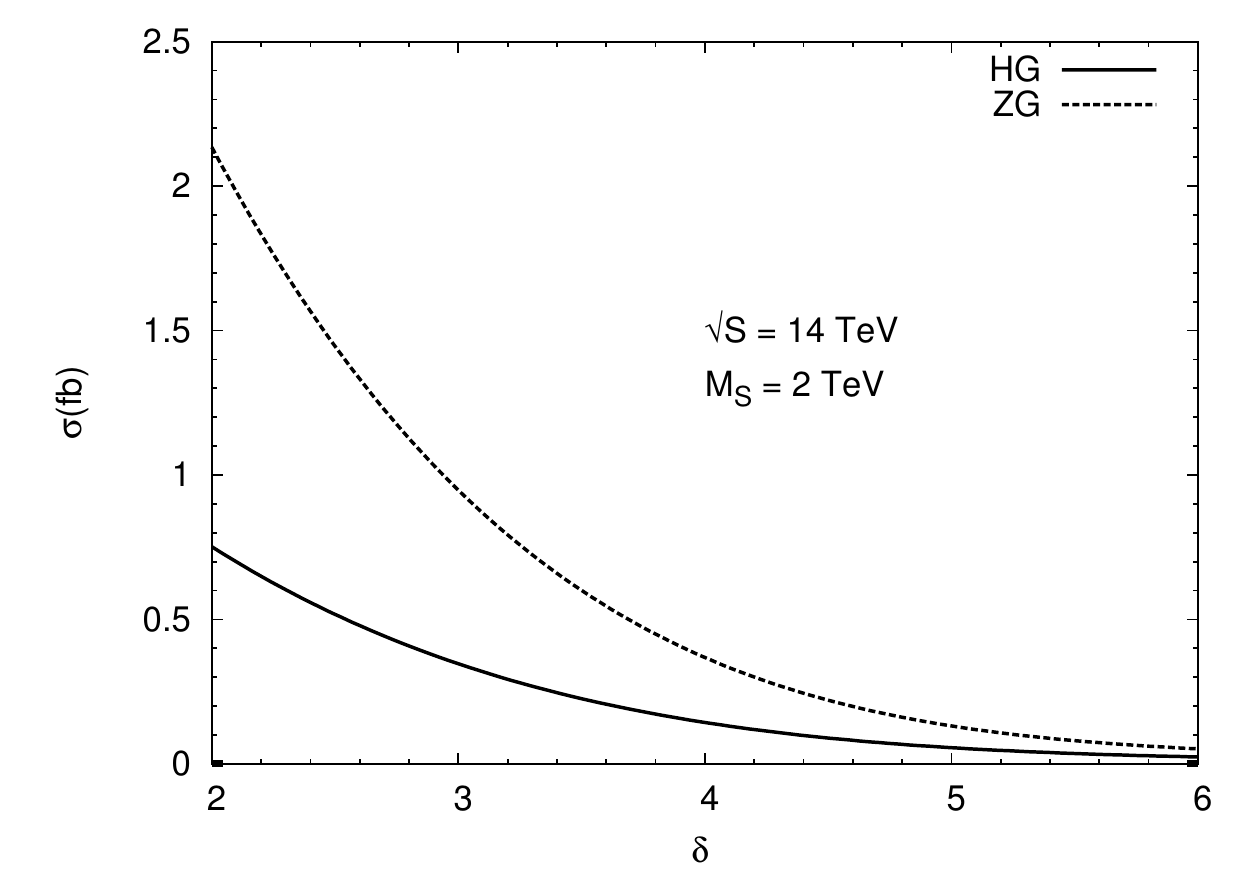}
\caption{Dependence of the cross sections on the number of extra dimensions $\delta$, for the scale $M_S = 2 $ TeV.}
\label{fig:sigma_nxd}
 \end{minipage}
 \hspace{0.5cm}
 \begin{minipage}[b]{0.5\linewidth}
\centering
 \includegraphics[width=\textwidth]{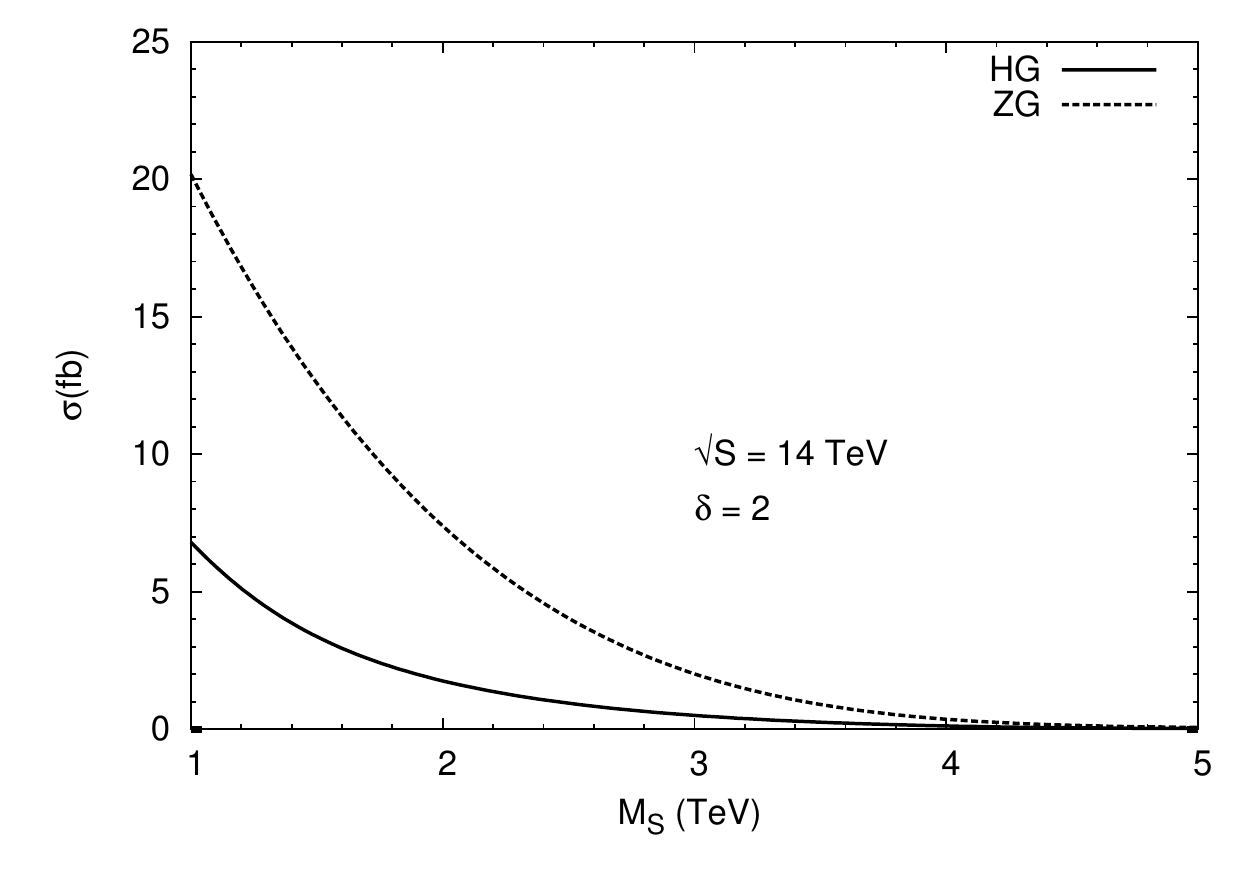}
\caption{Dependence of the cross sections on the scale $M_S$, for the number of extra dimension $\delta = 2$}
\label{fig:sigma_msc}
\end{minipage}
\end{figure}
The cut on the partonic centre-of-mass energy or equivalently on the $HG_{\rm KK}$/$ZG_{\rm KK}$ invariant mass 
is known as a {\it truncated scheme} in the literature. This cut is related to the fact that theoretical
 predictions within the ADD model, which is an effective field theory, can be valid only below the 
fundamental scale $M_S$. It will be interesting to probe the sensitivity of our predictions
on this kind of a constraint. We will further comment on the issue at the end of this section. 
Since our gluon fusion processes are finite, the partonic cross sections do not depend on
the factorization scale $\mu_f$. Also, their dependence on the renormalization scale $\mu_r$ is 
only through the strong coupling parameter $\alpha_s$. We have chosen the transverse energy 
($E_T\; =\sqrt{M^2 + (p_T)^2}$) of the weak bosons $H/Z$, as the common scale for the $\mu_f$ and 
$\mu_r$. In principle, we can work with both the LO and NLO PDFs.
We have used the LO CTEQ6L1 PDF, in the $HG_{\rm KK}$ case and the NLO CTEQ6M PDF, in $ZG_{\rm KK}$ 
case~\cite{Nadolsky:2008zw}. 
We note that at the 14 TeV LHC energy, the cross sections are 0.75 fb and 2.13 fb for the $HG_{\rm KK}$
and $ZG_{\rm KK}$ cases, respectively. The $HG_{\rm KK}$ cross section is much smaller than expected. The 
smallness of the cross section is due to two-orders of magnitude cancellation in the amplitude 
between the box and the triangle contributions. This destructive interference occurs due to the 
relative minus sign between the two contributions. However, the triangle and the box amplitudes are 
not separately gauge invariant. A similar cancellation is also seen in the $ZG_{\rm KK}$ 
amplitude. In the $HG_{\rm KK}$ 
case, we find that the bottom quark loop contribution to the cross section is less than a percent.
In the $ZG_{\rm KK}$ case, we have also calculated the hadronic cross section for 
a $p_T^Z >$ 400 GeV to avoid the SM background as suggested in \cite{Giudice:1998ck}.
We find that the cross section is about 0.2 fb and it is almost $10 \%$ of the NLO QCD correction 
calculated in \cite{Kumar:2010kv, Kumar:2010ca}. 
\begin{figure}[ht]
 \begin{minipage}[b]{0.5\linewidth}
\centering
\includegraphics[width=\textwidth]{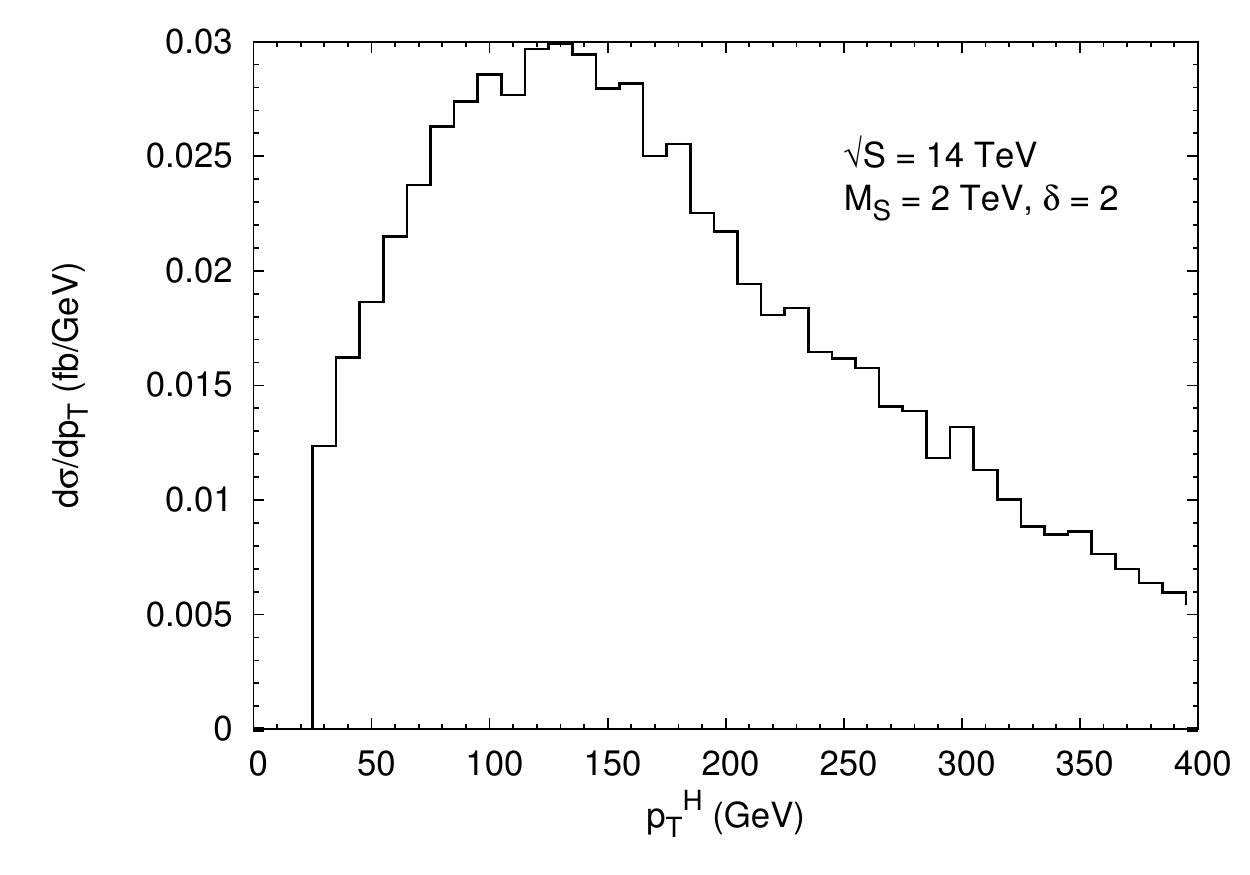}
\caption{Transverse momentum distribution of the Higgs boson for $M_S = 2$ TeV and $\delta = 2$.}
\label{fig:sigma_ptH}
 \end{minipage}
 \hspace{0.5cm}
 \begin{minipage}[b]{0.5\linewidth}
\centering
 \includegraphics[width=\textwidth]{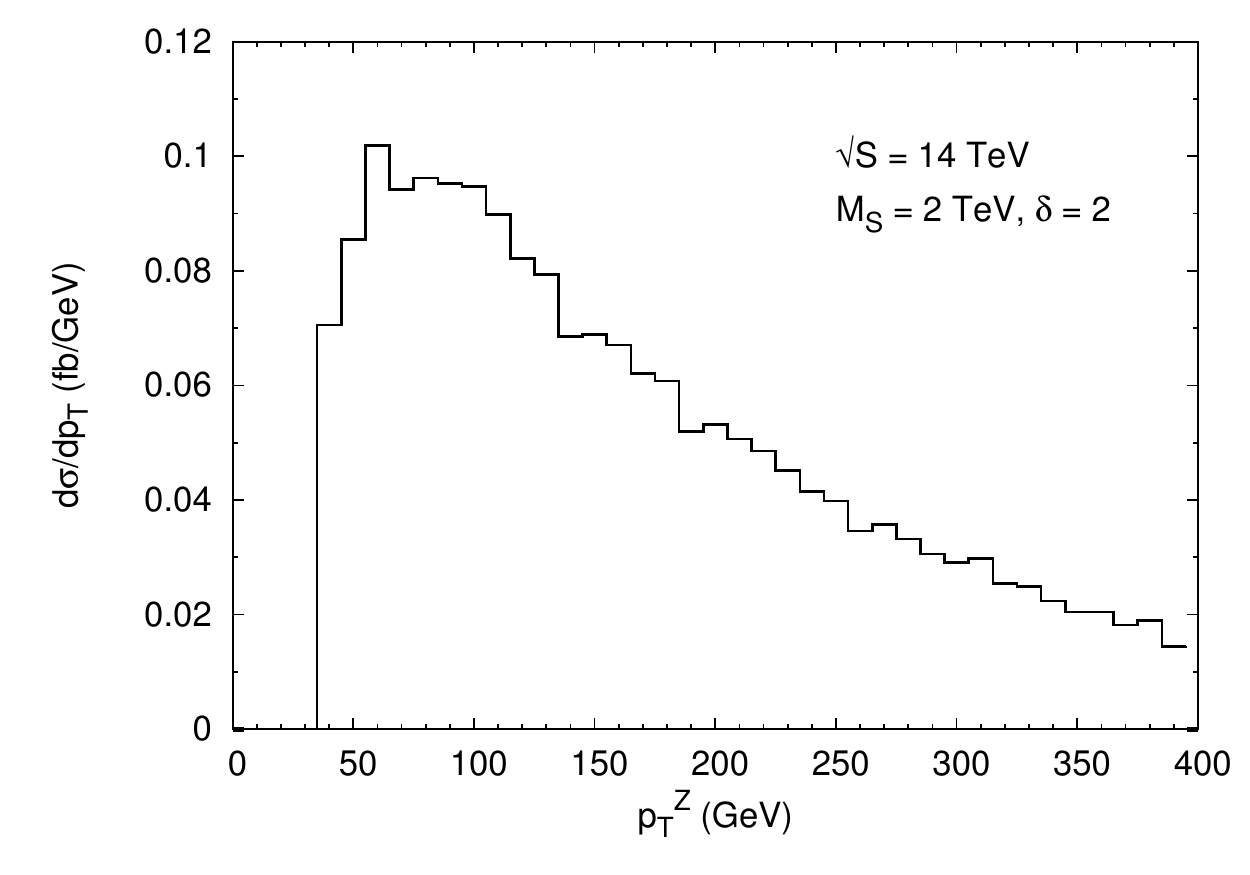}
\caption{Transverse momentum distribution of the $Z$ boson for $M_S = 2$ TeV and $\delta = 2$.}
\label{fig:sigma_ptZ}
\end{minipage}
\end{figure}

\begin{figure}[ht]
 \begin{minipage}[b]{0.5\linewidth}
\centering
\includegraphics[width=\textwidth]{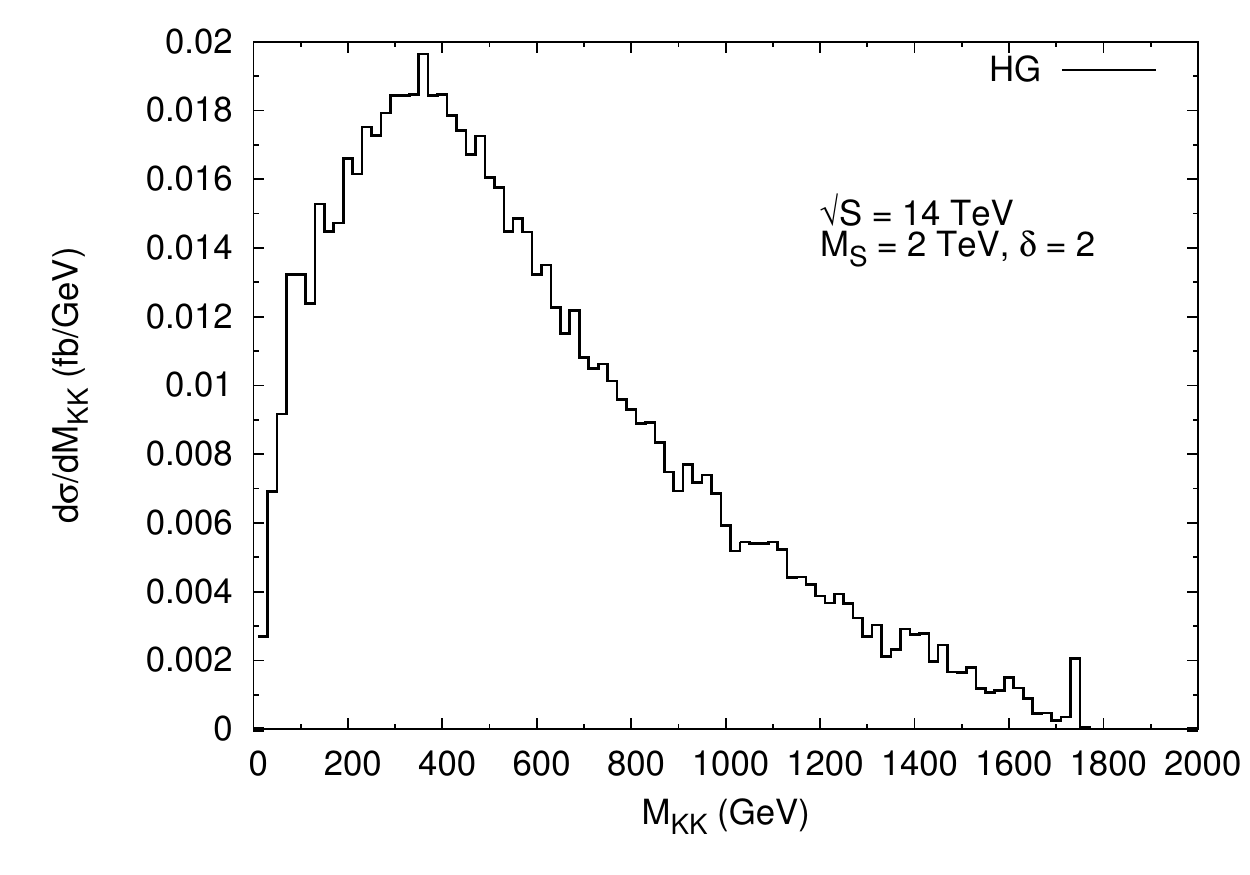}
\caption{KK-graviton mass distribution in $HG_{\rm KK}$ case, for $M_S = 2$ TeV and $\delta = 2$.}
\label{fig:sigma_mkkHG}
 \end{minipage}
 \hspace{0.5cm}
 \begin{minipage}[b]{0.5\linewidth}
\centering
 \includegraphics[width=\textwidth]{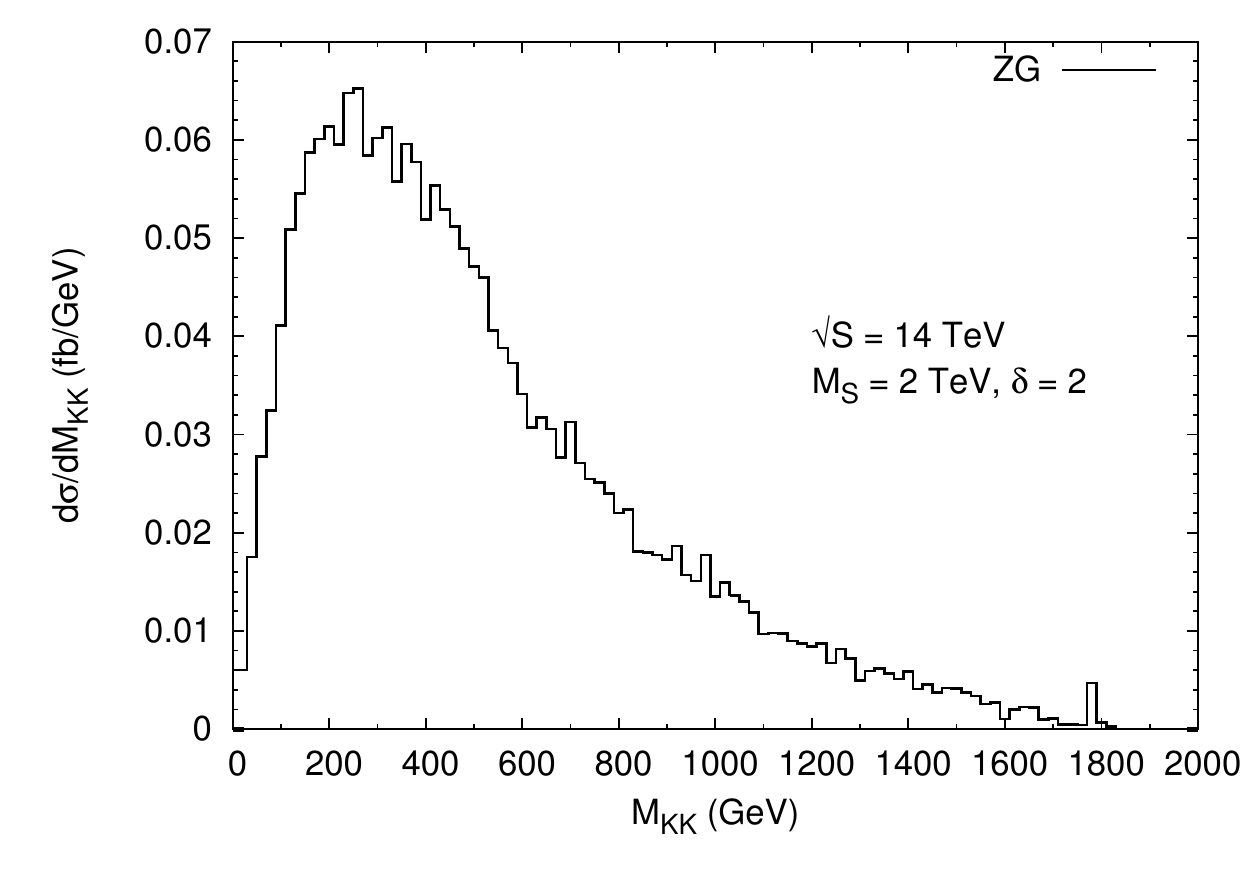}
\caption{KK-graviton mass distribution in $ZG_{\rm KK}$ case, for $M_S = 2$ TeV and $\delta = 2$.}
\label{fig:sigma_mkkZG}
 \end{minipage}
\end{figure}

We now illustrate various kinematic aspects of these processes at the 14 TeV centre-of-mass energy. 
In Figs. \ref{fig:sigma_nxd} and \ref{fig:sigma_msc}, we show the dependence of inclusive 
cross sections of the two processes on the ADD model parameters $\delta$ and $M_S$ . 
As $\delta$ or $M_S$ is increased, the density of states for KK-graviton modes falls (see Eq.\ref{eq:KK-density}), 
and therefore the 
cross sections also go down. The transverse momentum distributions of the Higgs and the 
$Z$ boson are plotted in Figs. \ref{fig:sigma_ptH} and \ref{fig:sigma_ptZ}, respectively.
The $p_T$ distributions are peaked about the masses of the weak bosons as one would expect. In the 
direct production processes of KK-graviton, all the kinematically allowed modes are produced.  
Figs. \ref{fig:sigma_mkkHG} and ~\ref{fig:sigma_mkkZG} show the KK-graviton mass distributions.
Since the density of KK-graviton modes increases with the KK-graviton mass $M_{\rm KK}$, the differential cross sections
also increase before the phase space suppression takes over. As the Eq.~\ref{eq:KK-density} suggests, the 
peak in the distributions will depend on the ADD model parameters $\delta$ and $M_S$.
Next, we study the scale dependence of the cross sections. We vary the common scale of the
factorization and renormalization around its central value, $\mu_0 = E_T^{H/Z}$. The cross sections
change by about $25-30\% $ by changing $\mu$ in the range between $\mu_0/2$ and $2\mu_0$. 
We find that the uncertainty in our calculations, due to the choice of different PDF sets, is 
in the range of $5-20 \% $ for the two processes.

\begin{figure}[h]
\begin{center}
\includegraphics[width=10cm]{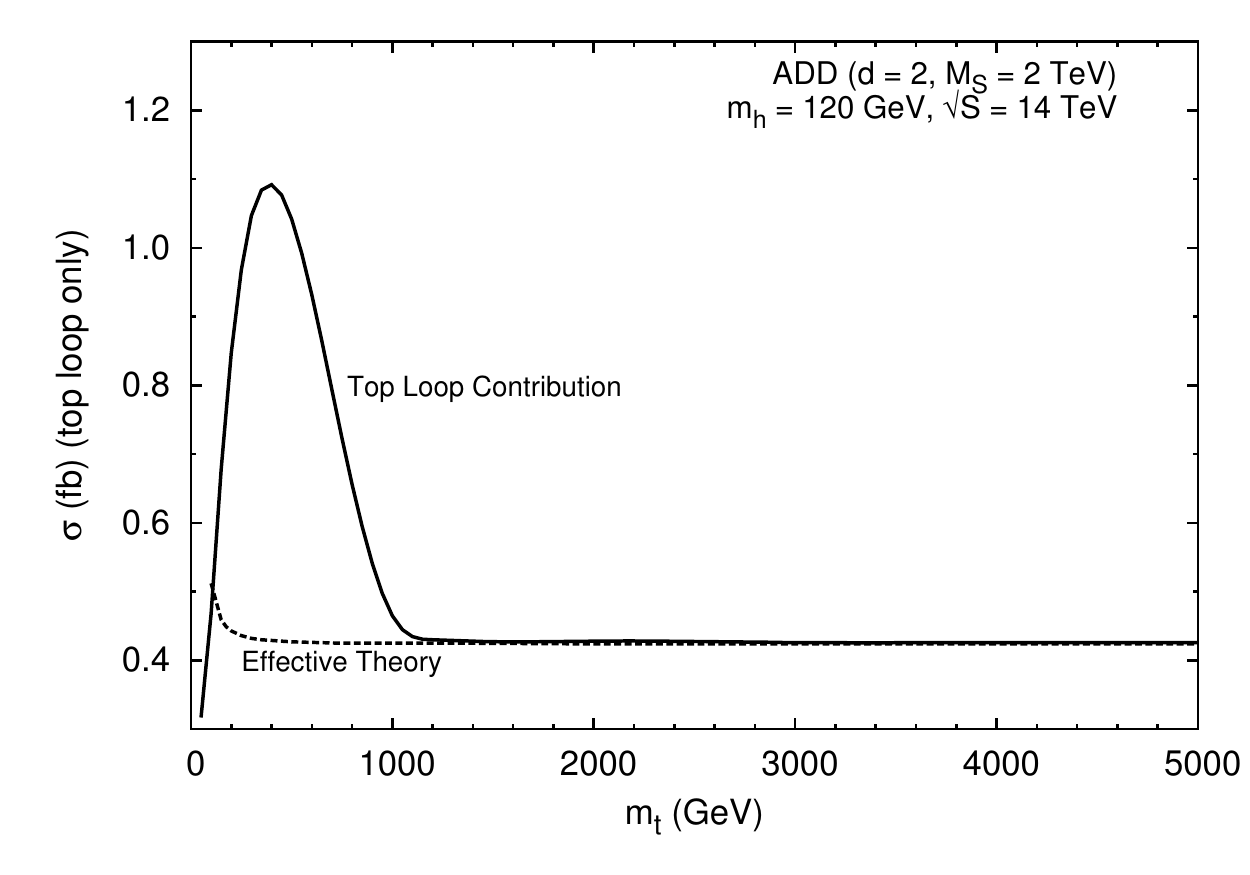}
\caption{A comparison of $\HG$ cross sections calculated in full theory and in the effective theory.}
\label{fig:sigma_mtHG}
\end{center}
\end{figure}

We discussed in the Sec.~\ref{section:HG-effective} that the process $\HG$ can also be calculated in an effective
theory of $ggH$-coupling.
The cross section calculations in the full theory and in the effective theory do not agree for 
$M_H=$120 GeV and $m_t = 175$ GeV. However, the two calculations agree very well for 
a very large top quark mass value ($m_t\ge 1.2$ TeV) as required. This is shown in Fig. ~\ref{fig:sigma_mtHG}. 
Note that unlike the
SM $gg\to H$ case, in our case, there is one extra scale present -- namely, the mass of the KK-graviton $M_{\rm KK}$.
We have seen that the  $M_{\rm KK}$ value is significantly larger than the mass of the top quark most 
of the time, see Fig.~\ref{fig:sigma_mkkHG}. Because of this, $\sqrt{s}$, which is larger than 
$M_H + M_{\rm KK}$, can go much beyond $2m_t$. Therefore,  one cannot expect the effective theory 
calculation to agree with the full calculation for $M_H=120$ GeV and the physical top quark mass $m_t = 175$ GeV. 
Finally, we comment on the UV sensitivity of our theoretical results. As mentioned before, 
we have presented all the results for which the partonic energy $\sqrt{s}$ is below the 
fundamental scale $M_S$, {\it i.e.}, the truncated scheme. It has been argued in Ref.~\cite{Giudice:1998ck} 
that, if relaxing this constraint does not change the results significantly, then results 
of the effective field theory can be trusted. We find that the cross sections in the truncated
and untruncated schemes 
(that is, with and without cut on $\sqrt{s}$) differ by $20 \%$ for $\delta=2$ and $M_S = 2$ TeV. 
This difference, as expected, increases for larger values of $\delta$ while it decreases with 
increasing $M_S$ \cite{Shivaji:2011re}.
%%%%%%%%%%%%%%%%%%%%%%%%%%%%%%%%%%%%%%%%%%%%%%%%%%%%%%%%%%%%%%%%%%%%%%%%%%%%%%%%%%%
%%%%%%%%%%%%%%%%%%%%%%%%%%%%%%%%%%%%%%%%%%%%%%%%%%%%%%%%%%%%%%%%%%%%%%%%%%%%%%%%
\section{A Discussion on $ZG_{\rm KK}$ Calculation}
The coupling of the KK-graviton with the quarks is such that the triangle and the box diagrams,
in the $ZG_{\rm KK}$ case, are linearly divergent in loop momentum and they have $VVA$ 
and $VVVA$ coupling structures respectively. The issue of chiral anomaly is well known in a linearly 
divergent fermion loop triangle diagram having $VVA$-structure. 
Because of the linear divergence, 
the $VVVA$-box diagrams also give anomalous contributions to the amplitude. This can be 
confirmed by checking the relations among charge-conjugated box diagrams, in the absence of a
suitable $n$-dimensional $\gamma^5$ prescription. The presence of anomaly affects the 
gauge invariance of 
the amplitude and for reliable predictions, our amplitude should be gauge invariant with 
respect to all the currents. 
We have already mentioned that if we do not regulate anomaly by using a suitable 
prescription for $\gamma^5$, the amplitude for the bottom/top quark in the loop, is not gauge 
invariant with respect to the gluons due to spurious anomalies. Even the relations among 
charge-conjugated diagrams and
the Bose symmetry of the amplitude do not hold. It turns out that the $n$-dimensional 
$\gamma^5$ prescription, given in Eq.~\ref{eq:g5-nd}, respects various symmetries of diagrams
and the amplitude. We have mentioned that, with this $\gamma^5$ prescription, both the bottom and top
quark contributions are separately gauge invariant with respect to the gluons and only the 
axial-vector current, corresponding to the $Z$ boson, is anomalous. The anomaly in axial-vector
current, being independent of the quark mass, also goes away in the full amplitude.  
See Eq.~\ref{eq:amp-ZG}. 
We have learned from our general discussion on the chiral anomaly that the 
anomalous contributions to the amplitude, including the spurious ones in 4 dimensions,
affect only the fermion mass independent rational part ${\cal R}$, see Sec.~\ref{section:rational}. 
The quark mass independence of the rational part, for an individual quark, is checked explicitly. 
In the full amplitude, shown in Eq.~\ref{eq:amp-ZG}, the rational terms will cancel between 
the bottom and the top quark contributions. Therefore, even if we work with 4-dimensional
$\gamma^5$, the full amplitude is going to be gauge invariant with respect to all the currents. Although,
the amplitude for an individual quark flavor in the loop will not be gauge invariant anymore, 
their difference in Eq.~\ref{eq:amp-ZG} is always gauge invariant. We have verified this in a 
separate calculation. All the numerical results,
presented above, agree with this way of doing calculation. It is definitely more economical because with 
4-dimensional $\gamma^5$ we require only one prototype box amplitude and two prototype triangle amplitudes
(one for each class) to generate the full amplitude. On the other hand, if we use $\gamma^5$ prescription 
given in Eq.~\ref{eq:g5-nd}, we need two prototype box amplitudes and four prototype triangle amplitudes
(two for each case) to generate the full amplitude. Also, with this $\gamma^5$ prescription, the 
trace calculation gives rise to a bigger expression for the amplitude of each diagram. 

%%%%%%%%%%%%%%%%%%%%%%%%%%%%%%%%%%%%%%%%%%%%%%%%%%%%%%%%%%%%%%%%%%%%%%%%%%%%%%%%%%%%%%%%%%%%%%%%%%%%%%%%%%%%%
%%%%%%%%%%%%%%%%%%%%%%%%%%%%%%%%%%%%%%%%%%%%%%%%%%%%%%%%%%%%%%%%%%%%%%%%%%%%%%%%%%%%%%%%%%%%%%%%%%%%%%%%%%%%%

% \newpage
% 
% \bibliographystyle{utcaps}
% \bibliography{thesis}
% 
% \end{document}

% Summary
\chapter{Summary}\label{chapter:summary}

In this thesis, we have considered two classes of gluon fusion processes which may be 
important at a high energy hadron collider such as the LHC. We 
have reported on the production of a pair of electroweak vector bosons with a jet via gluon 
fusion within the Standard Model. We have taken a model of large extra dimensions, the ADD model, as an 
example of new physics and have also considered the associated production of an electroweak boson  
and KK-gravitons. These gluon fusion processes receive contributions from the quark loop 
diagrams at the leading order and they are finite. The amplitude calculation is based on the 
traditional Feynman diagram approach. We have developed general purpose codes to perform 
the reduction of one-loop tensor integrals. All the basic scalar integrals which 
may appear in a one-loop amplitude are derived analytically and have been implemented in a 
FORTRAN routine. A flexible Monte Carlo integration routine based on the VEGAS 
algorithm is used to obtain the total as well as differential cross sections in these 
processes. To reduce the run time in the calculations of the SM processes, we have run 
the code in a parallel environment using the AMCI package.

We have verified our one-loop calculations by performing numerous 
checks on the amplitudes. We have checked the ultraviolet and infrared finiteness 
of the amplitudes. We 
have checked the structure of amplitudes by making gauge invariance checks. 
We have also verified the expected behavior of these amplitudes in the heavy 
quark mass limit. The general result regarding the infrared finiteness of an 
individual fermion loop diagram is also confirmed in these processes. 
We have shown  and verified that in a fermion loop amplitude, plagued from 
chiral anomaly, the anomaly is related to the rational part of the amplitude.
We have given a prescription for obtaining the correct rational part in a UV finite 
fermion loop amplitude utilizing the decoupling theorem.

In our study of the SM processes, we find that due to a large gluon flux available 
at the LHC, these processes are quite important. The typical hadronic cross 
section for the $\gpp$ is about 1 pb and it is about 10 $\%$ of the corresponding 
tree-level contribution. Like the $gg\to \gamma\gamma$ process, it is also important 
in the searches of a light Higgs boson. We find that the 
top quark loop contribution to the $gg \to \gamma\gamma g,\gamma Zg$ cross sections is negligible. 
For the processes $\gZZ$ and $\gWW$, we have kept only 
$\gamma Zg$-like contributions. Their cross sections 
are in the range of $4-15 \%$ of the corresponding tree-level cross sections. We have 
observed a qualitative similarity of these processes with the corresponding di-vector 
boson production cases. At the 14 TeV centre-of-mass energy, the cross sections of $\gVV$ 
processes are 20-30 $\%$
of those for the $\VV$ processes. In a detailed study of the $\gpZ$ process, we have  
compared this NNLO level contribution with the LO and the NLO predictions using the MCFM program. 
We note that the percentage contribution ($\sigma^{\rm NNLO}/\sigma^{\rm NLO}$) is about 
2-3$\%$ which can be enhanced by choosing
an appropriate set of kinematic cuts. The cross section of this process is dominated 
by the box contributions and therefore by the vector part of the amplitude. The axial-vector 
part of the amplitude contributes only about 10 $\%$ towards the cross section. 
Being leading order process, the 
scale uncertainty in the cross section calculation is governed by the strong coupling 
parameter $\alpha_s$ and the parton distributions, and it is quite large ($\sim$ 25-40 $\%$). 
The observability of this process at the LHC is discussed considering the decay of the
$Z$ boson into the charged leptons. We find that at 14 TeV and with 100 fb$^{-1}$ 
integrated luminosity, one can expect more than thousand events for 
$gg\to \gamma Z(\to l^+l^-)g$ process at the LHC. The issue of numerical instability 
in our calculations is dealt by systematically ignoring the contributions 
from the exceptional phase space points. We have adopted three different strategies to 
ignore their contributions, all in agreement within the allowed range of uncertainty. 
We have seen that such phase space points are very few and the contributions from 
such phase space points do not dominate the cross section.

In the ADD model processes, we have argued that the $gg\to\gamma G_{\rm KK}$ amplitude 
vanishes at the leading order due to an extension of Furry's theorem which includes 
the graviton. We find that, as expected, the gluon-gluon contribution to 
$pp\to HG_{\rm KK}+X$ dominates its cross section. For the model parameter values, 
$\delta=2$ and $M_S=$2 TeV, the hadronic cross section is only about 0.6 fb at 14 TeV LHC.
We have cross-checked our full one-loop calculation by working in an 
effective theory of gluon-Higgs coupling. In the effective theory, the process becomes 
a tree-level process. We find that for the physical top quark mass there is no agreement 
between the two calculations. This difference can be attributed to the fact that we have 
a large scale, the mass of the KK-gravitons, present in the theory. However, in the limit of 
a very heavy top quark the two results are in complete agreement as desired. In the case 
of $gg\to ZG_{\rm KK}$ process, we find that the amplitude gets contribution solely from the 
axial-vector part of the $Z$ boson coupling to a quark and the quarks of a massless/mass-degenerate 
generation do not contribute. Due to the nature of graviton-quark coupling, 
the box diagrams are also anomalous along with the triangle diagrams. We also learned that 
the anomaly should be regulated using a suitable $\gfive$ prescription in $n$ dimensions 
to ensure various symmetries of the diagrams and the amplitude. As expected, we find that the axial-vector current 
conservation in the amplitude holds only after including both the bottom and top quark contributions.
The typical cross section at 14 TeV is about 2 fb and it is about 10 $\%$ of the NLO cross section 
for $\delta=2$ and $M_S=$2 TeV. 
In both the $HG_{\rm KK}$ and $ZG_{\rm KK}$ cases, we find that there is a significant 
cancellation at the amplitude level between the triangle and the box contributions. The 
smallness of the cross sections for these processes, particularly in the $HG_{\rm KK}$ case, 
may be due to this cancellation.
The cross sections of these processes go down with increasing $\delta$ and/or $M_S$.
We see that the massive KK-graviton modes contribute significantly towards the inclusive cross section.
 Like in the case of the SM processes, we do observe a large uncertainty 
($\sim$ 5-20$\%$) due to the scale variation. To check the sensitivity of our results on the scale of 
Gravity $M_S$, we have calculated the cross section in both the truncated and untruncated schemes 
and the results differ by about $20 \%$.\\ 

% \emph{add an outlook}

 We have seen that at higher energies, the contributions of all these gluon fusion processes
increase. However, their cross sections suffer from large scale uncertainties as mentioned 
above. The scale uncertainties can be reduced by calculating radiative corrections to these
processes. The radiative corrections, which will also involve calculation of two-loop diagrams, 
are particularly important to our gluon fusion SM processes. Unlike the one-loop calculation, 
the calculation of two-loop amplitudes is not very common and its techniques are not 
yet standardized~\cite{Anastasiou:2001xx,Kosower:2011ty,Johansson:2012zv,Mastrolia:2012du}. 
Very few two-loop calculations of 
phenomenological importance, are available even for three and four-point functions. 
We have mentioned that beyond four-point function, the one-loop calculations are subject 
to numerical instabilities near exceptional phase space points. In the traditional approach of 
tensor reduction, this issue can be resolved to a certain degree by employing special expressions
for the reduction of higher point tensor integrals near such points~\cite{Denner:2005nn}. One 
may also use modern techniques (on-shell methods), based on the generalized unitarity cut, of calculating 
one-loop amplitudes~\cite{Berger:2006cz,Bern:2007dw,Ellis:2007br,Ellis:2011cr}. These techniques are results 
of many important ideas which 
have been developed over the years and due to them the automation of one-loop calculations, 
like that of tree-level calculations, seems 
feasible~\cite{Binoth:2005ff,Hirschi:2011pa,Cullen:2011ac,Bern:2012my,Rodgers:2012xx}. 
Like the case of $gg\to VV'$, 
the compact analytic expressions for $gg\to VV'g$ amplitudes can be calculated using these 
techniques~\cite{Dixon:1996wi,Bern:1997sc,Campbell:2011bn}. These analytic expressions will 
certainly reduce the computation time and can also be used for precision calculation at the LHC.
The SM gluon fusion processes $\gVV$ with soft jet form the real radiation 
part of the radiative correction to $\VV$ processes. Therefore, a full radiative 
correction of the LO $\VV$ processes can be a fruitful exercise~\cite{Bern:2001df}. I would 
like to conclude by quoting {\it L. D. Landau},
\begin{center}
{\it {\bf ``A method is more important than a discovery, \\
     since the right method will lead to new \\ 
     and even more important discoveries.''}}
\end{center}

 \setcounter{equation}{0}
 \setcounter{figure}{0}
\newpage
\begin{appendices}

%%%%%%%%%%%%%%%%%%%%%%%%%%%%%%%%%%%%%%%%%%%%%%%%%%%%%%%%%%%%%%%%%%%%%%%%%%%%%%%%%%%%%%%%%%%%%%%%%%%%%
\chapter{}
\section{Units, Conventions, Notations \& Definitions}\label{appendix:notation}
% 
% \begin{enumerate}
% 
%  \item 
{\bf Natural units and Conversion factors}
 \begin{eqnarray}
  \hbar &=& c = 1 \nn \\
  1\; {\rm GeV} &=& 1.783\times 10^{-24}\; {\rm g} \nn \\
  (1\; {\rm GeV})^{-1} &=& 0.1973\times 10^{-13}\;{\rm cm} \nn \\
  (1\; {\rm GeV})^{-1} &=& 0.658\times 10^{-24}\;{\rm sec} \nn \\
  1\;{\rm barn} &=& 10^{-24}\;{\rm cm^2} 
 \end{eqnarray}
%  
%   \item 
{\bf Metric}
 \begin{eqnarray}
  g^{\mu\nu} &=& g_{\mu\nu} = {\rm diag}(+1,-1,-1,-1) \nonumber \\ 
  p^\mu &=& (p^0,\vec{p}), \;\;\; p_\mu = g_{\mu\nu}p^\nu = (p^0,-\vec{p}) \nonumber \\ 
  p.k &=& p^\mu k_\mu = p^0k^0 - \vec{p}.\vec{k} \nn \\
  \partial^\mu &\equiv& \frac{\partial}{\partial x_\mu} = \left( \frac{\partial}{\partial t},-\vec{\nabla}\right),
 \end{eqnarray}
where $\vec{\nabla}$ is the gradient operator in three-dimensional space. 

{\noindent {\bf Pauli matrices}}
\begin{eqnarray}
\sigma^1 = \begin{pmatrix} 0 & 1 \\ 
                           1 & 0 \end{pmatrix},\;\;
\sigma^2 = \begin{pmatrix} 0 & -i \\ 
                           i & 0 \end{pmatrix},\;\;
\sigma^3 = \begin{pmatrix} 1 & 0 \\ 
                           0 & -1 \end{pmatrix}.
\end{eqnarray}

{\noindent {\bf Dirac matrices}}
\begin{eqnarray}
 \{ \gmu,\gnu \} &=& 2 g^{\mu\nu};\; \gmu = (\gamma^0,\gamma^i) \nn \\
  \{\gmu,\gfive\} &=& 0; \; \; \gfive = i \gamma^0\gamma^1\gamma^2\gamma^3 \nn \\
(\gmu)^\dagger &=& \gamma^0\gmu\gamma^0, \; (\gfive)^\dagger = \gfive  \nn \\
(\gamma^0)^2 &=& 1, \; (\gamma^i)^2 = -1, \; (\gfive)^2 = 1 \nn \\
 {\rm tr}(\gmu\gnu\gro\gsig) &=& 4(g^{\mu\nu}g^{\rho\sigma}-g^{\mu\rho}g^{\nu\sigma}
+g^{\mu\sigma}g^{\nu\rho}) \nn \\
 {\rm tr}(\gmu\gnu\gro\gsig\gfive) &=& -4i\;\veps^{\mu\nu\rho\sigma} \nn \\
 {\rm tr}(\gmu\gnu\gro\gsig.....) &=&  {\rm tr}(.....\gsig\gro\gnu\gmu) \nn \\
\gmu \gamma_\mu &=& n, \;
\gmu \gnu \gamma_\mu = -(n-2)\gnu 
\end{eqnarray}
where $\veps^{\mu\nu\rho\sigma}$ is completely antisymmetric tensor in 4 dimensions with following
properties
 \begin{eqnarray}
\veps^{0123} &=& -\veps_{0123} = +1, \nonumber \\
\veps^{\alpha\beta\mu\nu}\veps_{\alpha\beta\rho\sigma} &=& 
-2(\delta^\mu_\rho\delta^\nu_\sigma - \delta^\mu_\sigma\delta^\nu_\rho).
 \end{eqnarray}
{\bf Polarization sums}
\begin{eqnarray}
  {\rm spin\; \frac{1}{2}}:\;\; \sum_\lambda u^{\lambda}(p)\; \bar{u}^{\lambda}(p) &=& (\slashed{p} - m);\; 
\bar{u}^{\lambda}(p) = u^{\lambda}(p)^\dagger \gamma^0, \nn \\
 \sum_\lambda v^{\lambda}(p)\; \bar{v}^{\lambda}(p) &=& (\slashed{p} + m); \; {\slashed p} = \gmu p_\mu.
\end{eqnarray}
\begin{eqnarray}
 {\rm massless \; spin\;1}: \;\; \sum_\lambda \veps^{(\lambda)}_\mu(p) \; \veps^{*(\lambda)}_\nu(p) &\to& -g_{\mu\nu}, \nn \\
 {\rm massive \; spin\;1}: \;\; \sum_\lambda \veps^{(\lambda)}_\mu(p) \; \veps^{*(\lambda)}_\nu(p) 
                                &=& -g_{\mu\nu} + \frac{p_\mu\;p_\nu}{M^2}.
\end{eqnarray}
\begin{flalign}
 {\rm massive \; spin\;2}: \;\; \sum_\lambda \veps^{(\lambda)}_{\mu\nu}(p) \; \veps^{*(\lambda)}_{\rho\sigma}(p) 
                                =& \; \left(-g_{\mu\rho} + \frac{p_\mu\;p_\rho}{M^2}\right)
                                    \left(-g_{\nu\sigma} + \frac{p_\nu\;p_\sigma}{M^2}\right) \nn \\
                                &+ \left(-g_{\mu\sigma} + \frac{p_\mu\;p_\sigma}{M^2}\right)
                                    \left(-g_{\nu\rho} + \frac{p_\nu\;p_\rho}{M^2}\right) \nn \\  
                   &- \frac{2}{n-1}\left(-g_{\mu\nu} + \frac{p_\mu\;p_\nu}{M^2}\right)
                                    \left(-g_{\rho\sigma} + \frac{p_\rho\;p_\sigma}{M^2}\right).                              
\end{flalign}
% 
% \item 
{\bf Spin-1 propagator in $R_\xi$-gauge}
\begin{eqnarray}
 i D_{\mu\nu}(q^2,\xi) = \frac{i}{q^2-M^2} \left[ -g_{\mu\nu} +
                                                    \frac{q_\mu q_\nu}{q^2-\xi M^2} (1-\xi) \right].
\end{eqnarray}
Some of the very popular choices of the gauge-fixing parameter are
Landau gauge ($\xi = 0$), Feynman gauge ($\xi = 1$) and Unitary/Physical gauge ($\xi = \infty$).\\

% \item
%  \item 
{\noindent{\bf In the context of tensor reduction}}
\begin{eqnarray}\label{eq:appendixTR}
 \{v_i\} \; \rightarrow \; {\rm dual \; vectors \; to} \; \{p_i\},  \nn \\
 \{u_i\} \; \rightarrow \; {\rm dual \; vectors \; to} \; \{q_i\}, \nn \\
{\rm such \;that}\;\; v_i.p_j = u_i.q_j = \delta_{ij}, \;\; {\rm where} \;\; q_i = \sum_{j=1}^{i} p_i, \;\; q_0 = 0. 
\end{eqnarray}
% 
% \item
{\bf Gamma function}
\begin{eqnarray}
 \Gamma(z) &=& \int_0^\infty dt \; e^{-t}\; t^{z-1} \\
 \Gamma(1+z) &=& z\Gamma(z) \\
 \Gamma(\eps) &=& \frac{1}{\eps} - \gamma_E + {\cal O}(\eps)
\end{eqnarray}
where $\eps \to 0$ and $\gamma_E \simeq$ 0.5772157, is the Euler-Masheroni constant.

% \item 
{\noindent {\bf Integral parameterizations}}
\begin{eqnarray}
 \frac{1}{a^N} &=&  \frac{1}{\Gamma(N)}\int_0^\infty dt \; e^{-at}\; t^{N-1}
\end{eqnarray}
This is known as Schwinger parameterization. We can use this to obtain the Feynman parameterization,
\begin{eqnarray}
 \frac{1}{\prod_{i=1}^N a_i} &=&  \int_0^1 \prod_i dx_i \; \frac{(N-1)!}{(\sum_i a_ix_i)^N}\; 
\delta\left(\sum_ix_i-1\right).
\end{eqnarray}
{\bf $n$-dimensional integration}
\begin{eqnarray}
\int \frac{d^nl}{(2\pi)^n} \frac{1}{(l^2-M^2)^N} = i \frac{(-1)^N}{(4\pi)^{n/2}}
\frac{\Gamma(N-n/2)}{\Gamma(N)} \left(\frac{1}{M^2}\right)^{N-n/2}
\end{eqnarray}
% \item
{\bf $SU(N)$ algebra}\newline
The generators $T^a(a=1,2,..,N^2-1)$ of $SU(N)$ are hermitian, traceless 
matrices such that
\begin{eqnarray}
[T^a,T^b] &=& i f^{abc}\; T^c.
\end{eqnarray}
The numbers $f^{abc}$ are called the structure constants.
In the fundamental representation $T^a$ are $N$-dimensional matrices and they satisfy,  
\begin{eqnarray}
\{T^a,T^b\} &=& \frac{1}{N} \delta^{ab} +d^{abc}\; T^c, \\
{\rm with}, \;\; {\rm tr}(T^aT^b) &=& \frac{1}{2}\delta^{ab}.
\end{eqnarray}
Here $d^{abc}$ is totally symmetric in $a,b$ and $c$. Since 
${\rm tr}(T^aT^b) \propto \delta^{ab}$, it can be seen that $f^{abc}$ are 
totally antisymmetric in $a, b$ and $c$. Some useful identities are
\begin{eqnarray}
T^aT^b &=& \frac{1}{2N}\delta^{ab} + \frac{1}{2} d^{abc}T^c + \frac{1}{2} i f^{abc}T^c, \\
{\rm tr}(T^aT^bT^c) &=& \frac{1}{4}\left( d^{abc} + i f^{abc} \right), \\
f^{abc}f^{a'bc} &=& N\delta^{aa'}, \\
d^{abc}d^{a'bc} &=& \frac{(N^2-4)}{N}\delta^{aa'}, \;\; d^{aab} = 0.
\end{eqnarray}
In the adjoint representation, the generators $T^a$ are $(N^2-1)$-dimensional matrices and
\begin{eqnarray}
 \left( T^a \right)_{bc} = -if^{abc}.
\end{eqnarray}
% 
% \item 
{\bf Standard Model parameters}
\begin{eqnarray}
{\rm The\; Fermi\; constant},\; G_F &=& 1.166 \times 10^{-5}\;{\rm GeV}^{-2} \nn \\
M_Z &=& 91.1876\; {\rm GeV}, \;\; M_W = 80.403\; {\rm GeV} \nn \\
M_H &=& 125\; {\rm GeV}, \;\; m_b = 4.65\; {\rm GeV} \nn \\
m_t &=& 173.5\; {\rm GeV},\; {\rm sin^2{\theta_w}} = 0.23 \nn \\
\alpha_{em}^{-1}(M_Z^2) &=& 127.918,\; \alpha_s(M_Z^2) = 0.1176.
\end{eqnarray}

% 
% \end{enumerate}

%%%%%%%%%%%%%%%%%%%%%%%%%%%%%%%%%%%%%%%%%%%%%%%%%%%%%%%%%%%%%%%%%%%%%%%%%%%%%%%%%%%%%%%%%%%%%%%%%%%%%
\section{Logs and Di-Logs}\label{appendix:log-di-log}

The natural logarithm of a complex number is a multivalued function having a branch cut
along the negative real axis. For a complex number $z = r e^{i\theta}$,
\begin{eqnarray}
 {\rm ln}(z) &=& f_n(r,\theta) \nn \\
             &=& {\rm ln}(r) + i(\theta+2n\pi),
\end{eqnarray}
 with $r>0$ and $-\pi < \theta \le \pi$. The $n=0$ corresponds to the principal branch. The 
discontinuity of this function across the negative real axis can be seen by noting that,
\begin{eqnarray}
 \lim_{\eps \to 0} \left[f_n(r,\pi-\eps) - f_n(r,-\pi+\eps)\right]  = 2 i \pi.
\end{eqnarray}
Also, note that 
\begin{eqnarray}
 \lim_{\eps \to 0} \; f_n(r,\pi-\eps) =  \lim_{\eps \to 0} \;  f_{n+1}(r,-\pi+\eps), 
\end{eqnarray}
{\it i.e.}, the moment $\theta$ exceeds $\pi$ it goes to the next branch. Remember that
$f_n(r,\theta)$ is single valued on each branch with $-\pi < \theta \le \pi$. The rule for 
the logarithm of a product of two complex numbers $z_1$ and $z_2$ is,
\begin{eqnarray}
  {\rm ln}(z_1z_2) &=&  {\rm ln}(z_1) +  {\rm ln}(z_2) + 2i\pi\;\eta(z_1,z_2),
\end{eqnarray}
where
\begin{eqnarray}
  \eta(z_1,z_2) = \Theta(-{\rm Im}\;z_1)\; \Theta(-{\rm Im}\;z_2)\; \Theta({\rm Im}\;z_1z_2) \nn \\ 
                 - \Theta({\rm Im}\;z_1)\; \Theta({\rm Im}\;z_2)\; \Theta(-{\rm Im}\;z_1z_2).
\end{eqnarray} 

In the analytical evaluation of the one-loop scalar integrals, one often comes across the Spence function
or the di-logarithm. It is defined by~\cite{lewin:1981xx}
\begin{eqnarray}
 {\rm Sp}(z) &=& - \int_0^1 dt\; \frac{{\rm ln}(1-zt)}{t} = - \int_0^z dt\; \frac{{\rm ln}(1-t)}{t} \\
&=& z + \frac{z^2}{2^2} + \frac{z^3}{3^2} + ....\;\; {\rm for} \;\; |z| \le 1.
\end{eqnarray}
Since the branch cut of the logarithm is along the negative real axis, the branch cut of the
Spence function starts at $z=1$ along the positive real axis. Some of the special values are
\begin{eqnarray}
{\rm Sp}(-1) &=& -\frac{\pi^2}{12}, \; {\rm Sp}(0)=0, \nn \\
{\rm Sp}(1)&=&\frac{\pi^2}{6}, \; {\rm Sp}(2) = \frac{\pi^2}{4}.
\end{eqnarray}
We list few important identities related to Spence functions,
\begin{eqnarray}
- {\rm Sp}\left(1-z\right) &=& {\rm Sp}(z) + {\rm ln}(z) {\rm ln}(1-z) - \frac{\pi^2}{6} \\
 - {\rm Sp}\left(\frac{1}{z}\right) &=& {\rm Sp}(z) + \frac{1}{2} {\rm ln}^2(-z) + \frac{\pi^2}{6} \\
{\rm Sp}(z) + {\rm Sp}(-z) &=& \frac{1}{2}\; {\rm Sp}(z^2).
\end{eqnarray}

Some simple identities which could be useful in the evaluation of the one-loop scalar integrals are,
\begin{eqnarray}
{\rm ln}(a\pm \ieps) &=& {\rm ln}|a| \pm i\pi\;\Theta(-a)\\
{\rm ln}(a-\ieps\; a) &=& {\rm ln}(a+\ieps) \\
{\rm ln}(-a-\ieps\; a) &=& {\rm ln}(-a-\ieps )
\end{eqnarray}
\begin{eqnarray}
{\rm ln}(a+\ieps) + {\rm ln}(b-\ieps) &=& {\rm ln}\left[ab+\ieps\;{\rm sign}(b-a)\right]\nn \\
&=& {\rm ln}\left[ab+\ieps\;{\rm sign}(b)\right]\nn \\
&=& {\rm ln}\left[ab-\ieps\;{\rm sign}(a)\right] \\
% \end{eqnarray}
% \begin{eqnarray}
{\rm ln}\left(\frac{a}{b-\ieps}\right) &=& {\rm ln}\left(\frac{a}{b}+\ieps\; a\right) \nn \\
&=& {\rm ln}\left(\frac{a-\ieps}{b-\ieps}\right) \\
{\rm Sp}(a\pm \ieps\;a) &=& {\rm Sp}(a\pm \ieps) \\
{\rm Sp}(-a\pm \ieps\;a) &=& {\rm Sp}(-a\mp \ieps).
\end{eqnarray}
Here $a$ and $b$ are real numbers and $\eps$ is a vanishingly small positive number.

% \chapter{2 $\&$ 3-body Phase space}\label{appendix:phase-space}

%%%%%%%%%%%%%%%%%%%%%%%%%%%%%%%%%%%%%%%%%%%%%%%%%%%%%%%%%%%%%%%%%%%%%%%%%%%%%%%%%%%%%%%%%%%%%%%%%%
\chapter{}
\section{Simple analysis of infrared singularities at one-loop}\label{appendix:IR-structure}
Computation of virtual diagrams involve integration over undetermined loop momenta. These loop 
integrals may be ill-defined as they may diverge for certain limiting values of the momentum flowing 
in the loop. A well known case is that of the ultraviolet singularity which may arise as the loop momentum 
becomes very large. With massless particles present in the theory, these integrals may develop infrared (IR) 
or more appropriately mass singularities as well. Also, there are singularities which originate from 
specific phase space points such as physical and anomalous thresholds. A systematic study of the mass 
singularities and the threshold singularities can be done via Landau Equations 
\cite{Landau:1959fi,eden:2002xx,zuber:2005xx}. Kinoshita describes the mass singularities 
of Feynman amplitudes as pathological solutions of Landau Equations \cite{Kinoshita:1962ur}.

Here we limit ourselves to the IR singularity of one-loop diagrams, away from any threshold. 
Also we do not consider any exceptional phase space point, corresponding to the vanishing
of any partial sum of external momenta. According to Kinoshita, 
these are pure mass singularities, valid for all possible kinematic invariants made out of 
the external momenta. These mass singularities can be of soft or collinear type. 
Under certain circumstances, which we will discuss below, an overlapping of the
soft and collinear singularities may also occur.  
In Ref.~\cite{Kinoshita:1962ur}, this issue is discussed thoroughly, using a parametric form of 
the loop integrals. We will explore the structure of IR singularities of Feynman diagrams at 
the one-loop via naive power counting in the loop momentum. \\

The most general one-loop integral is of tensor type, in which the loop momentum appears in 
the numerator. One-loop diagrams with fermions in the loop are common examples of such 
integrals. Since any tensor integral at one-loop can be expressed in terms of scalar ones 
($i.e.$, the one-loop integrals in $\phi^3$-theory), it is sufficient to apply our analysis to
the scalar integrals only. 

%%%%%%%%%%%%%%%%%%%%%%%%%%%%%%%%%%%%%%
\subsection{Soft singularity}
\begin{figure}[h]
\begin{center}
\includegraphics [angle=0,width=0.4\linewidth] {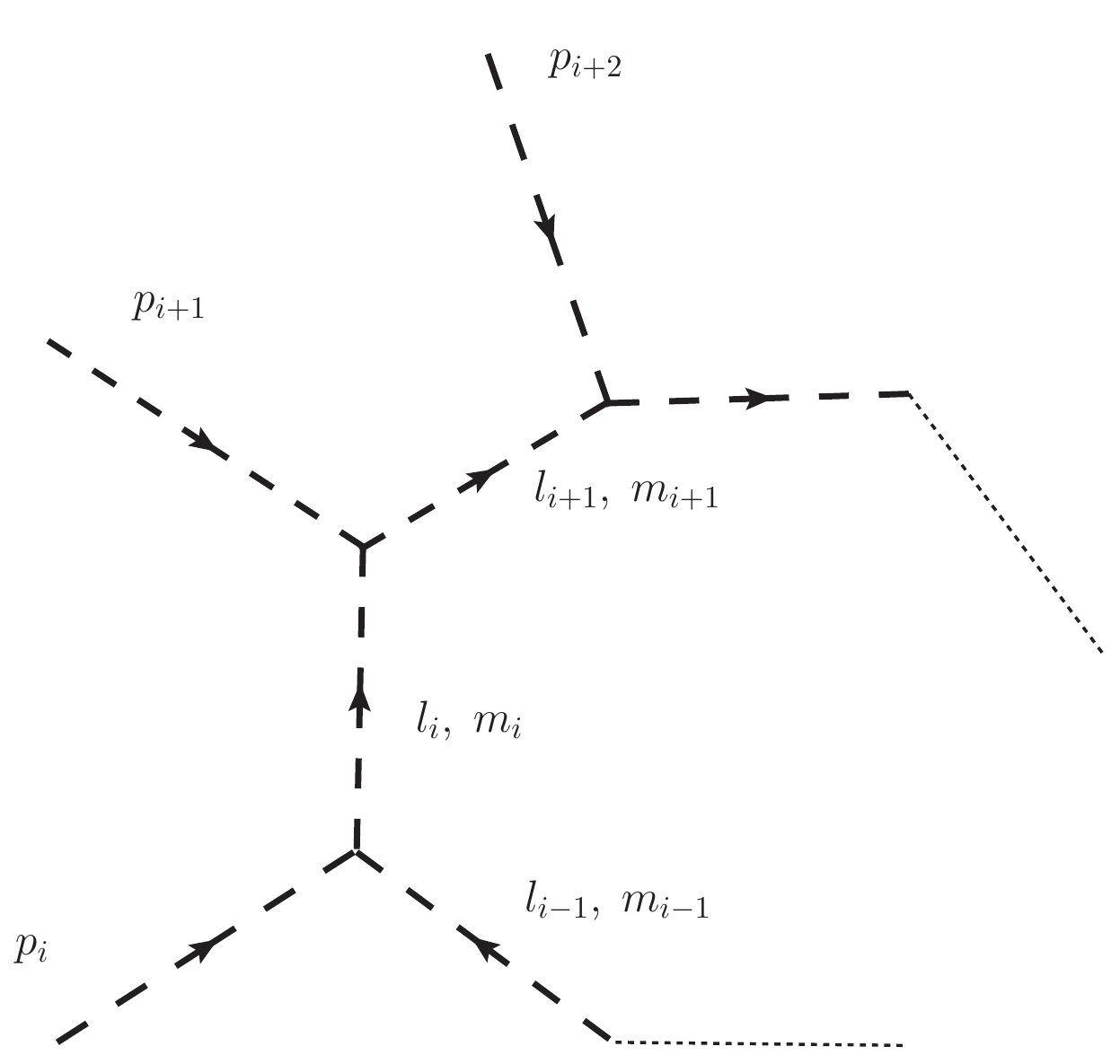}
\caption{ General scalar one-loop diagram with momentum assignment }
\label{fig:loopdiagram}
\end{center}
\end{figure}

\begin{figure}[h]
\begin{center}
\includegraphics [angle=0,width=0.3\linewidth] {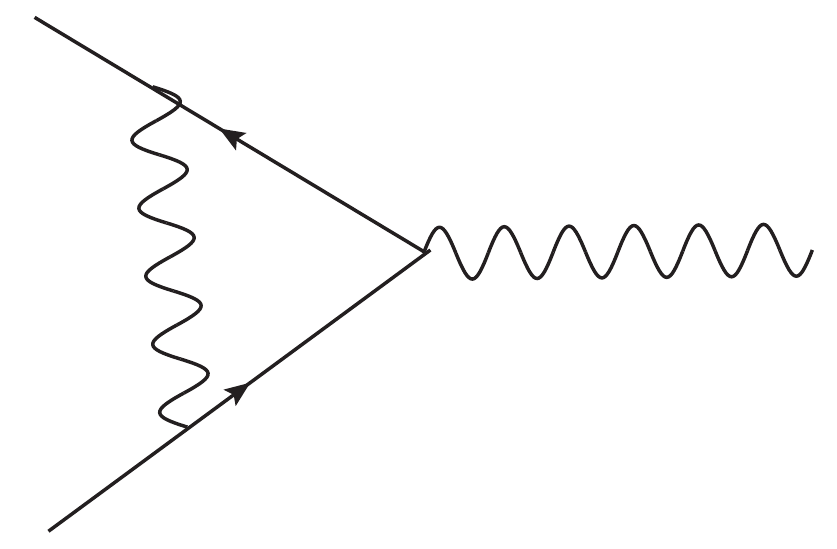}
\caption{ One-loop correction to QED vertex }
\label{fig:qed-vertex}
\end{center}
\end{figure}

These singularities appear as a result of any of the internal lines becoming soft, that is, 
its momentum vanishes. We consider the $N$-point scalar integral, Fig.~\ref{fig:loopdiagram}, 
in $n$ dimensions, given by
\begin{eqnarray}\label{eq:scalarloop}
  I^N (p_i,m_i;i=0\rightarrow N-1) = \int d^nl \; { \frac{1}{ d_0 d_1 ..... d_{N-1}}},
\end{eqnarray}
with following notations for simplification,
\begin{eqnarray}
 d_i &=& {l_i}^2 - {m_i}^2, \; l_i = l+q_i \;  \nonumber\\
 {\rm and} \;\; q_i &=& p_0+p_1+....+p_i.
\end{eqnarray}
In our notation, $p_0 = 0$, but to start with $m_0 \neq 0$. Now we 
wish to derive conditions under which above integral may diverge as one of the internal 
momenta in the loop, say $l_i$, becomes soft. We will take $l_i = \epsilon $, 
with the understanding that soft limit for $l_i$ is reached as $\epsilon \rightarrow 0$. 
With $l_i = \epsilon $, in our notation, relevant denominators take the following form 
\begin{eqnarray}
  d_{i-1} &=& p_i^2-2\epsilon \cdot p_i-m_{i-1}^2,\nonumber\\
  d_i &=& \epsilon^2-m_i^2   \nonumber\\
{\rm and} \;\; d_{i+1} &=& p_{i+1}^2+2\epsilon \cdot p_{i+1}-m_{i+1}^2.
\end{eqnarray}
We have dropped $\epsilon^2$ against $\epsilon\cdot p_i$ and $\epsilon\cdot p_{i+1}$, 
assuming $\epsilon$ is not orthogonal to $p_i$ and $p_{i+1}$. The above denominators 
vanish under the soft limit, if 
\begin{eqnarray}
 m_i=0, \; p_i^2 = m_{i-1}^2 \;\; {\rm and} \;\; p_{i+1}^2 = m_{i+1}^2.
\end{eqnarray}
Thus in the soft limit, the scalar integral in Eq.~\ref{eq:scalarloop}, behaves as
\begin{eqnarray}\label{eq:softloop}
 I^N \sim \int d^n\epsilon \; {1\over \epsilon\cdot p_i\; \epsilon^2 \; \epsilon\cdot p_{i+1} } \sim \epsilon^{n-4},
\end{eqnarray}
which diverges logarithmically in $n = 4$. In $m_i \rightarrow 0$ limit, the divergence appears as 
{\rm ln}($m_i$). Kinoshita called it $\lambda$-singularity. It is easy to check that 
no other denominator vanishes in the soft limit of $l_i$, in general. Thus, {\it the appearance of a soft 
singularity in one-loop diagrams is associated with the exchange of a massless particle between two 
on-shell particles}. The structure of the soft singularity in Eq.~\ref{eq:softloop} suggests that it can occur for 
$N\geq3$ point functions only. A text book example of soft-singular integral is the one-loop vertex 
correction in QED with massive fermions, as shown in Fig.~\ref{fig:qed-vertex}. 
%%%%%%%%%%%%%%%%%%%%%%%%%%%%%%%%%%%%%%%%
\subsection{Collinear singularity}
At one-loop, collinear singularity may appear when one of the internal momenta becomes 
collinear with a neighboring external leg. See Fig.~\ref{fig:loopdiagram}. We consider 
\begin{eqnarray}
 l_i = x p_{i+1} + \epsilon_\perp,
\end{eqnarray}
where $x\neq0,-1$ (since they correspond to the softness of $l_i$ and $l_{i+1}$, respectively) 
and $\epsilon_\perp\cdot p_{i+1}=0$. The collinear limit is obtained 
as $\epsilon_\perp \rightarrow 0$. In this case relevant denominators are,
\begin{eqnarray}
d_i &=& x^2 p_{i+1}^2 + \epsilon_\perp^2 - m_i^2  \nonumber\\
{\rm and}\;\; d_{i+1} &=& (x+1)^2 p_{i+1}^2 + \epsilon_\perp^2 - m_{i+1}^2 . 
\end{eqnarray}
The general conditions for these denominators to vanish are 
\begin{eqnarray}
 p_{i+1}^2=0, \; m_i = 0, \; m_{i+1}=0,
\end{eqnarray}
 that is, {\it one-loop diagrams in which a massless external leg meets two massless 
internal lines may develop a collinear divergence}. In fact, the first of the above three 
conditions is hidden in the assumption, $\epsilon_\perp\cdot p_{i+1}=0$. 
In $\epsilon_\perp \rightarrow 0$ limit, this equality implies
\begin{eqnarray}
 \cos\theta \simeq {p_{i+1}^0 \over \vert {\bf p_{i+1}} \vert} > 1,
\end{eqnarray}
for massive $p_{i+1}$. Thus for the collinear limit ($\theta \to 0 $ ), $p_{i+1}$ must be massless. 
No other denominator vanishes for non-exceptional phase space points. The $N$-point scalar integral 
in this limit goes as
\begin{eqnarray}
 I^N \sim \int d^n\epsilon_\perp \; {1\over \epsilon_\perp^2 \;\epsilon_\perp^2 } \sim \epsilon_\perp^{n-4},
\end{eqnarray}
which, like the soft singularity, is also logarithmically divergent. This singularity, sometimes 
referred as $m$-singularity, can be regularized by setting $m_{i+1} = m_i$ and taking 
$m_i \rightarrow 0$ limit. It appears as {\rm ln}($m_i$) term, as expected. Thus the one-loop QED
vertex correction diagram, shown in Fig.~\ref{fig:qed-vertex}, will also have a collinear type singularity
in the massless fermion limit.
% 
%%%%%%%%%%%%%%%%%%%%%%%%%%%%%%%%%%%%%%%
\subsection{Overlapping regions}
Since we have studied the structure of the soft and the collinear singularities of a one-loop diagram, 
it is desirable to seek the possibility of their overlap. A soft piece with all lines massless, 
contains two collinear pieces. Let this soft piece be the part of a scalar $N$-point function 
in $n$ dimensions. We set $l_i=\eps $, to write the $N$-point scalar integral for massless internal 
and external lines, keeping only (in general) potentially divergent denominators as 
$\eps \rightarrow 0 $.
\begin{eqnarray}\label{eq:overlap-loop}
 I^N \simeq \int d^n\eps \; {\frac{1}{(\eps^2-2\eps\cdot p_i)\; \eps^2 \;(\eps^2+2\eps\cdot p_{i+1})}}.
\end{eqnarray}
Instead of taking  $\eps \rightarrow 0 $ limit right away, which clearly corresponds to the soft limit, 
we wish to break the above integral into soft and collinear regions. This job is easily done in the 
light-cone coordinates as suggested in Ref.~\cite{Sterman:1995fz}. In this coordinate, a 4-vector is 
written as $ v \equiv (v^+,v^-,{\bf{v}_\perp}) $ where, $ v^\pm = \frac{1}{\sqrt{2}} (v^0\pm v^3)$ 
and ${\bf v_\perp} = (v^1,v^2)$ is a $2$-dimensional Euclidean vector. The dot product of two 4-vectors 
then takes following form
\begin{eqnarray}
 u\cdot v = u^+v^- + u^-v^+ - \bf{u_\perp}\cdot\bf{v_\perp}.
\end{eqnarray}
This can easily be generalized to $n$ dimensions. In the centre-of-mass frame of $p_i$ and $p_{i+1}$, 
using the light-cone variables, we may write
\begin{eqnarray}
 p_i &\equiv& \sqrt{2}\; \omega \;(1,0, {\bf0_\perp}), \;\;
 p_{i+1} \equiv \sqrt{2}\; \omega \;(0,1,\bf{0_\perp}).
\end{eqnarray}
Here $\bf{0_\perp}$ is a ($n-2$)-dimensional null vector in Euclidean space. Now the integral in 
Eq.~\ref{eq:overlap-loop}, reads
\begin{eqnarray}
 I^N \simeq \int  {\frac{ d\eps^+ \;d\eps^- \;d^{n-2}{\bf{\eps}_\perp}}
{(\eps^2-2\sqrt{2}\omega \eps^-) \; \eps^2 \;(\eps^2+2\sqrt{2}\omega \eps^+)}},
\end{eqnarray}
with $\eps^2 = 2\eps^+\eps^- - {\bf \eps_\perp}^2$. Notice that we can make $\eps$ and therefore $l_i$, 
collinear to $p_i$ by setting $\eps^-$ and $\bf{\eps}_\perp$ equal to zero. To do it more systematically, 
we choose $\eps^- = \lambda {\eps_\perp}^2 $ where $\lambda \ne 0$ and take $\eps_\perp \rightarrow 0$ 
limit in the above equation, so that the integral becomes
\begin{eqnarray}
  I^N \sim \int  {\frac{ d\eps^+ \;d\lambda \;{\bf{\eps_\perp}}^2 d^{n-2}{\bf{\eps}_\perp}}
{{\bf \eps_\perp}^2 {\bf \eps_\perp}^2 \eps^+}} =
\int { d\lambda\; \frac{d\eps^+}{\eps^+} \frac{d^{n-2}\eps_\perp}{{\eps_\perp}^2} }.
\end{eqnarray}
Once again we obtain the log-type collinear singularity in $n=4$ dimensions. Furthermore, 
if we take $ \eps^+ \rightarrow 0 $, 
we make $l_i$ soft and we see the overlapping of the soft and collinear singularities, which is also logarithmic 
in nature. Note that this singularity structure never gets worse for any other choice of $\eps^-$ made above.
It should be obvious that for non-exceptional phase space points there is no overlapping of two soft regions 
or two distant collinear regions in a one-loop diagram. Thus a one-loop IR divergent integral is written as 
sum of terms, each containing two large-log factors at the most. In dimensional regularization 
($n=4-2\eps_{IR}, \eps_{IR} \rightarrow 0^-$), IR singular terms at one-loop appear as coefficients of 
$ 1/ \epsilon_{IR}$ (soft and/or collinear case) and $1/\epsilon_{IR}^2$ (overlapping case) \cite{Ellis:2007qk}. 
In Fig.~\ref{fig:qed-vertex}, if we allow fermion lines to be massless, the one-loop correction to 
QED vertex exhibits the full structure of IR divergences that can appear in a general one-loop 
amplitude. \\

We have seen that for a one-loop diagram to have the IR singularity, at least one internal line must 
be massless. Diagrams with all the external legs off-shell are IR finite, even if all the internal lines 
are massless. 
The tensor integrals at most retain the IR-singular structure of the scalar integrals. Since, at maximum, three 
denominators of a general $N$-point scalar integral vanish at a time in infrared regions, one should 
expect the possibility of expressing IR-singular terms of any one-loop diagram ($N>3$), in terms of 
those of appropriate 3-point functions. In Ref.~\cite{Dittmaier:2003bc}, this expectation is achieved 
for the most general one-loop integral. We have also argued this in Sec.~\ref{section:IRfinite}. 
The above analysis also tells us that any one-loop diagram is IR finite in $n>4$ dimensions. 

%%%%%%%%%%%%%%%%%%%%%%%%%%%%%%%%%%%%%%%%%%%%%%%%%%%%%%%%%%%%%%%%%%%%%%%%%%%%%%%%%%%%%%%%%%%%%%%%%%%%%%%%
\section{One-loop scalar integrals}\label{appendix:scalars}

Evaluation of one-loop scalar integrals involves integration over undetermined loop 
momentum. We may encounter UV singularity in these integrals as loop momentum
becomes very large. Naive power counting suggests that only tadpole and bubble integrals
are UV divergent integrals. We use dimensional regularization ($n=4-2\eps$) to regulate UV 
singularities~\cite{collins:1986xx}. 
Although the tadpole has quadratic divergence while the bubble is log-divergent, in dimensional
regularization both type of UV singularities appear as $1/\eps$ pole. 
In this thesis, we are interested in fermion loop amplitudes. In many practical calculations, 
masses of fermions can be neglected. Presence of massless particles leads to another kind of 
singularity in the one-loop
scalar integrals, called the mass singularity. In 4 dimensions, these are of log-type as explained
in the previous section.
We have regularized these singularities by giving a small mass to the fermions. These scalar
integrals are derived in the special case of $m_i^2=m^2$, following 't Hooft and Veltman~\cite{'tHooft:1978xw}.
We maintain the full analytic continuation of these integrals.
Depending upon external virtualities there are many cases in each class of scalar integrals. 
\begin{enumerate}
 \item 
{\bf One-point (Tadpole) Scalar Integral}
\begin{eqnarray}
 A_0(m^2) &=& \frac{i}{16 \pi^2} \; m^2 \; \left[1 + \Delta_m\right], \nonumber \\
\Delta_m &=& \frac{1}{\eps} -\gamma_E + {\rm ln}\left(\frac{4\pi\mu^2}{m^2}\right) + \cal{O}(\eps).
\end{eqnarray}
Clearly for massless internal line, the tadpole scalar integral vanishes. 
 \item
{\bf Two-point (Bubble) Scalar Integral}
\begin{eqnarray}
  B_0(p^2;m^2) = \frac{i}{16 \pi^2} \left[ \Delta_m + 2 - z\;{\rm ln}\left(\frac{z+1}{z-1}\right)\right]
\end{eqnarray}
where $z = \sqrt{1-4(m^2-\ieps)/p^2}$. In the massless limit, this
integral becomes
\begin{eqnarray}
 B_0(p^2;m^2 \to 0) = \frac{i}{16 \pi^2} \; \left[\Delta_m + 2 - {\rm ln}\left(\frac{-p^2-\ieps}{m^2}\right) \right],
\end{eqnarray}
but there is no mass singularity.
These expressions cannot be used in the case $p^2=0$, and therefore the integral should be separately 
evaluated for this particular case. The result is,
\begin{eqnarray}
 B_0(p^2=0;m^2) = \frac{i}{16 \pi^2} \; \Delta_m. 
\end{eqnarray}
For the case of $p^2=0$, the integral develops a collinear singularity in $m^2\to0$ limit. 
However, the same integral can be shown to vanish in dimensional regularization, when 
$p^2=0, m^2=0$. In the special case of massless fermion loop amplitudes, one can see 
a correspondence between the mass regularization and the dimensional regularization of 
infrared singularities. Thus for our purposes, $A_0$ and $B_0$ integrals are never mass singular.\\
 
Since the triangle and box scalar integrals are UV finite, we can evaluate them in 4 dimensions. 
In $m^2\to0$ limit,
these and higher point scalar integrals have the following general structure,
\begin{eqnarray}
 I_0^{1-loop} = \frac{i}{16 \pi^2} \; \frac{1}{K} \left[f_2\;{\rm ln}^2(m^2) + f_1\;{\rm ln}(m^2) + f_0\right].
\end{eqnarray}
In 4 dimensions, the knowledge of $A_0, B_0, C_0$ and $D_0$ integrals is sufficient. 
The coefficients $f_i$, in various IR divergent scalar integrals are as follows:
 \item
{\bf Three-point (Triangle) Scalar Integrals} 

\underline{Case I: $p_1^2=0; p_2^2=0; p_3^2 \neq 0$}
\begin{eqnarray}
 K &=& 2 p_3^2, \nonumber \\
 f_2 &=& 1, \nonumber \\
 f_1 &=& -2\;{\rm ln}(-p_3^2-\ieps), \nonumber \\
 f_0 &=& {\rm ln}^2(-p_3^2-\ieps).
\end{eqnarray}
For finite $m^2$, $f_1 = f_2 = 0$ and $f_0 = {\rm ln}^2\left[(z+1)/(z-1)\right]$. Here 
$z = \sqrt{1-4(m^2-\ieps)/p_3^2}$. 
 
\underline{Case II: $p_1^2=0; p_2^2\neq0; p_3^2 \neq 0$}
\begin{eqnarray}
 K &=& 2 (p_3^2-p_2^2), \nonumber \\
 f_2 &=& 0, \nonumber \\
 f_1 &=& -2\;{\rm ln}(-p_3^2-\ieps) + 2\;{\rm ln}(-p_2^2-\ieps), \nonumber \\
 f_0 &=& {\rm ln}^2(-p_3^2-\ieps)-{\rm ln}^2(-p_2^2-\ieps).
\end{eqnarray}

\underline{Case III: $p_1^2\neq0; p_2^2\neq0; p_3^2 \neq 0$}
\begin{eqnarray}
 K &=& p_1^2 + p_2^2(1-2\alpha) - p_3^2 , \nonumber \\
 f_2 &=& 0, \nonumber \\
 f_1 &=& 0, \nonumber \\
 f_0 &=& \sum_{i=1,2,3} (-1)^i \left[ {\rm Sp}\left(\frac{y_i-1}{y_i+\ieps_i}\right) 
- {\rm Sp}\left(\frac{y_i}{y_i-1-\ieps_i}\right) \right],
\end{eqnarray}
where  
\begin{eqnarray}
y_1 &=& \frac{p_1^2(1-\alpha) + \alpha p_3^2}{\alpha K},\; y_2 = \frac{\alpha}{\alpha-1} y_1, \nn \\
y_3 &=& \frac{p_1^2-\alpha p_2^2}{K} \;\;
{\rm and} \;\; \eps_i = \eps \;{\rm sign}(p_i^2). 
\end{eqnarray}
Also, $\alpha$ satisfies
\begin{eqnarray}
p_2^2 \alpha^2-(p_1^2+p_2^2-p_3^2)\alpha + p_1^2 = 0. 
\end{eqnarray}
 \item
{\bf Four-point (Box) Scalar Integrals} 

\underline{Case I: $p_i^2 = 0, \forall \; i = 1,2,3,4$} 
\begin{eqnarray}
 K &=& s_{12} s_{23}, \nonumber \\
 f_2 &=& 2, \nonumber \\
 f_1 &=& -2\;{\rm ln}(-s_{12}-\ieps)-2\;{\rm ln}(-s_{23}-\ieps), \nonumber \\
 f_0 &=& 2\;{\rm ln}(-s_{12}-\ieps)\; {\rm ln}(-s_{23}-\ieps) - \pi^2.
\end{eqnarray}

\underline{Case II: $p_1^2=0; p_2^2=0; p_3^2 = 0; p_4^2 \neq 0$}
\begin{eqnarray}
 K &=& s_{12} s_{23}, \nonumber \\
 f_2 &=& 1, \nonumber \\
 f_1 &=& -2\;{\rm ln}(-s_{12}-\ieps)-2\;{\rm ln}(-s_{23}-\ieps)+2\;{\rm ln}(-p_4^2-\ieps), \nn \\
 f_0 &=& 2\;{\rm Sp}(1+\alpha s_{12}+\ieps) + 2\;{\rm Sp}(1+\alpha s_{23}+\ieps)
       - 2\;{\rm Sp}(1+\alpha p_4^2+\ieps) + \nn \\
        && {\rm ln}^2(-s_{12}-\ieps) + {\rm ln}^2(-s_{23}-\ieps) - {\rm ln}^2(-p_4^2-\ieps) - \frac{\pi^2}{3},
\end{eqnarray}
where $\alpha = (p_4^2 - s_{12} - s_{23})/K $ and $s_{ij} = (p_i+p_j)^2$.

\underline{Case III: $p_1^2=0; p_2^2\neq0; p_3^2 = 0; p_4^2 \neq 0$}
\begin{eqnarray}
 K &=& s_{12} s_{23} - p_2^2 p_4^2, \nonumber \\
 f_2 &=& 0, \nonumber \\
 f_1 &=& -2\;{\rm ln}(-s_{12}-\ieps)-2\;{\rm ln}(-s_{23}-\ieps)+2\;{\rm ln}(-p_2^2-\ieps)
         +2\;{\rm ln}(-p_4^2-\ieps), \nn \\
 f_0 &=& 2\;{\rm Sp}(1+\alpha s_{12}+\ieps) + 2\;{\rm Sp}(1+\alpha s_{23}+\ieps)- \nn \\
        && 2\;{\rm Sp}(1+\alpha p_2^2+\ieps) -2\;{\rm Sp}(1+\alpha p_4^2+\ieps)+ \nonumber \\
        && {\rm ln}^2(-s_{12}-\ieps) + {\rm ln}^2(-s_{23}-\ieps) - {\rm ln}^2(-p_2^2-\ieps) 
         - {\rm ln}^2(-p_4^2-\ieps),
\end{eqnarray}
where, $\alpha = (p_2^2+p_4^2 - s_{12} - s_{23})/K$. The above expressions of the box scalar integrals, 
can be compared with those derived in Ref.~\cite{Duplancic:2000sk} \footnote{ In Ref.~\cite{Duplancic:2000sk}, the 
IR divergent box integrals are derived in the dimensional regularization and their expressions contain terms
which are artifacts of the dimensional regularization. In an IR finite amplitude such terms do not contribute.}.
For the next two cases, the box scalar integrals that we have derived following 't Hooft and Veltman, 
have very complicated structure and we do not give those expressions here. The divergent coefficients $f_1$ and 
$f_2$ are derived using the identity given in Eq.~\ref{eq:IR-D0toC0}, while the expressions for $f_0s$ are 
taken from Ref.~\cite{Duplancic:2000sk}.
 
\underline{Case IV: $p_1^2=0; p_2^2=0; p_3^2 \neq 0; p_4^2 \neq 0$}
\begin{eqnarray}
 K &=& s_{12} s_{23}, \nonumber \\
 f_2 &=& \frac{1}{2}, \nonumber \\
 f_1 &=& -\;{\rm ln}(-s_{12}-\ieps)-2\;{\rm ln}(-s_{23}-\ieps)+ \;{\rm ln}(-p_3^2-\ieps)
        + \;{\rm ln}(-p_4^2-\ieps),\nn \\
  f_0 &=& 2\;{\rm Sp}(1+\alpha s_{12}+\ieps) + 2\;{\rm Sp}(1+\alpha s_{23}+\ieps) 
        - 2\;{\rm Sp}(1+\alpha p_4^2+\ieps) +\nn \\
      &&  2\;{\rm Sp}\left(1-\frac{p_3^2+\ieps}{s_{23}+\ieps}\right) 
        - 2\;{\rm Sp}\left(1-\frac{p_4^2+\ieps}{s_{12}+\ieps}\right) +\nonumber \\
      && {\rm ln}^2(-s_{12}-\ieps) + {\rm ln}^2(-s_{23}-\ieps) - {\rm ln}^2(-p_3^2-\ieps) 
       - {\rm ln}^2(-p_4^2-\ieps) +\nn \\
      && \frac{1}{2}\left[{\rm ln}(-p_3^2-\ieps) + {\rm ln}^2(-p_4^2-\ieps) -{\rm ln}(-s_{12}-\ieps) \right]^2,
\end{eqnarray}
where, $\alpha = (p_4^2 - s_{12} - s_{23})/K$.

\underline{Case V: $p_1^2=0; p_2^2\ne0; p_3^2 \neq 0; p_4^2 \neq 0$} 
\begin{eqnarray}
 K &=& s_{12} s_{23}-p_2^2p_4^2, \nonumber \\
 f_2 &=& 0, \nonumber \\
 f_1 &=& -\;{\rm ln}(-s_{12}-\ieps)-\;{\rm ln}(-s_{23}-\ieps)+\;{\rm ln}(-p_2^2-\ieps)
         +\;{\rm ln}(-p_4^2-\ieps),\nn \\
  f_0 &=& 2\;{\rm Sp}(1+\alpha s_{12}+\ieps) + 2\;{\rm Sp}(1+\alpha s_{23}+\ieps) 
        - 2\;{\rm Sp}(1+\alpha p_2^2+\ieps) -\nn \\
      &&  2\;{\rm Sp}(1+\alpha p_4^2+\ieps) + 2\;{\rm Sp}\left(1-\frac{p_3^2+\ieps}{s_{23}+\ieps}\right) 
        - 2\;{\rm Sp}\left(1-\frac{p_4^2+\ieps}{s_{12}+\ieps}\right) +\nonumber \\
      && {\rm ln}^2(-s_{12}-\ieps) + {\rm ln}^2(-s_{23}-\ieps) - {\rm ln}^2(-p_2^2-\ieps) 
       - {\rm ln}^2(-p_3^2-\ieps) -\nn \\
      && {\rm ln}^2(-p_4^2-\ieps) + \frac{1}{2}\left[{\rm ln}(-p_2^2-\ieps) + {\rm ln}^2(-p_3^2-\ieps) 
        -{\rm ln}(-s_{23}-\ieps) \right]^2 +\nn \\
      && \frac{1}{2}\left[{\rm ln}(-p_3^2-\ieps) + {\rm ln}^2(-p_4^2-\ieps) 
        -{\rm ln}(-s_{12}-\ieps) \right]^2,
\end{eqnarray}
where, $\alpha = (p_2^2 + p_4^2 - s_{12} - s_{23})/K$.
\end{enumerate}

The expressions of all the IR divergent 
scalar integrals, that we have derived in the mass regularization, can be compared with the complete list of 
divergent scalar integrals given in Ref.~\cite{Ellis:2007qk}. In this reference, the IR divergence of the scalar 
integrals is treated in the dimensional regularization ($n=4-2\eps$). We can see that there is a one-to-one 
correspondence (apart from terms that are artifacts of the dimensional regularization) between the results obtained 
in the two regularization schemes. The IR singularities, in the two regularization schemes are related as
\begin{eqnarray}
 \frac{1}{\eps} \leftrightarrow {\rm ln}(m^2),\;\;  \frac{2}{\eps^2} \leftrightarrow {\rm ln}^2(m^2).
\end{eqnarray}
We would also like to mention that the finite parts of the box scalar integrals which involve 
di-logs/Spence functions may lead to numerical instability in their present form, due to a large 
cancellation among di-log terms. This problem can be cured to a certain level by putting them in 
different forms using di-log identities. In this thesis, we have used only divergent pieces of 
these integrals to make finiteness checks on amplitudes. Finite numerical results are presented 
using the FF/ OneLOop library of scalar integrals \cite{vanOldenborgh:1990yc,vanHameren:2010cp}. 
 
For the case of massive internal lines, we are also interested in the large $m^2$ limit. In this limit, the basic 
scalar integrals are
\begin{eqnarray}
 B_0(m^2\to\infty) &=& \frac{i}{16 \pi^2} \;\left[\Delta_m + {\cal O}\left(\frac{1}{m^2}\right)\right], \\
 C_0(m^2\to\infty) &=&  \frac{i}{16 \pi^2} \; \left[-\frac{1}{2 m^2} + {\cal O}\left(\frac{1}{m^4}\right)\right], \\
 D_0(m^2\to\infty) &=&  \frac{i}{16 \pi^2} \; \left[\frac{1}{6 m^4} + {\cal O}\left(\frac{1}{m^6}\right)\right]. \label{eq:large-m-D0}
\end{eqnarray}
The non-leading terms will depend upon non-zero kinematic invariants of the scalars.

%%%%%%%%%%%%%%%%%%%%%%%%%%%%%%%%%%%%%%%%%%%%%%%%%%%%%%%%%%%%%%%%%%%%%%%%%%%%%%%%%%%%%%%%%%%%%%%%%%%%%%%%%%%%
\section{Derivation of pentagon scalar integral, $E_0$ }\label{appendix:E0}

We are interested in the case of equal masses for all the internal lines. This is not 
to simplify the derivation. The derivation can be carried out using 4-dimensional 
{\it Schouten identity}~\cite{vanNeerven:1983vr}. This identity is a result
of the statement that $\veps^{\mu_1\mu_2\mu_3\mu_4\alpha}$ is zero in 4 dimensions. Therefore 
antisymmetric combination of $\veps^{\mu_1\mu_2\mu_3\mu_4}$ and $l^{\alpha}$ should vanish. This implies,
\begin{eqnarray}
l^{\alpha} \veps^{\mu_1\mu_2\mu_3\mu_4} = l^{\mu_1} \veps^{\alpha \mu_2\mu_3\mu_4} + l^{\mu_2} \veps^{\mu_1\alpha \mu_3\mu_4}
                                      + l^{\mu_3} \veps^{\mu_1\mu_2\alpha\mu_4} + l^{\mu_4} \veps^{\mu_1\mu_2\mu_3\alpha} . 
\end{eqnarray}

Schouten identity can also be derived by evaluating, ${\rm tr}(\gamma^{\mu_1}\gamma^{\mu_2}\gamma^{\mu_3}\gamma^{\mu_4}\gamma^{\alpha}{\slashed l} \gamma^5)$
in 4 dimensions. In the case of 5-point function, we have four linearly independent momenta $p_1,p_2,p_3$
and $p_4$. In this derivation, it is more convenient to consider their linear combinations $q_1,q_2,q_3$ and $q_4$,
defined in chapter~\ref{chapter:oneloop}.
Multiplying the above equation by $q_1^{\mu_1}q_2^{\mu_2}q_3^{\mu_3}q_4^{\mu_4}$,
we get
\begin{eqnarray}
l^{\alpha} \veps^{q_1q_2q_3q_4} &=& l.q_1 \veps^{\alpha q_2q_3q_4} + l.q_2 \veps^{q_1\alpha q_3q_4}
                                      + l.q_3 \veps^{q_1q_2\alpha q_4} + l.q_4 \veps^{q_1q_2q_3\alpha} \nonumber \\
 {\rm or}\;\; l^\alpha &=& \sum_{i=1}^4 l.q_i\; u_i^\alpha \label{eq:l-alpha},
\end{eqnarray}
where $u_i$s are dual vectors of $q_i$s, satisfying $u_i.q_j = \delta_{ij}$ (clarify 
with Eq.~\ref{eq:v4}). This is the same expression 
as in Eq.~\ref{eq:l-mu}, with $m =4$ and $\omega^\mu_l = 0$ but with different basis vectors. 
Contracting with $l_\alpha$ both the sides in Eq.~\ref{eq:l-alpha}, we get
\begin{eqnarray}
 l^2 &=& \sum_{i=1}^4 l.q_i\; l.u_i \nonumber \\
\Rightarrow d_0 + m^2 &=& \frac{1}{2}\sum_{i=1}^4 l.u_i \;(d_i - d_0 - r_i) \nonumber \\
 {\rm or} \; \; 2 m^2 + (2 + \sum_{i=1}^4 l.u_i)d_0 +  l.w &=& \sum_{i=1}^4 l.u_i \;d_i \label{eq:E0-1},
\end{eqnarray}
where $r_i =  q_i^2$ and $w^\mu = \sum_{i=1}^4 r_iu_i^\mu$. Dividing both sides by $d_0d_1d_2d_3d_4$ and 
integrating over $d^4l$, we get
\begin{eqnarray}
 \int d^4l \; \frac{2 m^2 + (2 + \sum_{i=1}^4 l.u_i)d_0 +  l.w}{d_0d_1d_2d_3d_4} = 0. \label{eq:E0-2}
\end{eqnarray}

The contribution from the right hand side of Eq.~\ref{eq:E0-1} is zero, because 
$
\int d^4l \; \left(l^\mu \;d_i /d_0d_1d_2d_3d_4\right),
$
is a linear combination of those $q_i$s which are already present in the definition of the corresponding $u_i$s.
Further, note that
\begin{eqnarray}
\sum_{i=1}^4 u_i^\mu &=& (\veps^{\mu q_2q_3q_4} + \veps^{q_1 \mu q_3q_4} + \veps^{q_1q_2 \mu q_4} 
                       + \veps^{q_1q_2q_3 \mu})/\veps^{q_1q_2q_3q_4} \nonumber \\
 &=& \veps^{\mu (q_2-q_1)(q_3-q_2)(q_4-q_3)}/\veps^{q_1(q_2-q_1)(q_3-q_2)(q_4-q_3)} \nonumber \\
 &=& \veps^{\mu p_2p_3p_4}/\veps^{p_1p_2p_3p_4}  = v_1^\mu .
\end{eqnarray}
Also, shifting the loop momentum $l\rightarrow l-q_1$~\footnote{Note that the loop integral is UV finite.} 
in the second term of Eq.~\ref{eq:E0-2},
\begin{eqnarray}
\int d^4l \; \frac{(2 + \sum_{i=1}^4 l.u_i) }{d_1d_2d_3d_4} =
\int d^4l \; \frac{(1 + \sum_{i=1}^4 l.u_i) }{d_0d'_1d'_2d'_3} =
\int d^4l \; \frac{(1 +  l.v_1) }{d_0d'_1d'_2d'_3}. 
\end{eqnarray}
Here $d'_1 = (l+q_2-q_1)^2 - m^2, d'_2 = (l+q_3-q_1)^2 - m^2$ and $d'_3 = (l+q_4-q_1)^2 - m^2$. 
Once again, since 
$
\int d^4l \; (l^\mu /d_0d'_1d'_2d'_3)
$
is a linear combination of $p_2,p_3$ and $p_4$, the $l.v_1$ term in the above equation does not contribute. Finally,
we concentrate on the last term of the Eq.~\ref{eq:E0-2}. Using Eq.~\ref{eq:l-alpha}, we write
\begin{eqnarray}
 l.w &=& \sum_{i=1}^4 l.q_i\; u_i.w = \frac{1}{2} \sum_{i=1}^4 (d_i - d_0 -r_i)\; u_i.w \nonumber \\
 &=& \frac{1}{2} \sum_{i=1}^4 d_i(u_i.w) - d_0(v_1.w) -w^2 .
\end{eqnarray}

Substituting all these results in Eq.~\ref{eq:E0-2}, we get
\begin{eqnarray}
  \int d^4l \; \frac{4 m^2-w^2 + (2 - v_1.w)d_0 + \sum_{i=1}^4 (u_i.w)d_i }{d_0d_1d_2d_3d_4} = 0 \nn \\
\Rightarrow E_0 = \frac{1}{w^2-4 m^2} \left(  (2 - v_1.w)D_0^{(0)} + \sum_{i=1}^4 (u_i.w)D_0^{(i)} \right),
\end{eqnarray}
which is the desired relation mentioned in Eq.~\ref{eq:E0}. In $n=4-2\eps$ dimensions, this identity 
is expected to receive ${\cal O}(\eps)$ correction. This correction is proportional to 5-point
scalar integral in $n=6-2\eps$ dimensions \cite{Bern:1993kr}. It may appear that this ${\cal O}(\eps)$ piece
could be important in the $n$-dimensional reduction of 5-point tensor integrals of rank $m \ge2$ and we may
require its explicit form. It has been argued (and shown explicitly in some special cases) 
in Ref.~\cite{Bern:1993kr}, that the ${\cal O}(\eps)$ piece is dropped out in the reduction of five-point
tensor integrals with rank $\le 5$. A general proof is given in Ref.~\cite{Binoth:2005ff}. Thus, the tensor reduction of five-point functions can be carried 
out in $n=4$ dimensions without any ambiguity. In the case of five-point fermion loop amplitudes, considered
in this thesis, precise arguments can be given to justify the five-point tensor reduction in 4 dimensions; 
refer to Sec.~\ref{section:rational}. \\

In the above expression of $E_0$, We can also take the large $m^2$ limit. Using the large $m^2$ limit 
of the box scalars, given in Eq.~\ref{eq:large-m-D0}, we get
\begin{eqnarray}
 E_0(m^2\to\infty) &=& \frac{1}{(-4m^2)} 2\;D_0(m^2\to \infty) \nn \\ 
&=& \frac{i}{16 \pi^2} \; \left[-\frac{1}{12 m^6} + {\cal O}\left(\frac{1}{m^8}\right)\right].
\end{eqnarray}

%%%%%%%%%%%%%%%%%%%%%%%%%%%%%%%%%%%%%%%%%%%%%%%%%%%%%%%%%%%%%%%%%%%%%%%%%%%%%%%%%%%%%%%%%%%%%%%%%%%%%
\chapter{}
\section{Examples of fermion loop triangle amplitudes }\label{appendix:ggB}
We would like to give two specific and simple examples of fermion loop amplitudes. These
examples will complement the discussion presented in Sec.~\ref{section:rational}, on the special 
features of fermion loop amplitudes. 

\begin{figure}[h]
\begin{center}
\includegraphics [angle=0,width=0.8\linewidth] {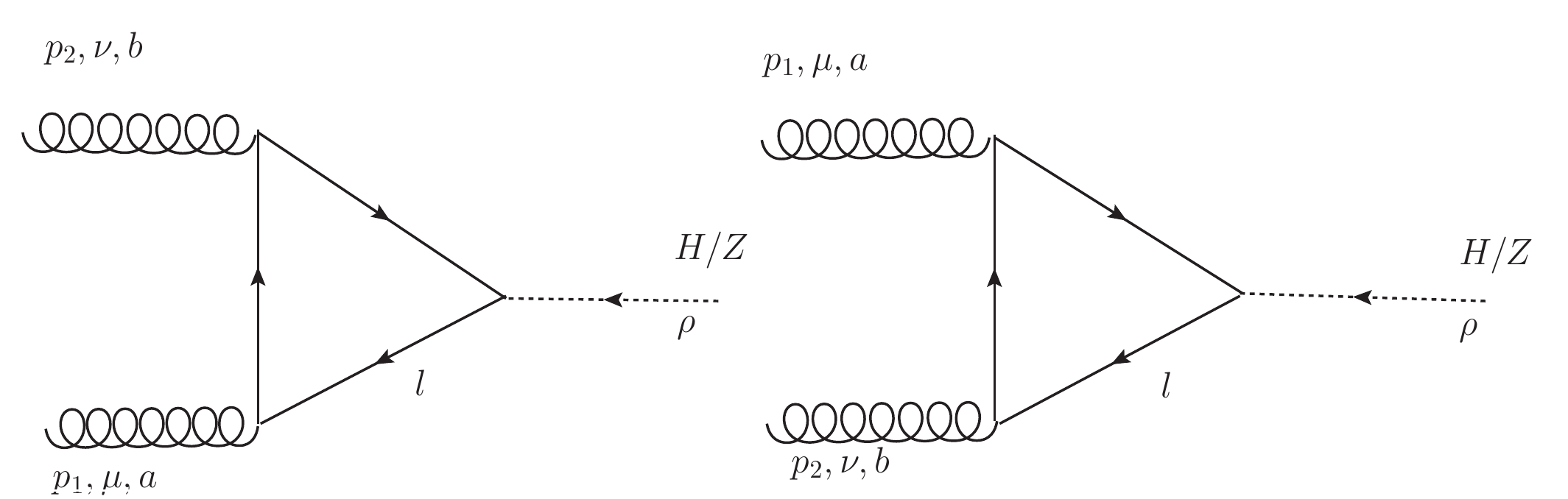}
\caption{ Feynman diagrams for $g g \to H/Z$. The dashed line may represent the Higgs boson or 
the $Z$ boson. The 4-vector index $\rho$ is applicable in the $gg\to Z$ case only.}
\label{fig:ggB} 
\end{center}
\end{figure}

\subsection{$ggH$ amplitude; $p_1^2 \ne 0, p_2^2 \ne 0$}\label{appendix:ggH}
Ignoring the color and coupling factors etc., the tensor structure of the 
amplitude may be written as
\begin{flalign}
{\cal M}^{\mu\nu}(p_1,p_2) =\;& m^2 \Big[F_0(p_1,p_2)\; g^{\mu\nu} + F_1(p_1,p_2)\; p_1^\mu p_1^\nu 
                           + F_2(p_1,p_2)\; p_2^\mu p_2^\nu\; + \nn \\
                           & F_3(p_1,p_2)\; p_1^\mu p_2^\nu + F_4(p_1,p_2)\;  p_2^\mu p_1^\nu \Big].
\end{flalign}
The form factors $\{F_i\}$ are obtained by calculating the triangle diagrams shown in Fig.~\ref{fig:ggB}. The
two diagrams are related by charge-conjugation and they are equal. The gauge invariance with respect to
the gluon currents, {\it i.e.}, ${\cal M}^{\mu\nu}p_{1\mu} = {\cal M}^{\mu\nu}p_{2\nu} = 0$, relates these 
form factors. The independent form factors are:
\begin{flalign}
 F_0(p_1,p_2) =\;& F_0(p_2,p_1) \nn \\
              =\;&  - 4\; \frac{p_1^2}{\Delta_2}\;(p_2^2 + p_1.p_2)\; R_1 -
                      4\; \frac{p_2^2}{\Delta_2}\;(p_1^2 + p_1.p_2)\; R_2 \;+ \nn \\
               & (4\; s\; p_1^2\; p_2^2/\Delta_2 + 16\; m^2 - 8\; p_1.p_2)\; C_0 + 8\; R,
\end{flalign}
\begin{flalign}
 F_1(p_1,p_2) &= (-p_2^2/p_1.p_2)\; F_3(p_1,p_2), \nn \\
 F_2(p_1,p_2) &= F_1(p_2,p_1),
\end{flalign}
\begin{flalign}
 F_3(p_1,p_2) =\;& F_3(p_2,p_1) \nn \\
              =\;&  
\big[-12p_1^2 p_2^2(p_1^2+p_1.p_2)/(\Delta_2)^2+4(3p_1^2 + 2p_1.p_2)/\Delta_2\big]\; R_1\;+ \nn \\
& \big[-12p_1^2 p_2^2(p_2^2+p_1.p_2)/(\Delta_2)^2+4(3p_2^2 + 2p_1.p_2)/\Delta_2\big]\; R_2\;+ \nn \\
               & \big[ -12p_1^2 p_2^2((p_1^2+p_2^2)p_1.p_2+2p_1^2p_2^2)/(\Delta_2)^2 + \nn \\
& 8((p_1^2+p_2^2+2m^2)p_1.p_2+4p_1^2 p_2^2)/\Delta_2 -8 \big]\; C_0 + 8\; p_1.p_2/\Delta_2\; R,
\end{flalign}
\begin{flalign}
 F_4(p_1,p_2) =\;& F_4(\p2,\p1) \nn \\
              =\;& F_3(p_1,p_2) -8\; (p_1^2 + p_1.p_2)/\Delta_2\; R_1 -8\; (p_2^2 + p_1.p_2)/\Delta_2\; R_2\;+ \nn \\ 
& \big[-8((p_1^2+p_2^2)p_1.p_2+2p_1^2p_2^2)/\Delta_2 + 16\big] \; C_0,
\end{flalign}
with
%  
% \begin{flalign}
\begin{eqnarray}
 R_1 &=& B_0(1) - B_0(2), \nn \\
 R_2 &=& B_0(1) - B_0(0).
\end{eqnarray}
% \end{flalign}
%  
In the above, $R\;(=1)$ is used as a flag to identify rational terms and $\Delta_2 = p_1^2p_2^2-(p_1.p_2)^2$. 
Note that all these coefficients are UV finite and they vanish in the decoupling limit. We see that,
after taking out the overall factor of $m^2$, the 
rational part of the amplitude is independent of the quark mass. Therefore, as we described in Sec.~\ref{section:rational}, the rational part can be calculated
utilizing the decoupling theorem even in amplitudes involving a Higgs boson.
For the special case of $p_1^2 = p_2^2 = 0$,
\begin{eqnarray}
 {\cal M}^{\mu\nu}(p_1,p_2) &=  m^2\;(g^{\mu\nu}-p_2^\mu p_1^\nu/\p1.\p2)\big[ 8\; R + (16 m^2-8p_1.p_2)\; C_0\big].
\end{eqnarray}

In the $m^2 \to \infty$ limit, this amplitude reduces to an effective $ggH$-vertex,
\begin{eqnarray}
 i {\cal M}^{\mu\nu}_{ab}(gg\to H)|_{m \to \infty} = \delta^{ab} \frac{\alpha_s}{3\pi v}\; (\p1.\p2g^{\mu\nu}-\p2^\mu \p1^\nu), 
\end{eqnarray}
where we have also included the couplings and the color factor. This is only the leading effect, obtained 
by expanding $C_0$ up to ${\cal O}(1/m^4)$ term.

\subsection{$ggZ$ amplitude; $p_1^2 \ne 0, p_2^2 \ne 0$}\label{appendix:ggZ}
Since the vector part of the amplitude vanishes due to Furry's theorem, 
the most general form of the amplitude, consistent with the vector current conservation, can 
be expressed as 
\begin{flalign}
 {\cal M}^{\mu\nu\rho}(p_1,p_2) =\;& F_1(p_1,p_2)\Big[p_1^\mu \veps^{\nu\rho p_1p_2} - p_1^2 \veps^{\mu\nu\rho p_2} \Big]
  + F_2(p_1,p_2)\Big[p_2^\nu \veps^{\mu\rho p_1p_2} - p_2^2 \veps^{\mu\nu\rho p_1} \Big] \nn \\
 & + F_3(p_1,p_2)\; (p_1+p_2)^\rho\; \veps^{\mu\nu p_1p_2} + F_4(p_1,p_2)\; (p_1-p_2)^\rho\; \veps^{\mu\nu p_1p_2}
\end{flalign}
where, due to symmetry
\begin{flalign}
 F_4(p_1,p_2)=0, \; \; F_1(p_1,p_2) = -F_2(p_2,p_1).
\end{flalign}
Here we are not concerned about the overall factors of coupling and color.
To obtain these form factors, we need to calculate the linearly divergent triangle diagrams, shown
in Fig.~\ref{fig:ggB}, in the presence of $\gamma^5$. We can use any of the $n$-dimensional $\gamma^5$ 
prescriptions, given in the next section, to perform the full calculation in $n=4-2\eps$ dimensions. 
The form factors thus obtained are:  
\begin{flalign}
F_2(p_1,p_2) =\;& \big[-6p_1^2 ((p_1^2+p_2^2)p_1.p_2+2p_1^2p_2^2)/(\Delta_2)^2+ 10p_1^2 /\Delta_2\big]\; R_1\;+ \nn \\
& \big[-6sp_1^2p_2^2/(\Delta_2)^2+ 2(p_1^2+s)/\Delta_2\big]\; R_2\;+ \nn \\
& \big[-6p_1^2 p_2^2(p_1^2(p_1^2+3p_2^2+3p_1.p_2)+p_2^2 p_1.p_2)/(\Delta_2)^2 + \nn \\
&  4(p_1^2(p_1^2+4p_2^2+2p_1.p_2)+2m^2(p_1^2+p_1.p_2)) /\Delta_2\big]\; C_0 \;+ \nn \\
& 4(p_1^2+p_1.p_2)/\Delta_2\; R,
\end{flalign}
\begin{flalign}
F_3(p_1,p_2) =\;& F_3(p_2,p_1) \nn \\
 =\; & \big[6p_1^2p_2^2 (p_1^2+ p_1.p_2)/(\Delta_2)^2-2p_1^2 /\Delta_2\big]\; R_1 \;+ \nn \\
& \big[6p_1^2p_2^2 (p_2^2+ p_1.p_2)/(\Delta_2)^2-2p_2^2 /\Delta_2\big]\; R_2 \;+ \nn \\
& \big[6p_1^2p_2^2 ((p_1^2+p_2^2)p_1.p_2+2p_1^2p_2^2)/(\Delta_2)^2- \nn \\
& 8(p_1^2p_2^2+m^2p_1.p_2) /\Delta_2\big]\; C_0 - 4p_1.p_2/\Delta_2\;R. 
\end{flalign}
Once again, we see that the rational terms are independent of the quark mass,
and the amplitude vanishes in the decoupling limit as a result of the cancellation 
between the rational part and the non-rational part of the amplitude.
We can calculate the anomalous contribution to the axial-vector current by 
dotting the amplitude with $(\p1+\p2)^\rho$,
\begin{eqnarray}
   {\cal M}^{\mu\nu\rho}(p_1,p_2)(\p1+\p2)_\rho = (16m^2\;C_0+8\;R)\;\veps^{\mu\nu p_1p_2}.
\end{eqnarray}
The first term on the right side of the above equation, shows the explicit breaking of the chiral symmetry due the 
quark mass and therefore non-conservation of the axial-vector current. Nevertheless, 
it goes away in the massless quark limit. The second term is the anomalous contribution and it is 
independent of the quark mass. The flag $R$, confirms that the anomaly affects only the rational 
part of the amplitude. The non-conservation of the axial-vector current, even in the massless
fermion limit, is the famous {\it chiral anomaly} of $VVA$-triangle diagrams~\cite{Adler:1969gk,Bell:1969ts}.
For on-shell gluons and the $Z$ boson,
\begin{eqnarray}
{\cal M}^{\mu\nu\rho}(p_1,p_2) &=& \frac{1}{\p1.\p2} (8m^2\;C_0+4\;R) \; \veps^{\mu\nu p_1p_2} (p_1+p_2)^\rho,\nn \\
\Rightarrow {\cal M}(Z \to g g)  &\equiv& {\cal M}^{\mu\nu\rho}(p_1,p_2) e_{1\mu} e_{2\nu} e_{3\rho} = 0.
\end{eqnarray}
Here $\{e_i\}$ are polarization vectors of the gauge bosons and we have used, $e_3.(\p1+\p2) = -e_3.\p3 =0$,
 for the $Z$ boson.
This is the statement of the Landau-Yang theorem, that is, a spin-1 massive particle cannot decay into two 
massless spin-1 particles~\cite{Landau:1948kw,Yang:1950rg}. Although we have proved it in a one-loop 
calculation, the Landau-Yang theorem is 
an exact quantum mechanical statement and it should be valid to all orders in perturbation theory. 
In the above, if the $Z$ boson is off-shell, the 4-current ($j^\rho$) attached to it can be treated, 
symbolically, as its polarization vector. Therefore, as long as this current is conserved, {\it i.e.}, 
$j^\rho(\p1+\p2)_\rho = 0$, the $ggZ$-amplitude with on-shell gluons, but off-shell $Z$ boson, 
will also vanish.

%%%%%%%%%%%%%%%%%%%%%%%%%%%%%%%%%%%%%%%%%%%%%%%%%%%%%%%%%%%%%%%%%%%%%%%%%%%%%%%%%%%%%%%%%%%%%%%%%%%%%
\section{$\gamma^5$ in $n$ dimensions}\label{appendix:g5-nd}
We discussed in Sec.~\ref{section:rational} that the chiral anomaly, in linearly divergent fermion loop
amplitudes, should be regulated by using an $n$-dimensional prescription for $\gamma^5$ in trace 
calculations, to 
ensure the conservation of vector currents and other symmetries of amplitudes. In 4
dimensions, $\gamma^5$ anticommutes with all other gamma-matrices, {\it i.e.}, 
$\{ \gamma^5, \gamma^\mu \} = 0$. In $n$ dimensions, the cyclic property of the trace of a string of 
gamma-matrices with a $\gamma^5$ is not compatible with this anticommutation relation~\cite{Jegerlehner:2000dz}. 
Noting that, in $n$ dimensions
\begin{eqnarray}
 \gal(\gmu\gnu\gro\gsig)\gamma_\alpha = 2 \gnu\gro\gsig\gmu + 2 \gmu\gsig\gro\gnu + (n-4)\gmu\gnu\gro\gsig,
\end{eqnarray}
we can write
\begin{eqnarray}
 {\rm tr}\left[\gfive\gal(\gmu\gnu\gro\gsig)\gamma_\alpha\right] &=& 
2\;  {\rm tr}\left[\gfive\gnu\gro\gsig\gmu\right] + 2\;  {\rm tr}\left[\gfive\gmu\gsig\gro\gnu\right] \nn \\
&&+ (n-4)\; {\rm tr}\left[\gfive\gmu\gnu\gro\gsig\right].
\end{eqnarray}
On the other hand, assuming the anticommuting property of $\gfive$ and using the cyclic property
of the trace, 
\begin{eqnarray}
 {\rm tr}\left[\gfive\gal(\gmu\gnu\gro\gsig)\gamma_\alpha\right] = 
-n\; {\rm tr}\left[\gfive\gmu\gnu\gro\gsig\right].
\end{eqnarray}
Here we have used $\gmu\gamma_\mu = n $. Combining the above two results, we get
\begin{eqnarray}
(n-4)\; {\rm tr}\left[\gfive\gmu\gnu\gro\gsig\right] = 0,
\end{eqnarray}
where we used, ${\rm tr}\left[\gsig\gro\gnu\gmu\gfive\right] = {\rm tr}\left[\gfive\gmu\gnu\gro\gsig\right]$.
The final result, clearly, makes sense only in 4 dimensions and therefore we must have a prescription for 
$\gfive$ in $n$ dimensions. The two very common prescriptions, in the literature, are:
\begin{enumerate}
 \item {\bf The 't Hooft-Veltman prescription:}~\cite{'tHooft:1972fi} \\
In this prescription, the $n$ dimensional $\gmu$-matrices are broken into 4-dimensional and 
$(n-4)$-dimensional parts such that $\gfive$ anticommutes with the 4-dimensional part while 
it commutes with the $(n-4)$-dimensional part, {\it i.e.},
\begin{eqnarray}
 \{\gmu_{(4)}, \gfive\} = 0, \;\; [\gmu_{(n-4)}, \gfive] = 0.
\end{eqnarray}
where $\gmu_{(4)} = \gmu; \mu=0,1,2,3$ and $\gmu_{(n-4)} = \gmu; \mu=4,5,...,n$. Furthermore, 
all the external momenta and polarizations are taken in 4 dimensions. However, the loop
momentum remains in $n$ dimensions. Thus,
\begin{eqnarray}
 \gamma.l = \gmu l_\mu = \gamma_{(4)}.l_{(4)} + \gamma_{(n-4)}.l_{(n-4)}.
\end{eqnarray}
\item {\bf The Larin's prescription:}~\cite{Larin:1992du} \\
This prescription is very straightforward and one uses
\begin{eqnarray}
 \gamma^5 &=& -\frac{i}{4!}\; \veps_{\mu\nu\rho\sigma}\;\gamma^\mu \gamma^\nu \gamma^\rho \gamma^\sigma,
% {\rm or}\;\; \gamma_\mu\gamma^5 &=& \frac{i}{3!}\; \veps_{\mu\nu\rho\sigma}\; \gamma^\nu \gamma^\rho \gamma^\sigma. 
\end{eqnarray}
in the fermion loop trace calculation. In this way of writing $\gfive$, its 4-dimensional 
nature is restricted to the fully antisymmetric $\veps$-tensor, which can be multiplied 
after calculating the trace in $n$-dimensions. The trace expression gets very lengthy with 
the above $\gfive$. We can use a simpler and more practical form,
\begin{eqnarray}
\gamma_\mu\gamma^5 &=& -\frac{i}{3!}\; \veps_{\mu\nu\rho\sigma}\; \gamma^\nu \gamma^\rho \gamma^\sigma. 
\end{eqnarray}
\end{enumerate}

We have used both these prescriptions to calculate the $ggZ$-amplitude, discussed above. Both these 
prescriptions generate correct
rational terms, consistent with the desired symmetries of the amplitude and the decoupling theorem. 
We would like to further comment that special prescriptions for $\gfive$ may be required, if there are 
odd number of $\gfive$-matrices in the trace. The 4-dimensional anticommuting property of $\gfive$
can be used safely, in presence of even number of $\gfive$-matrices in the trace. 

%%%%%%%%%%%%%%%%%%%%%%%%%%%%%%%%%%%%%%%%%%%%%%%%%%%%%%%%%%%%%%%%%%%%%%%%%%%%%%%%%%%%%%%%%%%%%%%%%%%%%
\chapter{}
\section{Feynman rules}\label{appendix:feynrules}
\begin{figure}[h!]
\begin{center}
\includegraphics [angle=0,width=1.0\linewidth] {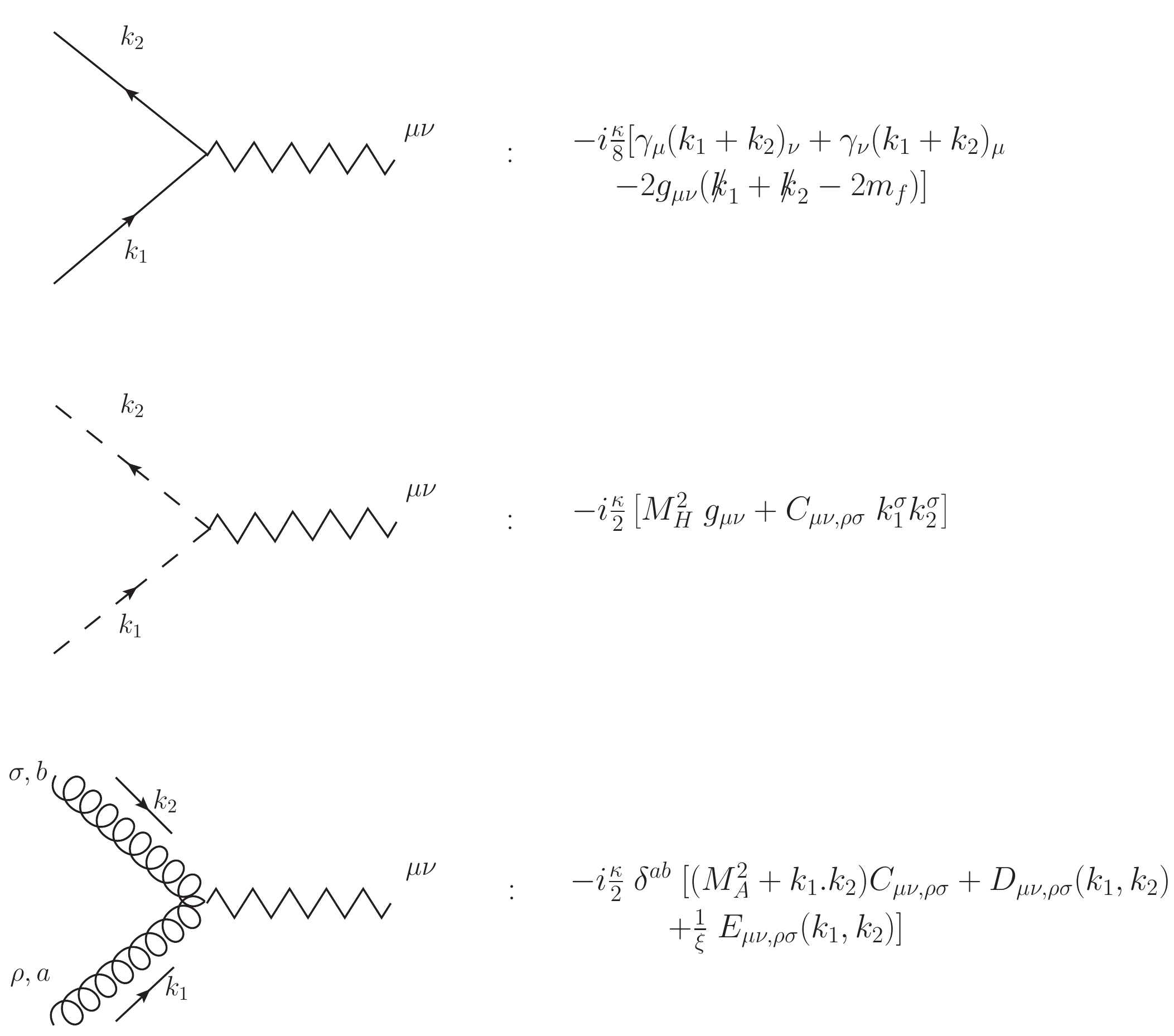}
\caption{ Feynman rules for 3-point vertices in the ADD model. The zig-zag line denotes the KK-graviton. The 
dashed line is for the Higgs boson while the curly line may denote any gauge boson with mass $M_A$. 
$\kappa = \sqrt{16\pi G_N} = \sqrt{2}/M_P$, and $\xi$ is the gauge-fixing parameter.}
\label{fig:ADD-3point} 
\end{center}
\end{figure}

\begin{figure}[h!]
\begin{center}
\includegraphics [angle=0,width=1.0\linewidth] {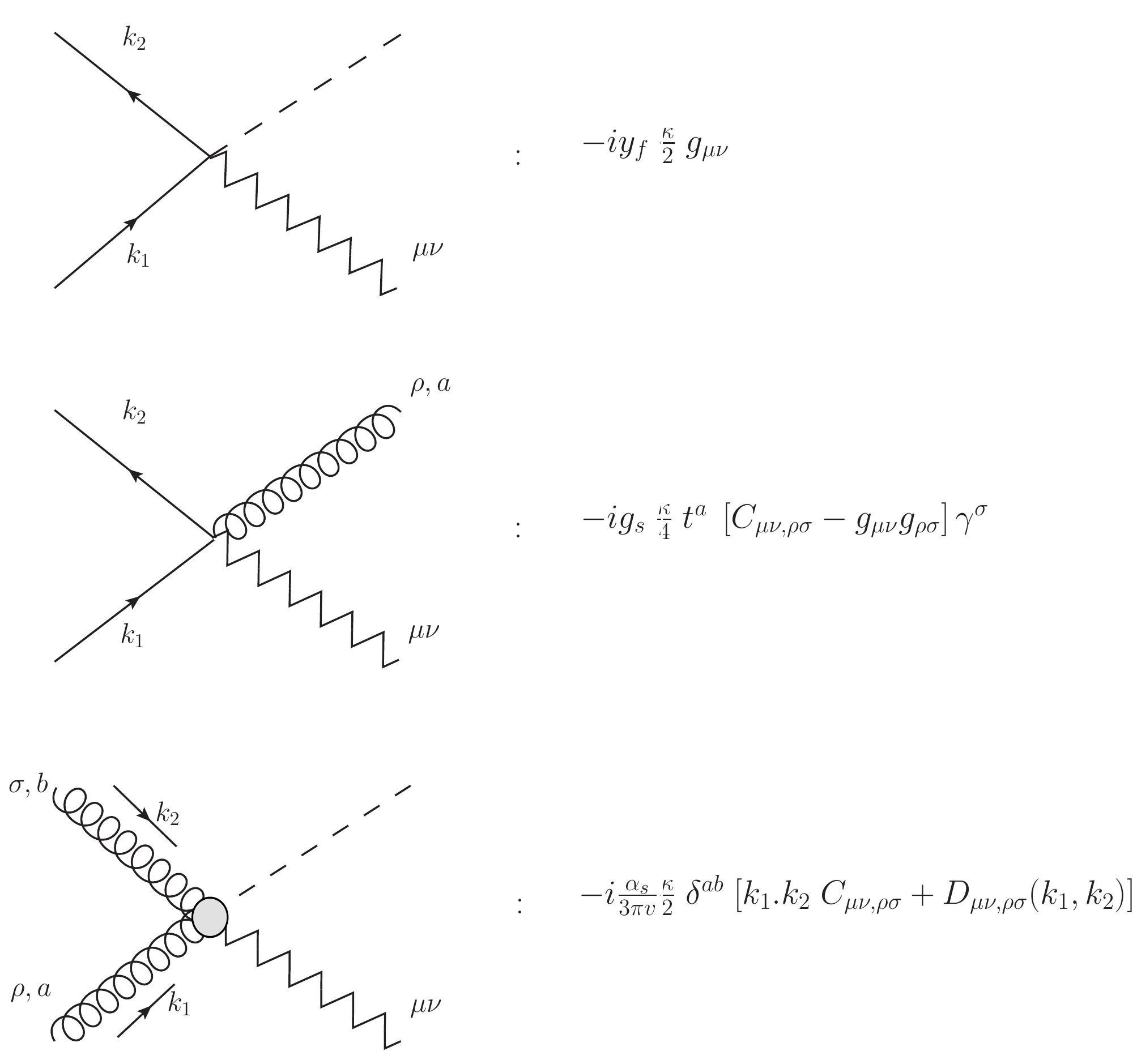}
\caption{ Feynman rules for some of the 4-point vertices in the ADD model. The Yukawa coupling, 
$y_f = \frac{1}{2} g_w(m_f/M_W)$. The last vertex is derived in an effective theory of $ggH$ coupling. }
\label{fig:ADD-4point} 
\end{center}
\end{figure}

We have given here the ADD model Feynman rules for those vertices which appear 
in our processes considered in chapter 4; see Figs.~\ref{fig:ADD-3point} and \ref{fig:ADD-4point}. 
These are taken from the Ref.~\cite{Han:1998sg}. The tensor structures which appear in these rules
are given by
\begin{flalign}
C_{\mu\nu,\rho\sigma} &= g_{\mu\rho}g_{\nu\sigma}+g_{\mu\sigma}g_{\nu\rho}-g_{\mu\nu}g_{\rho\sigma}, \\
D_{\mu\nu,\rho\sigma}(k_1,k_2) &= g_{\mu\nu}k_{1\sigma}k_{2\rho}- \Big[ g_{\mu\sigma}k_{1\nu}k_{2\rho}
+ g_{\mu\rho}k_{1\sigma}k_{2\nu} - g_{\rho\sigma}k_{1\mu}k_{2\nu} + (\mu \leftrightarrow \nu) \Big], \\
E_{\mu\nu,\rho\sigma}(k_1,k_2) &= g_{\mu\nu}(k_{1\rho}k_{1\sigma} + k_{2\rho}k_{2\sigma} + k_{1\rho}k_{2\sigma})
-\Big[\ g_{\nu\sigma}k_{1\mu}k_{1\rho} + g_{\nu\rho}k_{2\mu}k_{2\sigma} + (\mu \leftrightarrow \nu) \Big].
\end{flalign}
Note that for the on-shell production of the KK-graviton, the terms proportional to $g_{\mu\nu}$
do not contribute due to the traceless condition of the graviton polarization tensor. 
The SM Feynman rules used in thesis are taken from~\cite{peskin:2005xx}. 


\end{appendices}
% \chapter{List of Acronyms}
% 
% \newpage
% 
% \bibliographystyle{utcaps}
% \bibliography{thesis}
% 
% \end{document}

\newpage
 \addcontentsline{toc}{chapter}{Bibliography}
\bibliographystyle{utcaps}
\bibliography{mythesis}

\end{document}